\documentclass[preprint,12pt]{elsarticle}
\usepackage{tensor}	
\usepackage{caption}
\usepackage{mathrsfs}
\usepackage{color}
\usepackage[dvipsnames]{xcolor}
\usepackage{soul}
\usepackage{amssymb}
\usepackage{amsmath}
\usepackage{hyperref}
\usepackage{rotating}
\usepackage[utf8]{inputenc} 
\usepackage{multirow}

\def\l{\left}
\def\r{\right}
\def\f{\frac}

\def\nn{\nonumber}

\def\be{\begin{equation}}
\def\ee{\end{equation}}
\def\ba{\begin{eqnarray}}
\def\ea{\end{eqnarray}}
\def\gn{G_\mathrm{N}}

\def\muff{\mu_\mathrm{ff}}
\def\musc{\mu_\mathrm{sc}}
\def\fs{f\sigma_8}

\def\gsp{\eta}

\def\ak{\alpha_K}
\def\ab{\alpha_B}
\def\am{\alpha_M}
\def\at{\alpha_T}
\def\ah{\alpha_H}

\def\mc{M_\mathrm{C}}
\def\lc{\lambda_\mathrm{C}}
\def\bx{\beta^2_\xi}
\def\bb{\beta^2_B}

\def\mp{M_\mathrm{pl}}
\def\mps{M^2_\mathrm{pl}}

\def\om{\Omega_{\rm m}}
\def\omo{\Omega_{{\rm m},0}}
\def\omde{\Omega_{\rm DE}}
\def\omdeo{\Omega_{{\rm DE},0}}
\def\w{w_{\rm DE}}

\def\fg{\mathfrak{f}}
\def\cf{\mathfrak{F}}

\def\hiclass{\texttt{hi\_class} }
\def\eftcamb{\texttt{EFTCAMB} }
\def\coop{\texttt{COOP} }

\allowdisplaybreaks

\graphicspath{{figs/}}

\journal{Physics Reports}

\begin{document}

\begin{frontmatter}

\title{Effective Field Theory of Dark Energy: a Review}

\author{Noemi Frusciante$^a$}
\ead{nfrusciante@fc.ul.pt}
\author{Louis Perenon$^{b,c,d}$}
\ead{lperenon@uwc.ac.za}
\address[1]{Instituto de Astrof\'isica e Ci\^encias do Espa\c{c}o, Faculdade de Ci\^encias da Universidade de Lisboa, Edificio C8, Campo Grande, P-1749016, Lisboa, Portugal }
\address[b]{Department of Physics \& Astronomy, University of the Western Cape, \\Cape Town 7535, South Africa}
\address[c]{Cosmology and Gravity Group, Department of Mathematics and Applied Mathematics, University of Cape Town, Rondebosch 7701, Cape Town, South Africa}
\address[d]{Aix Marseille Univ, Universit\'e de Toulon, CNRS, CPT, Marseille, France}

\begin{abstract}
The discovery of cosmic acceleration has triggered a consistent body of theoretical work aimed at modeling its phenomenology and understanding its fundamental physical nature. In recent years, a powerful formalism that accomplishes both these goals has been developed, the so-called  effective field theory of dark energy. It can capture the behavior of a wide class of modified gravity theories and classify them according to the imprints they leave on the smooth background expansion history of the Universe and on the evolution of linear perturbations. The effective field theory of dark energy is based on a Lagrangian description of cosmological perturbations which depends on a number of functions of time, some of which are non-minimal couplings representing genuine deviations from General Relativity. Such a formalism is thus particularly convenient to fit and interpret the wealth of new data that will be provided by future galaxy surveys. Despite its recent appearance, this formalism has already allowed a systematic investigation of what lies beyond the  General Relativity landscape and provided a conspicuous amount of theoretical predictions and observational results. In this review, we report on these achievements.
\end{abstract}


\end{frontmatter}

\tableofcontents

%%%%%%%%%%%%%%%%%%%%%%
\section{Introduction}
%%%%%%%%%%%%%%%%%%%%%%

The Universe is a physical system whose dynamics is strongly affected by gravity and that is characterized, at very large scales, by invariance under spatial translations and rotations. Other \emph{bona fide} symmetries of our fundamental Lagrangians, time translations and Lorentz boosts, are spontaneously broken. One curious occurrence for the Universe is that its expansion rate has accelerated twice, once at very primordial times and high energies (inflation) and then through a second much later period that is still ongoing (``Dark Energy", DE). When trying to address these phenomena, it looks natural to speculate about the nature of gravity itself. Indeed, the necessity of invoking accelerating expansion twice might even cast doubts on the very basic geometrical description inherited from classical General Relativity (GR), when applied to the Universe as a whole. 

The standard cosmological model, $\Lambda$CDM, is based on GR and assumes the Universe is made of a DE component in the form of a cosmological constant ($\Lambda$), cold dark matter (CDM) particles and ordinary matter. Although $\Lambda$CDM gives an  astonishing description of the Universe~\cite{Aghanim:2018eyx,Troxel:2017xyo}, the model shows some shortcomings: the so-called cosmological constant problems. These arise from the mismatch between the naturally expected value of $\Lambda$ from quantum corrections and the observed one, and to the late time coincidence problem (see \cite{Weinberg:1988cp,Martin:2012bt,Joyce:2014kja} for reviews). Furthermore, some mild observational tensions among different datasets emerge in this model, for instance, on the value of the Hubble constant $H_0 \,(=100\,h$ km\,s$^{-1}$\,Mpc$^{-1})$ and amplitude of the matter power spectrum at present time and scale of 8 h$^{-1}$Mpc, denoted by $\sigma_{8,0}$. At the level of the background evolution of the Universe, comparing for example Cosmic Microwave Background (CMB) radiation data by Planck \cite{Adam:2015rua,Aghanim:2018eyx} and local measurements of $H_0$ based on the cosmic distance ladder leads to a tension of $4.4 \sigma$~\cite{Riess:2011yx,Riess:2016jrr,Riess:2019cxk}. On the other hand, Baryon Acoustic Oscillations (BAO) measurements from the Baryon Oscillation Spectroscopic Survey (BOSS)~\cite{Dawson:2012va} and the Sloan Digital Sky Survey (SDSS)~\cite{Abazajian:2008wr} show a $2.5 \sigma$ discrepancy in $H_0$ with Planck~\cite{Delubac:2014aqe}. Regarding matter perturbations, a discordance of about $2.3 \sigma$ on $\sigma_{8,0}$ can be spotted between the Kilo-Degree Survey (KiDS)~\cite{deJong:2015wca} and Planck data~\cite{Hildebrandt:2016iqg,deJong:2015wca,Kuijken:2015vca,Conti:2016gav,Joudaki:2019pmv}. This picture summarizes the motivations at the basis of speculations on the validity of the $\Lambda$CDM model and the search for new physics beyond the standard model.

When pondering about the theory of gravity, one immediately faces a general deep lesson: no other theory than GR is compatible with the basic requisites of a single massless spin two field (the graviton) and recovering Lorentz invariance. As corollary, Lovelock's theorem~\cite{Lovelock:1971yv,Lovelock:1972vz} implies any infrared departure from GR must bring in new degrees of freedom (DoFs). Usually one refers to these proposals as modified gravity (MG) theories~\cite{Joyce:2014kja,Lue:2004rj,Copeland:2006wr,Silvestri:2009hh,Nojiri:2010wj,Tsujikawa:2010zza,Capozziello:2011et,Clifton:2011jh,Bamba:2012cp,Koyama:2015vza,Avelino:2016lpj,Joyce:2016vqv,Nojiri:2017ncd,Ferreira:2019xrr,Kobayashi:2019hrl}. The simplest modification  at the basis of both DE and Inflationary models is to introduce an extra scalar DoF to GR thereby accessing the realm of scalar-tensor theories~\cite{Horndeski:1974wa,Fujii:2003pa,Deffayet:2009mn,Clifton:2011jh,Tsujikawa:2010zza,Gleyzes:2014dya,Joyce:2014kja,Koyama:2015vza,Langlois:2015cwa,Ferreira:2019xrr}. 
 It is possible to think of this scalar field as the Goldstone field of broken time translations characteristic of an expanding Universe~\cite{Piazza:2013coa}. This specific symmetry breaking pattern is well described within the unitary gauge which is equivalent to a choice of the time coordinate. Spacial diffeomorphisms are therefore left  unbroken. By setting the time coordinate proportional to the value of the scalar field, its fluctuation disappears from the dynamics and remains encoded in the metric DoFs. There comes an important advantage from this procedure and the unitary gauge: it allows to write down the most general action for  cosmological perturbations with relative ease, providing an effective framework that does not rely on any specific model, yet being a genuine description of departures from GR: a so-called effective field theory (EFT). Regarding expanding universe, this formalism was first applied to inflation~\cite{Creminelli:2006xe,Cheung:2007st,Bordin:2017hal}, then to quintessence models without conformal couplings~\cite{Creminelli:2008wc}, and finally to DE~\cite{Gubitosi:2012hu,Bloomfield:2012ff}. The latter case enables one to treat any DE/MG models with one additional scalar DoF in a model-independent approach described through a variety of geometrical operators compatible with the symmetry imposed and accompanied by time dependent functions, namely the EFT functions. By requiring that the linear perturbation equations contain derivatives up to second order, theories of the  ``Horndeski" type~\cite{Horndeski:1974wa,Deffayet:2009mn} can be recovered quite straightforwardly in the unitary gauge~\cite{Gleyzes:2013ooa,Bloomfield:2013efa}. Even more, working in the unitary gauge has made very natural to explore an entire new class of scalar-tensor theories beyond Horndeski, the so-called Gleyzes-Langlois-Piazza-Vernizzi (GLPV) theories~\cite{Gleyzes:2014dya}. The EFT approach can also encode the cosmology of Lorentz violating theories \cite{Gubitosi:2012hu,Frusciante:2015maa} such as Ho\v rava gravity~\cite{Horava:2008ih}. The landscape of theories that, despite the presence of higher derivatives in the equations of motions, avoid Ostrogradsky instabilities \cite{Ostrogradsky:1850fid}  has been exhausted by the Degenerate Higher Order Scalar-Tensor Theories (DHOST)~\cite{Langlois:2015cwa,Crisostomi:2016czh,Crisostomi:2016tcp,Achour:2016rkg,Langlois:2017mxy,Langlois:2018jdg}. In parallel, the original EFT formalism has been also extended to include those theories with second-order derivative equations of motion with a vector or tensor additional field \cite{Lagos:2016wyv,Lagos:2017hdr}, such as Generalised Proca \cite{Heisenberg:2014rta}, generalized Einstein-Aether \cite{Jacobson:2000xp} and massive bi-gravity \cite{Hassan:2011zd}. Finally, while the EFT formalism was originally developed to describe general patterns of DE/MG models at linear cosmological scales, there has also been great progress in extending the framework to include non-linear perturbative effects \cite{Bellini:2015wfa,Frusciante:2017nfr,Yamauchi:2017ibz,Cusin:2017mzw,Cusin:2017wjg,Kennedy:2019nie}. 

A key goal of future surveys such as Euclid \cite{Laureijs:2011gra}, Dark Energy Spectroscopic Instrument (DESI) \cite{Aghamousa:2016zmz}, Square Kilometre Array (SKA) \cite{Bacon:2018dui}, Stage-4 CMB experiment (CMB-S4) \cite{Abazajian:2016yjj}, Large Synoptic Survey Telescope (LSST) \cite{Abell:2009aa} is to test gravity on cosmological scales with an exquisite precision to shed light on the phenomenon of late time cosmic acceleration. Cosmological probes such as Galaxy Clustering (GC), CMB, Weak Lensing (WL), Redshift-Space Distortions (RSD), Supernovae Ia (SNIa) and BAO give access to precious physical information about the expansion history and growth of large-scale structures (LSS). The final aim is to confirm the cosmological standard model with even more accuracy, or, eventually, to single out signatures of DE/MG. The systematic exploration of DE/MG models against cosmological data is facilitated by the EFT approach. It offers indeed the possibility to relate the phenomenology of large classes of models directly to cosmological observables. This unified framework allows to explore the realm of gravitational theories in a coherent fashion identifying clear testable patterns. The synergy between theory and observations is further exploited thanks to Einstein-Boltzmann (EB) codes constructed on top of the EFT framework. The latter are \eftcamb \cite{Hu:2013twa,Raveri:2014cka}, \hiclass \cite{Zumalacarregui:2016pph}, \coop \cite{Huang:2015srv} and \texttt{EoS\_class} \cite{Pace:2019uow} and they have been used to obtain cosmological constrains using a wide selection of data sets.

In this review, we aim to provide a comprehensive overview of the many developments occurred in the field of MG when treated in the EFT of dark energy framework. The detailed structure of the review is as follows. Section \ref{sec:eftofde} is dedicated to its theoretical aspects which include: the construction of the EFT action, a description of background and linear perturbation equations, the {\it mapping} procedure to write any specific scalar-tensor theory in the EFT language and the theoretical stability requirements to guarantee the viability of a gravity model. We also discuss the so-called $\alpha$-basis which reshuffles the EFT functions according to physical properties of the Universe and the extension of the EFT framework to direct couplings between gravity and matter fields. Finally, we review the EB codes which implement the EFT framework. Section \ref{sec:novpred} describes the links between theory and observations where we highlight the novel predictions brought by the EFT formulation. We discuss the construction of a road map to select stable DE/MG models depending on crucial observations and their implications. We review the studies of large samples of DE/MG models within the EFT framework which identified clear patterns in observables such as the growth of LSS and power spectra. We also summarize the results about the impact of stability conditions on the parameter space. Section \ref{sec:constraints} is devoted to a discussion of cosmological constraints using present day data and forecasts from future surveys. We attempt to cover all the models analyzed in literature within the EFT framework, these include direct parameterizations of the EFT functions as well as the mapping of specific MG theories in the framework. Section \ref{Sec:Astro} contains an overview about the implications astrophysical constraints have on the parameter space identified by the EFT functions. In particular, we discuss constraints from massive astrophysical bodies such as dwarf, neutron stars, pulsars and galaxy clusters and the recent bounds derived from the joint detection of the GW170817 and GRB170817A events. Section \ref{sec:discussion} is devoted to final remarks and discussion on the prospect of future inputs into this field.

We provide a guide to acronyms and symbols used in this manuscript respectively in Table \ref{tab:acronym} and Table \ref{tab:symbols} (Appendix \ref{App:symbols}). We show the relations among the different EFT basis adopted in literature and a detailed list of the relevant EFT parameterizations used in the analyses in Appendix \ref{App:params}.  In Appendix \ref{summaryconstraints}, we give tables summarizing the observational constraints discussed in Section \ref{sec:constraints}.

%%%%%%%%%%%%%%%%%%%%%%%%%%%%%%%%%%%%%%%%%%%%%%%%%%%%%%%%%%%%%%%%%%%
\section{Effective Field Theory of Dark Energy} \label{sec:eftofde}
%%%%%%%%%%%%%%%%%%%%%%%%%%%%%%%%%%%%%%%%%%%%%%%%%%%%%%%%%%%%%%%%%%%

The EFT is a model-independent approach encompassing all single-field DE/MG models. It describes both the evolution of the cosmological background and linear perturbations. In the following, we review the building blocks of this approach and the {\it mapping} procedure which allows to write any specific scalar-tensor theories in the EFT language. We also discuss a phenomenological EFT basis, the so-called  $\alpha$-basis, and its extension to include the direct coupling between DE and matter fields. Furthermore, we present the stability requirements a gravity model has to satisfy to be considered theoretically viable. These are particularly needful when dealing with model-independent parameterizations. We conclude by reviewing the existing EB codes which implement the EFT framework. 

%--------------------------------------------------------
\subsection{Action and its formulation}\label{sec:action}
%--------------------------------------------------------

The EFT action is constructed in the unitary gauge and written in terms of operators compatible with the residual symmetries of unbroken spatial diffeomorphisms. The operators are organized in powers of the number of perturbations and spatial derivatives and these symmetries allow for time-dependent functions multiplying each operator. We review these features:
\begin{itemize}
\item \textit{The background:} in the EFT framework, the perturbations are assumed to evolve on a homogeneous and isotropic background, thus a Friedmann-Lema\^{i}tre-Robertson-Walker (FLRW) line element of the following form is considered:
\be
ds^2=-dt^2+a(t)^2\left(\frac{dr^2}{1-\bar{\kappa} r^2}+r^2d\bar{\Omega}^2\right)\,,
\ee
where $a(t)$ is the scale factor and $t$ is cosmic time, $\bar{\kappa}$ is the spatial curvature constant and $d\bar{\Omega}^2=d^2\theta+sin^2\theta d^2\varphi$. 

\item \textit{The unitary gauge:} the unitary gauge is also known as the \emph{velocity orthogonal gauge} ~\cite{Kodama:1985bj}. It corresponds to the choice of the basis in which the perturbation of the extra DoF, responsible for the spontaneous symmetry breaking, vanishes. In details, let us consider a scalar field, $\phi(t,\vec{x})$, and its decomposition in a perturbed FLRW Universe as follows
\begin{equation}\label{eq:decompphi}
\phi(t,\vec{x})= \bar{\phi}(t)+\delta\phi(t,\vec{x}),
\end{equation}
where $\vec{x}$ are the spatial coordinates, $\bar{\phi}(t)$ is the homogeneous background value of the scalar field and $\delta\phi$ its perturbation. In order to apply the unitary gauge, one has to choose the time coordinate such that $\delta \phi=0$, thus $t$ becomes a function of $\phi$, $i.e.$ $t=t(\phi)$ \footnote{This statement is valid at all perturbative orders as long as $\phi$ is a monotonic function of time.}. According to this choice $\phi$ defines a preferred time slicing ($\phi=const $) and constant time hypersurfaces coincide with constant scalar field hypersurfaces.  The gradient of the scalar field is thus assumed to be time-like. In a fluid description, the latter implies that the velocity of the scalar field is orthogonal to the constant time hypersurfaces, hence the name ``velocity orthogonal gauge''. The assumption of the unitary gauge implies the action will not show any explicit dependence on the scalar field, as such, the latter is said to be ``eaten" by the metric, making the manifest number of DoFs minimal. In conclusion, using this gauge implies two major consequences. Firstly, the EFT action will not be constructed in terms of $\phi$ and its perturbations, but will be written only in terms of the metric and geometrical quantities. Secondly, it breaks the full diffeomorphism invariance while leaving unbroken the subgroup of time-dependent spatial diffeomorphisms.
 
\item \textit{The operators:} according to the unitary gauge, the operators in the EFT action are constructed with all the invariants under the residual symmetries of unbroken spatial diffeomorphisms. The full EFT action is thus derived considering the metric $g_{\mu\nu}$ and the unit vector $n_\mu$, perpendicular to the time slicing, defined as \footnote{Note that in Ref.~\cite{Bloomfield:2012ff} $n_\mu$ is defined with the opposite sign. The different definition can change the sign of some operators in the EFT action, see for instance~\cite{Frusciante:2016xoj}.}
\begin{equation}
n_\mu = -\frac{\partial_\mu \phi}{\sqrt{-(\partial_\mu \phi)^2}} \rightarrow -\frac{\delta^0_\mu}{\sqrt{-g^{00}}}\;,
\end{equation}
where $g^{00}$ is the time-time component of the inverse metric and $\partial_\mu$ is the four dimensional derivative. The covariant derivatives of $n_\mu$ must be considered as well, or equivalently, its projection orthogonal to the constant time hypersurfaces, $i.e.$ the extrinsic curvature tensor and its trace
\begin{equation}\label{extrtensor}
K_{\mu\nu}= h_\mu^{\ \; \;\sigma} \nabla_\sigma n_\nu\;,\qquad K=\nabla^\nu n_\nu\,, 
\end{equation}
with the induced metric defined as $h_{\mu\nu}=g_{\mu\nu}+n_\mu n_\nu$ and $n^\sigma \nabla_\sigma n_\nu \propto h_\nu^{\ \; \;\mu} \partial _\mu g^{00}$ where $\nabla_\sigma$ is the four dimensional covariant derivative. Additional operators to  include  are also  the Ricci scalar $R$ and any curvature invariants and contractions of tensors with $n_\mu$, $g_{\mu\nu}$ and derivatives $\nabla_\mu$. Additionally, the symmetry of the action is still satisfied if each operator is accompanied by a time dependent function. These free functions of time which scale the perturbations on the FLRW background are called \textit{EFT functions}. Unlike standard covariant approaches, the EFT action is constructed using a perturbative approach thus each operator is expanded. For example, the operator $A$ will be written as $A(t,x_i)=\bar{A}(t)+\delta A(t,x_i)$, where $\bar{A}$ is the background value and $\delta A$ its perturbation. Note this perturbation scheme enables one to write down an action with operators expanded at any order. In a cosmological context, linear perturbations are generally the most focused upon and therefore only operators up to quadratic order are considered. Already at this order, a large number of operators and combinations can be constructed, especially if one considers an arbitrary number of spatial derivatives acting on the perturbed quantities. However, one has to keep in mind that additional spatial derivatives increase the scaling dimension of an operator and the higher the derivatives an operator contains, the less it becomes relevant on large linear scales.

\item \textit{Couplings with matter fields:} the first version of the EFT framework for dark energy assumes the validity of the weak equivalence principle (WEP) and hence the existence of a metric $g_{\mu\nu}$ universally coupled to the matter fields $\chi_m$, making the Jordan Frame the natural choice~\cite{Gubitosi:2012hu,Bloomfield:2012ff}. This assumption can be relaxed considering a frame where the gravitational interaction between the additional scalar DoF and the matter fields is explicit~\cite{Gleyzes:2015pma,Gleyzes:2014qga,Tsujikawa:2015upa,DAmico:2016ntq}. In this Section, we follow the historical path and we review the extension of the EFT framework to include direct couplings with matter fields in Section \ref{sec:couplings}.

\item \textit{Regime of validity and limitations:} the regime of applicability of the EFT framework spans from largest cosmological scales down to the ultraviolet (UV) cut-off, $i.e.$ the energy scale at which a classical description of gravity breaks down. In the cosmological framework, one can assume this cut-off $M_{\rm cut-off}$ must be larger than the Hubble parameter at present time $H_0$ in order to describe the background cosmology and observable perturbation modes \cite{Bloomfield:2011np}.  Extensions of the linear EFT framework to mildly non-linear scales are possible by including non-linear operators and  appropriate additional EFT functions to the action  \eqref{eftact} \cite{Frusciante:2017nfr,Yamauchi:2017ibz,Cusin:2017mzw,Cusin:2017wjg,Kennedy:2019nie}.  The  EFT approach to linear and non-linear scales is thus regulated by:  the strong coupling energy scale of the dark energy fluctuations and the screening scale.  Physically, the strong coupling scale corresponds to the scale at which the non-linear interactions of the EFT action exit perturbative unitarity \cite{Cusin:2017mzw} thus setting a cut-off above which the EFT looses predictability  and where UV completion becomes important. Such scale is characteristic of the specific model under consideration \cite{Luty:2003vm,Cusin:2017mzw}. While the screening scale  is associated to the non-linear scale of the extra scalar field.   If the screening is weak or  fails to operate (see Section \ref{Sec:Astro}), the linear EFT approach is still valid, on the other hand if the screening is strong non-linear terms dominate.
Other shortcomings of the EFT formulation exist. For example, it does not describe strong gravity nor higher-dimensional regimes. In parallel, the initial limitation of the EFT framework to include only DE/MG models based on a single scalar DoF was overcome recently by a proposal including additional vector and tensor fields \cite{Lagos:2017hdr}. We focus only on the original proposal encompassing DE/MG models with an extra scalar DoF in this review. 
\end{itemize}

Several versions of the EFT action exist in literature \cite{Gubitosi:2012hu,Bloomfield:2012ff,Gleyzes:2013ooa,Piazza:2013pua,Hu:2014oga,Tsujikawa:2014mba} which differ in notation, in number of operators and order of perturbations. In this Section, we consider the EFT action for DE/MG up to second order in perturbations, including the most relevant operators encompassing well known DE/MG models. We use the notation as presented for the first time in Ref.~\cite{Gubitosi:2012hu} and we discuss the other formulations in Appendix  \ref{App:params}. This EFT action reads~\cite{Gubitosi:2012hu} 
\begin{align} 
S = \frac12\int d^4x &\sqrt{-g}
\left[\mp^2  \fg(t) R \,  - 2 \Lambda(t)   - 2c(t)  g^{00} \right. \nn\\
&\left. +  M_2^4(t) (\delta g^{00})^2   - \bar m_1^3(t)\,  \delta g^{00} \delta K - \bar M_2^2(t)\,  \delta K^2 \right. \nn\\
&\left. - \bar M_3^2(t)\,  \delta K_{\mu}^{\ \nu} \delta K_{\ \nu}^\mu + \mu_1^2(t) \delta g^{00} \delta R+ m_2^2(t) h^{\mu \nu} \partial_\mu g^{00} \partial_\nu g^{00} \right. \nn\\
&\left.+ \dots \,\right]+ S_m[g_{\mu\nu},\chi_m]\,, 
	\label{eftact}
\end{align} 
where $\mp^2$ is the Planck mass, $g^{00}= -1+\delta g^{00}$, $g$ being the determinant of the metric, $\delta R$ and $\delta R_{\mu \nu} $ are the perturbations of the Ricci scalar and tensor, respectively. $S_m$ is the matter action for all matter fields $\chi_m$ and the EFT functions are $\fg$, $\Lambda$, $c$, $M_i$, $m_i$, $\bar M_i$, $\bar m_i$ and $\mu_i$. The action is organized in a specific way: the first line contains the operators contributing to both the background evolution of the Universe and the linear perturbation equations. The corresponding EFT functions are therefore denoted as {\it background} EFT functions. The operators in the second and third lines enter only in the linear perturbations equations. The ellipsis stand for additional second order operators which can be included, $e.g.$ $(\delta R)^2$ \cite{Gleyzes:2013ooa} or higher order terms in both derivatives and perturbations. For example Refs.~\cite{Kase:2014cwa,Frusciante:2015maa,Frusciante:2016xoj} include the operators with higher spatial derivatives to describe theories like high-energy Ho\v rava gravity~\cite{Blas:2009qj} in the EFT description. Higher order operators, such as $(\delta g^{00})^3$ and $(\delta g^{00})^2 \delta K$, are relevant for an extension of the action to non-linear scales~\cite{Frusciante:2017nfr,Cusin:2017mzw,Cusin:2017wjg}. Additional three operators need to be included~\cite{Langlois:2017mxy} in the above action to describe the recently found DHOST~\cite{Langlois:2018jdg,Langlois:2018dxi}. 

\begin{sidewaystable}[!].
\centering%
\begin{tabular}{ p{5cm} | c c c c c c c c c}
\hline\hline \\[-4mm]
& $\ \ \ \ \fg \ \ $ & $\ \ \Lambda \ \ $ &  $\ \ c  \ \ $ & $\ \ M^4_2 \ \ $ & $\ \ \bar{m}^3_1 \ \ $ & $\ \ \bar{M}^2_2 \ \ $ & $\ \ \bar{M}^2_3 \ \ $ & $\mu_1^2$ &$\ \ m^2_2 \ \ $ \\[0.7mm]   
\hline\hline \\[-4mm]
$\Lambda$CDM                                      & 1            & const.     & --         & --          & --          & --          & --          & --         & --          \\ \hline
Quintessence~\cite{Wetterich:1987fm,Ratra:1987rm,Caldwell:1997ii}                                     & 1/\checkmark & \checkmark & \checkmark & --          & --          & --          & --          & --         &--           \\ \hline
$K$-essence~\cite{ArmendarizPicon:2000dh}         & 1/\checkmark & \checkmark & \checkmark & \checkmark  & --          & --          & --          & --         &--           \\ \hline
Brans-Dicke~\cite{Brans:1961sx,Boisseau:2000pr}   & \checkmark   & \checkmark & \checkmark & --          & --          & --          & --          & --         &--           \\ \hline
$f(R)$~\cite{Sotiriou:2008rp,DeFelice:2010aj}     & \checkmark   & \checkmark & --         & --          & --          & --          & --          & --         &--           \\ \hline
Kinetic braiding~\cite{Deffayet:2010qz}           & 1            & \checkmark & \checkmark & \checkmark  & \checkmark  & --          & --          & --         &--           \\ \hline
DGP~\cite{Dvali:2000hr}                           & \checkmark   & \checkmark & \checkmark & $	\sharp$ & \checkmark  & --          & --          & --         & --          \\ \hline
$f(G)$-Gauss-Bonnet \cite{Nojiri:2005jg}          & \checkmark   & \checkmark & \checkmark & \checkmark  & \checkmark  & \checkmark  & $ \sharp$   & $\sharp$   & --          \\ \hline
Galileons~\cite{Nicolis:2008in,Deffayet:2009wt}   & \checkmark   & \checkmark & \checkmark & \checkmark  & \checkmark  & \checkmark  & $ \sharp$   & $	\sharp$  & --          \\ \hline
Horndeski~\cite{Horndeski:1974wa,Deffayet:2009mn} & \checkmark   & \checkmark & \checkmark & \checkmark  & \checkmark  & \checkmark  & $ \sharp$   & $\sharp$   & --          \\ \hline
GLPV \cite{Gleyzes:2014dya,Gleyzes:2014qga}       & \checkmark   & \checkmark & \checkmark & \checkmark  & \checkmark  & \checkmark  & $ \sharp$   & \checkmark & --          \\ \hline
low-energy  Ho\v{r}ava~\cite{Horava:2008ih}       & \checkmark   & \checkmark & \checkmark & \checkmark  & --          & \checkmark  & \checkmark  & --         & \checkmark  \\ \hline
\end{tabular}
\normalsize
\caption{Examples of well known DE/MG models described by the EFT action~(\ref{eftact}): \checkmark indicates that the EFT function is present, $\sharp$ means the corresponding EFT function is related to other EFT functions, -- indicates the EFT function is not present.}
\label{tab:eftth}
\end{sidewaystable}

As anticipated, the power of this model-independent construction of a gravitational action lies in the direct connection it has with most of the well known cosmological theories proposed in the last decades. It captures for example the physics at linear cosmological scales of theories such as $f(R)$-gravity~\cite{Sotiriou:2008rp,DeFelice:2010aj}, Horndeski~\cite{Horndeski:1974wa,Deffayet:2009mn}, GLPV theories~\cite{Gleyzes:2014dya,Gleyzes:2014qga}, low-energy Ho\v rava gravity~\cite{Horava:2008ih} and more, as illustrated in Table \ref{tab:eftth}. The four major sub-sets of theories encoded in the EFT action \eqref{eftact} are: 

\begin{enumerate}
\item Scalar-tensor theories  \textit{\`a la} Brans-Dicke, hereafter dubbed Generalized Brans-Dicke theories (GBD), which are described only by the background EFT functions $\{\fg, \Lambda, c\}$; 
\item Horndeski theories for which the relation $\bar M_2^2=-\bar M_3^2=2\mu_1^2$ (and $m_2^2=0$) must be imposed; 
\item GLPV theories which require the condition $\bar M_2^2=-\bar M_3^2$ (and $m_2^2=0$);
\item Lorentz violating theories, such as Ho\v rava gravity, for which $m^2_2\neq 0$.
\end{enumerate}
The second and third conditions allow to eliminate from the EFT action spatial derivatives higher than second order. 

The advantage of the EFT formalism is twofolds: on one hand it allows to perform model-independent explorations of DE/MG models without assuming any particular theory. This is done by selecting all the EFT functions in action \eqref{eftact} or a sub-set of them usually according to the above itemized list. This procedure is known as \textit{pure} EFT approach. The investigation of the above four sub-sets of theories in a {\it pure} EFT fashion allowed to identify quite general features of the gravity force as further discussed in Section \ref{sec:novpred}. On the other hand, the EFT can be used to investigate specific DE or MG models, $e.g.$ $f(R)$. In this case, the EFT functions assume characteristic forms~\cite{Gubitosi:2012hu,Bloomfield:2012ff,Bloomfield:2013efa,Gleyzes:2013ooa,Gleyzes:2014rba,Frusciante:2015maa,Frusciante:2016xoj}. We illustrate the mapping recipe to encode specific DE/MG theories in the EFT formulation in Section~\ref{sec:mapping}. One refers to this procedure as \textit{mapping} approach. The latter is particularly useful to test a specific theory against cosmological data as explained in Section \ref{sec:codes}.

%----------------------------------------------------------------------
\subsection{Modified background Friedmann equations}\label{sec:eftfried}
%----------------------------------------------------------------------

In this Section, we proceed further by illustrating the model-independent formulation of the background Friedmann equations for DE/MG models. In order to obtain such equations, one must vary the action \eqref{eftact} with respect to the metric. This procedure yields the modified Friedmann equations
\begin{align}
\label{eq:calC} c &= \mp^2 \fg \left( -  \dot H + \frac{\bar{\kappa}}{a^2} -  \frac12 \frac{\ddot \fg}{\fg} +  \frac{H}{2} \frac{\dot \fg}{ \fg} \right)  - \frac12 (\rho_m+p_m)  \;, \\
\label{eq:calL}\Lambda &= \mp^2 \fg \left(\dot H +3 H^2 +2 \frac{\bar{\kappa}}{a^2} + \frac12 \frac{\ddot \fg}{\fg}+\frac{5 H}{2}   \frac{\dot \fg}{\fg} \right)  - \frac12 (\rho_m-p_m)  \;,
\end{align}
where the dots correspond to derivatives with respect to cosmic time $t$, $H(t)$ is the Hubble function defined as $H\equiv \f{1}{a}\f{da}{dt}$ and $\rho_\mathrm{m}(t) $ and $p_\mathrm{m}(t)$ are respectively the background energy density and pressure of matter. To complete the set of background equations, one supplies
 the above system with the matter continuity equations as follows 
\begin{equation}\label{eq:conseft}
\dot \rho_\mathrm{m} + 3 H (\rho_\mathrm{m} + p_\mathrm{m}) =  0\; ,
\end{equation}
where the perfect fluid approximation is assumed. The modified Friedmann equations can be rewritten following the fluid description:
\begin{align}
 H^2 + \frac{\bar{\kappa}}{a^2}    &=  \frac1{3 \mp^2 \fg} (\rho_\mathrm{m} + \rho_{\rm DE}  ) \; ,\label{mfried1}\\
\dot H -  \frac{\bar{\kappa}}{a^2} &=  - \frac1{2 \mp^2 \fg} (\rho_\mathrm{m} + \rho_{\rm DE} +p_\mathrm{m} + p_{\rm DE}  ) \label{mfried2} \;,
\end{align}
where $\{\rho_{\rm DE}, p_{\rm DE}\}$ are the density and pressure of the dark fluid. Now, differentiating eq.~\eqref{mfried1} with respect to time and combining it with eqs.~\eqref{eq:conseft}-\eqref{mfried2}, one obtains a ``non-conservation" equation for the dark fluid, which reads
\begin{equation} 
\dot \rho_{\rm DE} + 3 H (\rho_{\rm DE} + p_{\rm DE}) = 3 \mp^2 \dot{\fg} \left(H^2 + \frac{\bar{\kappa}}{a^2}\right)\; .
\end{equation}
The non-conservation of the dark fluid is thus regulated by the time derivative of the EFT function $\fg$. Using eqs.~\eqref{mfried1}-\eqref{mfried2} in eqs.~\eqref{eq:calC}-\eqref{eq:calL} allows the simplification
\begin{align}
c \ & =  \  \frac12 ( - \ddot \fg + H \dot \fg ) \mp^2 + \frac12 (\rho_{\rm DE}+p_{\rm DE}) \;, \label{c2}\\
\Lambda \, & = \ \frac12 (  \ddot \fg + 5 H \dot \fg ) \mp^2 + \frac12 (\rho_{\rm DE}- p_{\rm DE})   \;. \label{L2}
\end{align}
One could equivalently define an effective density and pressure for the dark fluid $\rho_{\rm DE}^\mathrm{eff}$ and $p_{\rm DE}^\mathrm{eff}$ which have the forms 
\begin{align}
\rho_{\rm DE} & = \fg\rho_{\rm DE}^{\rm eff}+ (\fg-1) \rho_m \;, \\
 p_{\rm DE} & = \fg p_{\rm DE}^{\rm eff} + (\fg-1) p_m \;.
\end{align}
They allow to recover a more standard form of Friedmann equations
\begin{align}\label{eq:friedeft}
H^2 + \frac{\bar{\kappa}}{a^2}  &= \frac1{3 \mp^2} \left(\rho_\mathrm{m} + \rho^\mathrm{eff}_{\rm DE} \right) \; , \\
\dot H - \frac{\bar{\kappa}}{a^2} &= - \frac1{2 \mp^2} \left(\rho_\mathrm{m} + \rho^\mathrm{eff}_{\rm DE} +p_\mathrm{m} + p^\mathrm{eff}_{\rm DE} \right) \;,
\end{align}
where the dependency on the non-minimal coupling $ \fg$ has been hidden in the dark component. From the first line above, one recovers the usual constraint relation: $\om+\omde+\Omega_{\bar{\kappa}}=1$ where the density parameters $\Omega_i$ are defined as $\om=\rho_\mathrm{m}/3\mp^2H^2$, $\omde=\rho^\mathrm{eff}_{\rm DE}/3\mp^2H^2$ and $\Omega_{\bar{\kappa}}=-\bar{\kappa}/a^2H^2$. In the following we indicate with $\Omega_{i,0}$ their present day values.

Note that the EFT approach provides a general model-independent description to study the evolution of cosmological perturbations and, in this spirit, one can assume any background expansion. The modified Friedman equations contain three unknown EFT functions $\{\fg, \Lambda,c\}$ and the unknown expansion history, $H$. Thus, one has to fix two out of the four free functions, then the remaining can be obtained from the Friedman equations. Usually, the procedure adopted is to assume a chosen form for $\fg$ and an equation of state for the dark fluid, $i.e.$ $w_{\rm DE}$ ($p_{\rm DE}=w_{\rm DE}\rho_{\rm DE}$) in order to fix $H$. Then both $c$ and $\Lambda$ are fixed through eqs.~\eqref{c2}-\eqref{L2}. In this respect, the most common choice is to fix $H(t)$ to that of $\Lambda$CDM. An attempt to find an explicit form for $\fg$ was made in Ref.~\cite{Frusciante:2013zop}, where the study of the dynamical system associated to the background equations allowed to find appropriate ans$\ddot{\mbox{a}}$tz for $\fg$. One might claim that fixing $H$ to follow a specific expansion history might induce strong hypotheses on the behavior of the underlying gravity theory. Using too specific parametric forms could indeed bias the generality of predictions since one is already selecting a branch of models at the level of the background. To avoid such drawbacks, one can alternatively parameterize $\Lambda$ and $c$ and solve eqs.~\eqref{eq:calC}-\eqref{eq:calL} to derive $H$ and $\fg$~\cite{Raveri:2017qvt,Espejo:2018hxa}. Both the approaches are used in the phenomenological analysis presented in Section~\ref{sec:novpred} and cosmological constraints in Section~\ref{sec:constraints}. 

%---------------------------------------------------------
\subsection{The St\"uckelberg trick}\label{sec:stucktrick}
%---------------------------------------------------------

The unitary gauge used to write the action \eqref{eftact} is useful from a theoretical point of view to identify the main operators which introduce modifications at large scales. Furthermore, as presented before, it allows to select relevant sub-classes of models for cosmological purposes straightforwardly, yet it is not  convenient to study the evolution of the extra scalar DoF and  that of the metric perturbations separately. The reason is because the extra DoF is hidden inside the metric thus an explicit evolution equation for the scalar field cannot be obtained. It is possible to make such field appear explicitly in the action by restoring the full diffeomorphism invariance upon application of the \emph{St\"uckelberg trick} \footnote{The St\"uckelberg trick is a common procedure to study theories with broken gauge symmetries. The introduction of new fields is used to reveal a symmetry of the gauge-fixed theory.}\cite{Gubitosi:2012hu,Bloomfield:2012ff}. To do so, one has to force back the broken gauge transformation on the field in the Lagrangian by imposing the following time coordinate transformations
\begin{equation}
 t  \rightarrow \tilde{t} = t +\pi(x^\mu) \;, \qquad x^i \rightarrow \tilde{x}^i =x^i \;,
\end{equation}
where $\pi$ is the perturbation of the extra DoF. Time translation invariance is thereby restored. The above transformations induces time dependent functions in the action to transform as
\begin{equation}
\mathfrak{g}(t) \rightarrow \mathfrak{g}(t+\pi(t,x^i))= \mathfrak{g}(t) + \dot{\mathfrak{g}}(t)\pi(t,x^i) + \frac12\ddot{\mathfrak{g}}(t) \pi(t,x^i)^2+ ...\;,
\end{equation}
while scalars do not transform. The transformations of the quantities of interest in action (\ref{eftact}) are \cite{Gubitosi:2012hu,Bloomfield:2012ff}
\begin{align}
g^{00} &\to g^{00} + 2 g^{0 \mu} \dot \pi + g^{\mu \nu} \partial_\mu \pi \partial_\nu \pi \;, \\
\delta K_{ij} &\to \delta K_{ij} - \dot H \pi h_{ij} - \partial_i \partial_j \pi \;, \\
\delta K &\to \delta K - 3 \dot H \pi - \frac1{a^2} \partial^2 \pi \;, \\ 
\mathcal{R}_{ij} &\to \mathcal{R}_{ij} + H (\partial_i \partial_j \pi + \delta_{ij} \partial^2 \pi) \;, \\
\mathcal{R} &\to \mathcal{R} + \frac4{a^2} H \partial^2 \pi \;,
\end{align}
where $\mathcal{R}$ is the three dimensional Ricci scalar and $\mathcal{R}_{ij}$ is the corresponding tensor.

Once the above transformations have been inserted in the action (\ref{eftact}), one obtains the following action \cite{Gubitosi:2012hu,Bloomfield:2012ff}:
\begin{align}
S = \int & d^4x \sqrt{-g} \bigg[ \frac{\mp^2}{2} \fg(t + \pi) R - \Lambda(t + \pi) \nonumber \\
&- c(t + \pi) \left(-1+\delta g^{00} - 2 \dot{\pi} + 2 \dot{\pi}\,\delta g^{00} + 2 \nabla_i \pi \, g^{0i} - \dot{\pi}^2 + \frac{1}{a^2} \nabla^i \pi \nabla_i \pi \right) \nonumber \\
& + \frac{M_2^4 (t) }{2} \left(\delta g^{00} - 2 \dot{\pi} \right)^2 \nn\\
&- \frac{\bar{m}_1^3 (t )}{2} \left(\delta g^{00} - 2 \dot{\pi} \right) \left(\delta 
\tensor{K}{^\mu_\mu} + 3 \dot{H} \pi + \frac{\nabla_i\nabla^i \pi}{a^2}\right) \nonumber \\
& - \frac{\bar{M}_2^2 (t)}{2} \left(\delta \tensor{K}{^\mu_\mu} + 3 \dot{H} \pi + \frac{\nabla_i\nabla^i \pi}{a^2} \right)^2 \nonumber \\
& - \frac{\bar{M}_3^2 (t )}{2}\left[
 \left(\delta \tensor{K}{^i_j} + \dot{H} \pi \delta\indices{^i_j}
 + \frac{1}{a^2} \nabla^i \nabla_j \pi \right)
 \left(\delta \tensor{K}{^j_i} + \dot{H} \pi \delta\indices{^j_i}
 + \frac{1}{a^2} \nabla^j \nabla_i \pi \right) \right.\nonumber \\
& \left.\hspace{1.5cm}+ \left(\delta \tensor{K}{^0_0} \right)^2 + 2 \left( \delta \tensor{K}{^i_0} - \frac{H}{a^2} \nabla^i \pi \right) \left( \delta \tensor{K}{^0_i} + H \nabla_i \pi \right) \right] \nonumber \\
& + \frac{\mu^2_1 (t )}{2} \left(\delta g^{00} - 2 \dot{\pi} \right)\, \left(\delta\mathcal{R} +4H \frac{\nabla_i\nabla^i\pi}{a^2}\right) \nonumber \\
&+ \f{m_2^2(t)}{2}\left(g^{\mu\nu}+n^{\mu} n^{\nu}\right)\partial_{\mu}\left(g^{00}-2\dot{\pi}\right)\partial_{\nu}\left(g^{00}-2\dot{\pi}\right)\bigg] + S_{m} [g_{\mu \nu}, \chi_m]\,,
\end{align}
where $\nabla^i$ is the spatial covariant derivative.

It is now possible to obtain the dynamical equation for the extra DoF explicitly by varying the action with respect to $\pi$. This yields
\begin{align}\label{pieq}
A \ddot{\pi}+B\dot{\pi}+(C+k^2 D)\pi+ E=0,
\end{align}
where the spatial part has been Fourier transformed and $k$ is the wavenumber. $A,B,C,D,E$ are functions of time and scale $k$. $E$ also includes couplings with the metric linear perturbations. Their expressions can be found in~\cite{Hu:2014oga}. Let us note that in order to obtain observable predictions this equation needs to be coupled with the other linear perturbative equations found by varying the above action with respect to each metric component \cite{Bloomfield:2012ff,Gleyzes:2013ooa}. 

%--------------------------------------------------------
\subsection{Mapping: a general recipe}\label{sec:mapping}
%--------------------------------------------------------

The relevance of the EFT approach relies in its capability to encompass DE/MG models with a single additional scalar DoF.  Along the model-independent explorations of DE/MG (\textit{pure} EFT approach), it is possible to map specific theories with one additional scalar DoF in the EFT language (\textit{mapping} approach)~\cite{Gubitosi:2012hu,Bloomfield:2012ff,Bloomfield:2013efa,Gleyzes:2013ooa,Gleyzes:2014rba,Frusciante:2015maa,Frusciante:2016xoj}. Hereafter we discuss the latter. 

In order to map a theory in the EFT language one can follow two paths. The first consists in starting from the covariant action of a specific theory and then impose the unitary gauge. One has then to identify each term in the action with the corresponding one in EFT action. Let us present a simple example. We use the covariant quintessence Lagrangian and we impose the unitary gauge as follows 
\begin{equation}
\mathcal{L_{Q}}\sim-\frac{1}{2}(\partial \phi)^2-V(\phi) \xrightarrow[\mbox{unitary gauge}]{} -\frac{1}{2}\dot{\bar{\phi}}^2 g^{00}-V(\bar{\phi}) \;,
\end{equation}
where $\delta \phi=0$ is considered according to the definition of unitary gauge. Then, it is straightforward to identify the following correspondence:
\begin{equation}
c(t)=\frac{1}{2}\dot{\bar{\phi}}^2\,, \qquad \Lambda(t)=V(\bar{\phi})\,,
\end{equation}
from the EFT action \eqref{eftact}.
Deriving the mapping relations from a covariant Lagrangian requires to apply these steps for every theory and it can become cumbersome in more involved cases. This lack of generality is solved by the second option \cite{Kase:2014cwa,Gleyzes:2014rba,Frusciante:2016xoj}, $i.e.$ working out the mapping for a general Lagrangian written in unitary gauge with all the relevant operators. As a result each EFT function can be written in terms of this general Lagrangian. Once this general recipe is derived, any model can be translated in the EFT formalism. 
In the following we illustrate this general approach in details.  

For this purpose one introduces the Arnowitt-Deser-Misner (ADM) formalism \cite{Arnowitt:1959ah}, for which the line element can be written as: 
\begin{equation}
ds^2=-N^2 dt^2+h_{ij}(dx^i+N^i dt)(dx^j+N^j dt)\,,
\end{equation}
where $N(t,x^i)$ is the lapse function, $N^i(t,x^i)$ the shift and $h_{ij}(t,x^i)$ is the three dimensional spatial metric. Using the ADM formalism, a general Lagrangian describing scalar-tensor theories can be written as function of the following operators~\cite{Kase:2014cwa}:
\be\label{ADM_lagrangian}
L=L(N,{\mathcal R}, \mathcal{S}, K, {\mathcal Z}, {\mathcal U}, {\mathcal Z}_1, {\mathcal Z}_2, \alpha_1, \alpha_2, \alpha_3, \alpha_4, \alpha_5; t)\,,
\ee
where in details:
\ba\label{ADM_quantities}
&&\mathcal{S}=K_{\mu\nu}K^{\mu\nu}\,,\,\,{\mathcal Z}={\mathcal R}_{\mu\nu}{\mathcal R}^{\mu\nu}\,,\,\,{\mathcal U}={\mathcal R}_{\mu\nu}K^{\mu\nu}\,,\nonumber\\
&&{\mathcal Z}_1=\nabla_i{\mathcal R}\nabla^i{\mathcal R}\,,\,\,{\mathcal Z}_2=\nabla_i{\mathcal R}_{jk}\nabla^i{\mathcal R}^{jk}\,,\,\,\alpha_1=a^ia_i\,,\nonumber\\
&&\alpha_2=a^i\Delta a_i\,,\,\,\alpha_3={\mathcal R}\nabla_ia^i\,,\,\,\alpha_4=a_i\Delta^2a^i\,,\,\,\alpha_5=\Delta {\mathcal R}\nabla_ia^i,
\ea
with $a_\nu=n^{\mu}\nabla_{\mu}n_{\nu}$ being the acceleration of the normal vector and $\Delta=\nabla_k\nabla^k$. The operators considered in the Lagrangian~(\ref{ADM_lagrangian}) allow to describe gravity theories containing up to sixth order spatial derivatives. According to the EFT action \eqref{eftact}, in the following we consider only some of the operators introduced above, $i.e.$ $L=L(N,{\mathcal R}, \mathcal{S}, K, \alpha_1, t)$. However, an extended EFT action and the general mapping including the whole set of operators listed in eq. (\ref{ADM_lagrangian}) can be found in Refs. \cite{Kase:2014cwa,Frusciante:2016xoj}.

Expanding the Lagrangian \eqref{ADM_lagrangian} up to quadratic order in perturbations of these operators yields the following action \cite{Gleyzes:2013ooa}:
\begin{eqnarray} \label{actionexpanded}
S_{ADM}&=&\int{}d^4x\sqrt{-g}\left[\bar{L}+\dot{\mathcal{F}}+3H\mathcal{F}+(L_N-\dot{\mathcal{F}}) \delta N\right.\nn\\ 
&+&\left.\left(\dot{\mathcal{F}}+\frac{1}{2}L_{NN}\right)(\delta N)^2+L_{\mathcal{S}} \delta K_\mu^\nu\delta K_\nu^\mu+\frac{1}{2}\mathcal{A}(\delta K)^2 \right.\nn\\ 
&+&\left.\mathcal{B}\delta N \delta K+\mathcal{C}\delta K \delta\mathcal{R}+L_{N\mathcal{R}}\delta N\delta \mathcal{R}+L_{\mathcal{R}}\delta \mathcal{R} \right.\nn\\ 
&+&\left.\frac{1}{2}L_{\mathcal{R}\mathcal{R}}\delta \mathcal{R}^2+L_{\alpha_1}\partial_i\delta N\partial^i \delta N \right]\,,
\end{eqnarray}
where $\bar{L}$ is the background expression of the Lagrangian, $L_\mathcal{S}\equiv\partial L/\partial \mathcal{S}$ and equivalently for the others and 
\begin{align}\label{Coefficientdefinitions}
\mathcal{A}&=L_{KK}+4H^2L_{\mathcal{S}\mathcal{S}}+4HL_{SK},\qquad \mathcal{B}=L_{KN}+2HL_{\mathcal{S}N},\nn\\ 
\mathcal{C}& =L_{KR}+2HL_{\mathcal{S}R}, \qquad\qquad\qquad\quad \mathcal{F}=L_K+2HL_{\mathcal{S}}. 
\end{align}

The EFT action \eqref{eftact} has to be written in ADM form as well in order to be compared with eq. \eqref{actionexpanded}. After some manipulations (see~\cite{Gleyzes:2013ooa,Kase:2014cwa,Frusciante:2016xoj} for details) which include the use of the Gauss-Codazzi relation~\cite{Gourgoulhon:2007ue} and the transformation 
\be\label{linkg00N}
g^{00}=-\frac{1}{N^2}=-1+2\delta N-3(\delta N)^2+ ...\equiv-1+\delta g^{00}\,,
\ee
where $\delta N$ is the perturbation of the lapse function, from which one can deduce $(\delta g^{00})^2=4(\delta N)^2$ at second order, one obtains:
\begin{align}\label{EFTADM}
S_{EFT}&=\int d^4x\sqrt{-g}\left\{\frac{\mp^2}{2}\fg\mathcal{R}+3H^2\mp^2\fg+2\dot{H}\mp^2\fg+2\mp^2H\dot{\fg}+\mp^2\ddot{\fg} \right.\nonumber\\
&\left.+c-\Lambda+\left[H\dot{\fg}\mp^2-2\dot{H}\mp^2\fg-\mp^2\ddot{\fg}-2c\right]\delta N -(\mp^2\dot{\fg}+\bar{m}^3_1)\delta K\delta N \right.  \nonumber\\
&\left.+\f{1}{2}\left[\mp^2\fg-\bar{M}_3^2\right]\delta K^{\mu}_{\nu}\delta K^{\nu}_{\mu}-\f{1}{2}\left[\mp^2\fg+\bar{M}^2_2\right](\delta K)^2 +\mu_1^2\delta N\delta\mathcal{R} +\right.\nonumber\\
&\left.\left[2\dot{H}\mp^2\fg+\ddot{\fg}\mp^2-H\mp^2\dot{\fg}+3c+2M^4_2\right](\delta N)^2 +4m^2_2h^{\mu\nu}\partial_{\mu}\delta N\partial_{\nu}\delta N\right\}\,.
\end{align}
At this point, it is very easy to deduce the following mapping by identification:
\begin{align}\label{Map}
&\fg(t)=\frac{2}{\mp^2}L_{\mathcal{R}},\qquad c(t)=-\frac{1}{2}(L_N+\dot{\mathcal{F}})+(H\dot{L}_{\mathcal{R}}-\ddot{L}_{\mathcal{R}}-2L_{\mathcal{R}}\dot{H}),\nonumber\\
&\Lambda(t)=-\bar{L}+\dot{\mathcal{F}}+3H\mathcal{F}+2(3H^2L_{\mathcal{R}}+\ddot{L}_{\mathcal{R}}+2H\dot{L}_{\mathcal{R}}+2\dot{H}L_{\mathcal{R}})+c\,,\nonumber\\
&\bar{M}^2_2(t)=-\mathcal{A}-2L_{\mathcal{R}}, \qquad M_2^4(t)=\frac{1}{2}\left(L_N+\frac{L_{NN}}{2}\right)-\frac{c}{2}, \nonumber\\
&\bar{m}_1^3(t)=-\mathcal{B}-2\dot{L}_{\mathcal{R}},\qquad \bar{M}^2_3(t)=-2L_{\mathcal{S}}+2L_{\mathcal{R}},\nonumber\\ &m_2^2(t)=\f{L_{\alpha_1}}{4},\qquad \mu_1^2(t)=L_{N\mathcal{R}}.
\end{align}

Let us show a practical example by considering the $f(R)$-gravity theory~\cite{Sotiriou:2008rp,DeFelice:2010aj}. The mapping of the latter into the EFT language was first derived in Refs.~\cite{Gubitosi:2012hu,Gleyzes:2014rba}, but here we use the above recipe to find the mapping relations. The $f(R)$-gravity action is:
\be \label{fRaction}
S_f=\int{}d^4x\sqrt{-g}\f{\mp^2}{2}\l[R+f(R)\r],
\ee
where $f(R)$ is a general function of the four dimensional Ricci scalar. We expand this action around the background value of the Ricci scalar, $\bar{R}$:
\be\label{fRexpanded}
 S_f =\int{}d^4x\sqrt{-g}\f{\mp^2}{2}\left\{\l[1+f_R(\bar{R})\r]R+f(\bar{R})-\bar{R}f_R(\bar{R})\right\},
\ee
where $f_R\equiv \f{df}{dR}$. Now we use the Gauss-Codazzi relation~\cite{Gourgoulhon:2007ue} to write the above action in the ADM formalism:
\begin{align}
 S_f =\int{}d^4x\sqrt{-g}\f{\mp^2}{2}&\left\{ \l[1+f_R(\bar{R})\r]\l[\mathcal{R} +\mathcal{S}-K^2\r]\right. \nn\\
&\left. + \f{2}{N}\dot{f}_R K+f(\bar{R})-\bar{R}f_R(\bar{R})\right\}\,.
\end{align}
Finally, using eqs.~(\ref{Map}), one finds the following mapping relations
\be \label{fRmapping}
\fg(t)=1+f_R(\bar{R})\,, \qquad \Lambda(t)=\f{\mp^2}{2}f(\bar{R})-\bar{R}f_R(\bar{R})\,,
\ee 
 and the other EFT functions are zero.

The general recipe of eqs.~(\ref{Map}) is very handful when implementing a specific model in an EB code using the \textit{mapping} procedure as illustrated in Section~\ref{sec:codes}. 

%-------------------------------------------------------
\subsection{Stability conditions}\label{sec:eftofdestab}
%-------------------------------------------------------

We reviewed the construction of a very general framework enclosing many DE/MG models in the previous Sections. Because of the wide generality of the EFT approach, it is of crucial importance to ensure that the theory of gravity under consideration is free from pathological instabilities, such as ghosts, gradient and tachyonic instabilities~\cite{Sbisa:2014pzo}. When testing gravity models with cosmological data using statistical tools~\cite{Zhao:2008bn,Hu:2013twa,Raveri:2014cka,Zumalacarregui:2016pph}, these viability criteria can reduce the parameter space to explore~\cite{Piazza:2013pua,Salvatelli:2016mgy,Peirone:2017lgi} or even dominate over the constraining power of data~\cite{Raveri:2014cka,Frusciante:2015maa,Peirone:2017lgi}. 

Such instabilities are related to the evolution of the extra scalar DoF and when matter fields are involved, they can also contribute to the stability conditions and, as such, alter the viability space of the theory~\cite{Scherrer:2004au,Bertacca:2007ux,Bertacca:2007cv,Gergely:2014rna,Kase:2014cwa,Gleyzes:2014qga,DeFelice:2016ucp,Kase:2014yya}. Hence, the latter need to be consistently considered when analyzing the stability of the whole system. Fundamental then becomes the choice of the matter action. Recently, it has been shown~\cite{DeFelice:2015moy,DeFelice:2016ucp} that among the models describing the matter action~\cite{Schutz:1977df,Brown:1992kc,Scherrer:2004au,Bertacca:2007ux,Bertacca:2007cv,DeFelice:2011bh,Gergely:2014rna,Kase:2014cwa,Gleyzes:2014qga,Gleyzes:2015pma,DAmico:2016ntq,Kase:2014yya}, the more appropriate choice is the Sorkin-Schutz action~\cite{Schutz:1977df,Brown:1992kc}. This action describes general matter fluids and the canonical field characterizing the matter DoFs is the matter density perturbation, $\delta_m$. The latter allows to avoid the problem of a divergent action when including pressure-less matter fluids, such as baryon and dark matter. 

One can then construct an action $S$ made by the EFT action~(\ref{eftact}) ($S_{EFT}$) and the Sorkin-Schutz one ($S_m$), $S=S_{EFT}+S_m$. The action $S$ then includes one DoF for the gravity sector, namely $\zeta$, defined as the scalar perturbation of $h_{ij}$ \footnote{The  ADM metric perturbations for the scalar and tensor components reads: $ds^2=-(1+2\delta N)dt^2+2\partial_i\psi dtdx^i+a^2\l[(1+2\zeta)\delta_{ij}+h_{ij}^T\r]dx^idx^j$, where $\delta N(t,\vec{x})$ is the perturbation of the lapse function, $\partial_i\psi(t,\vec{x})$ is the scalar perturbation of the shift function, $\zeta(t,\vec{x})$ of the three dimensional metric  and $h_{ij}^T(t,\vec{x})$ are the perturbed metric components which contribute to tensor modes.}, as many DoFs, $\delta_{m,i}$, as matter fluids considered  and the tensor modes $h_{ij}^T$. As a result one obtain an action for scalar modes ($S^s$) and one for tensor modes ($S^T$). In Fourier space they have  the following compact forms:
\ba\label{actionshort}
S^s&=&\frac{1}{(2\pi)^3}\int{} d^3k\,dt\,a^3\l(\dot{\vec{\chi}}^t\textbf{A}
\dot{\vec{\chi}}-k^2\vec{\chi}^t\textbf{G}\vec{\chi}-
\dot{\vec{\chi}}^t\textbf{B}\vec{\chi}-\vec{\chi}^t\textbf{M}\vec{\chi}\r)\,,\\
\label{actiontensor}
S^T&=&\frac{1}{(2\pi)^3}\int{} d^3k\,dt\,a^3 \, \f{M^2(t)}{8}\l[(\dot{h}_{ij}^T)^2-c_t(t)^2\f{k^2}{a^2}(h_{ij}^T)^2\r]\,,
\ea
where $\vec{\chi}^t=(\zeta,\delta_i)$ is a dimensionless vector and $\textbf{A},\textbf{G},\textbf{B},\textbf{M}$ are matrices whose coefficients are combinations of EFT functions and some matrices also manifest a $k$ dependence  \footnote{A similar analysis for the no-ghost and no-gradient conditions can be performed by starting from the action \eqref{eftact} and restoring the broken symmetry by means of the St\"uckelberg trick (see Section \ref{sec:stucktrick}). In this case the propagating DoF associated to the gravity sector is $\pi$. The stability conditions have been derived with this approach for Horndeski-like models \cite{Bloomfield:2012ff,Piazza:2013pua}.}. We refer the reader to \cite{DeFelice:2016ucp} for further details and the complete expressions of the matrices.  Finally, $M^2(t)$ and $c_t^2$ are respectively  the   \textit{effective Planck mass} and the speed of propagations of tensor modes. In section \ref{sec:alternative} we provide their expressions in terms of EFT functions.

Let us now discuss the three main sources of instability related to the above  actions:
\begin{itemize}
\item The \emph{Ghost instability} corresponds to having modes with negative kinetic energy. In this case the high energy vacuum is unstable to the spontaneous production of particles~\cite{Cline:2003gs,Carroll:2003st}. Such a pathology is regulated by demanding for a positive kinetic term if only one field is involved, or a positive kinetic matrix if more fields define the system. In the   scalar modes action such condition corresponds to requiring $\textbf{A}$ to be positive definite, \textit{i.e.}  all eigenvalues must be strictly positive. The  condition is imposed only in the high energy regime because the ghost instability generated in the infrared regime corresponds to the physical phenomenon of the Jeans/tachyonic instability~\cite{Gumrukcuoglu:2016jbh}, which can be controlled demanding for specific conditions, as we will discuss in the following. Let us note that in the case of the EFT framework, one has to consider that such approach is valid up to a certain cut-off scale, namely $\Lambda_{\rm cut-off}$. Then, when performing high-$k$ expansions one has to assume the following relations between the scales involved, $H\ll k/a\ll \Lambda_{\rm cut-off}$.   The no-ghost condition when applied to the tensor modes action \eqref{actiontensor} reads  $M^2>0$.

\item The \emph{Gradient or Laplacian instability} occurs when the DoFs propagate with negative speeds, $i.e.$ $c_{s,i}^2<0$  and $c_t^2<0$. This signals the presence of exponentially growing modes. The regime in which the gradient instability manifests itself is in the high-$k$ regime. The speeds of propagation of each DoF can be identified by computing the field equations associated to the actions~(\ref{actionshort})  and \eqref{actiontensor} and considering their high-$k$ expansions. To avoid the gradient instabilities, which would be catastrophic for the system, one has to require that the speeds of propagation are positive, $i.e.$ $c_{s,i}^2>0$  and $c_t^2>0$. 

\item The \emph{Tachyonic and Jeans instabilities} are the less severe instabilities and they appear when the DoF has a negative mass squared. In particular, they arises when the perturbations are not computed about the true vacuum of the theory \cite{Joyce:2014kja}. In order to account for this pathology, one can look at the boundedness of the Hamiltonian at low momenta. These conditions are less explored with respect to the no-ghost/no-gradient conditions and a full and general derivation in the context of the EFT framework is done in Ref.~\cite{DeFelice:2016ucp}. Starting from the above action (\ref{actionshort}), it is possible to obtain the associated Hamiltonian, namely $\mathscr{H}(\Phi_i,\dot{\Phi}_i)$ of canonical fields $\Phi_i$, which in the case of one fluid assumes the form
\begin{equation}
\label{Hamiltonian}
\mathscr{H}(\Phi_i,\dot{\Phi}_i)=\frac{a^3}2\left[\dot{\Phi}_1^2 +\dot{\Phi}_2^2 +\mu_1(t,k)\,\Phi_1^2 +\mu_2(t,k)\,\Phi_2^2\right]\,,
\end{equation}
where $\mu_1$ and $\mu_2$ are the mass eigenvalues. The Hamiltonian is unbounded from below if the mass eigenvalues are negative, $i.e.$ $\mu_i(t,0)<0$. Requiring $\mu_i>0$ would result in a too stringent condition. A less severe request, if the $\mu_i$ are negative, is to demand they satisfy the condition $|\mu_i(t,0)|\lesssim H^2$. In this case the time scale of evolution of the instability is larger than the Hubble time so that it will not affect the stability of the system. Such condition will allow to have $\mu_i$ negative at some times. This behavior is known as Jeans instability and it is necessary in order for structures to form.
\end{itemize}
The pathologies listed above concern both the scalar propagating modes as well as the tensor ones. Let us note that  in the case of the scalar action~(\ref{actionshort}) a unique condition for each of the above points, encompassing all viable theories included in the EFT framework, cannot be computed. This is because of the large number of operators involved and the different $k$ dependence they carry, in particular the operators accompanying $\bar{M}^2_3,\bar{M}^2_2$ and $m^2_2$. Because the stability requirements discussed above demand for high-$k$ or low-$k$ limits, a certain number of sub-cases need to be considered to account for different powers of $k$ appearing in the action and equations~\cite{DeFelice:2016ucp}. The three most relevant cases are~\cite{DeFelice:2016ucp}: 1) the general case for which $m_2^2\neq0$ and $\bar{M}_2^3 + \bar{M}^2_2\neq 0$; 2) the GLPV case which includes Horndeski as sub-case; 3) the Ho\v rava gravity like case. Regardless of the sub-cases considered, the matter fields do not modify the ghost conditions. The same is not true for the case of gradient and tachyon instabilities where the combination of matter and gravity fields is non trivial. For example, in the case of GLPV, it has been found that the speed of propagation of the gravity DoF and that of the radiation fluid do not decouple even at high-$k$~\cite{Gleyzes:2014dya,Kase:2014cwa,DAmico:2016ntq,DeFelice:2016ucp}. Most of the investigations we review in the next Sections are limited to the case of Horndeski theories. For this case, the stability requirements concerning the absence of ghosts and positive speeds of propagation become very simple because the $k$ dependence in the matrices disappears and the matter fields do not affect any of them. On the contrary the mass eigenvalues are strongly modified by the presence of matter fields.  Regardless of the sub-case considered, if direct couplings between the extra scalar field and matter are included, the no-ghost and no-gradient conditions are modified~\cite{Gleyzes:2015pma,Tsujikawa:2015upa}.

Finally, it has been noticed that the conditions for the absence of ghost and gradient instabilities do not dependent on the gauge choice, on the contrary the expressions for the mass eigenvalues do~\cite{DeFelice:2017mwa}. The Hamiltonian \eqref{Hamiltonian} indeed is written in terms of fields which may not have a clear physical interpretation. Thus in order to look for the proper tachyonic conditions, it would be more appropriate to consider the following gauge invariant quantity describing the linear density perturbation of the DE field~\cite{DeFelice:2017mwa}:
\be
\delta_\phi\equiv \f{\delta \rho_\phi}{\bar{\rho}_\phi}+\f{\dot{\bar{\rho}}_\phi}{\bar{\rho}_\phi}\l[\psi-\dot{\gamma}+2H\gamma \r]\,,
\ee
where $\psi$ and $\gamma$ are the scalar perturbations respectively of the shift function and of the metric tensor of the three dimensional spatial slices. Standard bars stand for background values. This definition is very general and applicable both in the presence of matter fields and in the late time Universe. The action \eqref{actionshort} can be written in terms of $\delta_\phi$. The resulting stability requirements in the de Sitter limit, which arise by imposing the absence of ghost and gradient instabilities, do not change when considering the $\zeta$ field nor $\delta_\phi$. On the contrary, the mass term for the two fields is distinctively different~\cite{DeFelice:2017mwa}. Further analysis in this direction is still necessary in order to extend the results of~\cite{DeFelice:2017mwa} by adopting the procedure illustrated in~\cite{DeFelice:2016ucp}. This would allow to definitively identify the mass of the physical perturbation field $\delta_\phi$ in presence of matter fields.

The relevance of imposing physical motivated stability requirements is not limited to demanding a viable gravity theory but they also have power to constrain models parameters. We review and discuss some considerable results in Section~\ref{sec:impactstab}.

%--------------------------------------------------------------------------------------
\subsection{$\alpha$-basis: a phenomenological parameterization}\label{sec:alternative}
%--------------------------------------------------------------------------------------
 
An alternative parameterization of the EFT action, dubbed the $\alpha$-basis, was developed in~\cite{Bellini:2014fua} in order to describe specific physical properties of the Horndeski theory. In that case any departure from GR is described by four time dependent phenomenological functions, namely $\alpha_M(t)$, $\alpha_B(t)$, $\alpha_K(t)$, $\alpha_T(t)$. The original $\alpha$-basis was generalized later to include GLPV models by adding an additional function, $\alpha_H(t)$~\cite{Gleyzes:2014qga,Gleyzes:2014rba}, and finally it was further developed to include higher spatial derivatives operators accounting for Lorentz violation, $e.g.$ $\alpha_{K_2}(t)$, $\alpha_B^{GLPV}(t)$~\cite{Frusciante:2016xoj}.
The $\alpha$-basis has the benefit of relating the evolution of the coupling functions to clear physical effects, hence it is a more phenomenological approach. 

The quadratic action in the $\alpha$-basis encompassing Horndeski, GLPV and low-energy Ho\v rava gravity can be written in ADM formalism and Fourier space as follows~\cite{Frusciante:2016xoj}
\ba\label{alphageneralized}
S&=&\f{1}{(2\pi)^3}\int{}d^3kdt\,a^3\,\f{M^2}{2}\l\{\l(1+\alpha_H\r)\delta N\delta_1 \tilde{\mathcal{R}}+2H\alpha_B\delta N\delta \tilde{K}\r.\nn\\
&+&\delta \tilde{K}^\mu_\nu\delta \tilde{K}^\nu_\mu- (\alpha^{GLPV}_B+1)(\delta \tilde{K})^2+\l.\l(\alpha_K +\alpha_{K_2}\f{k^2}{a^2}\r)H^2(\delta N)^2\r.\nn\\
&+& \left.(1+\alpha_T)\delta_2 (\tilde{\mathcal{R}}\delta (\sqrt{h}))\r\}\,, 
\ea
where the geometrical quantities with tildes are the Fourier transforms, $\delta_2$ refers to taking the expansion at second order in perturbations, $h$ is the determinant of the spatial metric $h_{ij}$. The identification with the EFT basis reads 
\ba\label{alphageneralizeddef}
&&\alpha_B(t)=-\frac{\mp^2\dot{\fg}+\bar{m}^3_1}{H M^2}\,, \quad \alpha_T(t)=\frac{\bar{M}^2_3}{M^2}\equiv c_t^2-1\,,\quad
\alpha_K(t)=\frac{2c+4M_2^4}{H^2 M^2}\,, \nn\\
&& \alpha_{K_2}(t)=\frac{8m_2^2}{M^2H^2}\,,\quad \alpha_H(t)=\f{2\mu^2_1+\bar{M}^2_3}{M^2}\,,\quad \alpha^{GLPV}_B(t)=\f{\bar{M}^2_3+\bar{M}^2_2}{M^2}\,,
\ea
where $M^2(t)=\mp^2\fg-\bar{M}^2_3$ is the \textit{effective Planck mass} and $c_t$ is the speed of propagations of tensor modes or equivalently of gravitational waves (GWs). One can complement the above functions with the {\it running} of the effective Planck mass, 
\begin{equation}
\alpha_M=\frac{1}{H}\frac{d\ln M^2}{d\ln t}\;,
\end{equation}
which characterizes the evolution rate of the effective Planck mass.

Let us now discuss the physical interpretation of the above basis~\cite{Bellini:2014fua,Gleyzes:2014qga,Gleyzes:2014rba,Frusciante:2016xoj,Amendola:2016saw}:
\begin{itemize}
\item $\am$ is the \emph{running Planck mass}. As specified above, this function parametrizes the time evolution of the effective Planck mass. A running Planck mass modifies the growth of structures, introduces anisotropic stress and modifies the friction term in the GW equation \eqref{eq:gweq}.

\item $\{\alpha_B,\alpha_B^{GLPV} \}$: $\alpha_B$ is the \emph{braiding function}~\cite{Bellini:2014fua} \footnote{The definition of $\alpha_B$ presented here follows the one in Ref.~\cite{Bellini:2014fua}. We note that it differs by a minus sign and a factor 2 from the one defined in Ref.~\cite{Gleyzes:2014rba,Frusciante:2016xoj}.}. It describes the mixing between the metric and the DE field. $\alpha_B$ is different from zero for all the theories showing non-minimal coupling to gravity and/or possessing the $\delta N \delta K$ operator in the action, $i.e.$ $f(R), L_3^{H}, L_4^{H}, L_5^{H}, L_4^{GLPV}, L_5^{GLPV}$, where $L_i^H$ and $L_i^{GLPV}$ are respectively the Lagrangians of Horndeski and GLPV theories. This operator does not appear when one considers quintessence and K-essence models ($L_2^{H}$). The additional function $\alpha^{GLPV}_B$ extends the braiding effect to scalar-tensor theories beyond GLPV. Both braiding functions take place in the kinetic term and the speed of propagation of the scalar mode, hence impacting the clustering properties of DE. 

\item $\{\alpha_K, \alpha_{K2}\}$: $\alpha_K$ is called \textit{kineticity} and is purely a kinetic function and $\alpha_{K2}$ is its extension to Lorentz violating theories. They both enter the definition of the kinetic term. They affect the speed of propagation of the DE field hence the condition for the absence of a scalar ghost. In particular, large positive values of these functions suppress the sound speed of scalar perturbations. The $\alpha_K$ function is characteristic of theories belonging to GLPV models for which $\alpha_{K2}$ vanishes. The opposite holds for low-energy Ho\v rava gravity. $\ak$ is the only coupling present in quintessence or perfect-fluid DE models.

 \item $\alpha_T$ is the \textit{tensor speed excess} and describes the deviation of the speed of propagation of GWs from the speed of light. This function is present in Horndeski, GLPV and low-energy Ho\v rava gravity. It affects the evolution of the scalar gravitational potentials leading to anisotropic stress (see Section \ref{sec:gravcouplings}). 
 
\item $\alpha_H$ characterizes the departure from Horndeski theories. It contributes to the speed of propagation of the scalar DoF and couples the gravitational field to the velocity of matter~\cite{Gleyzes:2014dya}. This function is present in GLPV and low-energy Ho\v rava gravity models. In particular, in the latter case this function can be further extended with additional functions associated with higher order spatial derivatives terms in high-energy Ho\v rava gravity~\cite{Frusciante:2016xoj} for example.

\end{itemize} 

The above basis was carefully built to consider the different phenomenological aspects of the DE fluid. However, let us notice that the desired correspondence between the $\alpha$-functions and physical effects becomes weaker when going beyond the Horndeski class. 

%-------------------------------------------------------------
\subsection{Couplings with matter fields}\label{sec:couplings}
%-------------------------------------------------------------

The EFT formalism described through the action (\ref{eftact}) assumes the matter fields, $\chi_m$, are minimally coupled to gravity through a unique metric $L_m(g_{\mu\nu},\chi_m)$. This frame dubbed Jordan frame is the standard frame where the interpretation of cosmological measurements are performed. The reason for using the Jordan frame relies on the fact that stringent constraints exist on the couplings between the extra DoF and the standard matter species, $i.e.$ baryons and photons~\cite{Hui:2009kc,Creminelli:2013nua}. As a consequence such couplings are chosen to be minimal, $i.e.$ the matter fields are not coupled to the scalar curvature. However, in the case of dark matter and neutrinos, observational constraints are less severe and one has more freedom. One can consider therefore a frame where the gravitational interaction between the additional scalar DoF and the matter fields is explicit. The so-called Einstein frame. In this frame, the metric $\hat{g}_{\mu\nu}$ is related to the Jordan frame metric $g_{\mu\nu}$ by a conformal/disformal transformation as follows \footnote{Let us note that the transformation \eqref{couplings} can be more general, for instance $C_i$ can also depend on $X$ or even it can include an extended disformal term given by a rank-two symmetric tensor~\cite{Zumalacarregui:2013pma}. }:
\be\label{couplings}
\hat{g}_{\mu\nu}^{(i)}=C_i(\phi)g_{\mu\nu}+D_i(\phi,X)\partial_\mu\phi\partial_\nu\phi\,,
\ee
where $C_i$ and $D_i$ are respectively the conformal and disformal couplings for each matter species $i$ and $X=\partial_\mu\phi\partial^\mu\phi$. The condition $C_i>0$ is necessary to preserve the signature of the Jordan frame metric. Let us note that the transformation \eqref{couplings} preserves the structure of the Lagrangian in GLPV theories~\cite{Gleyzes:2014qga}, while in the case of Horndeski theories the structure is preserved if and only if $D_i(\phi)$~\cite{Bettoni:2013diz,Zumalacarregui:2013pma}. 

Couplings with matter fields have been investigated in the EFT framework for Horndeski theory~\cite{Gleyzes:2015pma} and later generalized to GLPV~\cite{Gleyzes:2014qga,Tsujikawa:2015upa,DAmico:2016ntq}. We treat hereafter the general case of GLPV~\cite{DAmico:2016ntq}. Using the unitary gauge and the ADM formalism, the disformal transformation in eq.~(\ref{couplings}) reads
\be
\hat{g}_{\mu\nu}^{(i)}=C_i(t)g_{\mu\nu}+D_i(t,N)\delta^0_\mu\delta^0_\nu\,.
\ee 
Then, one can add to the $\alpha$-basis in action (\ref{alphageneralized}) for each matter species three coupling functions defining the conformal and disformal interactions~\cite{DAmico:2016ntq}
\be
\alpha_{C,i}=\f{1}{2H}\f{dlnC_i}{dt}\,,\qquad \alpha_{D,i}=\f{D_i}{C_i-D_i}\,,\qquad \alpha_{X,i}=\f{1}{2C_i}\f{\partial D_i}{\partial N}\,.
\ee
 Note that for the case of Horndeski theories $\alpha_{X,i}=0$. One has a total of $3N_i + 5$ time-dependent functions if $N_i$ matter species are present where the +5 are the standard coupling functions characterizing GLPV theories. However, the arbitrariness in the choice of the gravitational metric used to define the gravitational and matter sectors makes three of these functions redundant. The structure of the action is preserved under transformations of the reference metric thus the number of physically relevant functions of time reduces to $3N_i + 2$~\cite{DAmico:2016ntq}.

The inclusion of disformal couplings has been shown to impact the no-ghost and no-gradient stability requirements~\cite{Gleyzes:2015pma,Tsujikawa:2015upa}. The disformal couplings indeed contribute to the kinetic term of the scalar modes, thus the condition for the absence of ghost is modified and in turn the speed of propagation as well. Furthermore, the corresponding conditions are frame independent~\cite{Gleyzes:2015pma,Tsujikawa:2015upa}.

%-----------------------------------------------------
\subsection{Einstein-Boltzmann codes}\label{sec:codes}
%-----------------------------------------------------

In order to perform explorations of cosmological observables and constrain cosmological and model parameters, one option is to modify existing EB codes based on $\Lambda$CDM. The latter allow to numerically evaluate the linear evolution of relevant perturbed quantities in DE/MG scenarios. As a result, general purpose codes have been developed using the EFT formalism in order to encompass a wide range of DE/MG models and to allow for model-independent exploration of their properties (\textit{pure} EFT approach) as well as for a full implementation of specific models once the mapping has been worked out (\textit{mapping} approach). In the \textit{pure} EFT approach, the user has to specify the expansion history, $i.e.$ choose the functional form of two out of the four background functions ($\fg$, $\w(a)$, $c$ and $\Lambda$) and the functional forms of the other EFT functions. In the \textit{mapping} approach, the user has to implement a background solver to find the expansion history, $H$, for the desired model and provide the mapping relations for the EFT functions as discussed in Section \ref{sec:mapping}. A variant in the background solver is the so-called \textit{designer} approach, $i.e.$ the expansion history is fixed ($\w(a)$) and the dynamical equation for the gravity field is solved. The EFT functions are then specified using the mapping recipe. Such EB codes allow to evolve the full set of linear perturbative scalar and tensor equations without relying on any Quasi Static (QS) approximation. The QS approximation consists in neglecting the time derivatives of linear perturbations and it is usually applied within the sound horizon of the DE mode, $k/aH>c_s$~\cite{Sawicki:2015zya,Frusciante:2018jzw}. However, the latter is a necessary condition for the QS approximation to hold but not sufficient to exploit the full dynamics of the extra DoFs~\cite{Peirone:2017ywi}. 

A collection of EB solvers based on the EFT framework has been compared and cross-calibrated recently~\cite{Bellini:2017avd,Pace:2019uow}. Among them, there are: the Effective Field Theory for \texttt{CAMB} (\texttt{EFTCAMB}) \cite{Hu:2013twa,Raveri:2014cka}, Horndeski in \texttt{CLASS} (\texttt{hi\_class}) \cite{Zumalacarregui:2016pph}, Cosmology Object Oriented Package (\texttt{COOP}) \cite{Huang:2015srv} and Equation of State for \texttt{CLASS} (\texttt{EoS\_class}) \cite{Pace:2019uow}. All are publicly available \footnote{ \eftcamb webpage: \url{www.eftcamb.org}; \texttt{hi\_class}: webpage: \url{www.hiclass-code.net}; \texttt{COOP} webpage: \url{www.cita.utoronto.ca/~zqhuang}; \texttt{EoS\_class} webpage: \url{https://github.com/fpace}.}. Here we briefly describe their main features:

\begin{itemize}

\item \eftcamb is a patch for the public EB solver \texttt{CAMB}~\cite{Lewis:1999bs} which implements the EFT approach using the formulation presented in~\cite{Bloomfield:2012ff}. The set of EFT functions used in \eftcamb is in Appendix  \ref{App:params} (we refer to it as \eftcamb basis hereafter). The code has built-in models which include specific theories such as {\it designer} $f(R)$-gravity~\cite{Raveri:2014cka}, $f(R)$-Hu Sawicki model~\cite{Hu:2016zrh}, minimally coupled quintessence~\cite{Hu:2014oga}, low-energy Ho\v rava gravity~\cite{Frusciante:2015maa}, covariant Galileon~\cite{Peirone:2017vcq}, K-mouflage~\cite{Benevento:2018xcu}, Galileon Ghost Condensate \cite{Peirone:2019aua}, beyond Horndeski model \cite{Peirone:2019yjs} as well as several model-independent parameterizations of the DE equation of state and choices for the EFT functions. It allows to use the $\alpha$-basis as well~\cite{Bellini:2014fua}. A novelty introduced in the \eftcamb patch is the built-in stability module which checks for the viability of the underlying theory of gravity by imposing the full set of physical conditions discussed in Section~\ref{sec:eftofdestab}, $i.e.$ no-ghost and no-tachyonic conditions and a positive speed of propagation. The resulting viable parameter space is supplied as prior when using the Markov Chain Monte-Carlo (MCMC) code named \texttt{EFTCosmoMC}~\cite{Raveri:2014cka}. 

\item \texttt{hi\_class} is a modified version of the public EB solver \texttt{CLASS}~\cite{Lesgourgues:2011re} which implements Horndeski theory by using the formalism of the $\alpha$-basis, $i.e.$ $\alpha_M,\alpha_K,\alpha_T,\alpha_B$~\cite{Bellini:2014fua}. It has been recently extended to include also $\ah$ which is characteristic of GLPV models~\cite{Traykova:2019oyx}. The code includes the \eftcamb basis as well. The code is comprised of several built-in specific theories such as quintessence, with different choices for the potential, Brans-Dicke theory~\cite{Alonso:2016suf}, covariant Galileon~\cite{Renk:2017rzu} and also allows for a model-independent exploration of the $\alpha$-basis with built-in models. A stability check ensures the viability of the theory by implementing the no-ghost and positive speed conditions.  \texttt{hi\_class} is interfaced with \texttt{MontePython} \cite{Audren:2012wb,Brinckmann:2018cvx} to compute cosmological constraints. 

\item \texttt{COOP} is an EB code which solves the linear cosmological perturbations for GLPV theories and sub-classes by implementing the formalism of the $\alpha$-basis. It also includes a stability module verifying the absence of ghost and gradient instabilities. The likelihoods to perform cosmological constraints are embedded within the code itself.

\item \texttt{EoS\_class} is a modified version of the public EB solver \texttt{CLASS}~\cite{Lesgourgues:2011re}. It implements the $\alpha$-basis description of the Horndeski theory~\cite{Bellini:2014fua}. Unlike the previous EB codes, \texttt{EoS\_class} is based on the equation of state approach \cite{Battye:2013aaa}. This formalism encloses the modifications to GR in an effective fluid described by a non-trivial stress-energy tensor and the coefficients multiplying the fluid perturbations are written in terms of $\alpha$-functions. The code also includes as specific theory the {\it designer} $f(R)$-model (\texttt{CLASS\_EOS\_FR})~\cite{Battye:2015hza}.
 
\end{itemize}

Authors in Ref. \cite{Bellini:2017avd} compared the shapes of the CMB angular power spectrum and of the dark matter power spectrum predicted by  \texttt{EFTCAMB},  \texttt{hi\_class} and  \texttt{COOP}. The results agree at a level of $0.1\%$ for the matter power spectrum at all scales and for the TT, EE, TE spectra of CMB for angular scales $\ell>100$. Deviations of up to $0.5\%$ arise for $\ell<100$ due to known lack of convergence issues already detected when comparing results of \texttt{CAMB} and \texttt{CLASS} in $\Lambda$CDM. The analysis of  \texttt{COOP}, instead,  shows that \texttt{COOP} achieves the required precision only for $k<1h{\rm Mpc}^{-1}$. \texttt{EoS\_class} was cross-checked with \hiclass for the $\alpha$-basis \cite{Pace:2019uow} and with \eftcamb for the {\it designer} $f(R)$-model \cite{Bellini:2017avd} showing a sub-percent agreement. These results strengthened the confidence on these numerical codes making them efficient for precision constraints on cosmological and gravitational parameters. 

The initial conditions (ICs) for N-body codes can be fixed by using the linear EB codes or using the Zel'dovich' approximation \cite{Crocce:2006ve}. Due to the complexity of modeling the physical phenomena approximations are sometimes employed to set ICs, which for example do not account for dynamical perturbations of DE that instead might be significant. In this regard, ICs for N-body simulations were discussed within the EFT framework \cite{Valkenburg:2015dsa}. The model considered for the analysis is the $f(R)$-gravity, modeled using the \textit{designer} approach~\cite{Pogosian:2007sw}, $i.e.$ the expansion history is fixed and the dynamical equation is then solved for $f(R(a))$. In this specific case the expansion history is chosen to closely mimic the $\Lambda$CDM and $w$CDM. As result, ICs set at early time leave imprints up to 5\% at Mpc scales. This notably implies that one must go beyond the $\Lambda$CDM ICs for a proper implementation of $N$-body simulations in MG theories. A public code for the generation of ICs exists: \texttt{FalconIC}~\cite{Valkenburg:2015dsa} \footnote{\texttt{FalconIC} webpage: \url{http://falconb.org}}. It can be linked to any version of both EB codes \texttt{CAMB} and \texttt{CLASS}, including  \texttt{EFTCAMB}.

Many of the phenomenological investigations and cosmological constraints we present in the next Sections are obtained using the above linear EB packages.

%%%%%%%%%%%%%%%%%%%%%%%%%%%%%%%%%%%%%%%%%%%%%%%
\section{Novel predictions} \label{sec:novpred}
%%%%%%%%%%%%%%%%%%%%%%%%%%%%%%%%%%%%%%%%%%%%%%%

Ongoing and future cosmological surveys offer an unprecedented insight into gravity on cosmological scales. The EFT framework revealed to be a powerful theoretical tool  to systematically identify clear patterns and predictions of MG and DE proposals. In this Section, we review the novel predictions obtained with this approach, namely by using the systematic enforcement of stability conditions and the straightforward computation of cosmological predictions and observables. Let us note that until now the widely investigated class of models within the EFT framework is Horndeski gravity, which encloses a large class of well known scalar-tensor theories. Thus, most of the results presented in this Section apply to Horndeski models even though in some cases they can be straightforwardly extended.

%-----------------------------------
\subsection{Gravitational couplings}\label{sec:gravcouplings}
%-----------------------------------

Cosmological probes can be schematically divided into two subsets: surveys observing the smooth expansion of the Universe, such as SNIa and BAO, will constrain the Hubble rate $H(t)$ or the DE equation of  state $\w$; GC, RSD, CMB and WL surveys scrutinize the clumpy nature of Universe at large scales leading to measurements of  gravitational potentials, matter density and temperature fluctuations  power spectra. The latter data can be seen as  hybrid probes  since they contain information both on the evolution of the background and perturbations. These cosmological probes can be used to test DE and MG proposals at cosmological scales. Regarding the evolution of perturbations in scalar-tensor theories, the two powerful and handy phenomenological functions $\mu(t,k)$ and $\Sigma(t,k)$ prove useful to interpret theoretical predictions in light of observations~\cite{Amendola:2007rr,Bean:2010zq,Silvestri:2013ne,2010PhRvD..81j4023P,Amendola:2019laa}. The former, known as the {\it effective gravitational coupling} or the effective Newton constant, characterizes the modifications of gravity on the clustering of matter. The latter, called the {\it light deflection parameter}, describes the modifications of gravity on null geodesics, $i.e.$ how light travels on cosmological distances. To define these phenomenological functions, let us consider linear scalar perturbations for which the line element in Newtonian gauge reads
\be
ds^2=-(1+2\Psi)dt^2+a(t)^2(1-2\Phi)\delta_{ij}dx^idx^j\,,
\ee
where $\{\Psi(t,x^i), \Phi(t,x^i)\}$ are the gravitational potentials. The $\mu(t,k)$ and $\Sigma(t,k)$ functions are defined in Fourier space as
\be\label{muSigma}
 -\frac{k^2}{a^2} \Psi = 4 \pi \gn \,\mu(t, k)  \rho_\mathrm{m} \Delta_{\rm m}\,, \qquad
 -\frac{k^2}{a^2} (\Psi+\Phi)=8\pi G_N\Sigma(t,k) \rho_\mathrm{m}\Delta_{\rm m}\,,
\ee
where $G_N$ is the Newton gravitational constant and the comoving density contrast is defined as $\Delta_{\rm m}= \delta_{\rm m} +3Hv/k$, where $\delta_{\rm m}=(\rho_{\rm m}-\bar{\rho}_{\rm m})/\bar{\rho}_{\rm m}$ is the density contrast and $v$ is the irrotational component of the peculiar velocity. GC and RSD data, being statistics of the matter perturbation stochastic field, are direct probes of $\mu$. $\Sigma$ measuring the deviation in the Weyl potential ($\Phi+\Psi$) can be probed with measurements sensitive to the lensing of light. A third quantity, although its connection to observations is less obvious, called the {\it gravitational slip parameter}, is often considered 
\be\label{slip}
\gsp(t,k) = \frac{\Phi}{\Psi} \,.
\ee
The three phenomenological functions are thus linked by the relation
\be\label{eq:linkphenofunc}
\Sigma(t,k) = \frac{\mu(t,k)}{2} \left(1+\gsp(t,k) \right) \,.
\ee
Note that the GR limit is recovered when the three phenomenological functions are equal to unity. 

Finding analytical forms of any of the above functions is generally not possible but requires numerical solving. However, a direct connection between them and a specific theory of gravity can be derived if the QS approximation is assumed. This approximation amounts to neglecting terms involving time derivatives in the Einstein equations for linear perturbations. It has been proved to be a valid assumption within the sound horizon of the DE mode \cite{Sawicki:2015zya,Frusciante:2018jzw}. The advantage offered by the EFT approach is twofold. On the one hand, it grants the possibility to explore modification of gravity beyond the simplifying assumption that is the QS approximation. In this case $\mu$, $\Sigma$ and $\eta$ can be computed numerically using EB codes which evolve the full linear perturbative equations of the EFT formulation. On the other hand, it allows to obtain explicit and algebraic expressions in the QS limit for each phenomenological function, enabling a direct and neat connection between observables and the relevant couplings of the gravitational interaction, $i.e.$ the EFT functions. In this case the analytical expressions are obtained for large classes of models ($e.g.$ Horndeski, GLPV, Ho\v rava gravity) instead of being derived for any specific theory such as $f(R)$-gravity. As we will see further on, most of the discussions on the phenomenology of $\mu$, $\Sigma$ and $\eta$ in literature deals with Horndeski models and sub-classes. We thus focus on these cases here while we refer the reader to \cite{Gleyzes:2015pma,Langlois:2017mxy,Cusin:2017mzw,DAmico:2016ntq,Traykova:2019oyx} for the expression of the phenomenological functions in scenarios beyond this landscape. For the case of Horndeski models and sub-classes, the analytical expressions of the phenomenological functions in the QS approximation can be written as follows 
\ba\label{QSmueta}
&&\mu   = \dfrac{\mp^2}{M^2}\; \dfrac{1+\mc^2 \dfrac{a^2}{k^2}}{\dfrac{1}{2}f_1 f_3 M^2 +\dfrac{\mc^2}{1+\at}\dfrac{a^2}{k^2}}\;,\\[2mm]
&&\gsp  = \dfrac{\dfrac{f_5}{f_1}+\dfrac{\mc^2}{1+\at}\dfrac{a^2}{k^2}}{1+\mc^2\dfrac{a^2}{k^2}}\;,\\[2mm]
\label{QSsigma}
&&\Sigma =  \dfrac{\mp^2}{2M^2}\; \dfrac{1+\dfrac{f_5}{f_1}+\mc^2\left(1+\dfrac{1}{1+\at}\right)\dfrac{a^2}{k^2}}{\dfrac{1}{2}f_1 f_3 M^2 +\dfrac{\mc^2}{1+\at}\dfrac{a^2}{k^2}}\;,
\ea
where $f_i$ and $\mc^2$ depend on the EFT functions (see \cite{Pogosian:2016pwr} for details). $\mc$ sets a crucial transition scale: the one below which the scalar field mediates a fifth force \cite{Joyce:2014kja}. A fifth force is characteristic of MG theories with extra DoFs and the transition scale depends on the dynamical mechanism which screens the strength of the scalar fifth force in local environments \cite{Joyce:2014kja}. Astrophysical scales are  typical examples of screened environments.
In other words, the transition scale relates to the Compton wavelength of the scalar field, $\lc$, as $\mc \propto \lc^{-1}$. Theories of the chameleon type \cite{Khoury:2003aq,Khoury:2003rn} display a small Compton wavelength $\lc \lesssim 1\; \rm{Mpc}$, whereas models exhibiting self acceleration, and in a more general sense, models where the extra DoF sources cosmic acceleration, bare a very large Compton wavelength $\lc \propto H^{-1}$. Let us look into both regimes in more details. 

On super-Compton scales, $i.e.$ $k/a\ll \mc$, which we identify with subscript ``${\rm sc}$'', the eqs. \eqref{QSmueta}-\eqref{QSsigma} reduce to \cite{Pogosian:2016pwr}
\be
\label{eq:obs0}
\mu_{\rm sc}  = \dfrac{\mp^2}{M^2} \left(1+\at\right)\;, \quad
\gsp_{\rm sc}  = \dfrac{1}{1+\at}\;,\quad
\Sigma_{\rm sc} = \dfrac{\mp^2}{M^2}\left(1+\dfrac{\at}{2}\right)\;.
\ee
The above are thus representative of the modification of gravity which remain in a screened environment. For scalar-tensor theories such environment amounts to a medium where the scalar field is decoupled from the matter fields. The only mediators left to transmit long-range interactions are the tensors modes. 

On the other hand, in the sub-Compton regime, $i.e.$ $k/a \gg \mc$, which we denote with subscript ``$\infty$'', the eqs. \eqref{QSmueta}-\eqref{QSsigma} yield \cite{Pogosian:2016pwr}
\ba
\label{eq:obsinfinity}
\mu_\infty  &=& \dfrac{\mp^2}{M^2} \left(1+\at+\bx\right)\;, \nn\\[1mm]
\gsp_\infty  &=& \dfrac{1+\beta_B\beta_\xi/2}{1+\at+\bx}\;,\nn\\[1mm]
\Sigma_\infty &=& \dfrac{\mp^2}{M^2}\left(1+\dfrac{\at}{2}+\dfrac{\bx}{2}+\dfrac{\beta_B\beta_\xi}{4}\right)\;,
\ea
where
\ba 
\bb   &=& \dfrac{2}{c_s^2\alpha} \ab^2\;,\nn\\[1mm]
\bx   &=& \dfrac{2}{c_s^2\alpha} \left(\dfrac{\ab}{2}(1+\at)+\am-\at \right)^2\;,\nn\\[1mm]
\alpha &=& \ak +\dfrac{3}{2} \ab^2 \;,\label{ghostalpha}
\ea
with $\alpha$ being the no-ghost condition and $c_s^2$ the speed of propagation of the scalar mode. These represent then the full modifications of gravity on large cosmological scales. 

The phenomenological functions written as in eq. \eqref{eq:obsinfinity} allows a deeper understanding of the origins of modifications of gravity. For example the effective gravitational coupling $\mu$ can be understood as
\be \label{effectivemu}
\mu_\infty = \musc \left(1 + \muff\right)\;,
\ee
where $\muff = \beta_\xi^2/(1+\at)$ characterizes directly the strength of the fifth-force mediated by the scalar field \cite{Perenon:2015sla}. This contribution must be positive for a viable model since the gravitational interaction induced by a healthy spin-0 field is always attractive. This is indeed the case thanks to the stability conditions: $\{c_s^2, \alpha\}>0$ for scalar modes and $1+\at>0$ for tensor modes. 

We must emphasize that care must be taken regarding the normalization of the effective Planck mass. Defining a screened version of the phenomenological functions amounts implicitly to deciding the normalization scheme adopted for the effective Planck mass. Experiments measuring the value of the Newton constant are performed in a screened environment and thus probe the gravitational coupling which remains in the EFT action once the scalar field is decoupled from the gravitational potentials: $G_{\rm sc}= (1+\at)/M^2$ \cite{Perenon:2015sla}. Therefore, at present time, $t_0$, its value must coincide with the Newton constant, $i.e.$ $G_{\rm sc}(t_0) \equiv \gn$, which implies $\musc$ must be normalized as \cite{Perenon:2015sla}
\be\label{eq:normmusc}
\musc= \frac{G_{\rm sc}(t)}{G_{\rm sc}(t_0)} = \dfrac{M^2(t_0)\left(1+\at\right)}{M^2\left(1+\at(t_0)\right)}\;.
\ee
Upon identification with eq. \eqref{eq:obs0} one deduces $\mps=M(t_0)^2/\left(1+\at(t_0)\right)$. Importantly, this also implies that $\musc(t_0)=1$. Note that writing a simple and analytical definition of $\musc$ might not be straightforward when the QS approximation is not applied. In this respect, this normalization is not always adopted in literature. In the rest of the review, if not stated otherwise, we assume the convention to set the effective Planck mass today to the Planck mass is the one adopted\footnote{The previous being a matter of definition, one might be led to believe a simple rescaling of results in one normalization into the other to be sufficient to establish a fair comparison. This however has some caveats because of the underlying stability conditions. Let us consider an illustrative example. In the context of Horndeski theories, stability conditions and observations generically push towards $M^2(t)\gtrsim \mps$. Assuming this strict prior and considering the normalization $M^2(t_0)=\mps$, an MCMC analysis would therefore produce models satisfying $dM^2(t_0)/dt \gtrsim 0$ only. On the contrary, if the normalization of $M^2$ is left free, provided $M^2(t_0) \neq \mps$, both outcomes $dM^2(t_0)/dt \gtrsim 0$ and $dM^2(t_0)/dt \lesssim 0$ would appear and be stable. As a result, the posterior distribution on $M^2$, if not all parameters of the analysis, will be different than in the former case. Both the size and the shape thus depend on the normalization and a simple rescaling would not suffice.}.

\begin{figure}[!]
\begin{center}
 \includegraphics[scale=0.18]{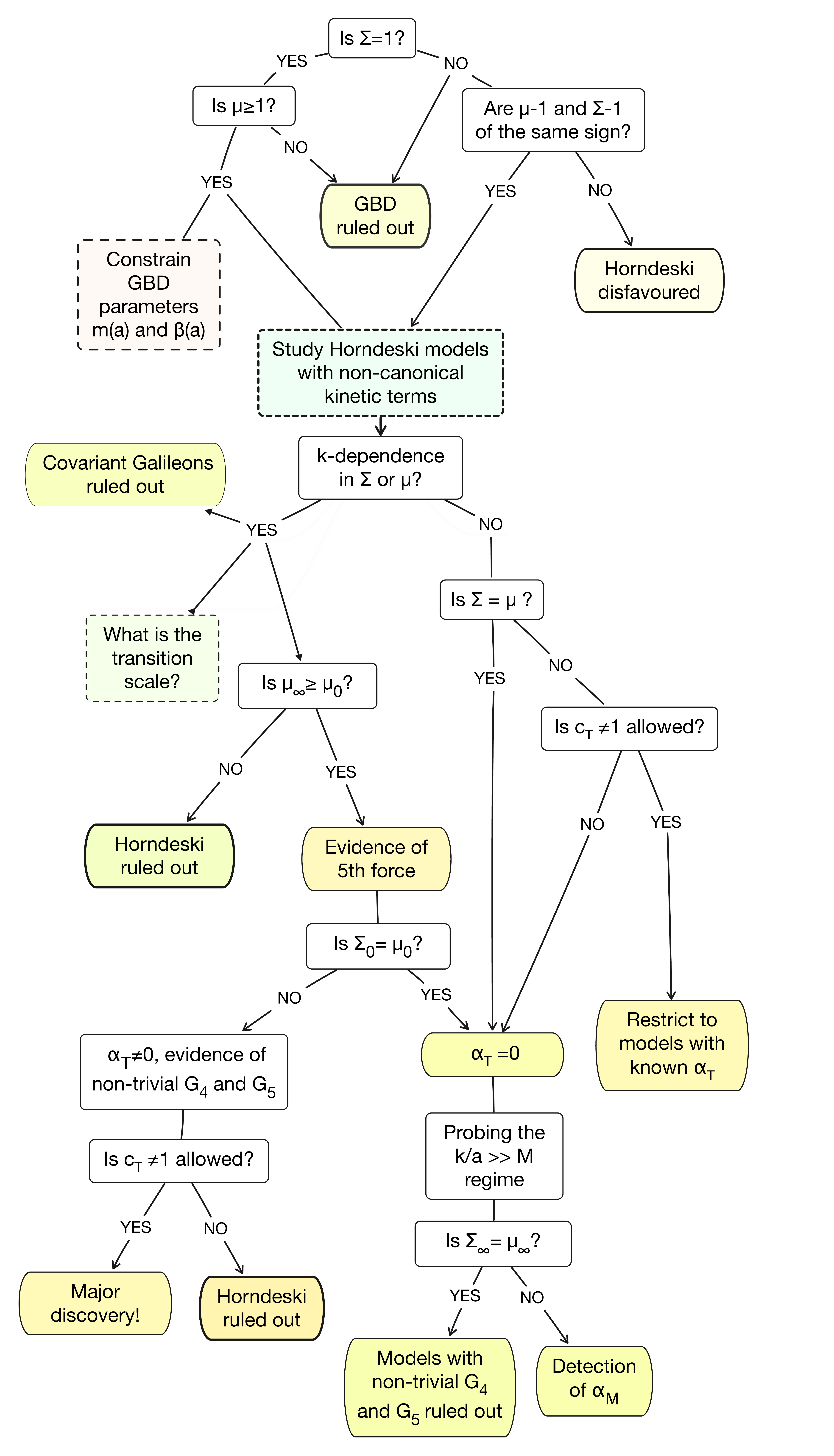}
 \caption{Figure 1 in Ref. \cite{Pogosian:2016pwr}. The diagram summarizes systematic interpretations of the phenomenological functions $\mu$ and $\Sigma$ according to their potentially measured values with the purpose of constraining/ruling out Horndeski models. $G_i$ functions are the free functions in Horndeski theory and $\mu_0, \Sigma_0, c_T$ are respectively $\mu_{\rm sc}$, $\Sigma_{\rm sc}$ and $c_t$ in this review.}
 \label{fig:horncon}
\end{center}
\end{figure}

%---------------------------------------------------------
\subsection{Phenomenology from $\mu$, $\Sigma$ and $\eta$}\label{sec:phenomenology}
%---------------------------------------------------------

The phenomenological functions $\mu, \Sigma, \eta$ introduced in the previous Section can be used for a systematic interpretation of observations with the aim of constraining or ruling out classes of DE/MG models. Analyzing their expressions in terms of EFT functions already gives an insight about the phenomenology of models. Furthermore, numerical investigations of these phenomenological functions give also precious information since the planes identified by $\mu$ - $\gsp$ and $\mu$ - $\Sigma$ have been proven to be sound benchmarks to highlight peculiar features of DE/MG models. In this regards the use of a Monte-Carlo approach to systematically generate large samples of {\it pure} EFT models ($\sim 10^4$) under stability requirements enables to draw conclusions of wide applicability.

Figure \ref{fig:horncon} \cite{Pogosian:2016pwr} is a practical example of how model-independent measurements can help in the diagnostic of Horndeski models. Self-accelerating models \cite{Neveu:2013mfa,Barreira:2013xea} have a very small scalar mass, comparable to the Hubble rate, then a detection of $k$-dependence by LSS surveys in either of the phenomenological functions would rule these models out. The condition $\mu_{\infty}>\mu_{\rm sc}$ follows from the presence of attractive fifth force in Horndeski models, hence a detection of the opposite, $\mu_\infty <\musc$, would rule out the whole class of models. In the super-Compton limit, $\at$ regulates the deviation in the slip parameter. Thus an observation of $\eta_{\rm sc}=1$ or $\musc=\Sigma_{\rm sc}$ would imply $\at=0$. The opposite also holds, a measurement of $\at=0$ requires $\eta_{\rm sc}=1$ otherwise Horndeski theories would be ruled out. 
 GWs are weakly constrained on cosmological scales with CMB but recently the Laser Interferometer Gravitational-Wave Observatory (LIGO)/Virgo and the INTernational Gamma-ray Astrophysics Laboratory (INTEGRAL)/Fermi collaborations constrained the deviation in the speed of GWs to be smaller than $10^{-15}$~\cite{TheLIGOScientific:2017qsa,Monitor:2017mdv}. This has severe implications for MG and in particular on the form of $G_4$ and $G_5$ in the Horndeski Lagrangian \cite{Creminelli:2017sry,Baker:2017hug,Ezquiaga:2017ekz,Creminelli:2018xsv,Amendola:2017orw} (see Section \ref{Sec:GW} for a more detailed discussion). Assuming therefore that $\at$ is negligible, a measurement of $\mu_{\infty}\neq\Sigma_{\infty}$ would give a signature of $\am \neq 0$. On the contrary it would constrain the form of $G_4$ and $G_5$ in the Horndeski Lagrangian \cite{Pogosian:2016pwr}. 

\begin{figure}[!]
\begin{center}
\includegraphics[scale=0.27]{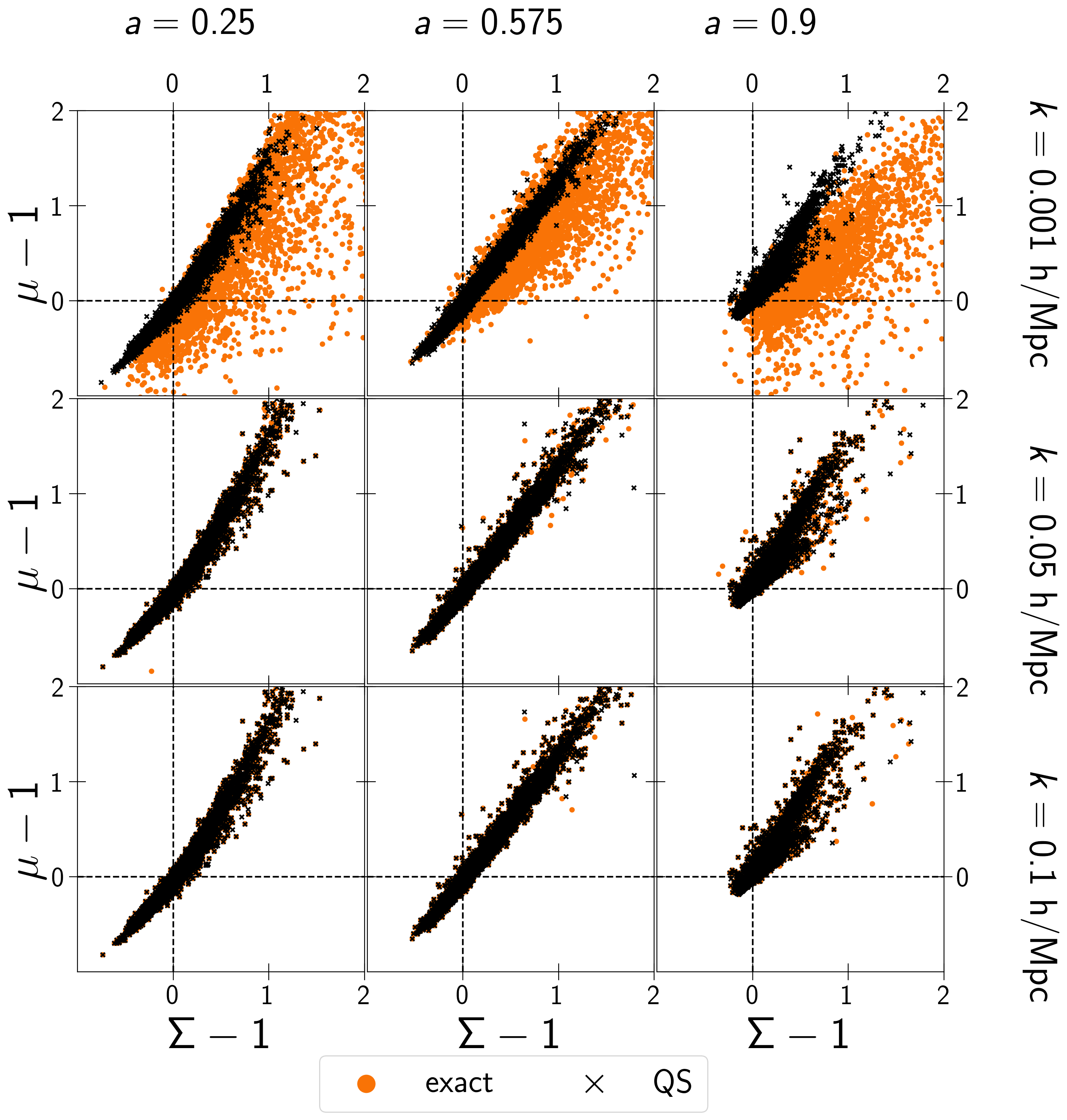}
\caption{Figure 4 in Ref. \cite{Peirone:2017ywi}. The distribution of the $\mu$ and $\Sigma$ from the Monte-Carlo sampling of viable {\it pure} Horndeski models with $c_t(z=0)=1$ as function of the scale factor $a$ and for three fixed values of $k$. The black crosses represent models obtained in the QS approximation while orange dots are models computed exploiting the full dynamics.}
\label{fig:hornlssobs}
\end{center}
\end{figure}

The QS relations of $\mu$ and $\Sigma$ discussed in Section \ref{sec:gravcouplings} allowed to deduce another important feature characterizing the Horndeski models, the so-called $\mu$-$\Sigma$ conjecture \cite{Pogosian:2016pwr}
\begin{equation}
(\mu-1)(\Sigma-1) \geqslant 0\;,
\end{equation}
which states that measurements of $\mu-1$ and $\Sigma-1$ of opposite signs at any redshift, $z$, and scale would strongly disfavor Horndeski models. A value of $\musc<1$ is predicted by self-accelerating models due to an increasing effective Planck mass. Then one does not expect to observe $\Sigma_{\rm sc}>1$. The latter is extremely unlikely since it would require a large positive $\at$ to change the trend of the effective Planck mass only in $\Sigma_{\rm sc}$. In the sub-Compton regime by comparing $\mu_\infty$ and $\Sigma_\infty$ from eqs. \eqref{eq:obsinfinity}, it is straightforward to deduce that the conjecture breaks down if the following inequalities hold \cite{Peirone:2017ywi}
\begin{align}
& 1 +\dfrac{1}{2} \left(\at +\bx+\beta_\xi \beta_B\right) < \dfrac{M^2}{\mps\left(1+\at \right)} < 1+\at+\bx \;, \\
& 1+\at+\bx < \dfrac{M^2}{\mps\left(1+\at \right)} < 1 +\dfrac{1}{2} \left(\at +\bx+\beta_\xi \beta_B\right) \;.
\end{align}
The sign agreement between $\mu_\infty-1$ and $\Sigma_\infty-1$ thus depends on the competition between the EFT functions $\lbrace M^2,\,\at \rbrace$ and the quantities $\lbrace \beta_\xi,\,\beta_B \rbrace$. This physically translates into the competition between screened and fifth force contributions. The failure of the conjecture on sub-Compton scales would then require a significant fine-tuning. The sign agreement in $\mu_\infty-1$ and $\Sigma_\infty-1$ is more obvious because they have the same pre-factor and the additional contributions from $\at$ and the fifth force are of the same order \cite{Pogosian:2016pwr}. The existence of a correlation between $\mu$ and $\Sigma$ was also found using a Monte-Carlo approach to generate a large sample of viable Horndeski model with a fixed $w$CDM background \footnote{$w$CDM is  the cosmological model where $\w=w_0$ and  $w_0$ is a free constant parameter.} and the QS approximation on sub-Compton scales to model $\mu_\infty$ and $\Sigma_\infty$ \cite{Perenon:2016blf}. This diagnostic suggests that Horndeski theories would be strongly disfavored if \cite{Perenon:2016blf}: $i)$ $\mu_\infty$ and $\Sigma_\infty$ are observed to have opposite sign for $z \gtrsim 1.5$; $ii)$ $\mu_\infty<1$ at $z=0$. A complementary analysis allowed to investigate the validity of the conjecture not only at different redshift but also at different scales as shown in Figure \ref{fig:hornlssobs} \cite{Peirone:2017ywi}. The results showed that the conjecture holds very well within the QS approximation, but the exact behaviors of $\mu$ and $\Sigma$ obtained with \eftcamb violate the conjecture at $k = 0.001$ h/Mpc, where the full dynamics of the Horndeski models allows $\Sigma > 1$ and $\mu < 1$ \cite{Peirone:2017ywi}. The validity of the conjecture at all times  is in tension with the previous work where the sign agreement is lost for $z<1.5$ \cite{Perenon:2016blf}. The reasons stem from how viable are the models considered: using only stability conditions (ghost and gradient) leads to the violation of the conjecture at small redshifts \cite{Perenon:2016blf}, while requiring additional priors based on observational constraints allows for the conjecture to be restored \cite{Pogosian:2016pwr}. In particular, Solar System bounds on the variation of the Newton's constant translate into a stringent prior on $\fg$ at present time \cite{Perenon:2019dpc}. Importantly, the breaking of the conjecture at $k = 0.001$ h/Mpc  was shown to occur despite such scale being well within the sound horizon of the scalar field, highlighting the fact that time derivatives of the scalar field and the metric can no longer be neglected. Thus the condition of validity of the QS approximation within the sound horizon of the dark field has to be seen as a necessary condition but not sufficient \cite{Peirone:2017ywi}.

\begin{figure}[t!]
\begin{center}
\includegraphics[scale=0.5]{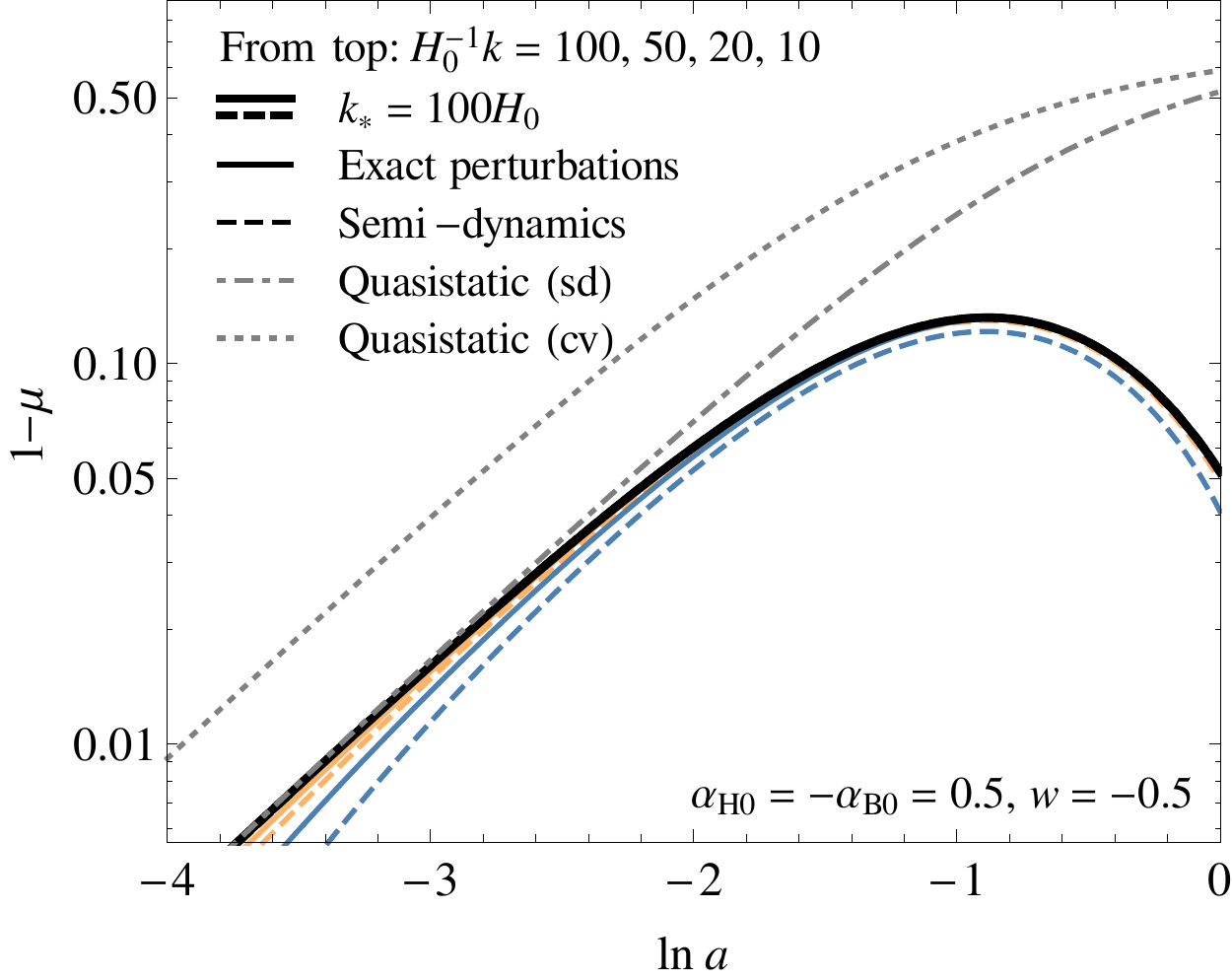}
\includegraphics[scale=0.5]{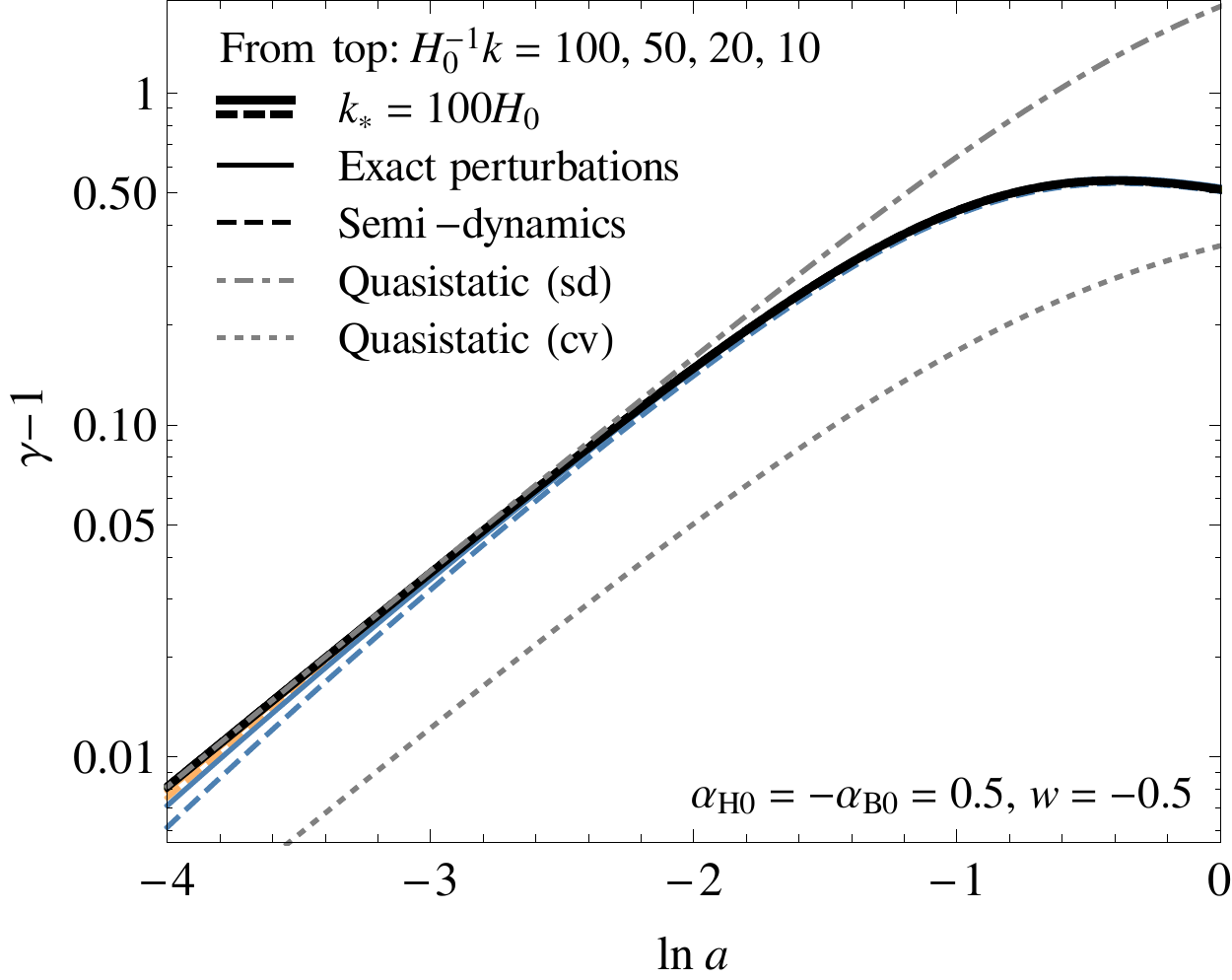}
\caption{Figure 4 in Ref. \cite{Lombriser:2015cla}. The predictions of $\mu$ and $\gamma$ ($=\eta$ in this review) in GLPV models using the semi-dynamical treatment of perturbations, the exact solution of linear perturbations, the QS approximation within the semi-dynamical approach (sd) and the QS approximation (cv).}
\label{fig:hornlssobs2}
\end{center}
\end{figure}

A way to go beyond the QS approximation at large scales while keeping analytical expressions of the phenomenological functions is to adopt a semi-dynamical treatment \cite{Lombriser:2015cla}, which amounts to consider time derivatives of the metric potentials and velocity fields on large scales \cite{Lombriser:2015cla}. This extension of the QS approximation was designed by evolving the perturbations at a given pivot scale and extrapolating the relations between the perturbations to other scales. Doing so a more precise treatment of large scales is encapsulated while the connection between theory and observables remains thus straightforward thanks to analytical expressions of $\mu$ and $\eta$. They read \cite{Lombriser:2015cla}:
\ba
\mu &=& \f{\mps}{M^2}\f{\mu_{+2}k_H^2+\mu_{+4}k_H^4+\mu_{+6}k_H^6}{\mu_{-0}+\mu_{-2}k_H^2+\mu_{-4}k_H^4+\mu_{-6}k_H^6}\,,\nn\\[1mm]
\eta &=& \f{\eta_{+0}+\eta_{+2}k_H^2+\eta_{+4}k_H^4}{\mu_{+2}+\mu_{+4}k_H^2+\mu_{+6}k_H^4}\,,
\ea
where $\mu_{\pm i}$ and $\eta_{+i}$ are expressed in terms of EFT functions and corrective terms arising from the semi-dynamical treatment.   Corrections near the Hubble scale, $k_H=k/aH$, up to order $k_H^6$ in $\mu$ and up to $k_H^4$ in the gravitational slip parameter  appear. In the small scales limit ($k\rightarrow \infty$), the semi-dynamical expressions for $\eta_\infty$ and $\mu_\infty$ for Horndeski models match naturally the ones in the QS approximation. However in the same limit, the velocity field and time derivative of the spatial metric potential in GLPV theories should not be neglected. The semi-dynamical expressions differ from the QS ones by factors proportional to $\ah$. Thus for these theories a semi-dynamical treatment should be preferred over the QS one even at small scales. The validity of this result is shown in Figure \ref{fig:hornlssobs2} \cite{Lombriser:2015cla}.

\begin{figure}[t!]
\begin{center}
\includegraphics[scale=0.28]{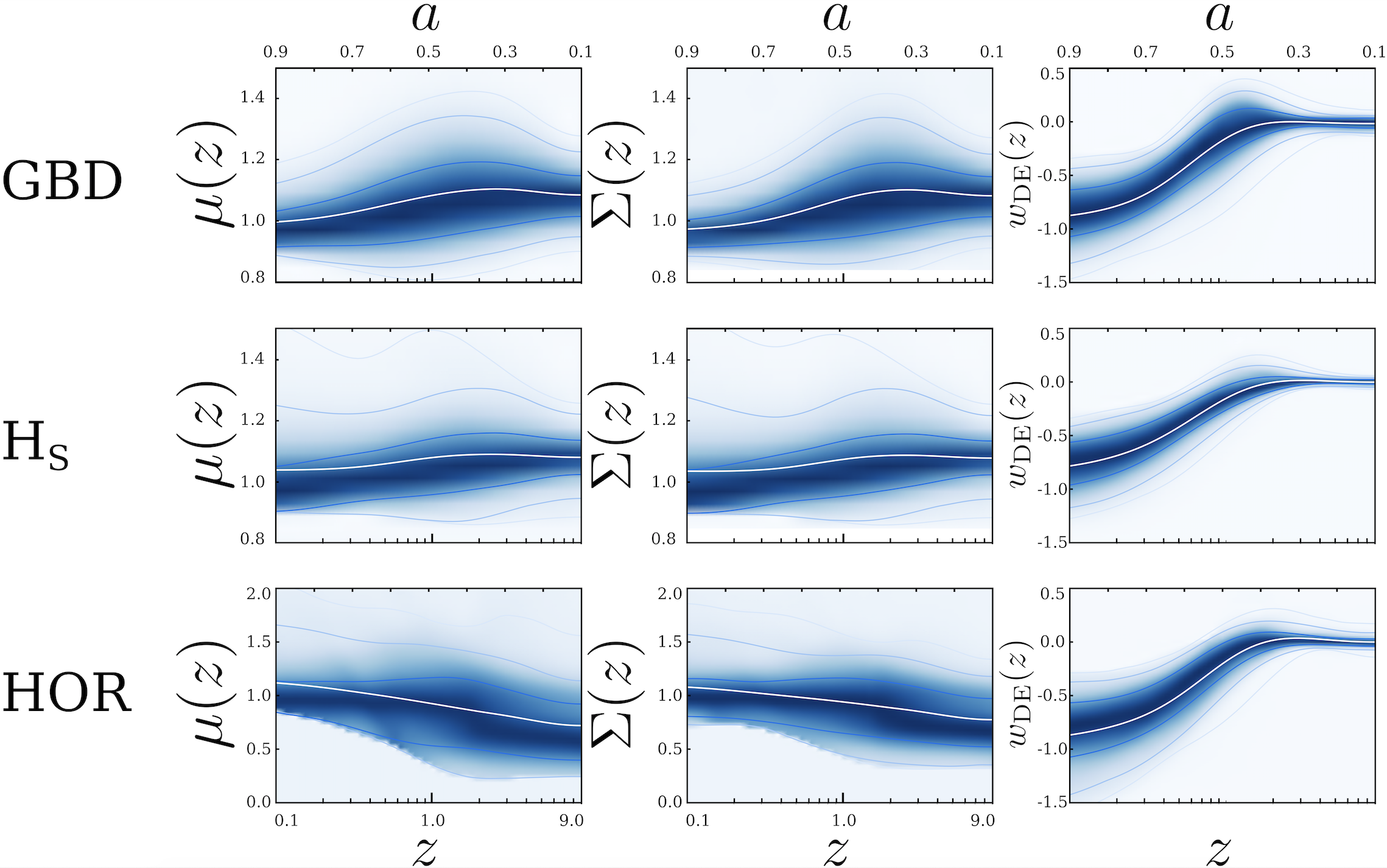}
\caption{Figure 1 in Ref. \cite{Espejo:2018hxa}. The blue density depicts the probability distribution function (at 68\%, 95\% and 99\% C.L., the white being the mean) of $\mu$, $\Sigma$ and $\w$ versus z of the Monte-Carlo sample for GBD (top row), Horndeski models with $c_t=1$ (H$_{\rm S}$, middle row) and Horndeski models with $c_t(z=0)=1$ (HOR, bottom row). }
\label{fig:hornlssobs3}
\end{center}
\end{figure}

With the perspective of large amount data from future missions, one might also tailor efficient procedures to reconstruct the time scaling of the phenomenological functions. As such, one can capture the evolution of $\mu$, $\Sigma$ and $\w$ in joint prior covariance matrices \cite{Espejo:2018hxa}. These matrices can hence be considered as priors which encapsulate the definite trends on $\w$, $\mu$ and $\Sigma$ of Horndeski theories and sub-classes that are shown in Figure \ref{fig:hornlssobs3}. The Monte-Carlo procedure highlighted that for instance these models can display $\mu(z=0)<1$ as soon as the normalization of the effective Planck mass today is left free. In parallel, this analysis  shows quantitatively how the redshift evolution of $\mu$ and $\Sigma$ is strongly correlated in scalar-tensor theories, while their correlation with $\w$ is progressively washed out when going beyond GBD models. The correlations to use in future model reconstructions can be effectively mapped into the CPZ parameterization \cite{Crittenden:2005wj} where for $\mu$ and $\Sigma$ they are shown to scale as $|a - a^\prime|$ and for $\w$ as $|\mbox{ln}a - \mbox{ln}a^\prime|$, where $a$ and $a'$ are two bins in the scale factor \cite{Espejo:2018hxa}. 

%----------------------------------------------------------------
\subsection{Impact of stability conditions}\label{sec:impactstab}
%----------------------------------------------------------------

\begin{figure}[!]
\begin{center}
\includegraphics[trim = 2.75cm 6.5cm 2.75cm 6.5cm, scale=0.25]{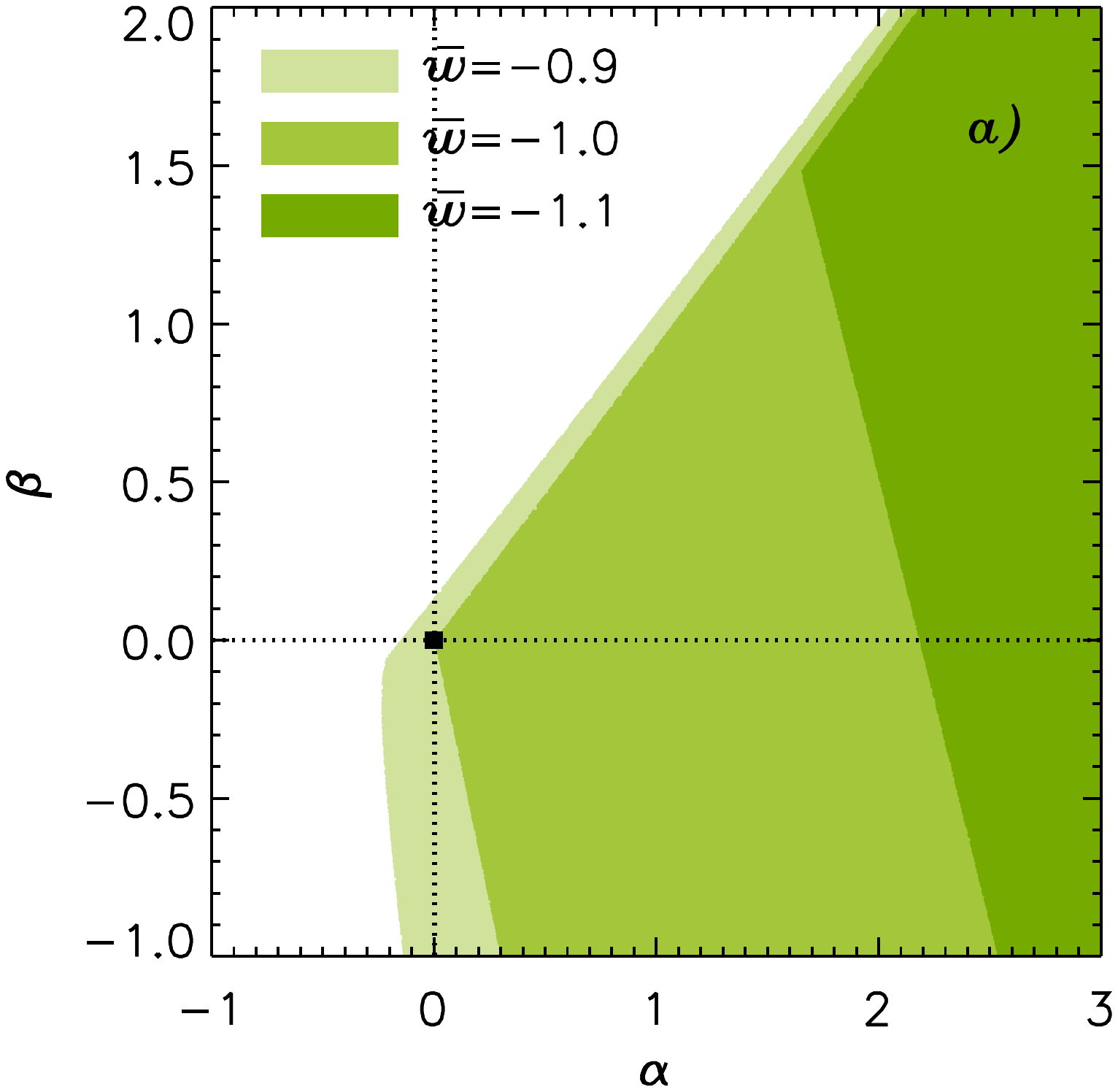}
\includegraphics[trim = 2.75cm 6.5cm 2.75cm 6.5cm, scale=0.25]{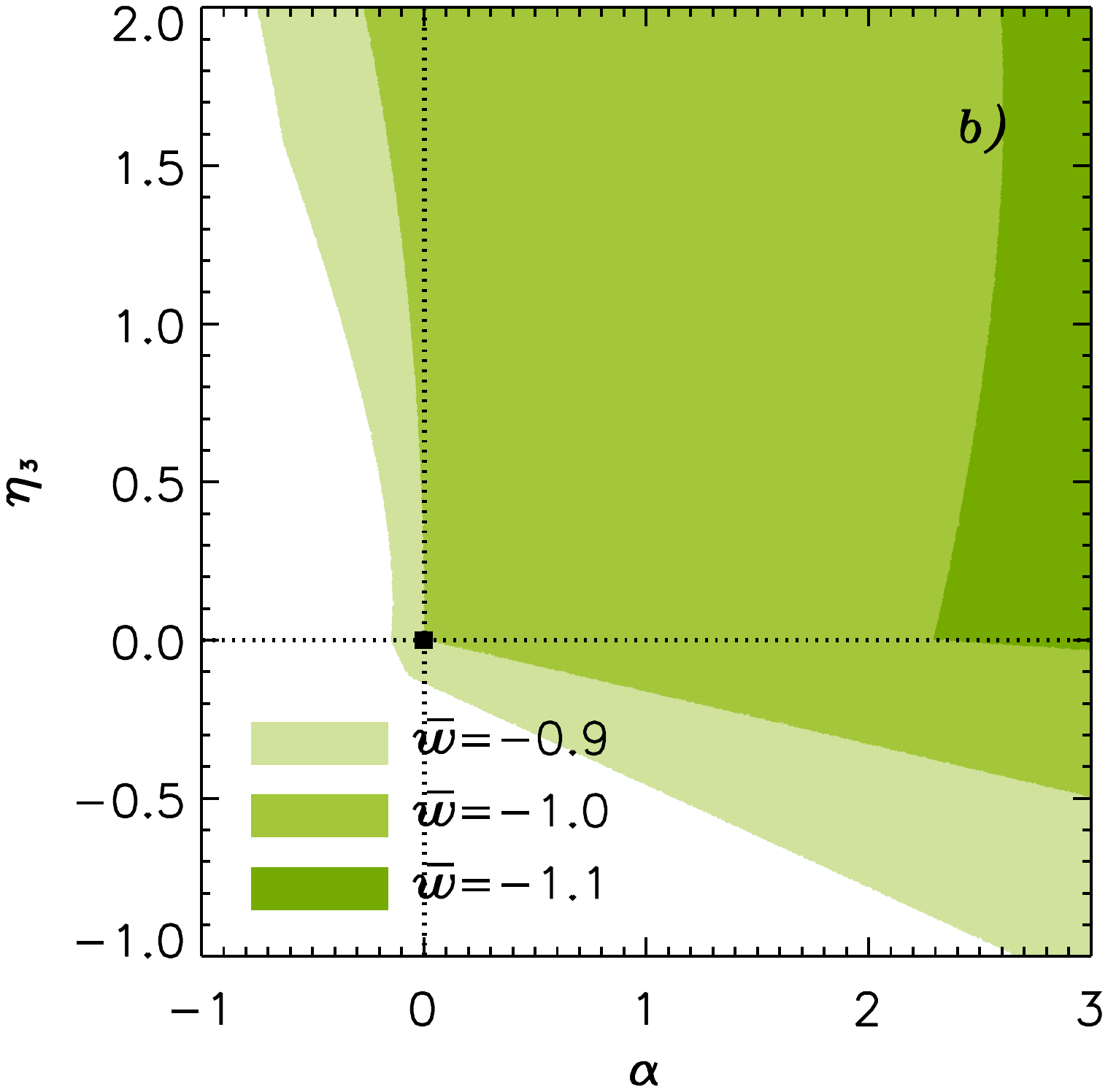}\vskip-2mm
\includegraphics[trim = 2.75cm 6.5cm 2.75cm 6.5cm, scale=0.25]{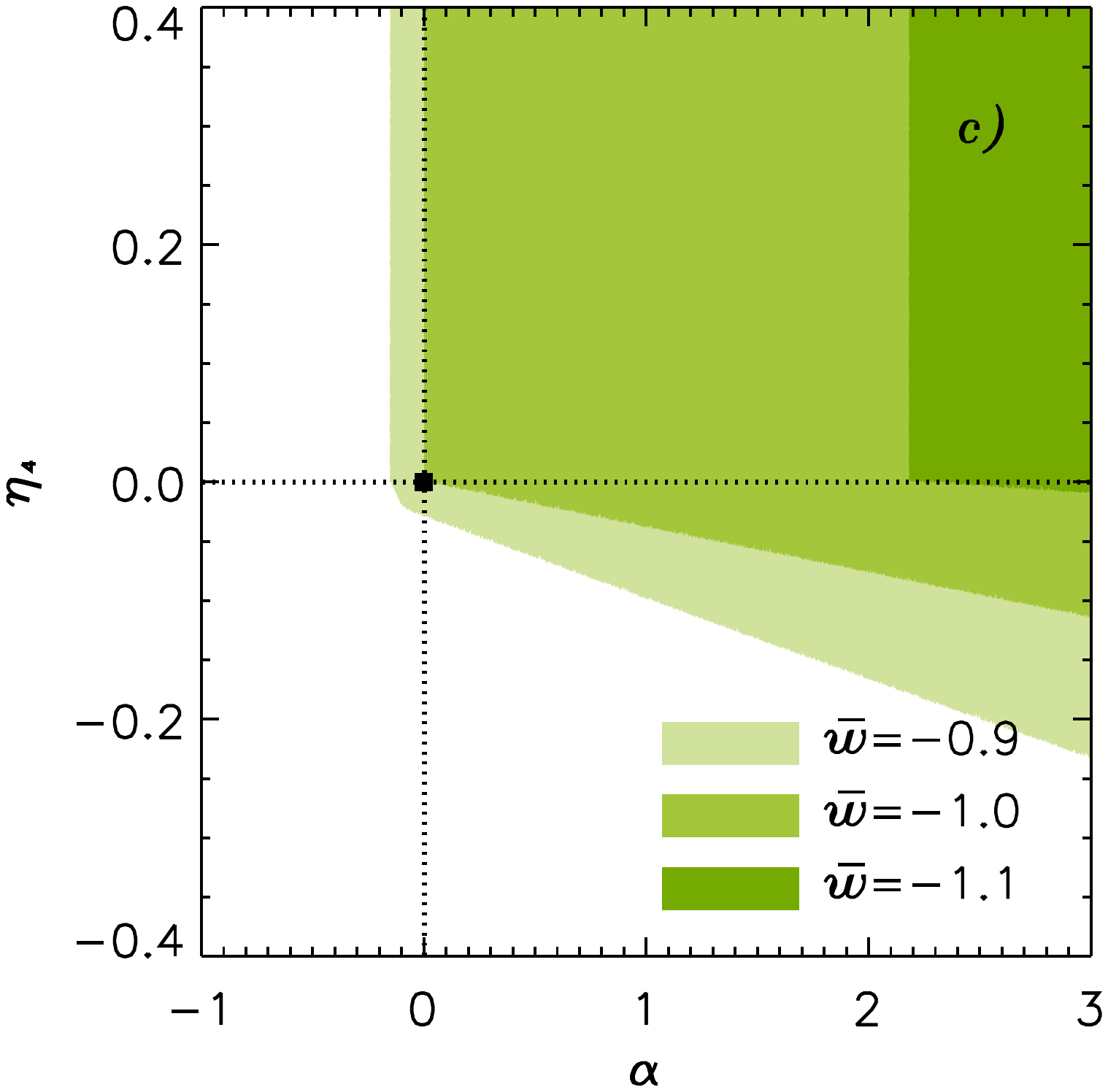}
\includegraphics[trim = 2.75cm 6.5cm 2.75cm 6.5cm, scale=0.25]{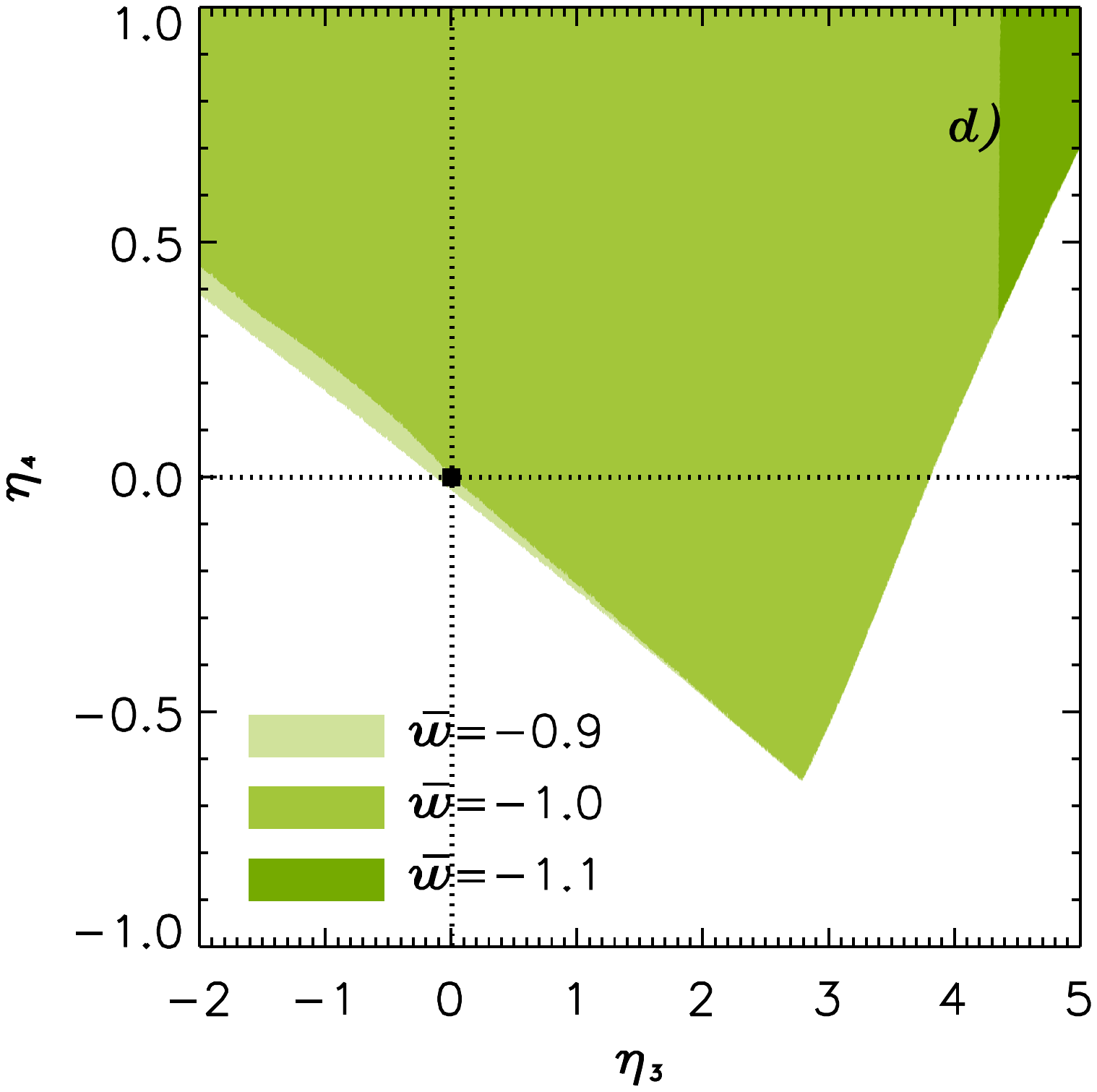}
\vskip4mm
\includegraphics[trim = 2.75cm 6.5cm 2.75cm 6.5cm, scale=0.25]{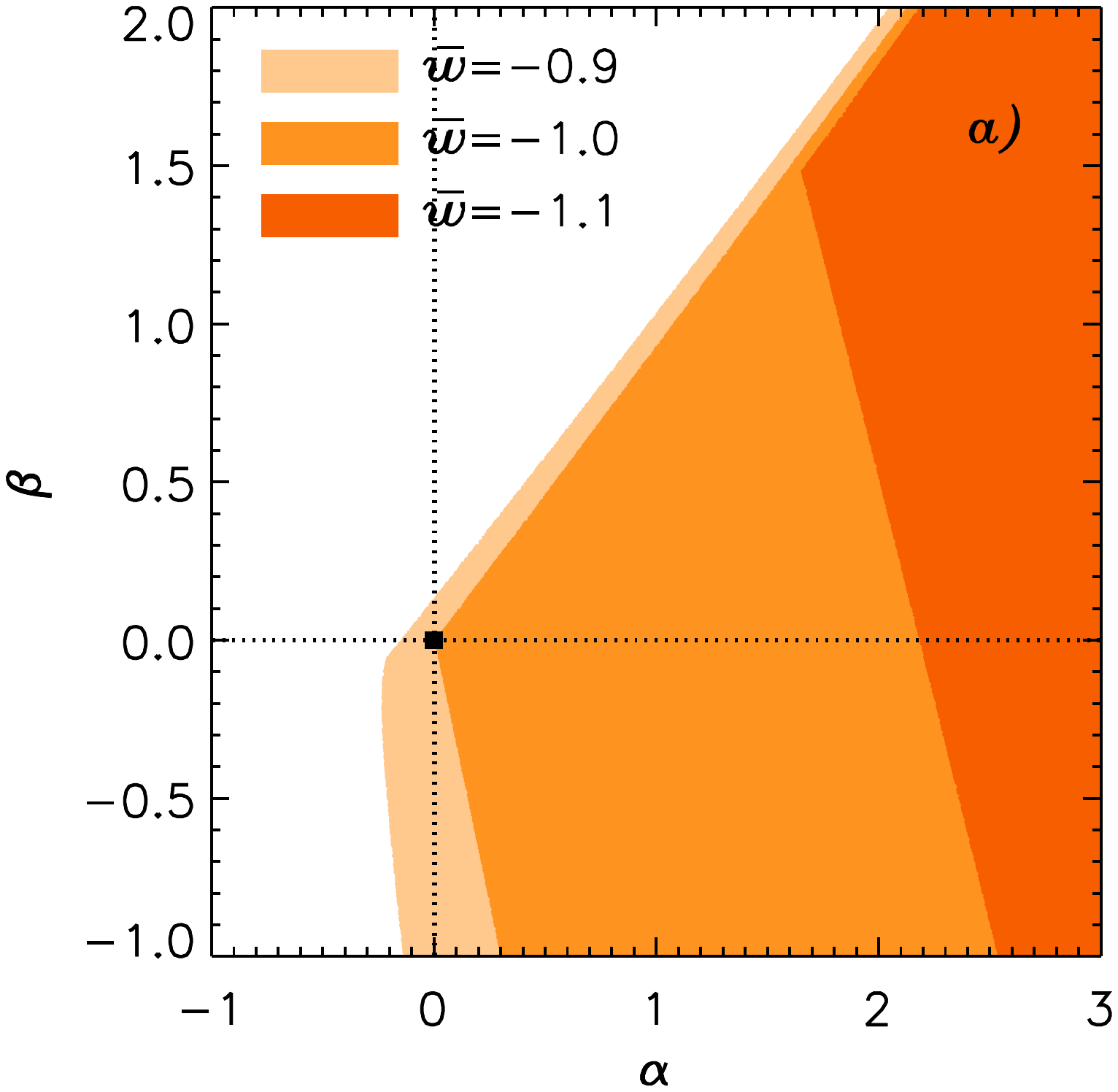} 
\includegraphics[trim = 2.75cm 6.5cm 2.75cm 6.5cm, scale=0.25]{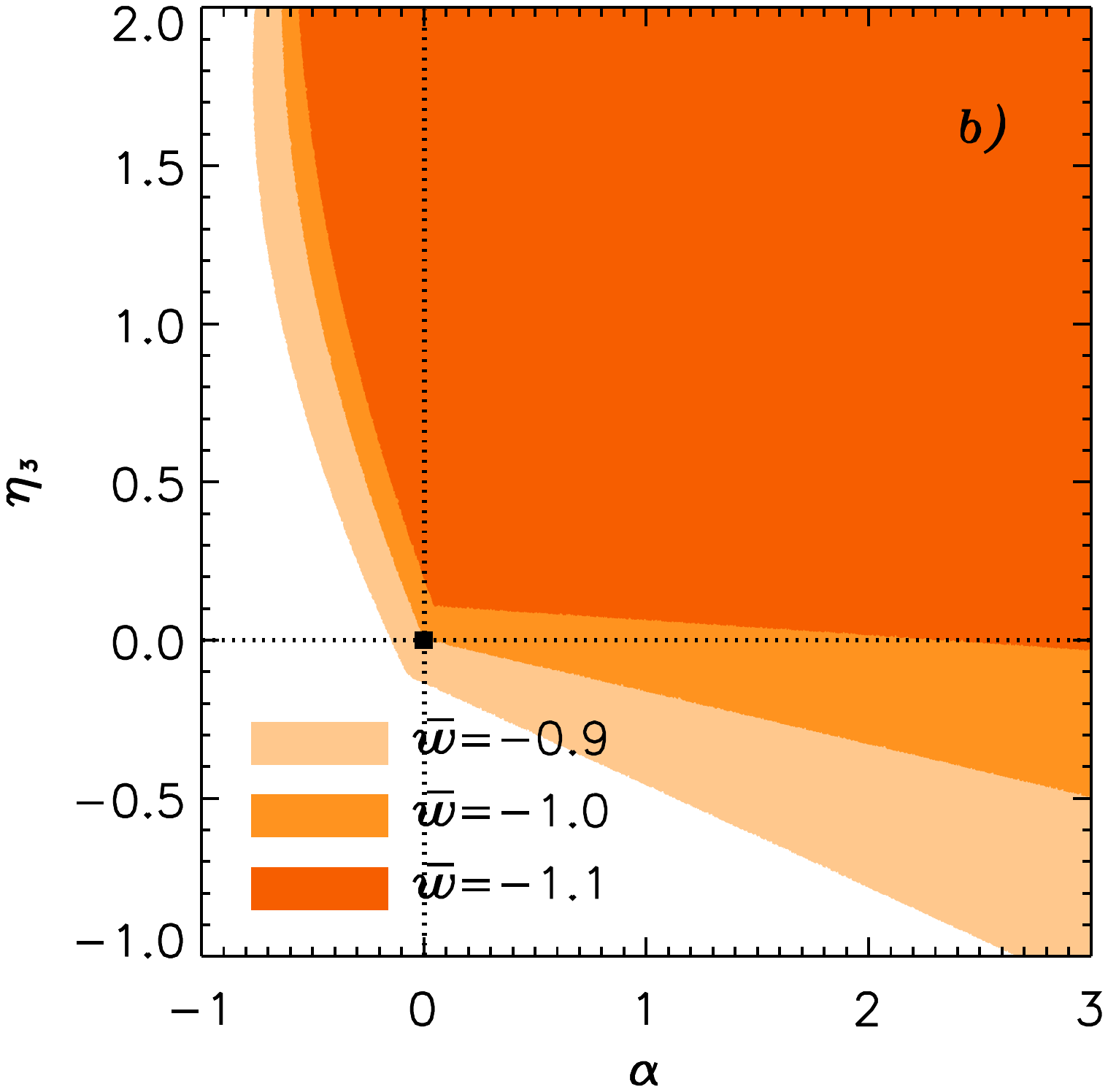}\vskip-2mm
\includegraphics[trim = 2.75cm 6.5cm 2.75cm 6.5cm, scale=0.25]{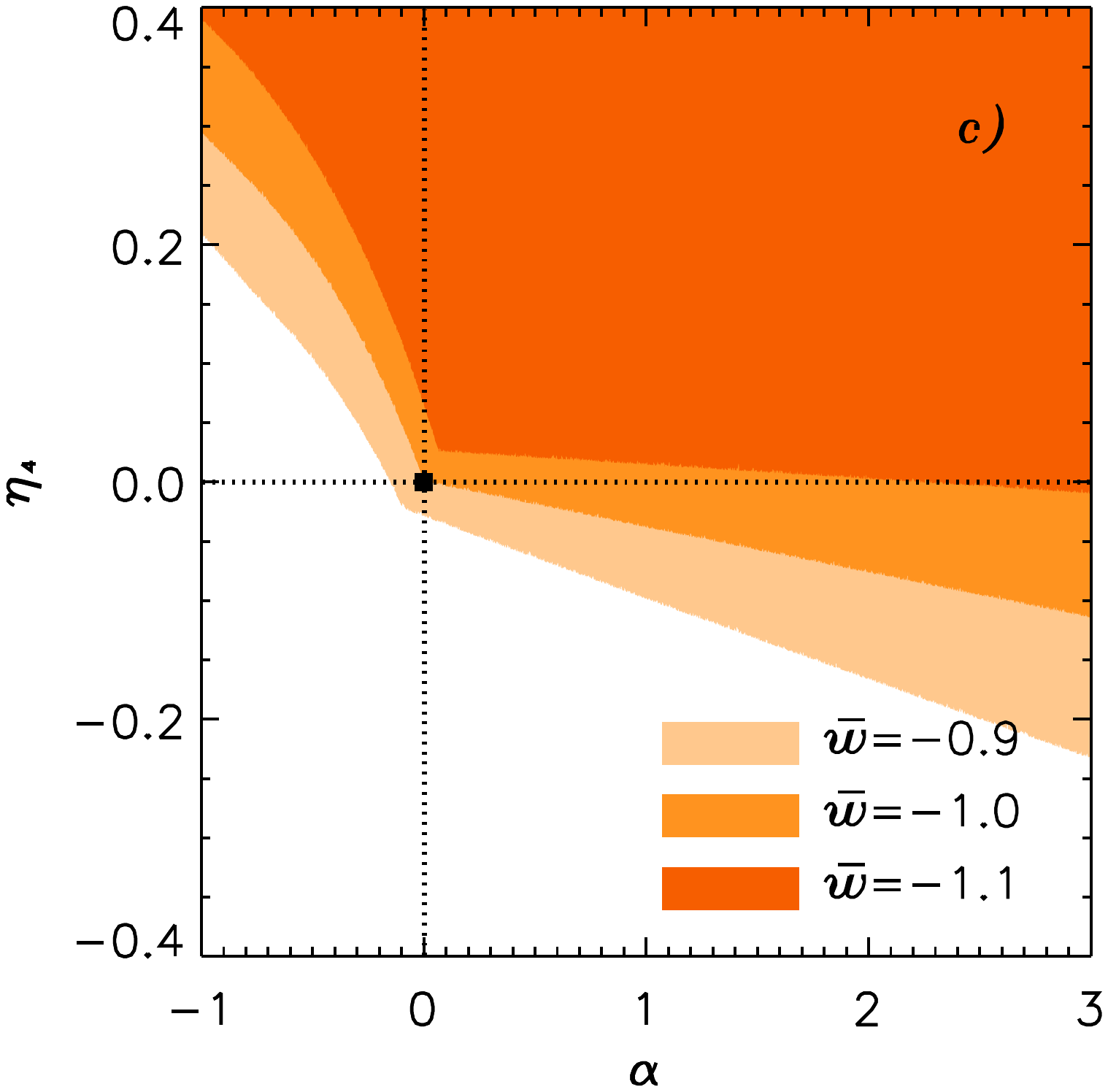}
\includegraphics[trim = 2.75cm 6.5cm 2.75cm 6.5cm, scale=0.25]{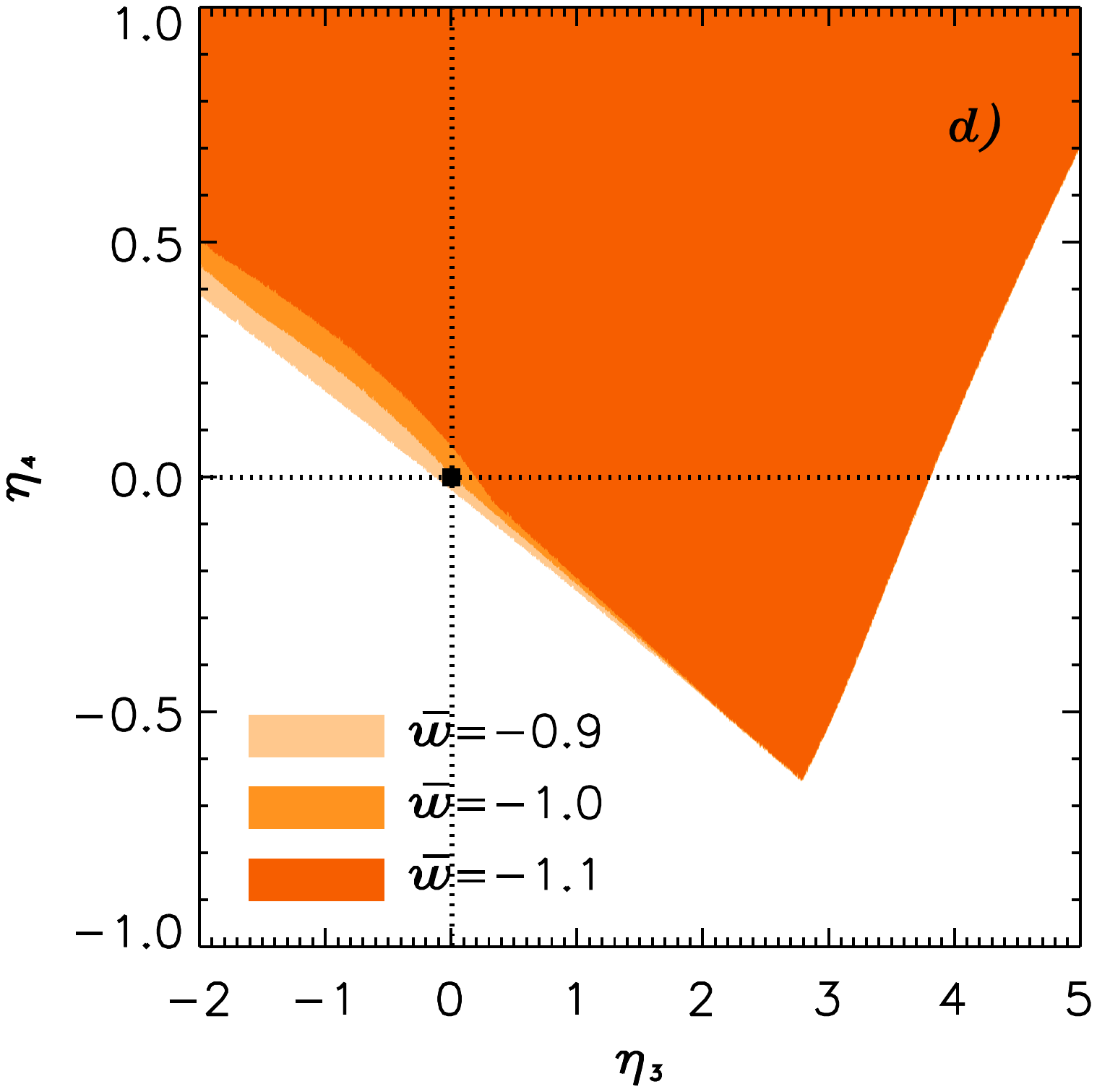}
\caption{Figure 1 and 2 in Ref. \cite{Piazza:2013pua}. The stability regions of {\it pure} Horndeski models for different choices of $\bar{w}$ ($=\w$ in this review). The results are obtained using the basis in Ref. \cite{Piazza:2013pua} and the 6-parameter form (eq. \eqref{6form}) where $\alpha$, $\beta$ are the free parameters characterizing the variation of the effective Planck mass, $\eta_2$ the kineticiy, $\eta_3$ the cubic Galileon coupling and $\eta_4$ the speed of GWs. The \emph{Top 4 panels} in green display the results for $\eta_2=0$, hence $M_2^4=0$, and \emph{bottom 4 panels} in orange for $\eta_2=10^6$.}
\label{fig:stabregion1}
\end{center}
\end{figure}

In the EFT formulation, the conditions which guarantee the stability of a  theory are expressed in terms of the EFT functions. This allows to gauge straightforwardly whether a gravitational model is physically viable. We have discussed in Section \ref{sec:eftofdestab} the required physical stability conditions a model has to satisfy, namely, the avoidance of ghosts, gradient and tachyonic instabilities. Here, we review the implications of such conditions on the parameter space of models. 

From this stability perspective, the ``naturalness'' of the $\Lambda$CDM scenario among Horndeski theories can be addressed \cite{Piazza:2013pua}. In this case the Horndeski class of models is parametrized within the EFT framework using the {\it pure} EFT approach. We refer to them as {\it pure} Horndeski models hereafter. As displayed in Figure \ref{fig:stabregion1}, the ghosts and gradient stability conditions shrink the allowed space of EFT parameters up to the point where $\Lambda$CDM corresponds to a corner of the stable region. This happens quite naturally since $\Lambda$CDM corresponds to a non-propagating extra DoF which follows from a vanishing kinetic term and speed of propagation. In other words, the realm of Horndeski theories with an acceleration of the background expansion compatible with current observational constraints has $\Lambda$CDM at its border. Interestingly, this analysis also revealed the parameter space location of models which violate the null energy condition in a stable manner, $i.e.$ ``super-accelerating'' (or ``phantom'') models with $\w<-1$. The volume of the parameter space occupied by the stable {\it pure} Horndeski models gradually shrinks as the background equation of state decreases (see Figure \ref{fig:stabregion1}). In particular, the $\Lambda$CDM model does not fall within the set of phantom models, highlighting the fact that at least one non-minimal coupling must be invoked for models to violate the null energy condition in a stable way. In parallel, there is one EFT function which does not play a role in the expression of $\mu$ nor $\gsp$ when the QS approximation is considered: $ M_2^4$ or equivalently $\alpha_K$. This coupling acts solely on the ghost stability condition and regulates hence the speed of sound of the scalar perturbations. Figure \ref{fig:stabregion1} shows also how large values of this coupling open up the space of stable models and thereby helps to oppose the shrinking effect due to considering $\w<-1$. This feature allowed to identify a large set of viable null energy condition violating theories.

\begin{figure}[!]
\begin{center}
\includegraphics[clip, trim = 0cm 0cm 7.2cm 0cm, scale=1]{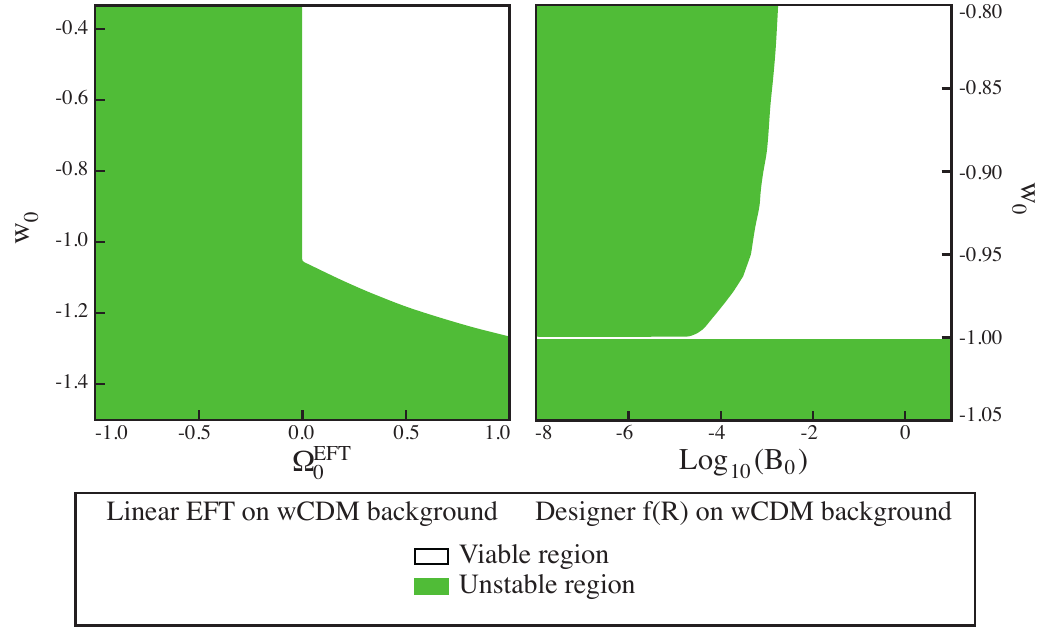}
\vskip6mm
\includegraphics[clip, trim = 0cm 0cm 0cm 7.9cm, scale=0.8]{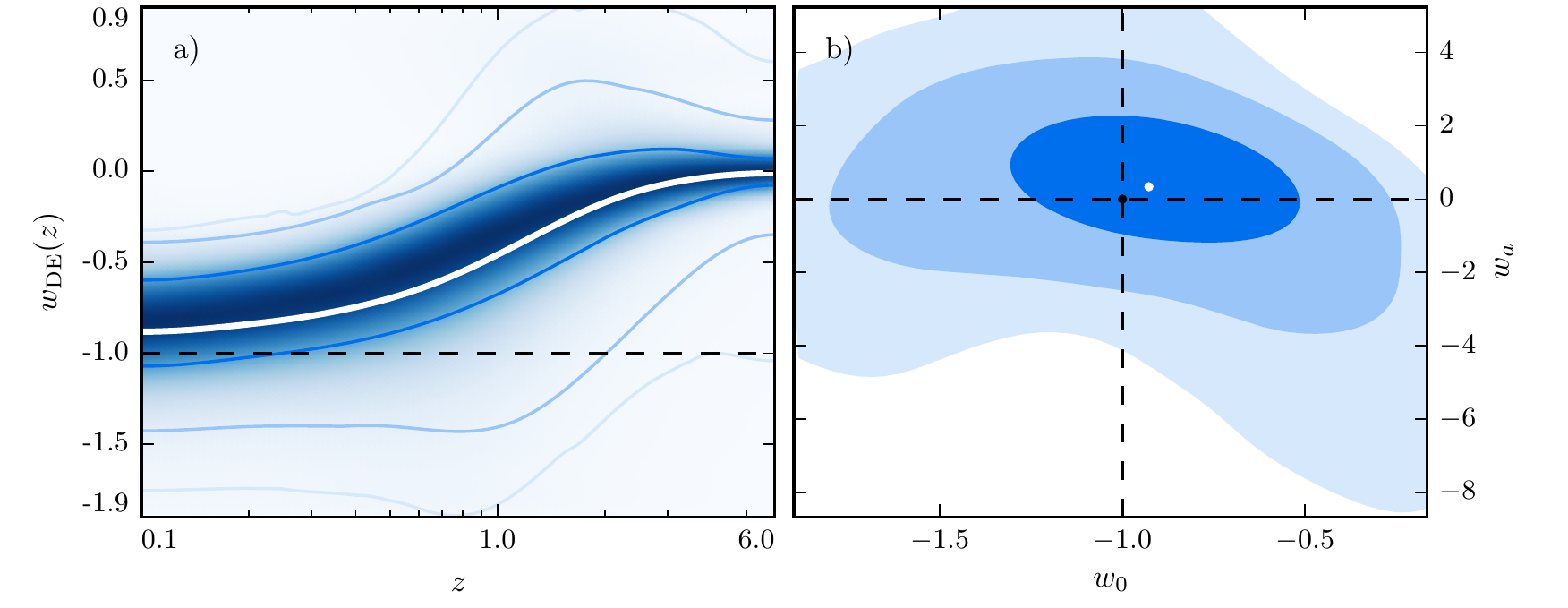}
\caption{\emph{Top panel}: Figure 1 in Ref. \cite{Raveri:2014cka}. The stability regions in the linear EFT and $f(R)$ {\it designer} models. $w_0$ corresponds to the constant value of $\w$. \emph{Bottom panel}: Figure 2 in Ref. \cite{Raveri:2017qvt}. The probability density, indicated by the shades of blue, of $\w(z)$ (a) and its projection in the $w_0$-$w_a$ plane of the CPL parameterization (b) are shown. The white lines/points correspond to the mean, the blue lines/contours to the 68\%, 95\%, 99\% C.L. intervals and the dotted lines to the $\Lambda$CDM predictions.}
\label{fig:stabregion2}
\end{center}
\end{figure}

The impacts of ghost and gradient conditions on GBD class of models were investigated using the \textit{pure} EFT approach with an effective Planck mass defined by the so-called ``linear EFT model'' \cite{Raveri:2014cka}:
\begin{equation}\label{eq:lineft}
\fg (t)= \frac{1}{2}\left(1+ \Omega_0^\mathrm{EFT}\,a(t)\right) \,,
\end{equation}
where $\Omega_0^\mathrm{EFT}$ is a constant.
As a consequence, all MG effects are captured by the parameter $\Omega_0^\mathrm{EFT}$ and the constant DE equation of state parameter $\w$ defining a $w$CDM expansion history. This analysis was the first to highlight how the stability conditions act as a strong prior on the constraints of cosmological parameters. In the top left panel of Figure \ref{fig:stabregion2}  one can notice for instance that a stable linear EFT model implies $\Omega_0^\mathrm{EFT}>0$. Once the effective equation of state parameter becomes smaller than $-1$, the likelihood contours produced by a statistical analysis would therefore be significantly tightened. This result independently confirms the shrinking theory space effect in the super accelerating regime discussed previously \cite{Piazza:2013pua}. 

Another specific model with a running Planck mass is $f(R)$-theory. In this case, one can study the impact of stability conditions on the parameters of the model using a \textit{designer} approach \cite{Pogosian:2007sw}. The resulting family of models are parametrized by the present value of the Compton wavelength of the \textit{scalaron}, $f_R=df/dR$, \cite{Starobinsky:2007hu}, defined as \cite{Song:2006ej}
\begin{equation}
\label{ComptonWave}
B=\frac{f_{RR}}{1+f_R}\frac{H \dot{R}}{\dot{H}}\,.
\end{equation} 
As shown in the top right panel of Figure \ref{fig:stabregion2}, the stability conditions which include the no-ghost, no-gradient and a positive mass ($m^2=f_{RR}>0$) induce a stringent cut of the parameters space \cite{Raveri:2014cka}. One can note that as opposed to the linear EFT model, the {\it designer} $f(R)$ model does not allow stable regions beyond $\w<-1$ as expected theoretically, because this amounts to a minimally coupled scalar field for which violating the null energy condition can only be at the expense of it being a ghost. These viable parameter spaces have been used to put stringent constraints on the parameters of both linear EFT model and $f(R)$-theory showing in some cases to have a higher constraining power than that of data \cite{Raveri:2014cka}. We will further discuss this point in Section \ref{sec:constraints}.

\begin{figure}[!]
\begin{center}
\includegraphics[scale=0.48]{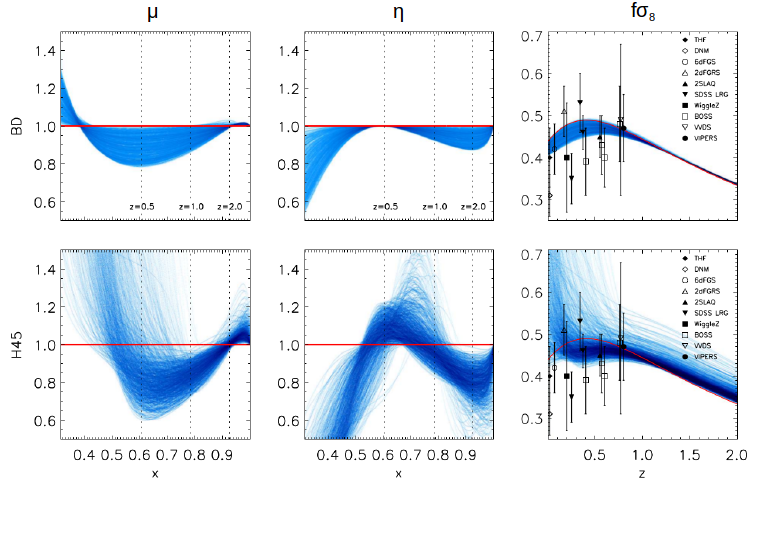}\vskip-2mm
\caption{Figures in Ref. \cite{Perenon:2015sla}. Predictions of $\mu$, $\eta$ ($\mu_{\infty}$, $\eta_{\infty}$ in this review) as function of the reduced background matter density $x=\om$ and $\fs$ versus $z$ for viable {\it pure} Horndeski (H45) and Generalised Brans-Dicke (BD) models obtained using a Monte-Carlo approach. The shades of blue define the density of curves and the red line correspond to $\Lambda$CDM.}
 \label{fig:muetatrend}
\end{center}
\end{figure}

The selection of the viable parameter space induced by the stability conditions has non trivial implications for  EFT  predictions.  We already discussed in Section \ref{sec:phenomenology} the $\mu-\Sigma$ conjecture \cite{Pogosian:2016pwr} and its implications \cite{Perenon:2016blf,Peirone:2017ywi}. Other general behaviors can be identified using a Monte-Carlo approach to generate a vast sample of healthy {\it pure} Horndeski models and explore the phenomenology of $\mu$, $\gsp$, $\Sigma$ and $\fs$ across $z$.  $\fs$ is the so-called growth function, $i.e.$ the product of $\sigma_8$ and the linear growth rate of matter density perturbations $f$ (see eq. \eqref{eq:linf}). For instance, definite features can be found for models exhibiting sub-luminal propagation of scalar and tensor perturbations and DE contributes only at low redshifts \cite{Perenon:2015sla}. As shown in Figure \ref{fig:muetatrend}, the effective gravitational coupling displays a characteristic $S-shape$ pattern, a regular alternating succession of epochs where gravity is stronger and weaker than predicted by GR \cite{Perenon:2015sla}. This behavior significantly affects the growth function $\fs$. The amplitude of this observable is generically suppressed, compared to the value expected in $\Lambda$CDM models, at intermediate redshifts ($0.5 \lesssim z \lesssim 1$), the opposite being true at all other cosmic epochs, when a phenomenon of super growth is highlighted. The gravitational slip parameter $\gsp$ is, instead, predicted to be bounded from above. Notably, for GBD theories, the gravitational slip parameter is at most unity and a larger deviation from GR is found only at intermediate redshifts for the Horndeski case. 

\begin{figure}[!]
\begin{center}
\includegraphics[scale=0.17]{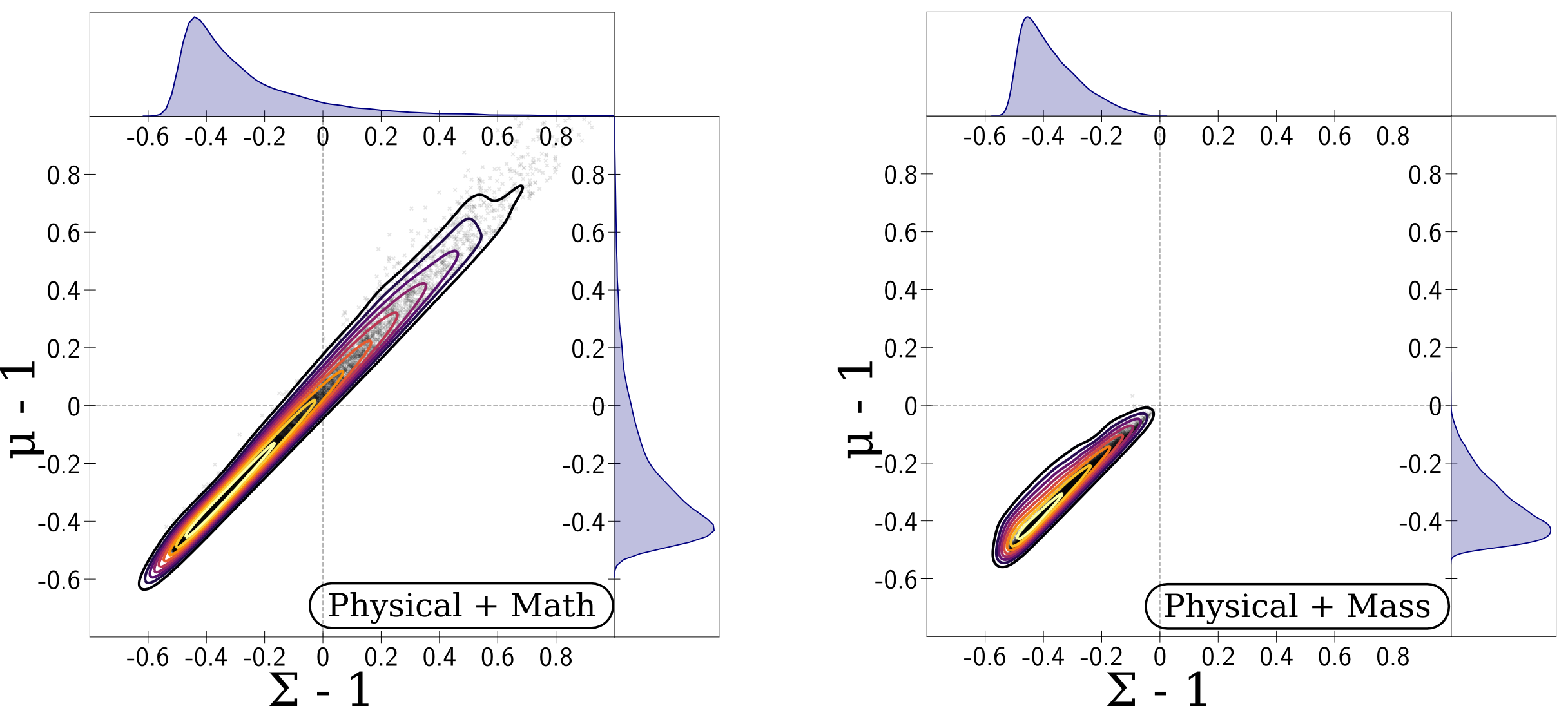}
\vskip5mm
\includegraphics[scale=0.17]{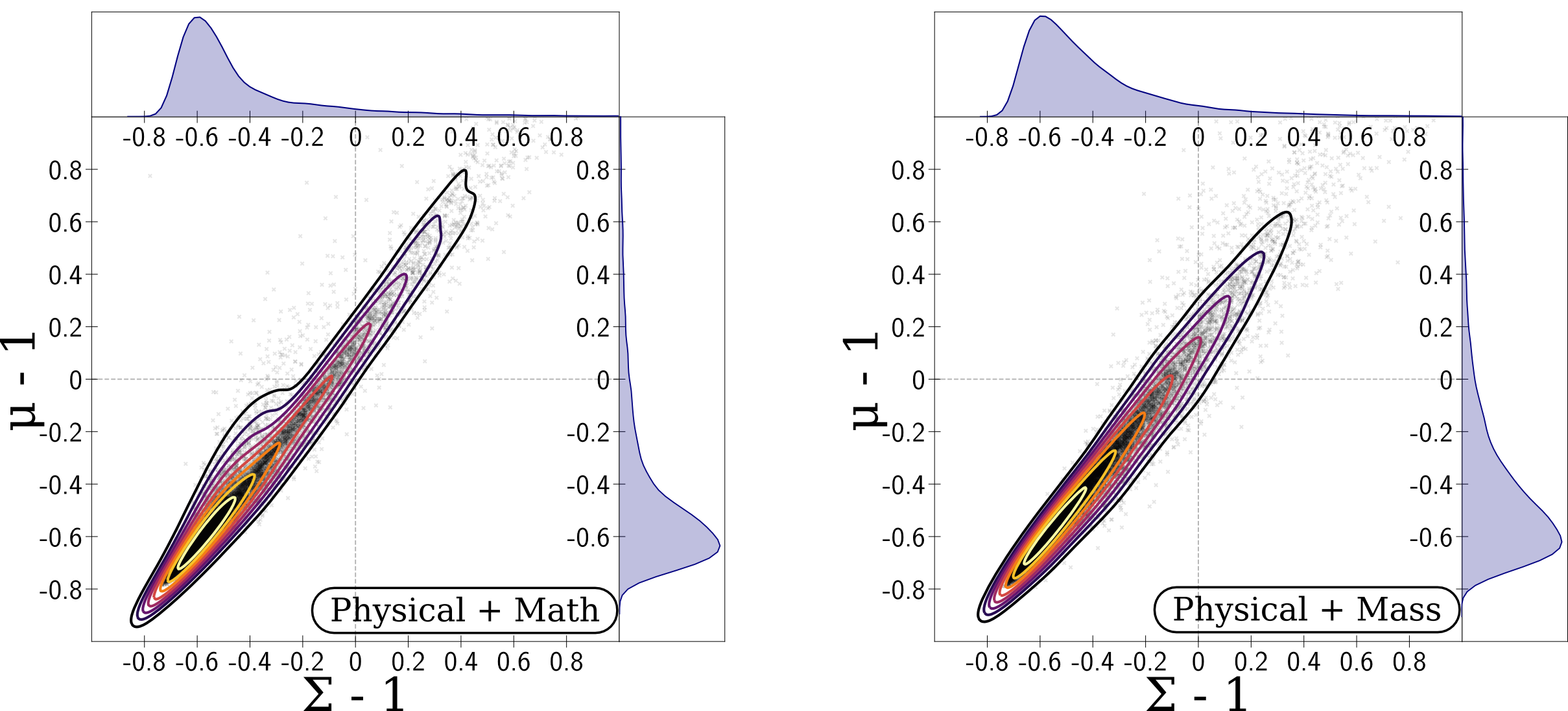}
\caption{Figures 3 and 4 in Ref. \cite{Frusciante:2018vht}. Marginalized 2D and 1D distributions of $\mu$ and $\Sigma$ at $z=0.1$ for GBD (\emph{top doublet}) and Horndeski models (\emph{bottom doublet}) generated in a Monte-Carlo fashion for several combinations of the stability conditions.}
\label{fig:tachyon}
\end{center}
\end{figure}

Historically only the no-ghost and no-gradient conditions are employed when exploring the stability of a gravity theory in literature. As discussed in Section \ref{sec:eftofdestab}, such requirements are high-$k$ dependent statements thus they cannot guarantee stability on the whole range of cosmic scales. In EB codes, to prevent exponentially growing modes at low-$k$, ad-hoc mathematical conditions are implemented to complete the set of physical conditions. In the EFT formulation, these additional requirements are worked out at the level of the perturbation equation for the scalar field $\pi$ \cite{Hu:2013twa,Hu:2014oga}. The stability module of \eftcamb  includes the tachyon conditions as a function of the mass eigenvalues $\mu_i$, $i.e.$ the mass condition \cite{Frusciante:2018vht}. The evaluation of the impacts of switching on this additional condition  reveals the mass condition to be very efficient for substituting the mathematical condition in practically all cases of cosmological interest \cite{Frusciante:2018vht}. We display the results obtained thanks to the Monte-Carlo generation of healthy {\it pure} Horndeski and GBD models in the $\mu-\Sigma$ plane at $z \approx 0.1$ in Figure \ref{fig:tachyon} \footnote{Note that in this investigation the effective Planck mass today was not normalized to $\mps$.}. One can notably observe that as opposed to the general case of {\it pure} Horndeski, the action of the mass condition becomes very peculiar in the {\it pure} GBD case and cuts out models from the $ \Sigma\,,\,\mu>1$ region at $z \approx 0.1$. The tail of models that instead are present when only the no-ghost, gradient and mathematical stability were active are thus singled out. This cut translates directly in a lower bound of the effective Planck mass which has to be positive at all redshifts. From these investigations, the mass requirements together with the no-ghost and no-gradient conditions prove a reliable and complete set of robust and physically motivated theoretical priors that guarantees the soundness of the theories on all linear scales. 

A common thread linking a lot of the analyses presented throughout this review is the processing of the background expansion rate as a fixed prior when studying the effects of MG theories in the perturbed sector. As anticipated in Section \ref{sec:eftfried}, a complementary approach can be investigated. Using the free background approach, where $H(t)$ is computed from the time evolving $\w$ expressed in terms of the background EFT functions $\Lambda$ and $c$, reveals that universal behaviors could also be found on quantities characterizing the background expansion \cite{Raveri:2017qvt}. The Monte-Carlo procedure employed shows that the stability conditions favor the emergence of a tracking behavior in the scaling of $\w$ of Horndeski models with the value of this function becoming close to zero deep in matter domination while approaching the value $\w=-1$ at present time (see Figure \ref{fig:stabregion2}). This renders the possibility to map this evolution onto the Chevallier-Polarski-Linder (CPL) parameterization \cite{Chevallier:2000qy,Linder:2002et}, $i.e.$ $\w=w_0+w_a(1-a)$ where $w_0$ and $w_a$ are constants. The same exploration technique for quintessence models reveals a tracker behavior with however $\w<-1$ excluded by stability, as expected theoretically. The definite features of the background evolution are translated into a theoretical prior covariance matrix for $\w$, correlating its values at different redshifts to simplify the reconstructions of $\w$ from future data. This has been recently used also for the reconstruction of $\w$ through Gaussian process in quintessence and Horndeski theories \cite{Gerardi:2019obr}.

Stability conditions are usually used on top of a chosen parameterization for the EFT functions. However it is possible to construct a {\it stable EFT basis} which evades ghost and gradient instabilities or accommodates further theoretical priors such as a luminal or sub-luminal scalar propagation speed \cite{Kennedy:2017sof,Kennedy:2018gtx,Lombriser:2018olq}. Such basis instead of using the usual EFT functions (or $\alpha$-basis) employs directly the stability functions $\alpha, M^2, c_s^2$ with a boundary condition $\alpha_{B0}=const.$. An advantage of this basis is that it avoids $\Lambda$CDM to be confined to a narrow corner of the stability space. This corner would be difficult to sample and could lead to false evidence against the $\Lambda$CDM model. The enforcement of theoretical stability conditions when performing cosmological constraints is definitively an advantage as they ensure the resulting observational bounds lay in a physical parameter space. This is now common practice and most of the available codes have built-in modules which automatically check for the stability of the considered model. A further benefit in using the {\it stable EFT basis} would be secured when running MCMC algorithms: Markov chain would no longer randomly explore unstable regions of the parameter space making the whole process much faster. We note however  that the {\it stable EFT basis} does not include the tachyonic condition discussed in Section \ref{sec:eftofdestab},  thus the stability of the underlying theory of gravity might not be guaranteed at all scales.

%-------------------------------------------------------------
\subsection{Linear growth rate of matter density fluctuations}\label{Sec:LGR}
%-------------------------------------------------------------

\begin{figure}[!]
\begin{center}
\includegraphics[scale=0.4]{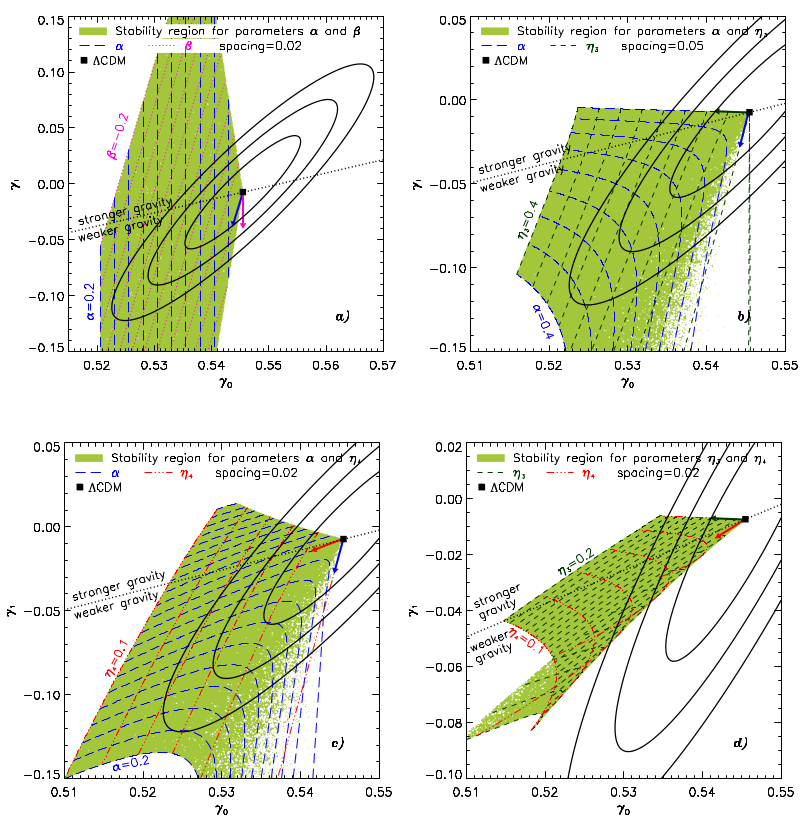}
\caption{Figure 3 in Ref. \cite{Piazza:2013pua}. Stable regions of a {\it pure} Horndeski model in the $\gamma_0-\gamma_1$ plane. The dotted line corresponds to the boundary between stronger and weaker gravity predictions with respect to $\Lambda$CDM. The 68\%, 95\% and 99\% likelihood contours correspond to the constraints for a Euclid-like survey.}
 \label{fig:gammasteig}
\end{center}
\end{figure}

The EFT formulation contributed to understand how linear structures are likely to grow in scalar-tensor theories of gravity and what theoretical patterns a viable theory should have to better accommodate cosmological observations. RSD, GC and WL independent  measurements all consistently detected a lower growth rate of matter density perturbations than predicted by the  $\Lambda$CDM model. This is at variance with results extrapolated from CMB studies~\cite{Hildebrandt:2016iqg,deJong:2015wca,Kuijken:2015vca,Conti:2016gav,Abbott:2017wau,Abbott:2017smn,Joudaki:2019pmv}. Scalar-tensor theories are not expected to produce lower growth relative to $\Lambda$CDM because of the fifth force they exhibit. The EFT has contributed to showing that this is not necessarily the case \cite{Piazza:2013pua,Tsujikawa:2015mga,Perenon:2015sla}. 

The effective gravitational coupling modifying the Poisson equation \eqref{muSigma} alters the growth of matter perturbations $\delta_m$ as follows \footnote{Let us note that in GLPV models, an additional modification proportional to $\alpha_H$ is present in the friction term in eq. \eqref{eq:evoldeltam}. This is due to the fact that terms involving matter velocity cannot be neglected on sub-horizon scales, see for instance \cite{Traykova:2019oyx}.}
\begin{equation}\label{eq:evoldeltam}
\ddot{\delta}_{\rm m} +2H \dot{\delta}_{\rm m} -4\pi \gn \mu \rho_{\rm m} \delta_{\rm m} =0\;.
\end{equation}
The linear growth rate $f$ is then defined by \cite{1980lssu.book.....P}
\begin{equation}\label{eq:linf}
f=\frac{d \ln \delta_m}{d \ln a}\;.
\end{equation}

In the case of Horndeski models, using the sub-Compton limit one can deduce from eq. \eqref{effectivemu} that in order to have lower growth the condition $\mu_\infty<1$ has to be satisfied. One necessary condition is to require $\musc<1$ which originates only from the tensor part, in particular it depends on the ratio between $c_t^2/M^2$, $i.e.$ the tensor speed and the effective Planck mass ($M^2$) \cite{Tsujikawa:2015mga}. However, the $\musc<1$ condition is not sufficient to guarantee a  weaker gravity because of the interaction between the scalar field and matter which always enhances $\mu_\infty$, $i.e.$ the fifth force. Numerically the growth of structure in Horndeski models was investigated using a large samples of Monte-Carlo simulated healthy Horndeski models and sub-classes with sub-luminal propagation of scalar and tensor perturbations \cite{Perenon:2015sla}. The results are displayed in Figure \ref{fig:muetatrend}  where it is clear that the majority of models satisfy $M^2>\mps$ and $c_t^2<1$, hence producing lower growth relative to $\Lambda$CDM at low redshifts. At higher redshifts instead it is possible to note a period of super-growth with a linear growth rate larger than that predicted in the standard model. GLPV theories are even more flexible in describing the time dependence of the linear growth rate because $\ah$ provides extra freedom on top of $M^2$ and $c_t$ to modulate the stronger/weaker gravity pattern \cite{Tsujikawa:2015mga}. In particular, for GLPV it is possible to obtain a suppression in the matter power spectrum with respect to $\Lambda$CDM \cite{Traykova:2019oyx,DAmico:2016ntq}.

In the standard model of cosmology, the linear growth rate can  be efficiently  parametrized as $f=\om(a)^{\gamma(a)}$  \cite{1980lssu.book.....P} where the parameter $\gamma(a)$ is  the so-called growth index \cite{Linder:2005in}.  MG models can predict a slight deviation from $\Lambda$CDM parameterization yet detectable. For example,  \cite{Steigerwald:2014ava} showed that the growth rate of a  large class of MG models can be accurately described by 
\begin{equation}\label{eq:fgamma}
f=\Omega_\mathrm{m}(a)^{\gamma_0+\gamma_1 \ln(\Omega_\mathrm{m}(a))}\;,
\end{equation} 
where $\gamma_0$ and $\gamma_1$ are parameters which depend, in a predictable way,  on the adopted gravitational  theory. The EFT basis can be directly mapped onto the $\gamma_0-\gamma_1$ parameterization \cite{Piazza:2013pua}. Thus once a specific parameterization of the EFT functions is chosen and a background is fixed one can obtain constraints on $\gamma_i$ parameters. This phenomenological approach allows to gain further  insights on the mechanisms that tend to  suppress the growth of structures on large scales. Assuming a $\Lambda$CDM background evolution and no-ghost and no-gradient conditions, the space of theories leading to weaker gravity in the $\gamma_0-\gamma_1$ plane is found to be much larger than that of stronger gravity \cite{Piazza:2013pua} (see Figure \ref{fig:gammasteig}). In other words, statistically, Horndeski theories are more likely to produce models which exhibit weaker gravity, in the local Universe, as compared to $\Lambda$CDM. In particular, once the cosmic expansion rate is fixed to mimic $\Lambda$CDM, no viable theory can show a value for $\gamma_0$ which is larger than that of standard scenario \cite{Piazza:2013pua}. Figure \ref{fig:gammasteig} displays also forecast for a Euclid like survey in this approach. 

\begin{figure}[!]
\begin{center}
\includegraphics[scale=0.39]{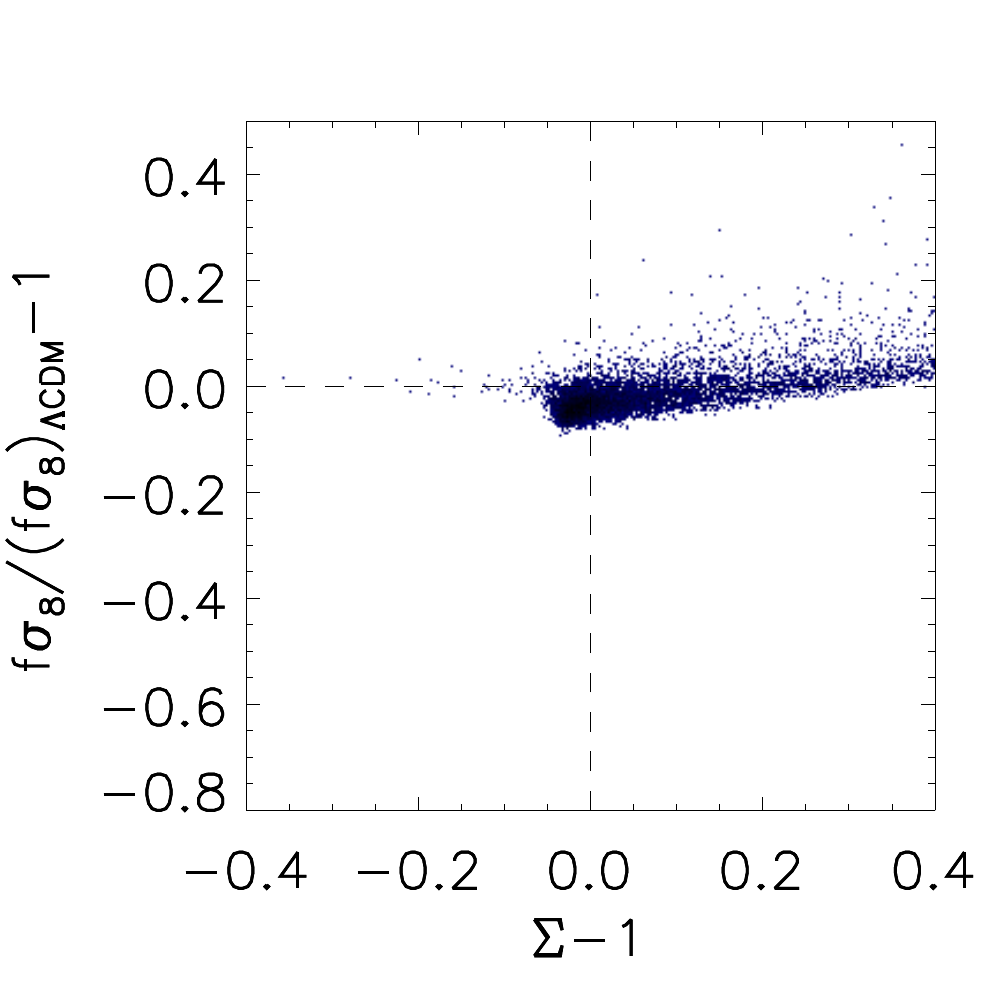}\hskip-7mm
\includegraphics[scale=0.39]{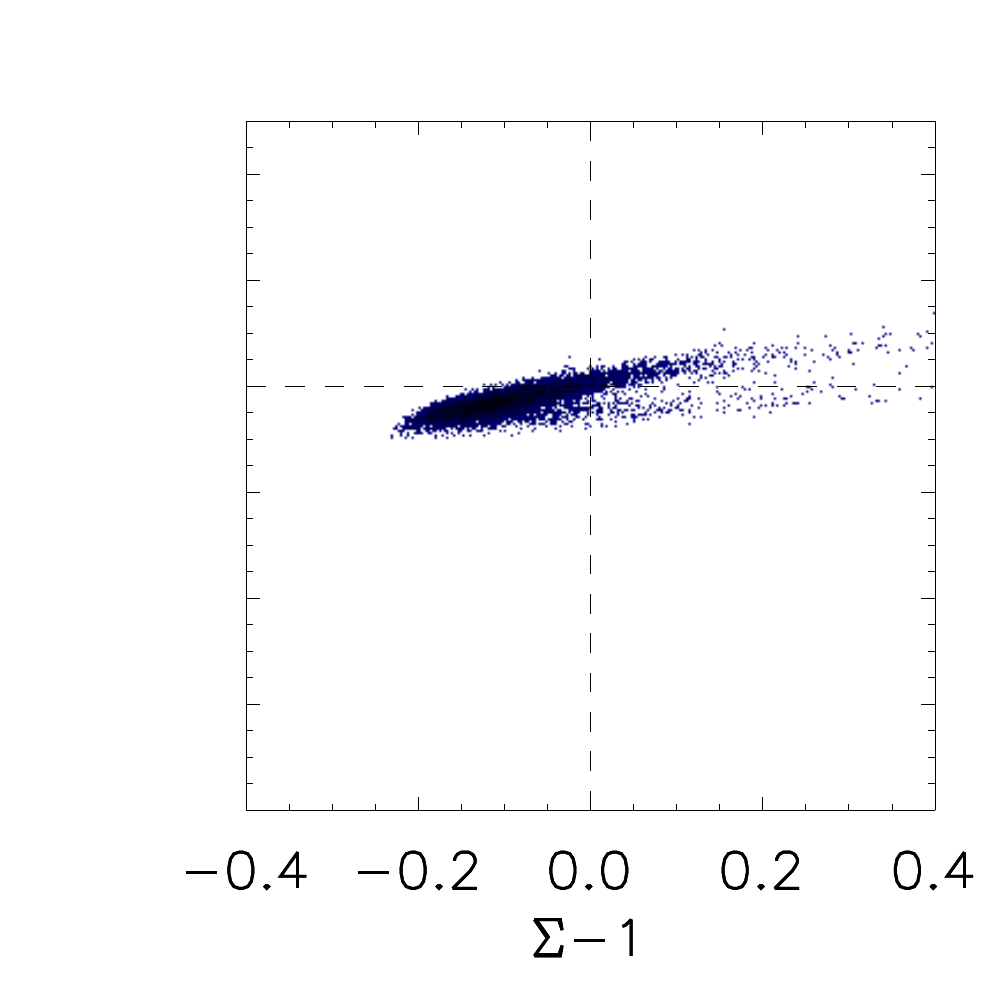}\hskip-7mm
\includegraphics[scale=0.39]{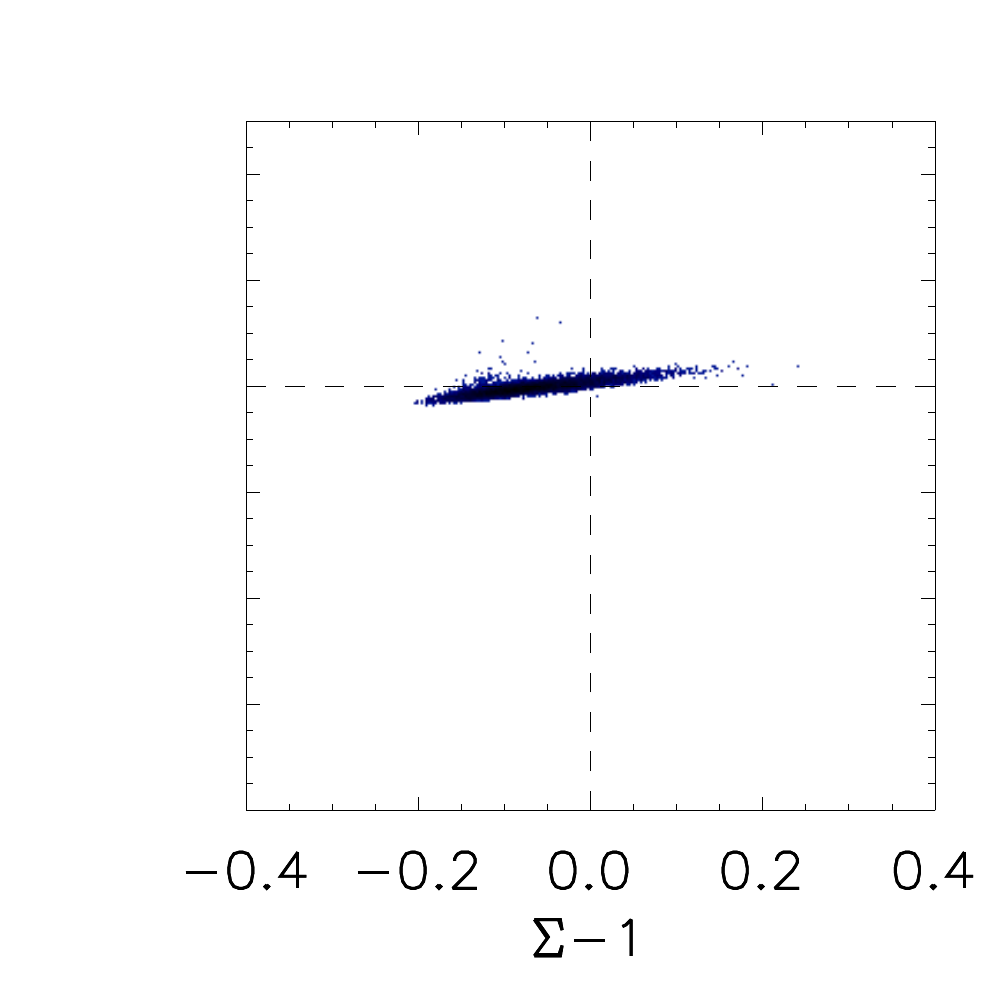}\hskip-7mm
\includegraphics[scale=0.39]{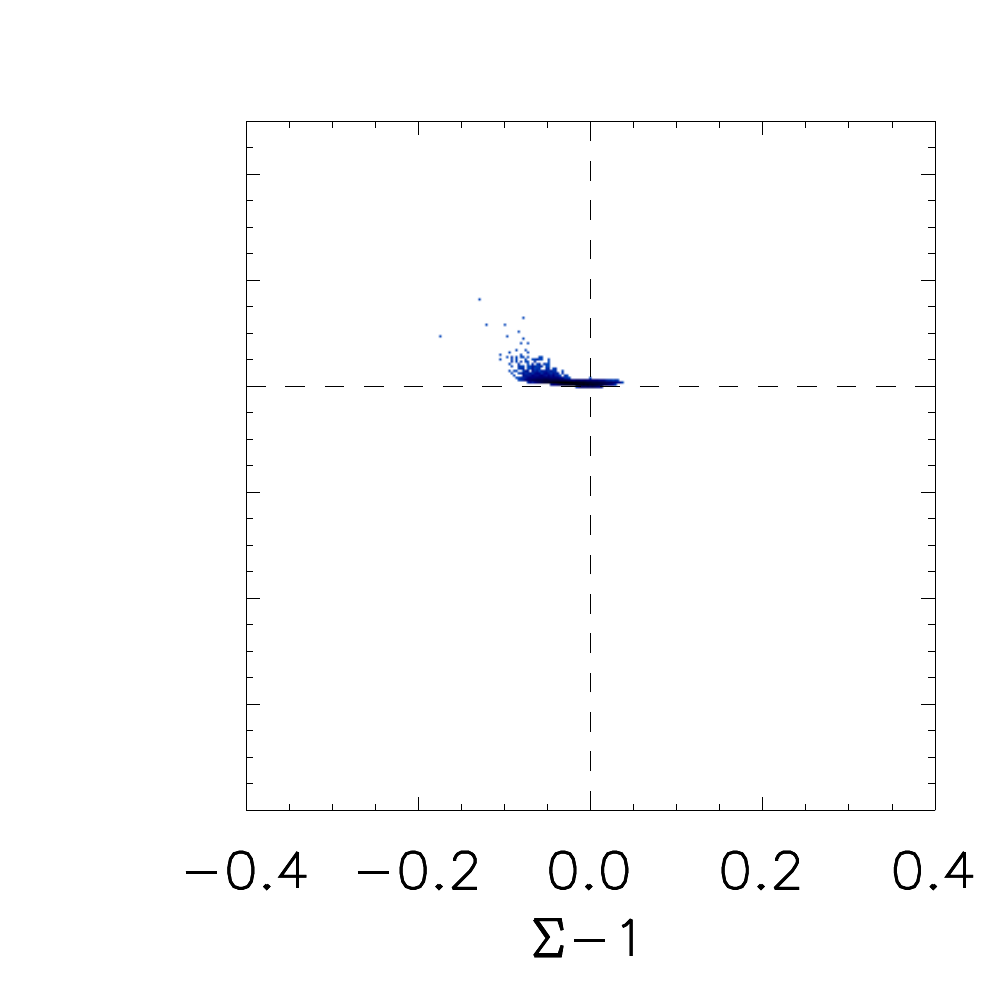}
\vskip-5mm
\includegraphics[scale=0.39]{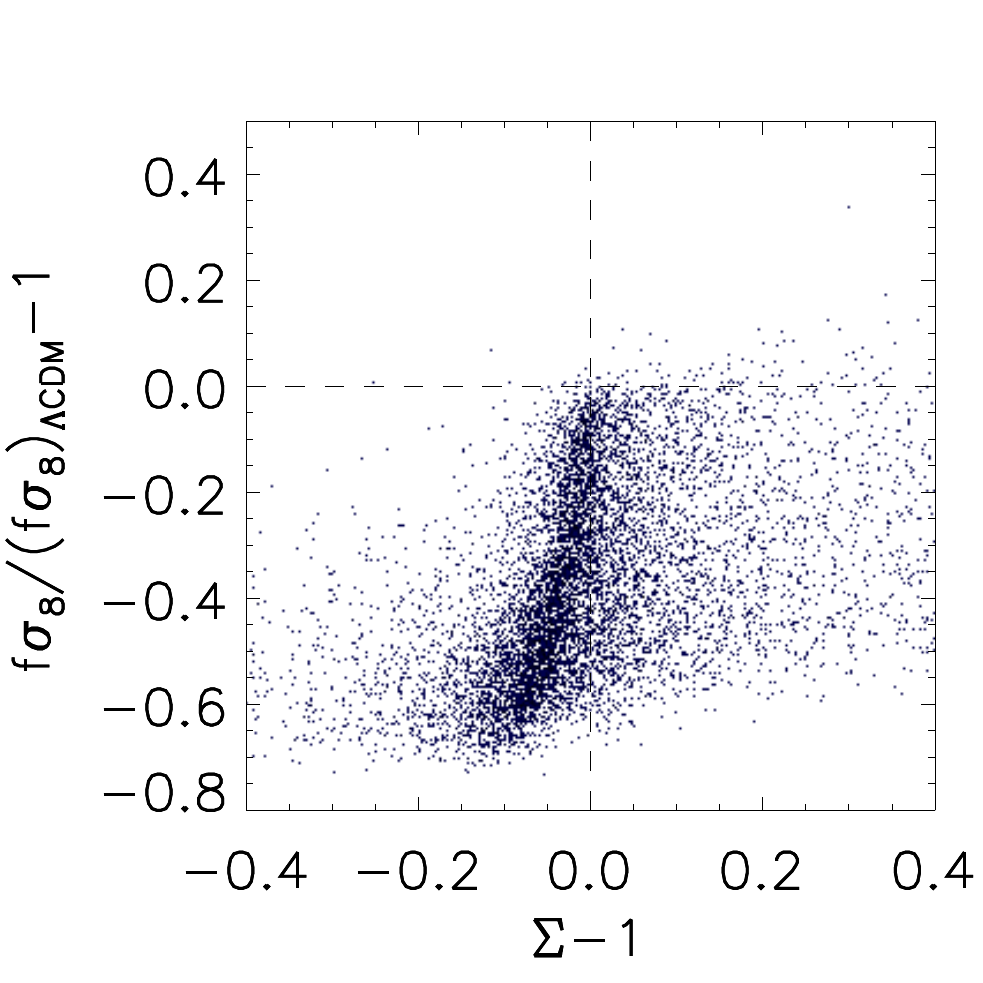}\hskip-7mm
\includegraphics[scale=0.39]{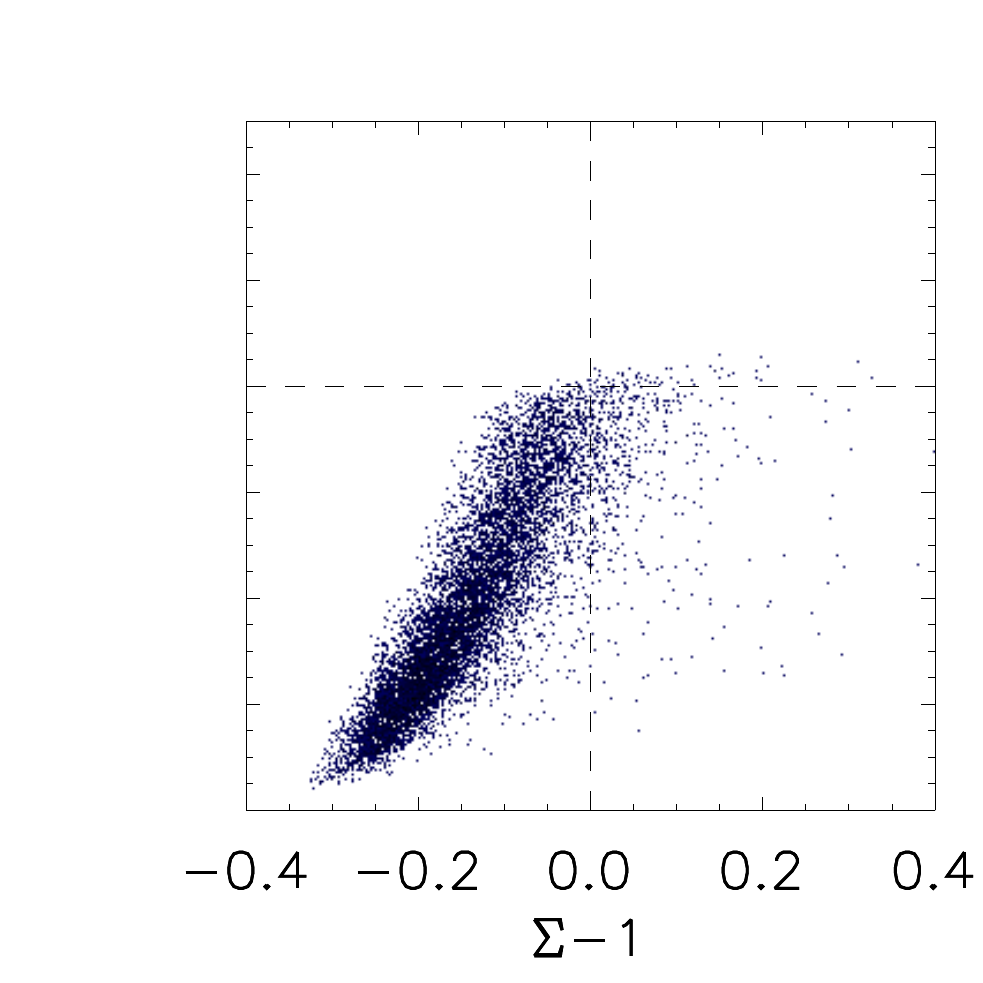}\hskip-7mm
\includegraphics[scale=0.39]{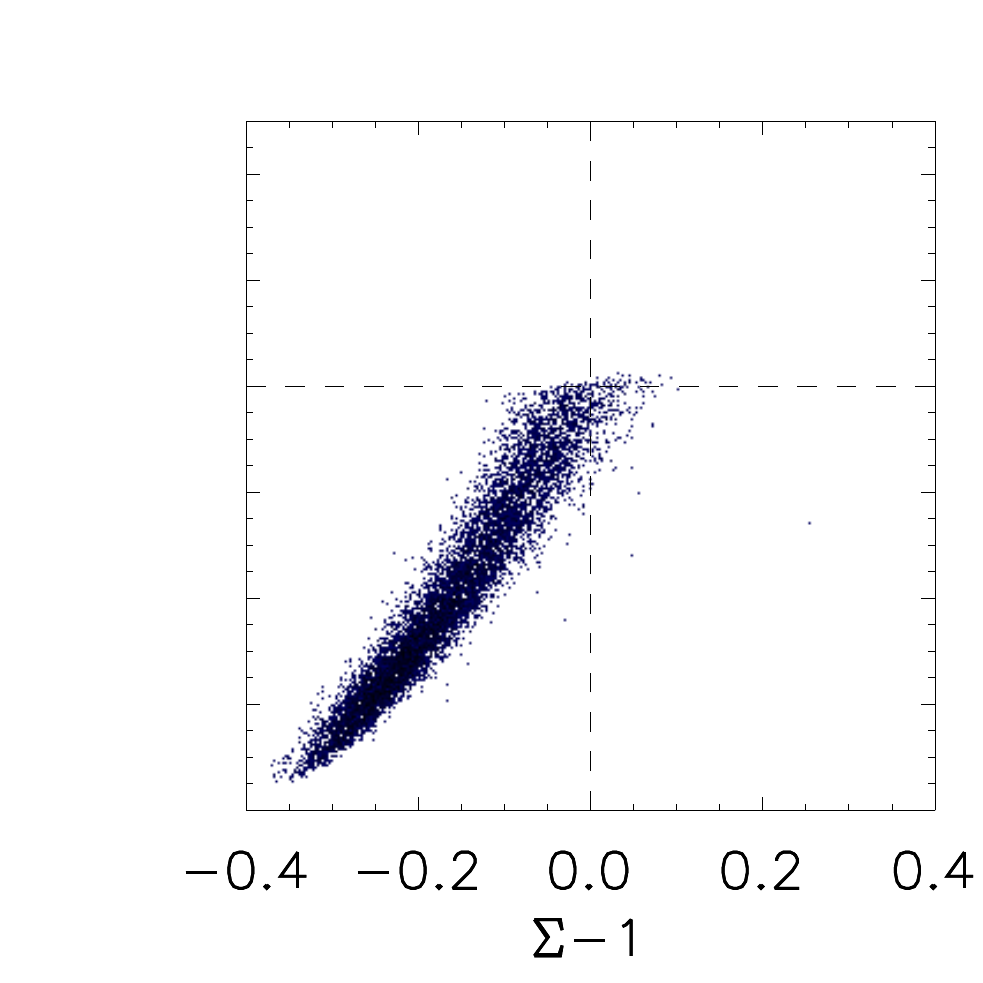}\hskip-7mm
\includegraphics[scale=0.39]{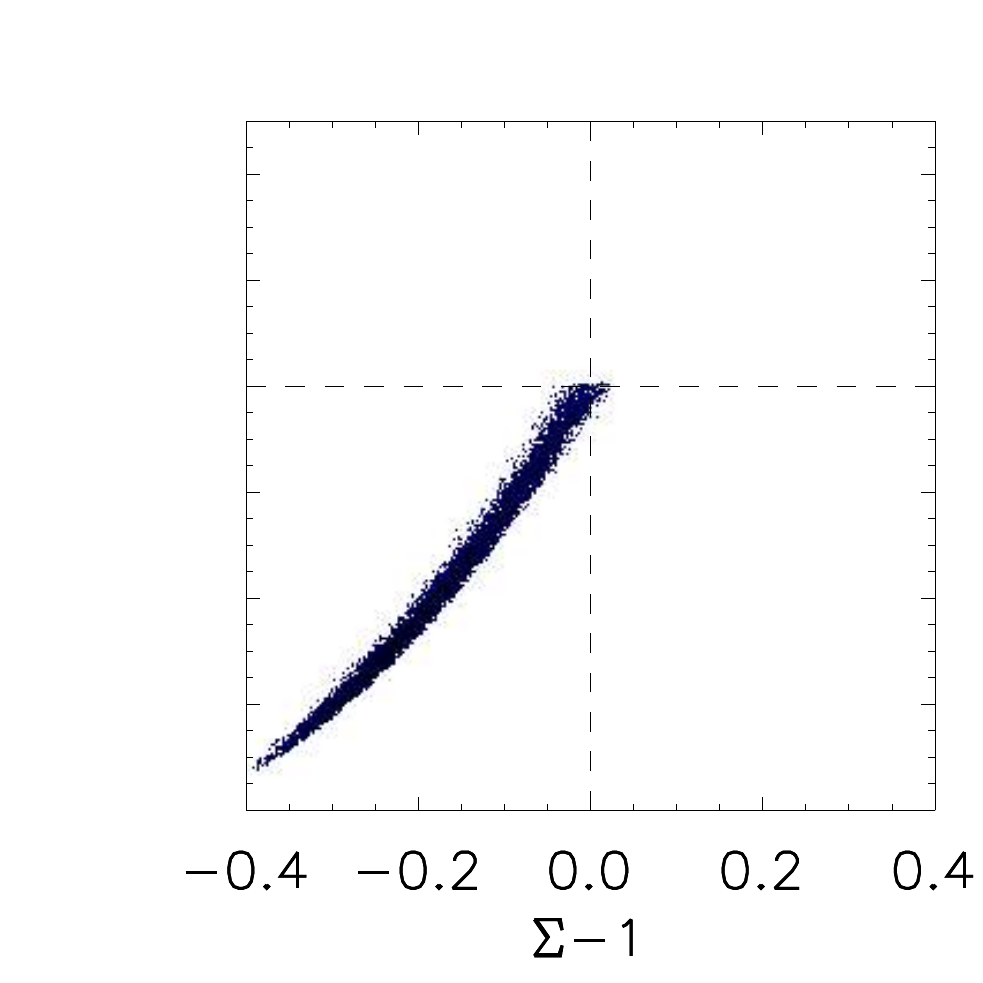}
\vskip-5mm
\includegraphics[scale=0.39]{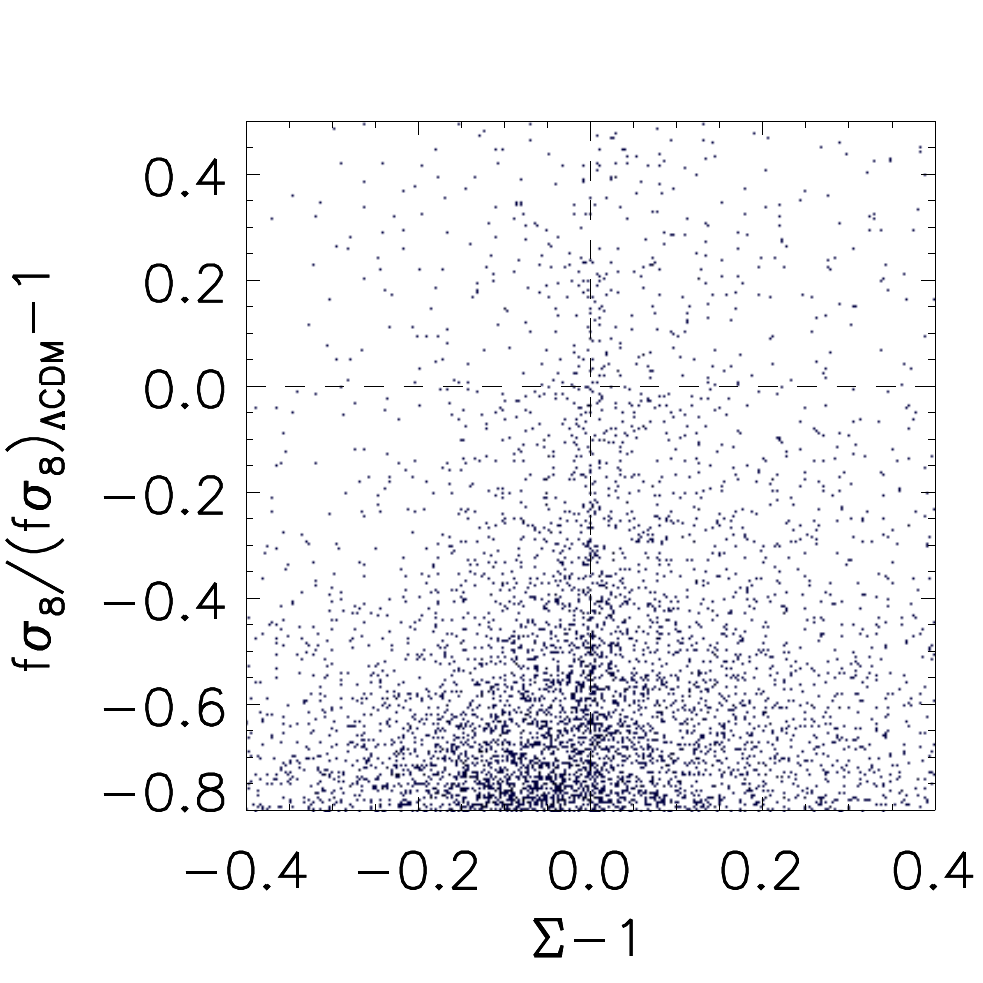}\hskip-7mm
\includegraphics[scale=0.39]{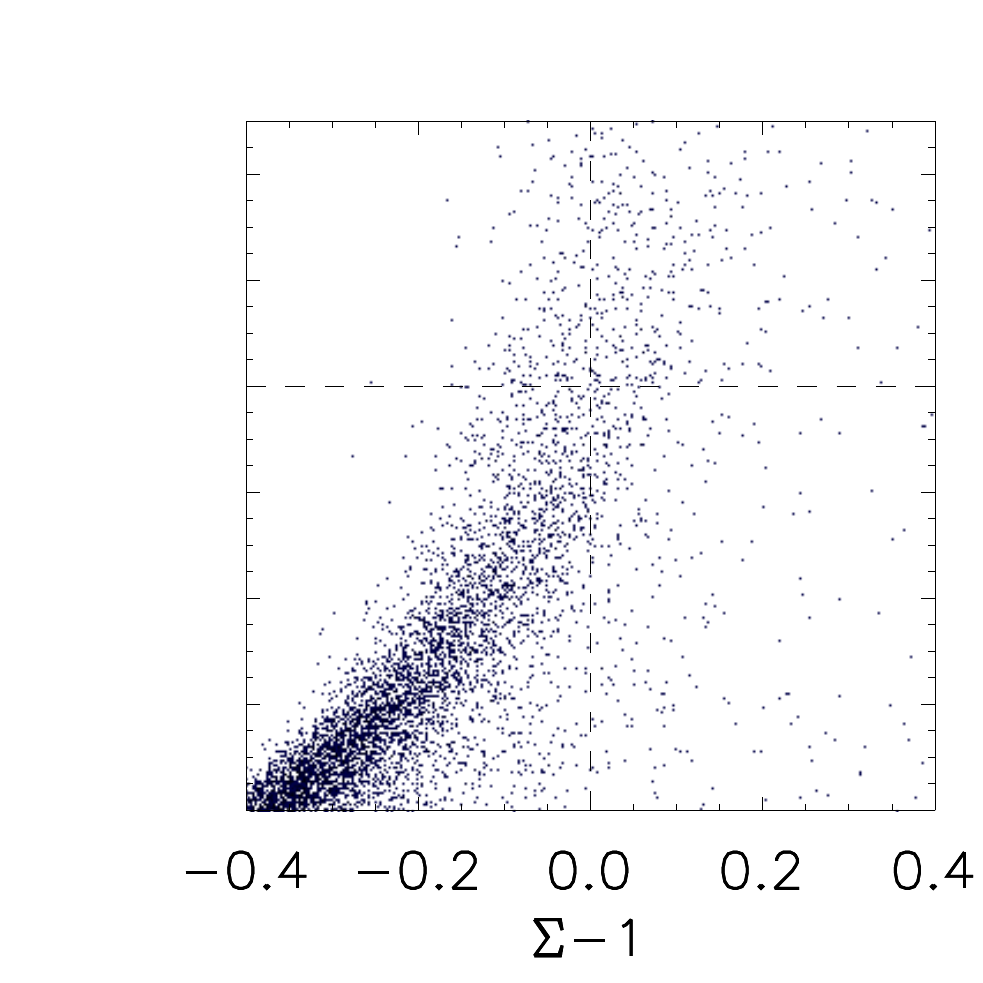}\hskip-7mm
\includegraphics[scale=0.39]{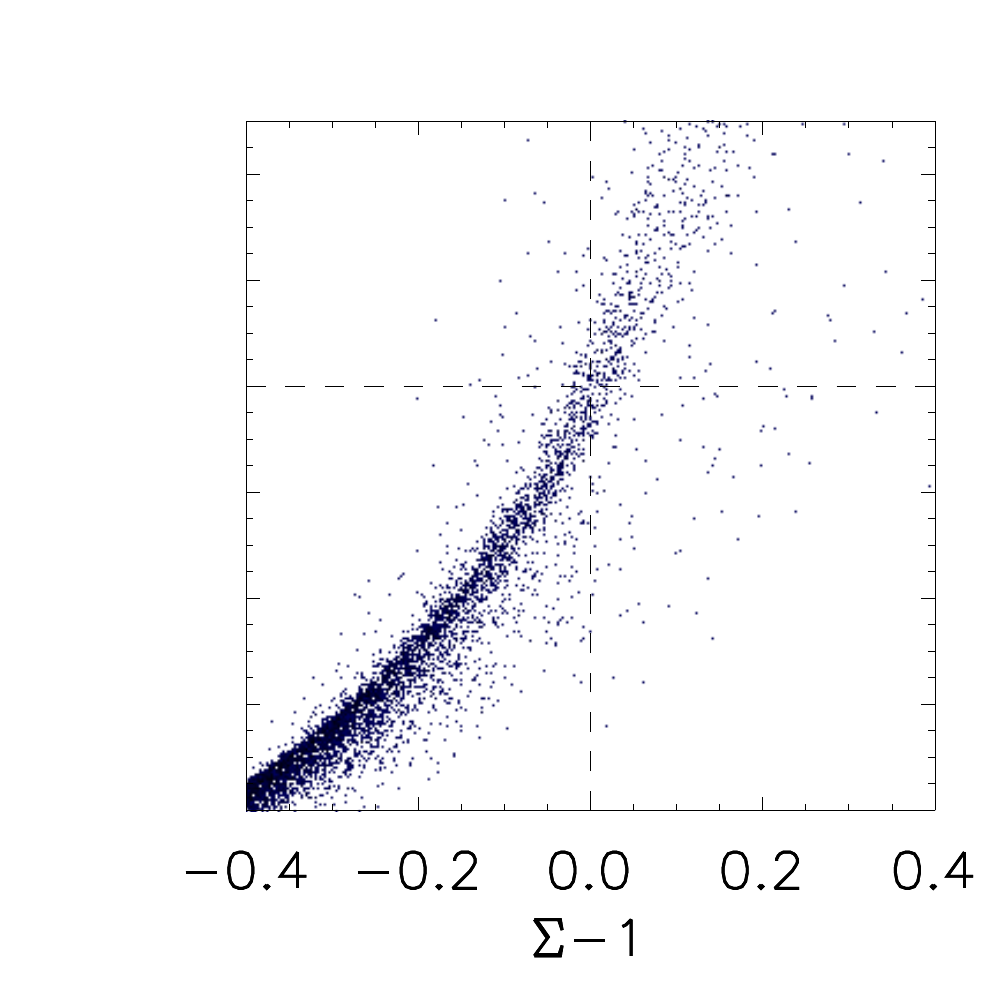}\hskip-7mm
\includegraphics[scale=0.39]{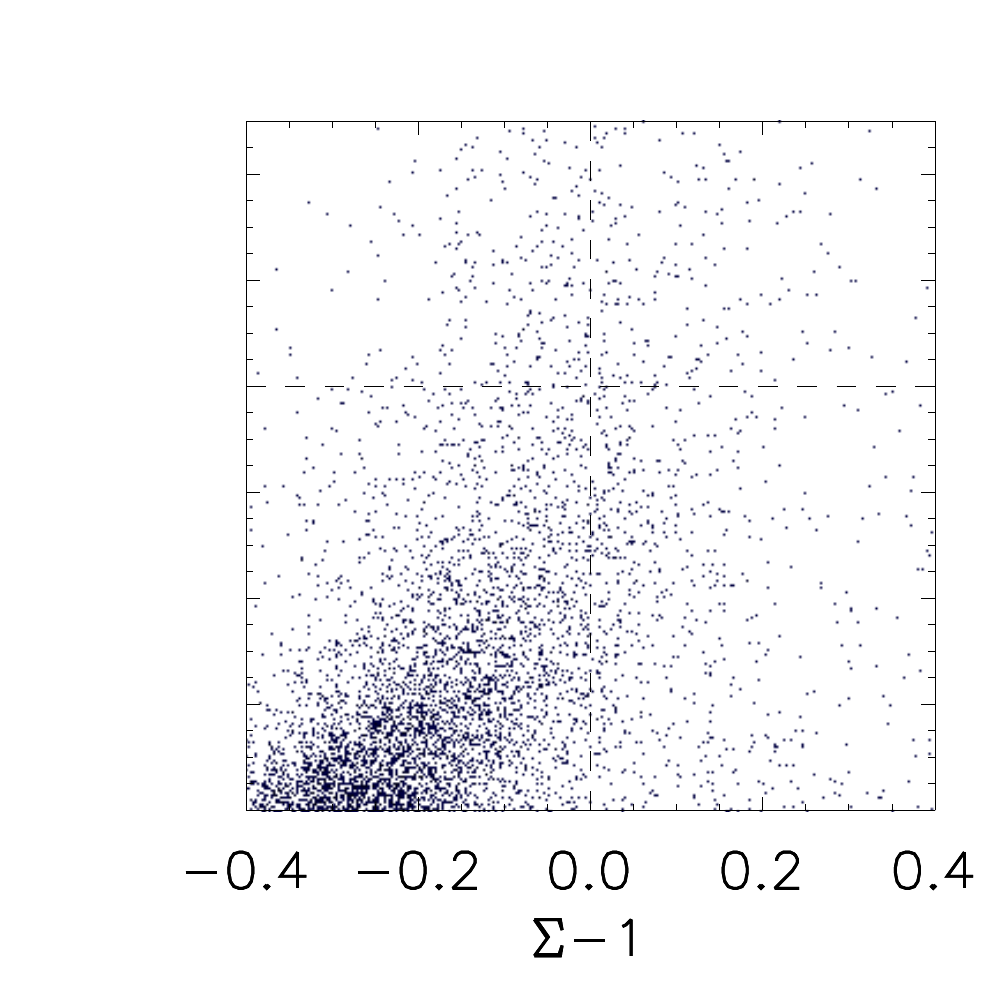}
\caption{Figures from \cite{Perenon:2016blf}. Monte-Carlo samples of viable {\it pure} Horndeski models in the $\fs/(f\sigma_8)_{\Lambda{\rm CDM}}-1$ - $\Sigma_\infty$ plane. The \textit{top panels} corresponds to LDE models, \textit{middle panels} to EDE and \textit{bottom panels} to EMG models. For each row, the redshift spans from left to right as $z=\lbrace 0,\,0.5,\,1,\,2 \rbrace$. The blue gradient scale gives a measure of the density of models.}
 \label{fig:fsigSigmaplane}
\end{center}
\end{figure}

The tendency of viable Horndeski theories to produce lower growth relative to $\Lambda$CDM is not necessarily limited to late-time DE (LDE) models but it is also present in scenarios where DE contributes throughout matter domination \cite{Perenon:2016blf}. The former are models for which the GR limit at early times is recovered, the latter are distinguished between early DE (EDE) models, $i.e.$ models with the effective Planck mass contributing to the total energy moment tensor even at early time, and early MG (EMG) models, $i.e.$ models where on top of the previous the EFT function do not vanish at early times. These scenarios are constructed using different asymptotic behaviors of the EFT functions. The $\fs$ - $\Sigma_\infty$ plane proves to be instrumental in discriminating between the aforementioned behaviors embedded within Horndeski theories and their capability to generate a lower growth. The diagnostic reveals that LDE models are strongly disfavored if $\fs < (f \sigma_{8})_{\Lambda CDM}$ at $z \gtrsim 1.5$; EDE models  if $\fs > (f \sigma_{8})_{\Lambda CDM}$ at $z\gtrsim 1.5$ or, simultaneously, $\fs <(f \sigma_{8})_{\Lambda CDM}$ and $\Sigma_\infty>1$ at $z\gtrsim 1.5$. The only possibility to display a large enhancement of growth is to use the EMG model.  The redshift evolution of these models in the $\fs$ - $\Sigma_\infty$ plane is displayed  in Figure \ref{fig:fsigSigmaplane}.

\begin{figure}[!]
\begin{center}
\includegraphics[scale=0.26]{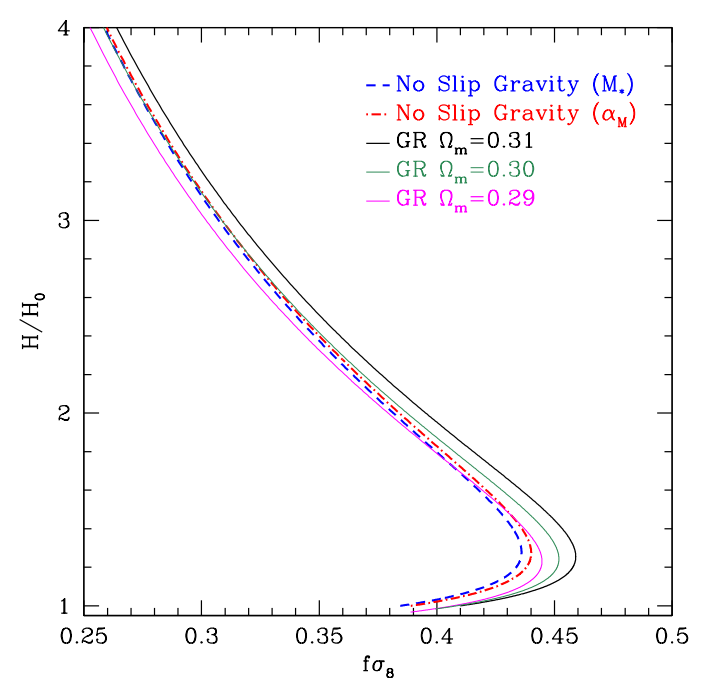}
\hskip4mm
\includegraphics[scale=0.275]{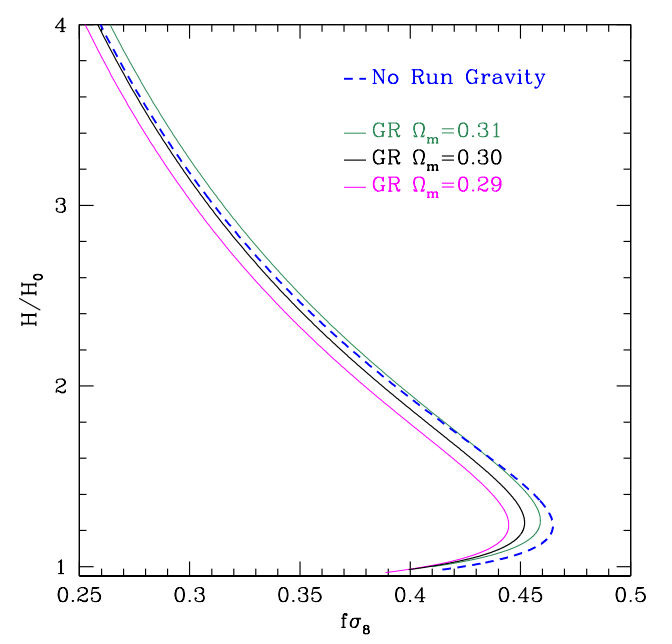}
\caption{\emph{Left panel}: Figure 4 in Ref. \cite{Linder:2018jil}. Predictions of the {\it No Slip Gravity} model in the $\fs-H/H_0$ plane. Two parameterizations are used to characterize the model: either the running of the Planck mass is parametrized (red) or the effective Planck mass (blue). They are modeled with the {\it e-fold} form (see eq. \eqref{param_efold}). \emph{Right panel}: Figure 5 in Ref. \cite{Linder:2019bqp}. Predictions of the {\it No Run Gravity} model with $\Omega_{\rm m}=0.3$ displayed against the standard model.} 
\label{fig:noslipgravity}
\end{center}
\end{figure}

Within the class of Horndeski theories, a specific phenomenological model known as the \textit{No Slip Gravity} model \cite{Linder:2018jil}  produces lower growth than $\Lambda$CDM at low redshifts as shown in Figure \ref{fig:noslipgravity} (left panel). The model is characterized by a gravitational slip parameter equal to unity, an unmodified speed of propagation for tensor modes and by the relation $\ab=-2 \,\am$. As consequence the effective gravitational coupling $\mu$ is lower than GR in the past inducing a lower growth of structure.

On the other hand, suppose future observational data do not allow: $i)$ lower growth of structure relative to the standard model; $ii)$ a gravitational slip different from unity; $iii)$ changes in the damping ($\am \neq 0$) and speed ($\at\neq 0$) of GWs. This does not induce all alternative scenarios to GR to be ruled out. Indeed, modifications of gravity due to a non-vanishing braiding function are still possible. An example of such a model is the {\it No Run Gravity} model characterized only by the $\ak$ and $\ab$ functions \cite{Linder:2019bqp}. Consequently, the deviation in the effective gravitational coupling depends only on the braiding term. In particular, stability conditions allow only positive values of $\ab$ and the model therefore always predicts an enhancement in the growth of structures relative to $\Lambda$CDM as depicted in Figure \ref{fig:noslipgravity} (right panel) \cite{Linder:2019bqp}.

%-------------------------------------------------
\subsection{Imprints on cosmological power spectra}\label{sec:obsersignatures}
%-------------------------------------------------

Non-standard gravitational scenarios are best inspected by analyzing the shape of large scale observables, such as the matter density power spectrum, the CMB angular power spectrum, the lensing spectrum, lensing B-mode contribution and the bi-spectrum. They can for instance change the lensing potential ($\Phi+\Psi$) when additional perturbative terms are included \cite{Acquaviva:2005xz}, change the growth of structure modifying the Poisson equation and affect the shape of the temperature-temperature CMB power spectrum at low multipole-$\ell$ through the integrated Sachs-Wolfe (ISW) effect sourced by $\dot{\Phi}+\dot{\Psi}$ \cite{Sachs:1967er,Kofman:1985fp}, shift its high-$\ell$ peaks due to a modified expansion history \cite{Hu:1996vq} and change the ratio between odd and even peaks in models where DE is coupled to dark matter \cite{Amendola:2011ie}.

\begin{figure}[!]
\begin{center}
\includegraphics[scale=0.55]{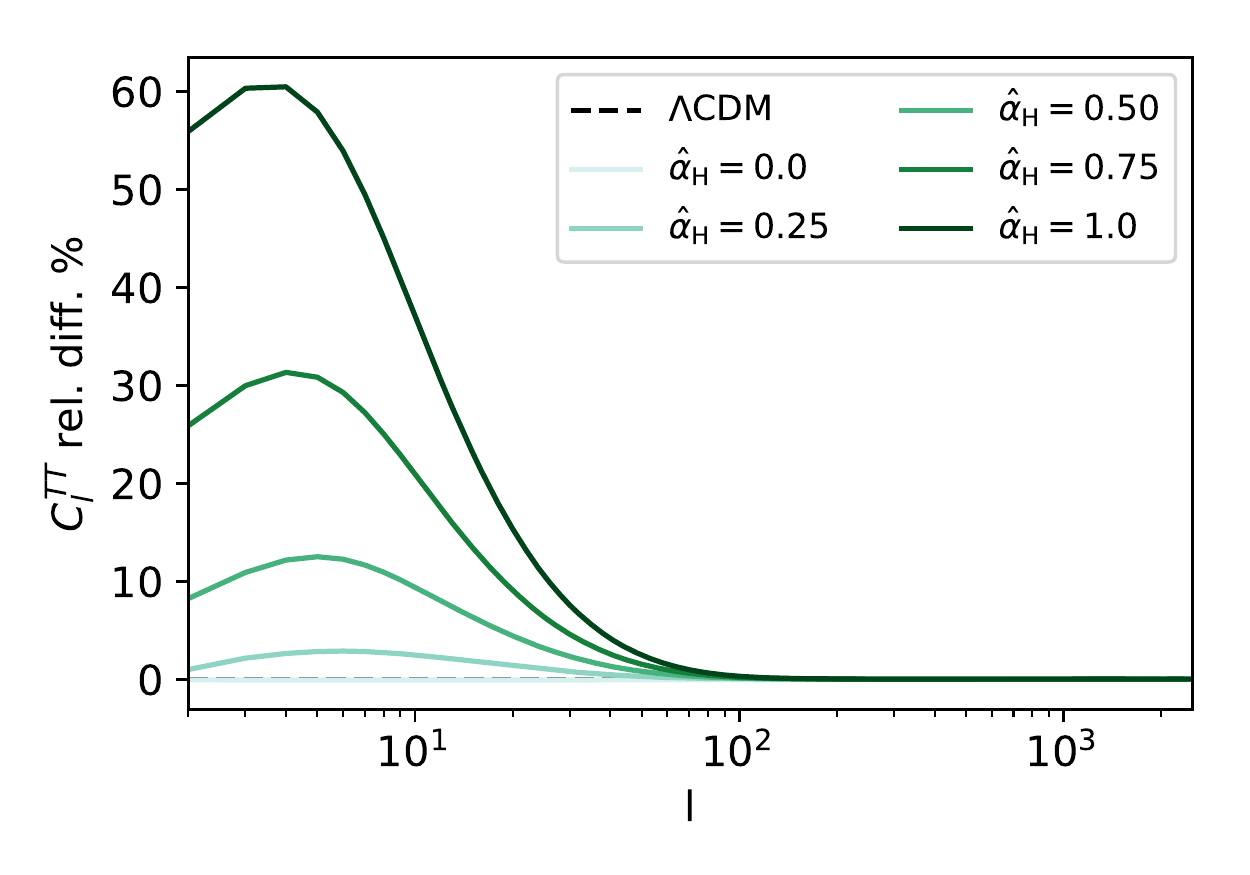}
\hskip2mm
\includegraphics[scale=0.34]{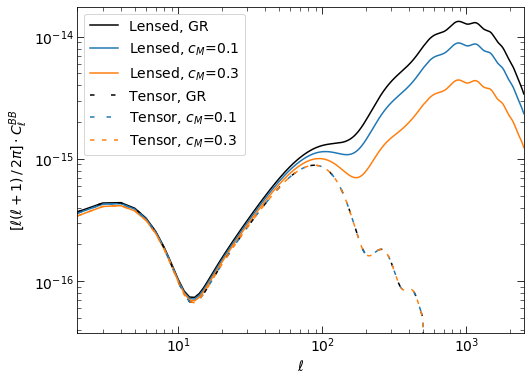}
\caption{{\it Left panel}: Figure 6 in Ref. \cite{Traykova:2019oyx}. The effects of different values of $\hat{\alpha}_H$ (the free constant parametrizing $\alpha_H$ in the {\it linear-de} form \eqref{param_de0})  on the relative difference of the angular TT-power spectrum with respect to the standard model. \emph{Right panel}: Figure 3 in Ref. \cite{Brush:2018dhg}. The effects of the {\it No Slip Gravity} model on the CMB B-mode power spectrum for different values of $c_M$ (the free constant parametrizing $\am$ in the {\it e-fold} form \eqref{param_efold}). The dotted lines represent the contribution from the tensor modes only.}
 \label{fig:powerspectra}
\end{center}
\end{figure}

The EFT approach allows thus to investigate the phenomenology of these effects at linear scales for large class of models \cite{Amendola:2014wma,Bellini:2015wfa,Salvatelli:2016mgy,Renk:2016olm,Zumalacarregui:2016pph,DAmico:2016ntq,Yamauchi:2017ibz,Brush:2018dhg,Garcia-Garcia:2018hlc,Frusciante:2018jzw,Hirano:2018uar,Traykova:2019oyx,Duniya:2019mpr,Pace:2019uow}. In this case the connection with a specific operator in the action \eqref{eftact} or a physical effect enclosed in the $\alpha$-basis can be easily identified by switching on/off single EFT functions per time. In the following we report on some of them. We note that the results in literature might depend on the chosen parameterization. Values of the effective Planck mass larger than $\mp^2$ are shown to suppress the lensing power spectrum with respect to the standard scenario and to modulate the shape of the TT power spectrum at low multipoles $l\lesssim 30$ giving rise to an enhanced ISW tail \cite{Salvatelli:2016mgy}. The deviations in the speed of propagation of GWs ($\at$) are responsible for changing the location of the inflationary peak of the BB spectrum \cite{Amendola:2014wma}. The kineticity coupling, $\ak$, has been found to modulate the low-$\ell$ CMB TT power spectrum due to the late-time ISW effect, however this effect is unmeasurable being dominated by the cosmic variance \cite{Frusciante:2018jzw}. A non-vanishing positive braiding term, $\ab$, leads to both an enhanced lensing auto-correlation function and matter power spectrum and, depending on its magnitude, it can generate either a suppressed ISW tail or an enhanced one \cite{Pace:2019uow}. The GLPV function, $\alpha_H$, is responsible for the damping of the matter power spectrum. Increasing positive values of $\alpha_H$ enhances the CMB TT power spectrum at low-$\ell$ \cite{Traykova:2019oyx,Duniya:2019mpr,DAmico:2016ntq} as it can be noticed in Figure \ref{fig:powerspectra} (left panel). On the same line the lensing potential decreases as a function of $\alpha_H$ \cite{Traykova:2019oyx,DAmico:2016ntq}. For the particular case of the \textit{No Slip Gravity} model, the running of the Planck mass, $\am$, is shown to mildly affect B-mode reionization and recombination bumps on low multipoles $l \lesssim 10 $, while at $l \gtrsim 100 $ its impacts on the lensing B-mode polarization are more significant (of order four times the maximum value of $\am$) \cite{Brush:2018dhg}, see Figure \ref{fig:powerspectra} (right panel).

\begin{figure}[!]
\begin{center}
\includegraphics[scale=0.5]{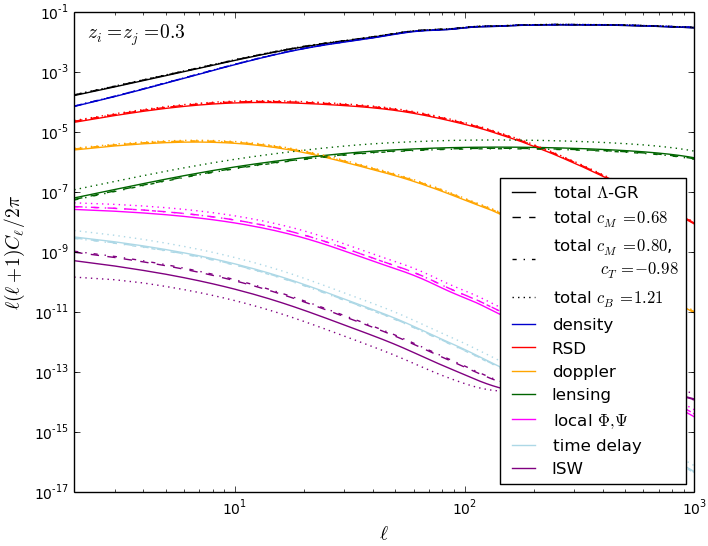}
\vskip2mm
\includegraphics[scale=0.38]{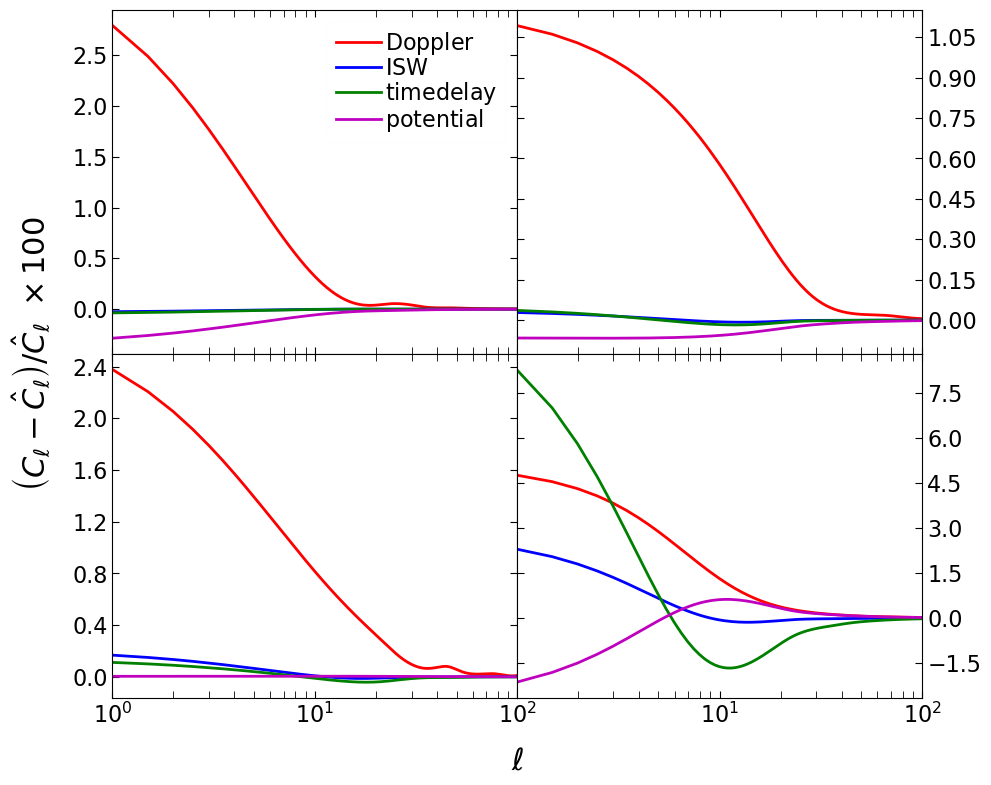}
\caption{{\it Top panel}: Figure 3 in Ref. \cite{Renk:2016olm}. The effects of the EFT functions on several relativistic contributions to the autocorrelation of the galaxy number count angular power spectrum at $z=0.3$. $c_M,c_T, c_B$ are the constants parametrizing respectively $\am,\at,\ab$ in the {\it linear-de} form \eqref{param_de0}. {\it Bottom panel}: From Figure 6 in Ref. \cite{Duniya:2019mpr}. Redshift evolution of the ultra-large scale relativistic effects for a given GLPV model, $\am = 0.03$, $\ah = 0.085$ and $\ak = 0$, on the relative difference of the angular power spectrum with respect to $\Lambda$CDM (hat). The panels correspond to $z = 0.1 $ (top left), $z = 0.5$ (top right), $z = 1$ (bottom left) and $z = 3$ (bottom right).}
 \label{fig:powerspectra2}
\end{center}
\end{figure}

Given the ever-increasing precision and  scales probed by surveys, integrated effects on ultra-large scales such as ISW, Doppler, Shapiro time-delay, etc, might  soon enter the observational window.  It is thus interesting to explore their effective potentiality in probing departures from standard gravity. Such effects have been scrutinized within the EFT formulation for Horndeski and GLPV models \cite{Renk:2016olm,Duniya:2019mpr}. For instance, the amplitude of the ISW signal when forecasted for Horndeski theories may deviate by up to $\mathcal{O}(1000\%)$ from standard GR expectations, while the same models only induce $\mathcal{O}(10\%)$ modifications of the amplitude of local observable such as $\fs$ \cite{Renk:2016olm}. In parallel, taking into consideration the lensing convergence or the Shapiro time delay leads to additional contributions highlighting different sensitivities to the EFT functions as displayed for the galaxy number count angular power spectrum in Figure \ref{fig:powerspectra2} (top panel) \cite{Renk:2016olm}. The ISW contribution remains the most sensitive probe to MG effects, yet including other integrated effects can therefore improve the total sensitivity of cosmological spectra to measure any gravity departures from GR. Note that the contribution of each ultra-large scale relativistic effects changes across redshift. For instance, the Doppler effect is shown in Figure \ref{fig:powerspectra2} (bottom panel) to be significant at all epochs in GLPV theories while the ISW, the time-delay and velocity potential effects become significant for $z\gtrsim 3$ \cite{Duniya:2019mpr}.

%%%%%%%%%%%%%%%%%%%%%%%%%%%%%%%%%%%%%%%%%%%%%%%%%%%%%%%%%
\section{Cosmological constraints}\label{sec:constraints}
%%%%%%%%%%%%%%%%%%%%%%%%%%%%%%%%%%%%%%%%%%%%%%%%%%%%%%%%%

In this Section, we review the cosmological constraints on DE/MG models obtained using the EFT approach. The key characteristic of these analyses is the use of MCMC procedures or Fisher forecasts to constrain gravity at large scales combined often with the theoretical priors exposed in the previous Section. We distinguish between models described by a \textit{pure} approach and specific gravity models implemented in EB codes using the \textit{mapping} approach (see definitions in Section \ref{sec:codes}). We include in Appendix  \ref{App:params} the list of the \textit{pure} EFT parameterizations used in this Section and tables  summarizing their  observational constraints in Appendix  \ref{summaryconstraints}.

%--------------------------------------------------
\subsection{Running Planck mass}\label{sec:running}
%--------------------------------------------------

Modifications of gravity induced by a running Planck mass have several distinct effects on observables. In the scalar sector, a Planck mass varying across time has been shown to impact, for instance, the background evolution, the growth of matter perturbations, lensing potential and the ISW tail of the CMB power spectrum. In the tensor sector, it alters the friction term in the GW equation (eq.~\eqref{eq:gweq}) affecting the amplitude of the primordial polarization peak in B-modes. We reviewed these features in Section \ref{sec:novpred}. In the EFT formalism, the running of the Planck mass is encoded by  $\am$ in the $\alpha$-basis and a linear combination of the derivatives of $\fg(t)$ and $\bar{M}^2_3(t)$ in the EFT basis (see Section \ref{sec:alternative} for details).
 
Generally, it is not possible to consider only $\am$ and set all other EFT functions to zero. This choice would give a vanishing kinetic term (see eq. \eqref{ghostalpha}) leading to ghost instabilities and strong coupling problems. One way to proceed is to tweak the function $\ak$ and fix it to a positive value to prevent the appearance of any pathological behavior. Note that in such a configuration any noticeable effects on cosmological observables are only due to $\alpha_M$ because $\alpha_K$ has no detectable impact \cite{Bellini:2015xja,Kreisch:2017uet,Frusciante:2018jzw}. We comment further about this aspect in Section \ref{sec:horncons}. 

\begin{figure}[!]
\begin{center}
\includegraphics[scale=0.45]{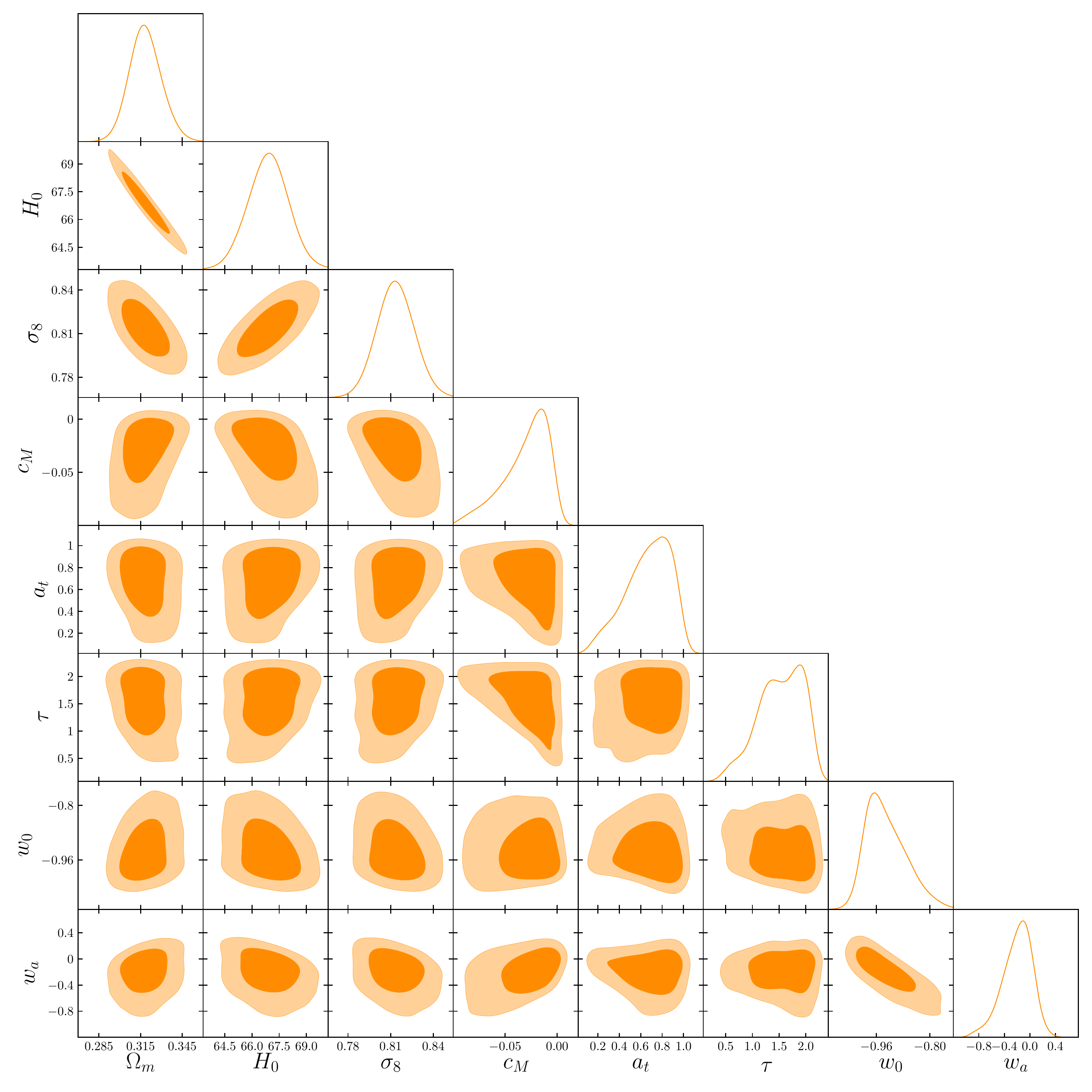}
\caption{Figure 7 in Ref. \cite{Brando:2019xbv}. Constraints obtained on the \textit{No Slip Gravity} model with $\am$ parametrized with the {\it hill} form (eq. \eqref{param_hill}) on a CPL background. The parameters controlling the time and sharpness of the \textit{hill} form are fixed to $a_t=0.5$ and $\tau=1$.}
\label{fig:running}
\end{center}
\end{figure}

One choice for the form of $\alpha_M$ often adopted is to model it as proportional to the DE density parameter, $i.e$ $\alpha_M = \alpha_{M,0}\omde(a)$, where $\alpha_{M,0}$ is a constant  (eq. \eqref{param_de0}). Using this model with a $\Lambda$CDM background, the combination of Planck CMB data and the $H_0$ prior from Riess \emph{et al.} \cite{Riess:2011yx} is found to favor a positive $\alpha_M$ at $\sim 2\sigma$, however, such preference is weakened when geometrical probes are included \cite{Huang:2015srv}. 

The effects of a running Planck mass have been further investigated when a specific link with the braiding function $\ab$ is retained. In particular, the relation $\am=-\ab$ is typical of conformally coupled models such as $f(R)$-gravity, Brans-Dicke and chameleon theories. Investigations of this class of models has been performed by the Planck collaboration in 2015 \cite{Ade:2015rim}. There, $\am$ is modeled as a power law in the scale factor with amplitude $\alpha_{M,0}$ and scaling $\beta$ with a $\Lambda$CDM background. The combination of datasets involving CMB, WL, BAO, RSD measurements leads to stringent constraints on the amplitude of the running Planck mass, $\alpha_{M,0}<0.097\,\, (95 \% \mbox{C.L.})$, while the statistical power to constrain its scaling is much weaker, $0.92^{+0.53}_{-0.25}\,\, (95 \% \mbox{C.L.})$. For this model, WL data prefer higher values of the expansion rate than CMB ones and because of that WL data lead to weaker constraints \cite{Ade:2015rim}. Using the latest Planck data (2018), the results favor $\alpha_{M,0}$ to be negative when  only the CMB data are  employed, while the inclusion of  WL, BAO, RSD measurements give the bounds $\alpha_{M,0}=-0.015^{+0.019}_{-0.017}$ at $68\% \mbox{C.L.}$ (with CMB lensing) \cite{Aghanim:2018eyx}. 

Following on the constraints where a relation between $\am$ and $\ab$ is imposed, \textit{No Slip Gravity} models are representatives with $\ab=-2\am$ \cite{Linder:2018jil}. For this class of models, stability conditions require $\am \ge 0$. Using the \textit{e-fold} form eq. \eqref{param_efold} to parameterize alternatively $\am$ or $M^2$ and a $\Lambda$CDM background, Planck 2015 data favor values of the running Planck mass much lower than what found with LSS data \cite{Linder:2018jil,Brush:2018dhg}. This is justified by $\am>0$ which induces a higher CMB lensing power and the loss of power on large scale in the matter power spectrum. A time dependent DE equation of state allows to explore larger portions of the  parameter space. In this case, for example,  the running Planck mass can even assume negative values \cite{Brando:2019xbv}. CMB, BAO, SNIa and RSD datasets constrain indeed $\am$ to be negative as shown in Figure \ref{fig:running}. Remarkably a negative running of the Planck mass at $z\gtrsim 1$ does not prevent the model to produce lower growth relative to the standard model nor a positive ISW-galaxy cross-correlation \cite{Brando:2019xbv}.

\begin{figure}[!]
\begin{center}
\includegraphics[clip, trim = 0cm 0cm 8.9cm 0.5cm, scale=1]{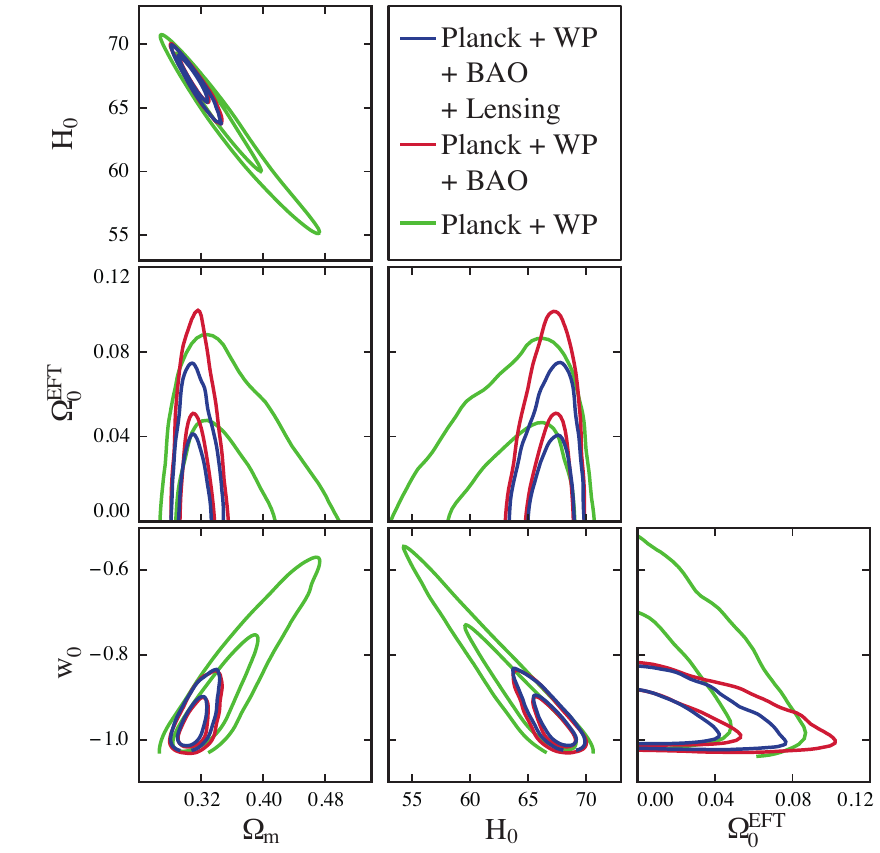}
\caption{Figure 2 in Ref. \cite{Raveri:2014cka}. 68\% and 95\% confidence regions on the parameters of the linear EFT model (eq.\eqref{eq:lineft}) with a $w$CDM background shown for several combination of observables. WP stands for WMAP low-$\ell$ polarization data.}
\label{fig:running2}
\end{center}
\end{figure}

The EFT basis (action \eqref{eftact}) offers a natural environment for  constraining the running of the Planck mass as well.  For example using the linear EFT model eq. \eqref{eq:lineft}, the cosmological constraints shown in Figure \ref{fig:running2} yield the bound $\Omega_0^\mathrm{EFT} < 0.061$ at 95\% C.L. for a $\Lambda$CDM background and $\Omega_0^\mathrm{EFT} < 0.058$ for a $w$CDM background \cite{Raveri:2014cka}, in agreement with what is found later by the Planck collaboration (2015) \cite{Ade:2015rim}. The linear EFT model has also been explored to gauge whether this proposal could reconcile the tension between Planck CMB estimation of the lensing amplitude parameter $A_L$ and the result from the lensing reconstruction \cite{Ade:2015xua}. Simulations of CMB anisotropy and CMB lensing spectra, assuming Planck 2015's best-fit values and Planck blue book on beam and noise specifications, show that models with an effective Newton constant stronger than $G_\mathrm{N}$ can have a modulating effect similar to that of $A_L$ \cite{Hu:2015rva}. Nevertheless, this induces higher values of $\sigma_{8,0}$ making  the tension with WL surveys more severe~\cite{deJong:2015wca,Kuijken:2015vca,Hildebrandt:2016iqg,Conti:2016gav,Abbott:2017wau,Abbott:2017smn}. The $A_L$ tension remains an open issue as confirmed by the latest Planck 2018 results \cite{Aghanim:2018eyx}. 

%--------------------------------------------------------------------
\subsection{{\it Pure} Horndeski and GLPV models}\label{sec:horncons}
%--------------------------------------------------------------------

The {\it pure} Horndeski and GLPV models are characterized in the EFT language by a selected set of free functions of time beyond $H(t)$. For {\it pure} Horndeski models they are $\{\fg,\bar{m}^3_1,M^4_2,\bar{M}^2_2\}$ in the EFT basis and $\{\am,\ak,\ab,\at\}$ in the $\alpha$-basis. An additional function is required to select {\it pure} GLPV models, i.e $\mu_1^2$ or $\alpha_H$ depending on the basis. In this description, instead of choosing the functional form of the $G_i$ or $F_i$ functions respectively in the Horndeski Lagrangians~\cite{Kobayashi:2011nu} and GLPV ones~\cite{Gleyzes:2014dya}, one can fix the form of each EFT function and describe the linear part of the theory. These sets of functions allow to explore and constrain the primary features of Horndeski and GLPV theories in a \textit{pure} EFT fashion. It is noteworthy to say that choosing an  appropriate model  for the EFT functions is  a challenging task. Sometimes the chosen forms of the EFT functions result in simplified behaviors of relevant physical quantities, such as $\mu(t,k)$ or $\Sigma(t,k)$, if compared to the complex behavior they have when the \textit{mapping} approach is considered \cite{Linder:2015rcz,Linder:2016wqw}. The consequence is that one might underestimate the modification of the gravity force or even miss its signatures. In this regard one has to be careful in generalizing the results obtained in the EFT framework recalling that they might be strictly related to the chosen parameterization. On the other hand, it is worth to note that relevant results have been obtained and common trends have been identified using this approach with different parameterizations. We review them in the following. 
 
\begin{figure}[!]
\begin{center}
\includegraphics[scale=0.28]{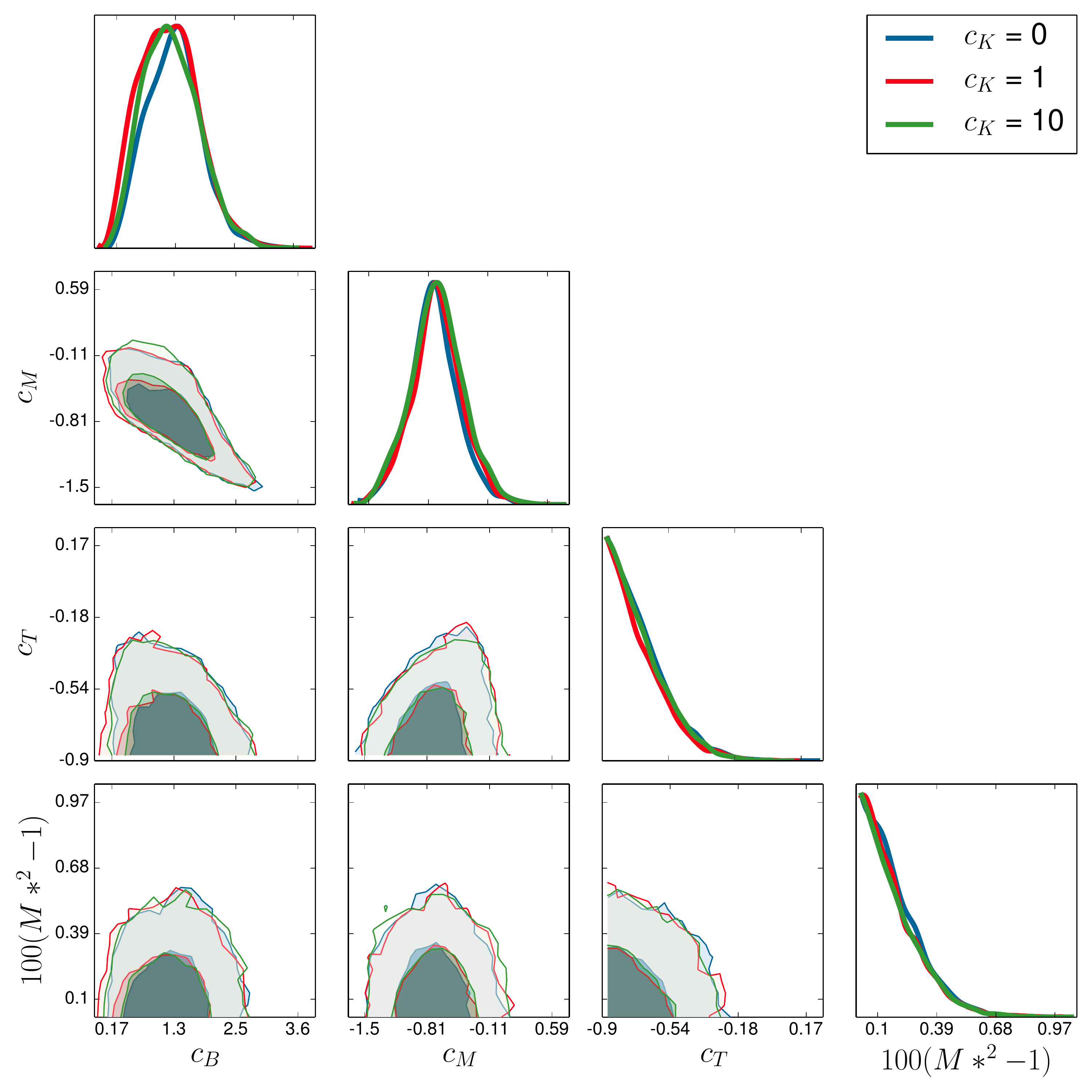}
\caption{Figure 3 in Ref. \cite{Bellini:2015xja}. Constraints on a {\it pure} Horndeski model in the $\alpha$-basis where each function is parametrized with the {\it linear-de} form (eq. \eqref{param_de0}) on a $\Lambda$CDM background.
\label{fig:bellcons}}
\end{center}
\end{figure}

The first observational constraints of {\it pure} Horndeski models have been derived in the $\alpha$-basis, where each $\alpha$-function is parametrized using the {\it linear-de} form (eq. \eqref{param_de0}). The constraints on the $\alpha_i$ parameters using cosmological datasets, such as CMB, BAO, RSD and the power spectrum of galaxies from the WiggleZ survey, are shown in Figure \ref{fig:bellcons} \cite{Bellini:2015xja}. Notably, the data favor models with the effective Planck mass $M^2$ larger than the Planck mass and stability conditions induces a hard prior on the initial value of the effective Planck mass, $i.e.$ $M^2_{\rm ini}\geq \mps$. The constraints display a neat preference for a positive value for the braiding, negative for the running Planck mass and a sub-luminal propagation of tensor modes. This parameterization was further investigated in light of KiDS+GAMA data considering luminal propagation of tensor modes \cite{SpurioMancini:2019rxy}. While the constraints on cosmological parameters are compatible with $\Lambda$CDM, a preference for positive values of $\am$ and $\ab$ is found. Furthermore, the clustering quantity $S_8=\sigma_8\sqrt{\Omega_\mathrm{m}/0.3}$ is highlighted to be better in agreement with the Planck estimate when considering Horndeski theories than $\Lambda$CDM.  Just as considering stability or sub-luminal priors improves observational constraints significantly, so does the consideration of ``positivity bounds'' \cite{Melville:2019wyy}.  The latter arise from requiring basic principles such as a unitary, causal, local UV completion \cite{Adams:2006sv,Nicolis:2009qm,Bellazzini:2016xrt,deRham:2017zjm}. Such positivity bounds when applied to Horndeski theory imply indeed some constraints on the $G_i$ functions which can be rewritten in terms of EFT functions \cite{Melville:2019wyy}. The inclusion of these additional bounds as theoretical priors when performing parameter estimation analysis, led to a reduction of over 60\% of the allowed parameter space for the sub-classes Horndeski models parametrized with the {\it linear-de} form (eq. \eqref{param_de0}). The additional constraint of sub-luminal propagation of GWs in the same models  narrowed down further the  viable parameter space to less than a 1\%.

\begin{figure}[!]
\begin{center}
\includegraphics[scale=0.53]{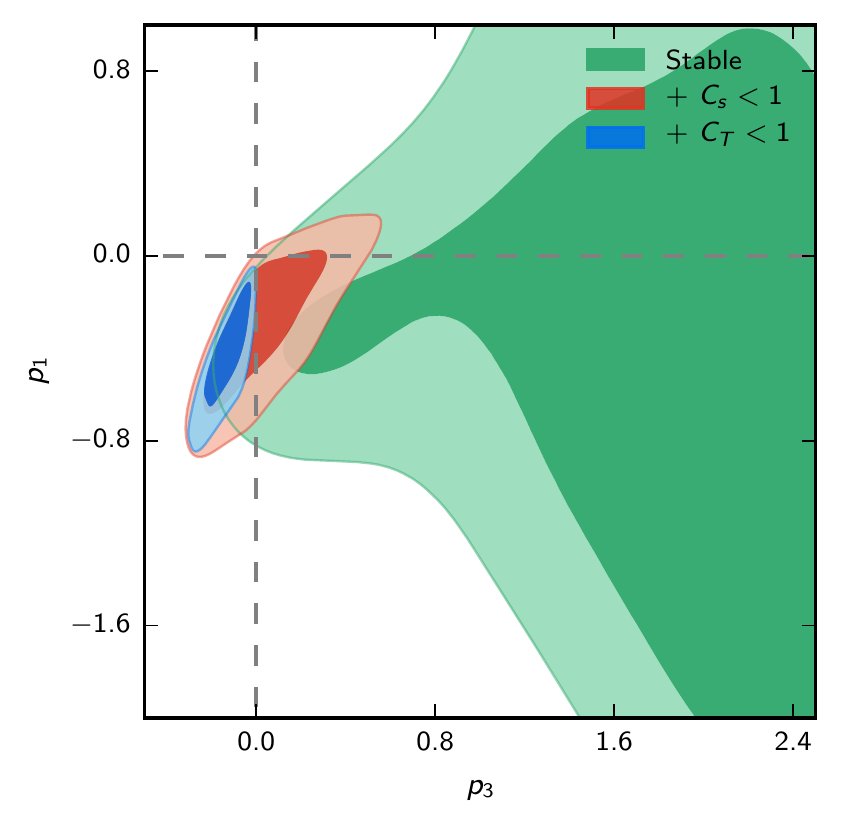}\hskip-1mm
\includegraphics[scale=0.53]{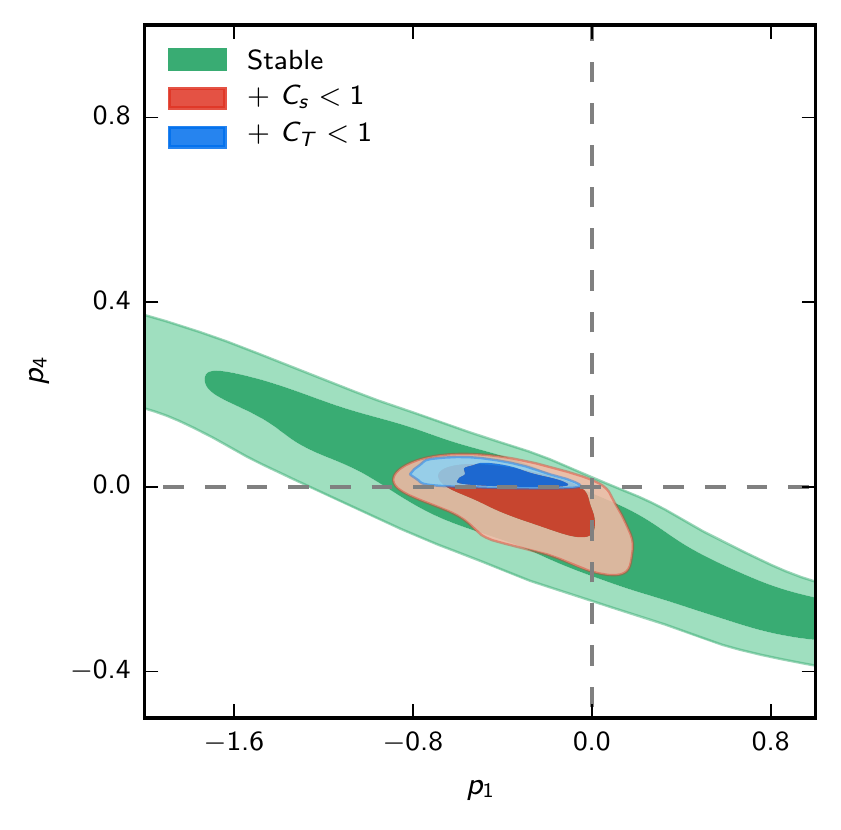}\hskip-1mm
\includegraphics[scale=0.53]{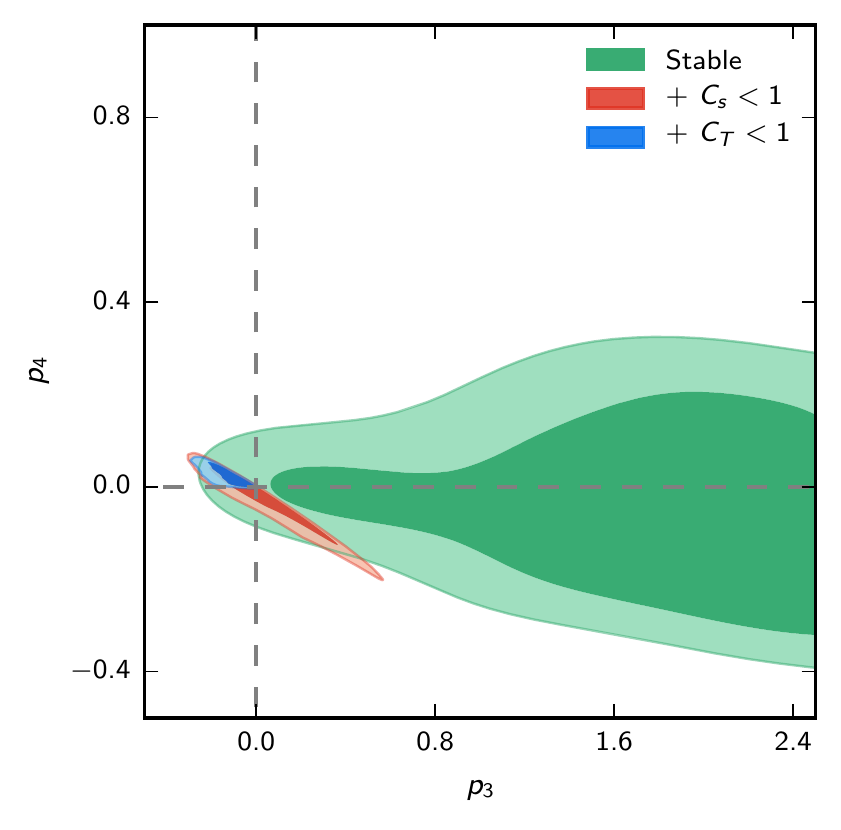}
\vskip5mm
\includegraphics[scale=0.45]{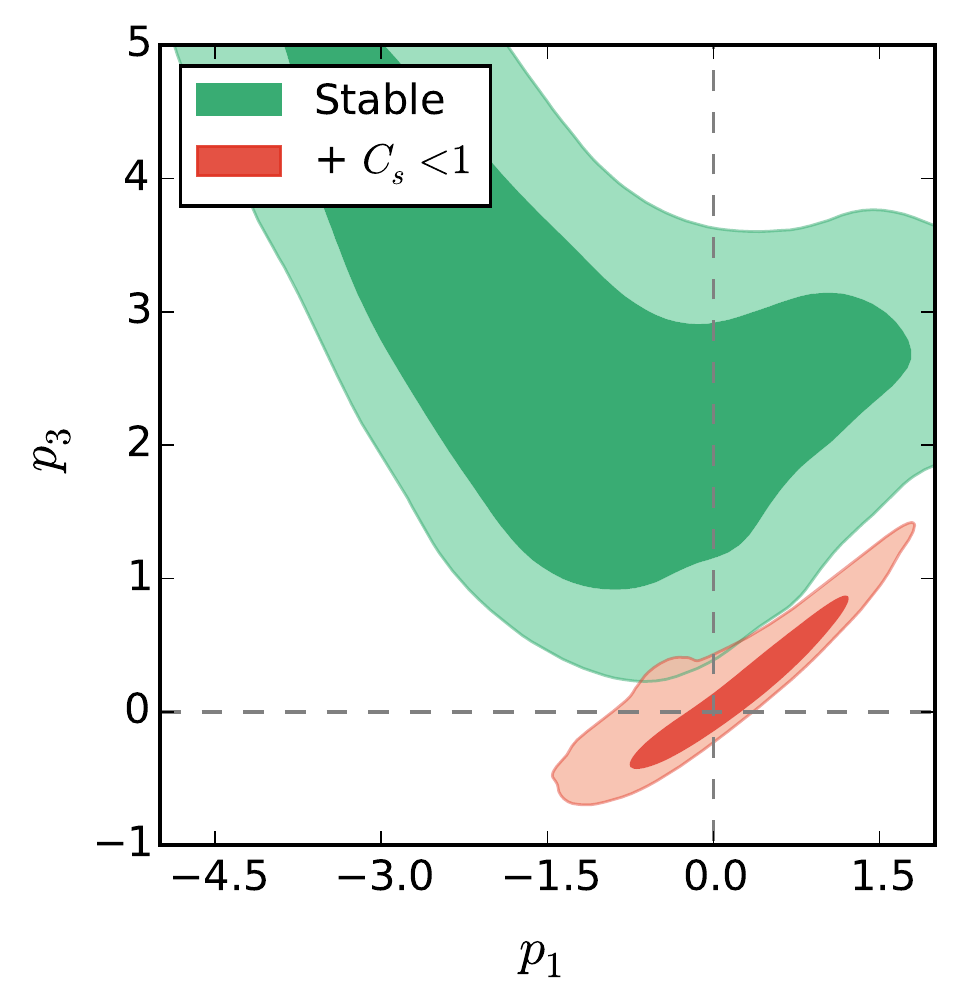}\hskip-1mm
\includegraphics[scale=0.45]{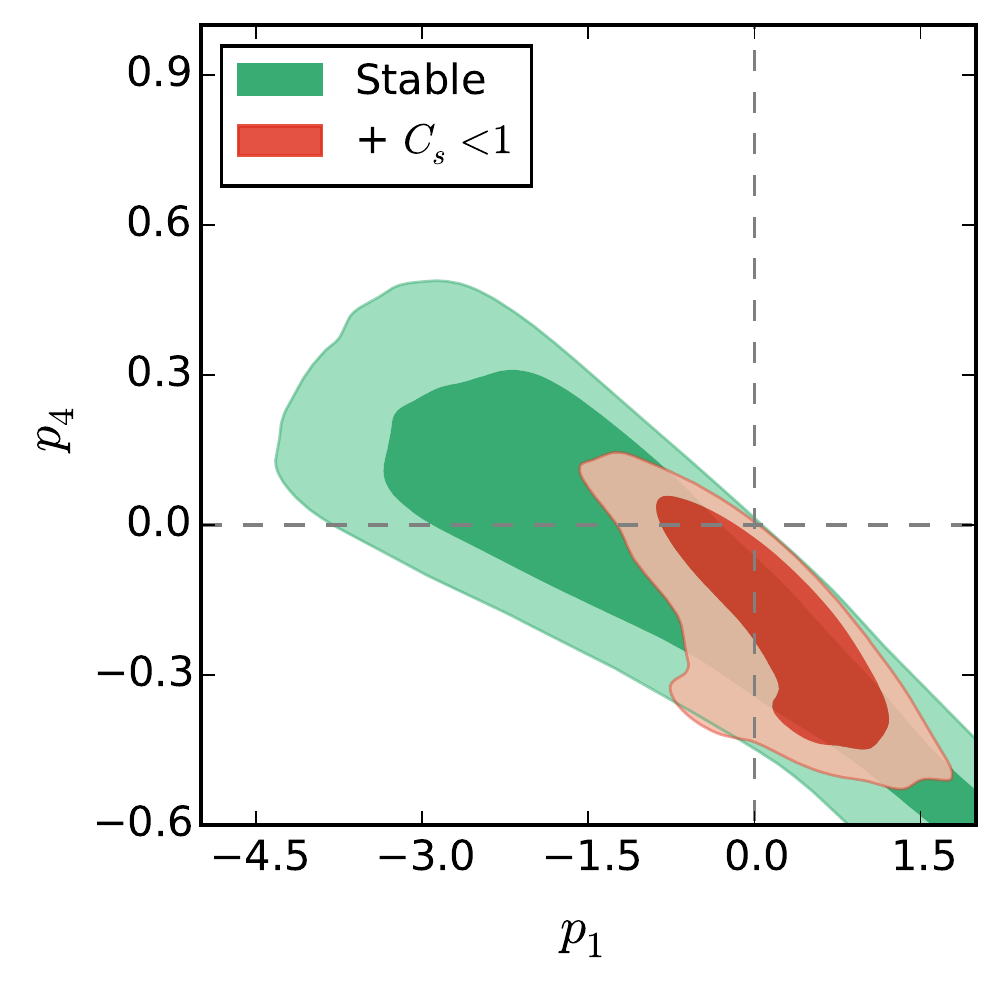}\hskip-1mm
\includegraphics[scale=0.45]{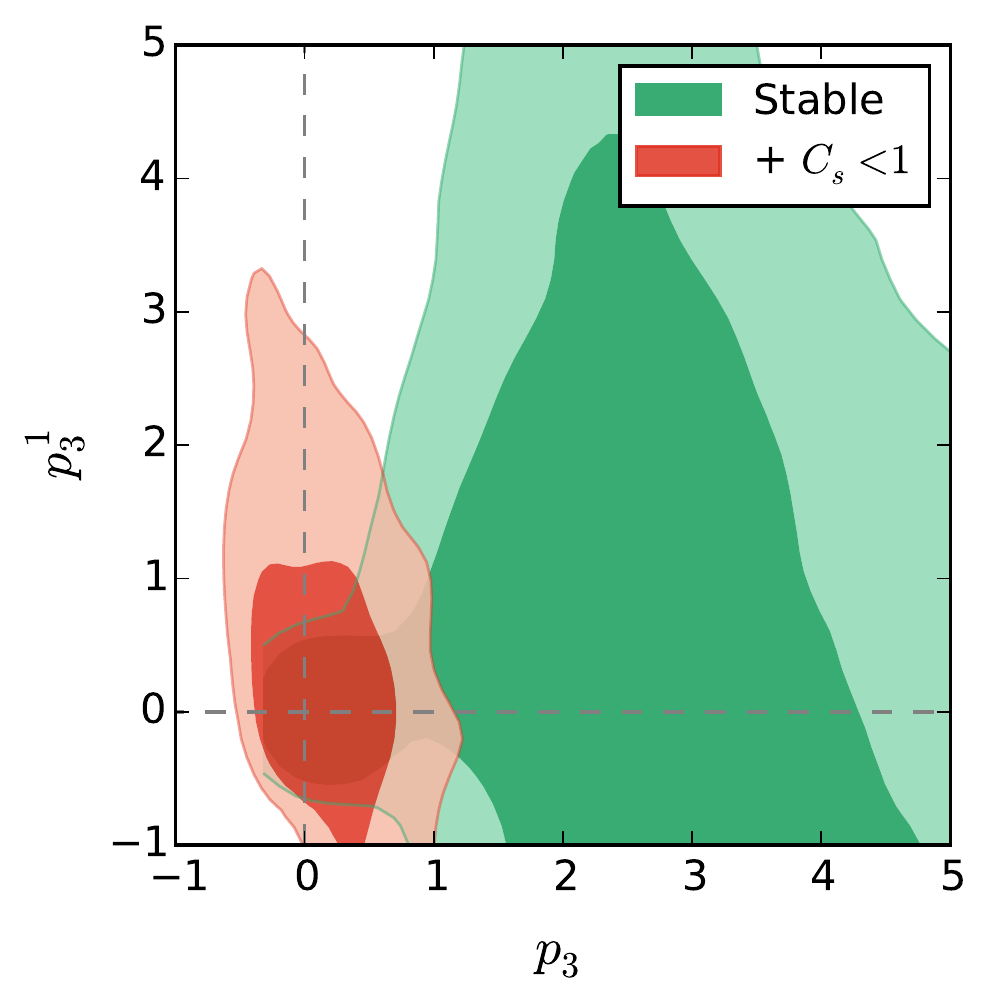}\hskip-1mm
\includegraphics[scale=0.45]{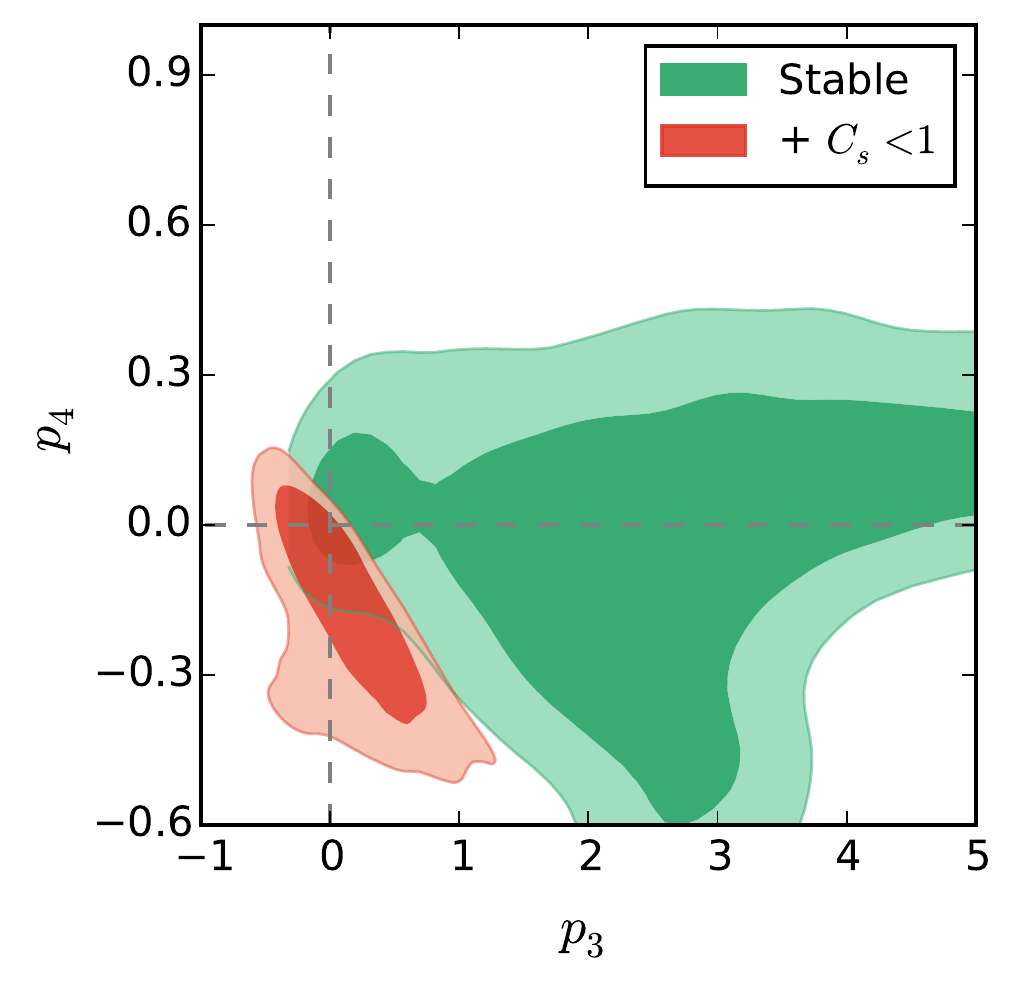}\hskip-1mm
\includegraphics[scale=0.45]{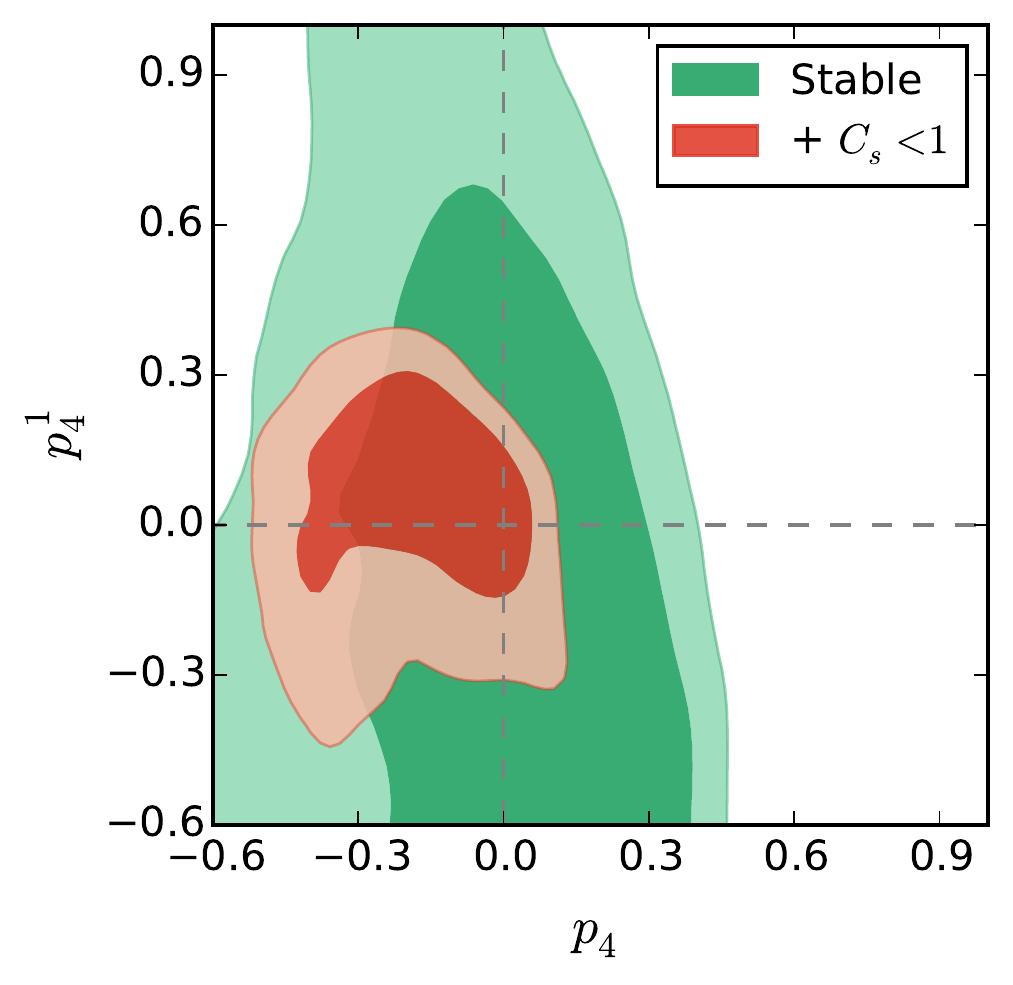}
\caption{\emph{Top panels}: Figure 4 in Ref. \cite{Salvatelli:2016mgy}. The CMB constraints are shown on the parameters characterizing {\it pure} Horndeski models where the EFT functions are parametrized as the {\it linear-de} form (eq. \eqref{param_de0}) in the  basis of Ref. \cite{Piazza:2013pua}. \emph{Middle and Bottom panels}: Figure 6 in Ref. \cite{Salvatelli:2016mgy}. The same class of models as in the top panels but here the EFT functions are parametrized in the {\it de-1} form (eq. \eqref{param_de}). In all these plots, subscript(s) 1 corresponds to the free parameter(s) describing the evolution of effective Planck mass, 2 kineticity, 3 cubic Galileon coupling and 4 the speed of GWs. As shown in the labels, the green posteriors correspond to the CMB constraints where only ghost and gradient stability conditions are assumed, the red contours have the additional prior $c_s<1$ and blue correspond to the addition of both $c_s<1$ and $c_T<1$ ($=c_t$ in this review) priors.}
\label{fig:salvcons}
\end{center}
\end{figure}

A parameterization of the EFT basis also in terms of {\it linear-de} form (eq. \eqref{param_de0}) has been considered with a $\Lambda$CDM background. For this model \cite{Salvatelli:2016mgy}, the constraining power of viability priors with CMB data are put in perspective. Setting the kineticity to zero so as to be in the most restrictive setup in terms of stability, the posterior distributions of the EFT parameters are understood to be mostly driven by the stability priors (see Figure 2 in Ref. \cite{Salvatelli:2016mgy}). On top of this, asking for sub-luminal propagation of scalar and tensor perturbations reduces drastically the marginalized contours of the parameters (see Figure \ref{fig:salvcons} top panels). Despite the different parameterization with respect to the model previously discussed \cite{Bellini:2015xja,SpurioMancini:2019rxy}, the tendency of the data to favor a lower effective Planck mass and a lower propagation speed  for tensor modes is recovered. Furthermore, with the aim to be more general, one can be led to increase by one the number of free parameters characterizing each EFT function. This is done expanding in powers of $\Omega_{\rm m}-\omo$ for instance and retaining up to two free parameter per EFT function following the {\it de-1} parameterization (eq. \eqref{param_de}). This has the consequence of loosening the posterior constraints as depicted in Figure \ref{fig:salvcons} (middle and bottom panels).  From this figure it is also clear that the sub-luminal propagation of perturbations can impact the data constraints and even limit the loosening of the constraints when more free parameters are considered.

\begin{figure}[!]
\begin{center}
\includegraphics[scale=0.35]{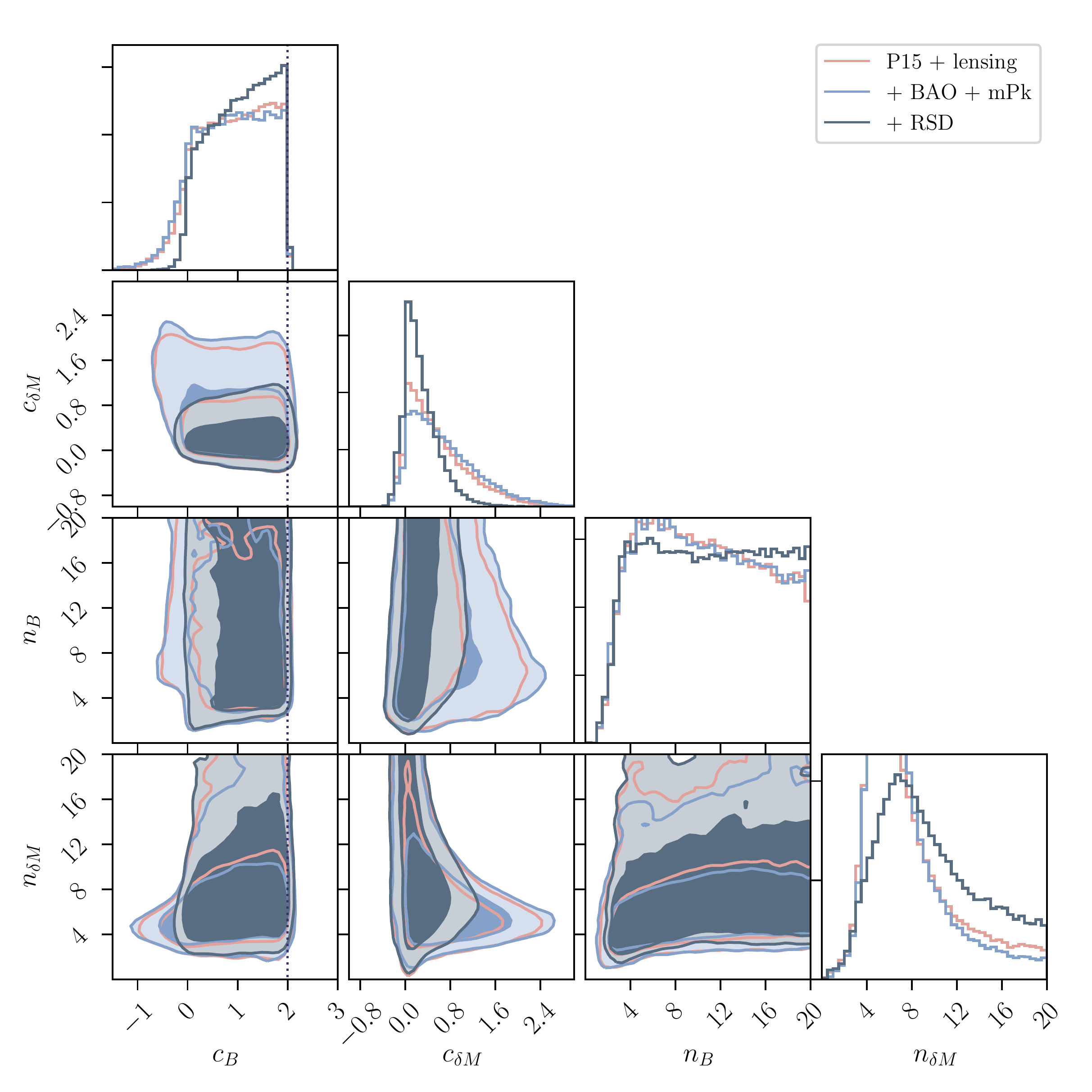}
\caption{Figure 2 in Ref. \cite{Noller:2018wyv}. Constraints on a {\it pure} Horndeski model with $c_t^2=1$ on a $\Lambda$CDM background and EFT functions modeled with the {\it scaling-a} form (eq. \eqref{param_scalinga}). P15 and mPk stand respectively  for Planck 2015 data and  data on the shape of the matter power spectrum from SDSS DR4.\label{fig:nollercons}}
\end{center}
\end{figure}

Constraining {\it pure} Horndeski models with a large set of cosmological probes tends to show that CMB data is only next to RSD data in terms of constraining power \cite{Kreisch:2017uet,Noller:2018wyv}. Within CMB data itself, the constraints on the EFT functions are mostly driven by the ISW effect while the inclusion of RSD has the important effect of breaking degeneracies between $\am$ and $\ab$ (when $\alpha_T=0$) \cite{Noller:2018wyv}. This result is illustrated in Figure \ref{fig:nollercons} \cite{Noller:2018wyv} where the constraints are obtained using the {\it scaling-a} parameterization (eq.\eqref{param_scalinga}) of the $\alpha$-functions. Assuming luminal propagation of tensor modes, RSD data sets tight bounds on the evolution of the running of the Planck mass disfavoring large positive values. Interestingly, the integrated components of the CMB probe tend to provide more sensitivity to the scaling of the EFT functions while combining with RSD data ameliorates the stringent bounds on their present value. When $\alpha_T \neq 0$, data prefer sub-luminal values for the speed of GWs and a larger viability region for the running Planck mass is possible. The intricate link between $\am$ and $\ab$ can be further constrained when considering the radiative stability of Horndeski theories \cite{Noller:2018eht}.

Focusing further on RSD constraints on MG, the recent release of measurements of $f$ and $\sigma_8$ separated thanks to galaxy-galaxy lensing by VIPERS \cite{delaTorre:2016rxm} and SDSS \cite{Shi:2017qpr}  give constraints on {\it pure} Horndeski models as competitive as the full set of $\fs$ data \cite{Perenon:2019dpc}. The models investigated  display $c_t^2=1$ and are modeled using the {\it de-1} form (eq. \eqref{param_de}) with a background expansion set to $\Lambda$CDM. The combination of this new set of data with the full $\fs$ recollection yields a gain on the precision of the constraints on the EFT parameters of at least 20\% with respect to the $\fs$ set alone. In parallel, a stringent bound from Solar System tests on the evolution of the Newton constant, $|\dot{G}_{\rm N}/\gn|< 0.002\;H_0$, is shown to translate into a prior on the unscreenable contribution to gravity, $\musc$, or equivalently the running Planck mass, which in turns increases the constraining power of $f$ and $\sigma_8$. In parallel, this prior stands as drastic cut on the weaker gravity the effective Planck mass can induce. Nevertheless, {\it pure} Horndeski theories are still able to produce lower growth relative to $\Lambda$CDM with $\sigma_{8,0}$ found to be lower than the Planck $\Lambda$CDM cosmology at more than 2$\sigma$. 

Regardless of the parameterization chosen, the kineticity $\alpha_K$ or equivalently $\gamma_1$ (see Appendix  \ref{App:params}) has been shown to have a weak impact on observables. This coupling is thus mostly unconstrained even though it bears a fundamental role for the viability of the model considered \cite{Bellini:2015xja,Kreisch:2017uet,Frusciante:2018jzw,Perenon:2019dpc}. The kineticity does not affect the constraints on the other parameters even when considering a large combination of data sets as displayed in Figure \ref{fig:bellcons} \cite{Bellini:2015xja}. The difficulty in constraining the kineticity coupling has been further investigated \cite{Kreisch:2017uet,Frusciante:2018jzw}. It has been shown in particular that it cannot be constrained by data directly because its contribution to observables is below the cosmic variance \cite{Frusciante:2018jzw}. In principle, a way to overcome the cosmic variance limitation could be to use sophisticated multi-tracer techniques \cite{Camera:2016cpr}. Despite this feature, the kineticity coupling has a significant role in defining the stable parameter space as it enters in the definition of the no-ghost condition for the scalar sector and thus regulates the speed of propagation of scalar modes. As consequence, it changes the viable parameter space used in the MCMC explorations and it cannot be simply discarded \cite{Kreisch:2017uet,Frusciante:2018jzw}.  Furthermore, bounds on $\ak$ or $\gamma_1$ can be obtained using on top of stability conditions the prior on $\gn$, $f$, $\sigma_8$ and $\fs$ data sets \cite{Perenon:2019dpc}. These significantly restrict the space of viable models which in turn put indirect constraints on $\ak$ or $\gamma_1$.

Moving beyond the Horndeski landscape, constraints on {\it pure} GLPV models with $c_t^2=1$ are derived using the {\it linear-de} form (eq. \eqref{param_de0}) with both $\Lambda$CDM and CPL backgrounds \cite{Traykova:2019oyx}. A combination of CMB, BAO and RSD datasets shows that the GLPV parameter $\alpha_H$ is degenerate with the braiding parameter $\alpha_B$ and the running Planck mass $\alpha_M$ but not with either the cosmological parameters nor the CPL ones. The marginalized distributions of $\hat{\alpha}_H$ (the constant parameter of $\alpha_H$) excludes the GR limit ($\hat{\alpha}_H=0$) in most of the cases analyzed, the data constrain $\hat{\alpha}_H$ to be of $\mathcal{O}(1)$ and favoring generally positive values. However, no statistically significant preference of this {\it pure} GLPV model over $\Lambda$CDM is found. 

The previous constraints are \textit{de facto} parameterization dependent. Interestingly,  it is possible  to minimize such dependence  by exploiting  data-driven reconstruction techniques. In these approaches indeed the number of parameters increases largely, but the constraining power of data can be explored more faithfully. Such an approach is used to constrain Horndeski models and sub-classes (GBD, Quintessence, K-essence and Kinetic Gravity Braiding) by making use of the EFT formulation \cite{Raveri:2019mxg}. The EFT functions are modeled on a fixed time grid and interpolated thanks to piece-wise fifth order spline. The information from the specific theories are encoded thanks to a Gaussian smoothing kernel defined by the CPZ correlation \cite{Crittenden:2011aa}. Doing so, the EFT functions identifying the models are reconstructed across cosmic times using cosmological data and could thus be used to derive specific model properties.  This data-driven approach notably highlights that the constraining power of present cosmological probes such as CMB, WL, BAO, SNIa and local measurements of $H_0$, is not only high for Horndeski but is enhanced as the complexity of the theory considered increases. In other words, the more EFT functions are considered, the more freedom is allowed, the more the data can express their constraining efficiency. In parallel, the analysis also shows that Horndeski theories might alleviate the tension in $H_0$ at low redshifts while the sub-classes do not. 

%-------------------------------------------------
\subsection{Forecasts with next generation surveys }\label{sec:forecasts}
%--------------------------------------------------

\begin{figure}[!]
\begin{center}
\includegraphics[scale=0.42]{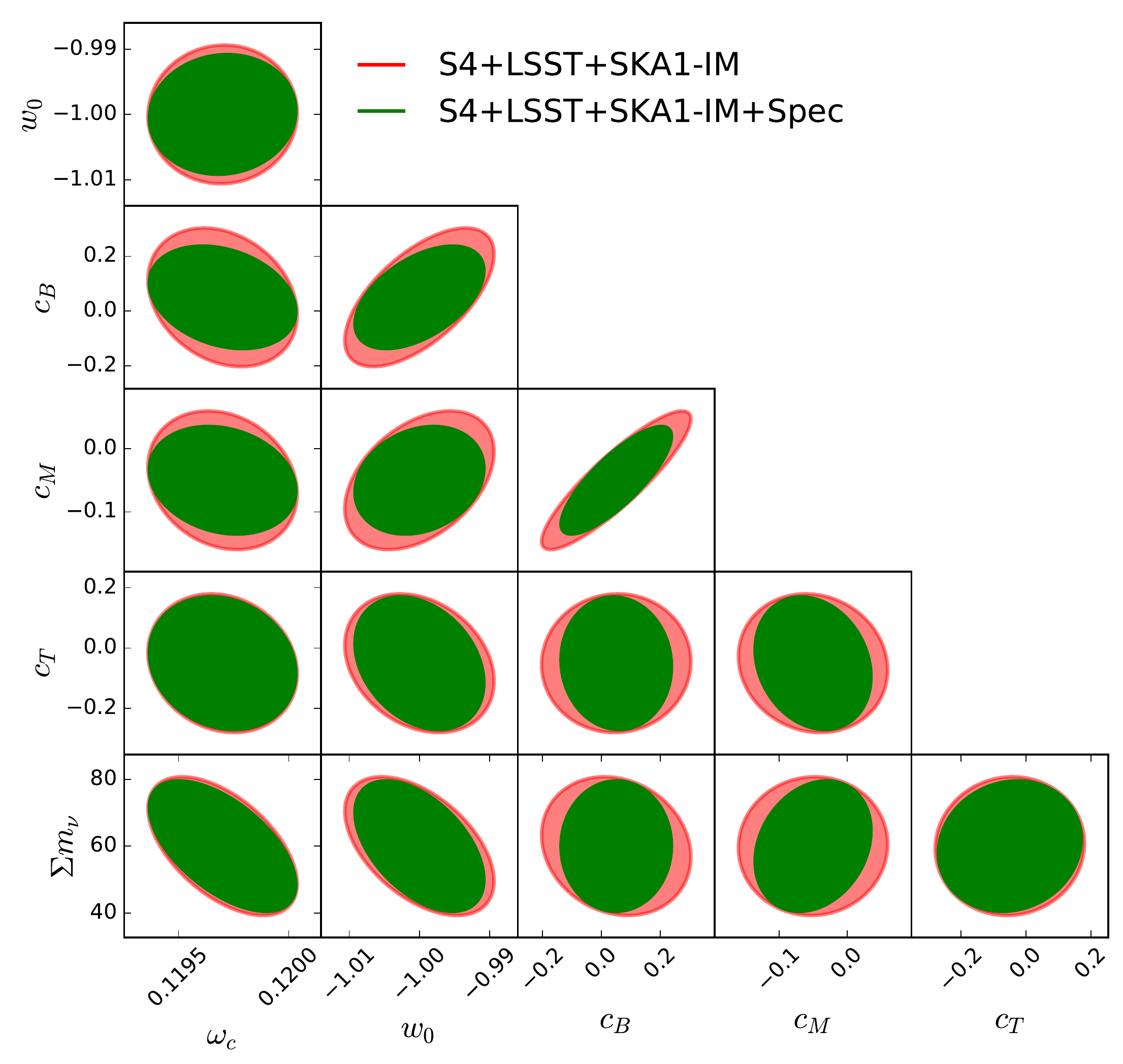}
\includegraphics[scale=0.65]{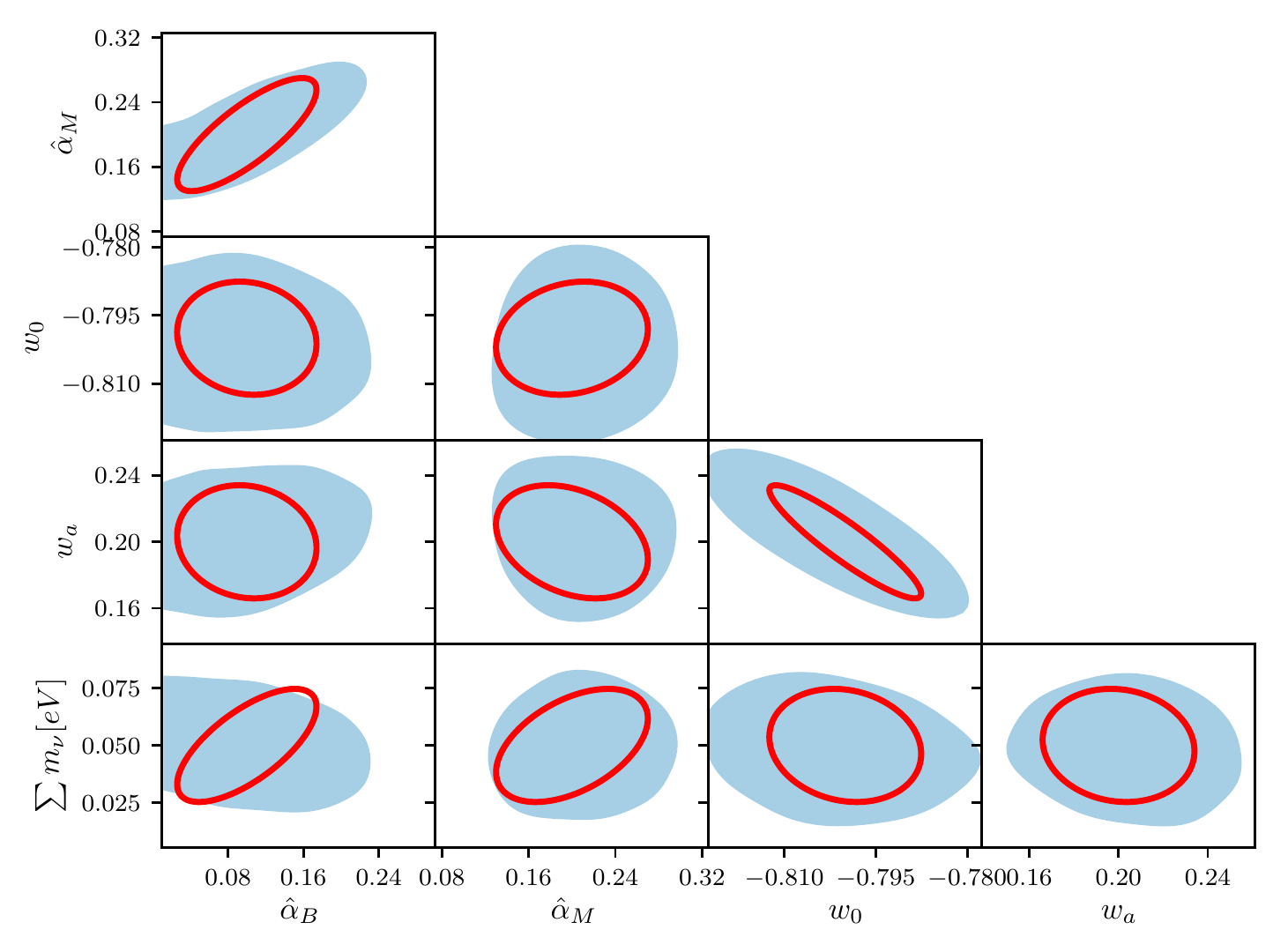}
\caption{\emph{Top panel}: Figure 1 in Ref. \cite{Alonso:2016suf}. The cosmological constraints on {\it pure} Horndeski models parametrized with the {\it linear-de} form (eq. \ref{param_de0}) with a $w$CDM background for the combination of CMB-S4, LSST (galaxy clustering and shear) and an intensity mapping experiment by SKA Stage 1 are shown in pink. Adding BAO and growth rate measurements from a DESI-like experiment produces the green contours. \emph{Bottom panel}: Figure 4 in Ref. \cite{Reischke:2018ooh}. 68\% constraints region on the free parameters of {\it pure} Horndeski models with $c_t=1$ modeled with the {\it linear-de} form (eq. \ref{param_de0}) and a CPL background obtained from a MCMC analysis (blue contours) and Fisher analysis (red ellipses).}
\label{fig:hornforecast1} 
\end{center}
\end{figure}

Next generation surveys will probe the Universe at extended and complementary redshifts and scales delivering highly accurate data and offering an unprecedented insight into gravity on cosmological scales. Exploiting their potentiality and investigating the improvement in constraining cosmological parameters is one of the main goals of current investigations. Since the EFT framework does not rely on a specific model but allows to make general prediction on large classes of models, it provides a powerful benchmark for  forecasting the cosmological signals to which the future missions will give access to. In the following, we review the cosmological forecasts analyses performed using the \textit{pure} EFT approach \cite{Gleyzes:2015rua,Alonso:2016suf,Leung:2016xli,Abazajian:2016yjj,Reischke:2018ooh,Mancini:2018qtb,Frusciante:2018jzw}. Also in this case, the class of models that has been largely explored belongs to the {\it pure} Horndeski models.

\begin{figure}[!]
\begin{center}
\includegraphics[scale=0.8]{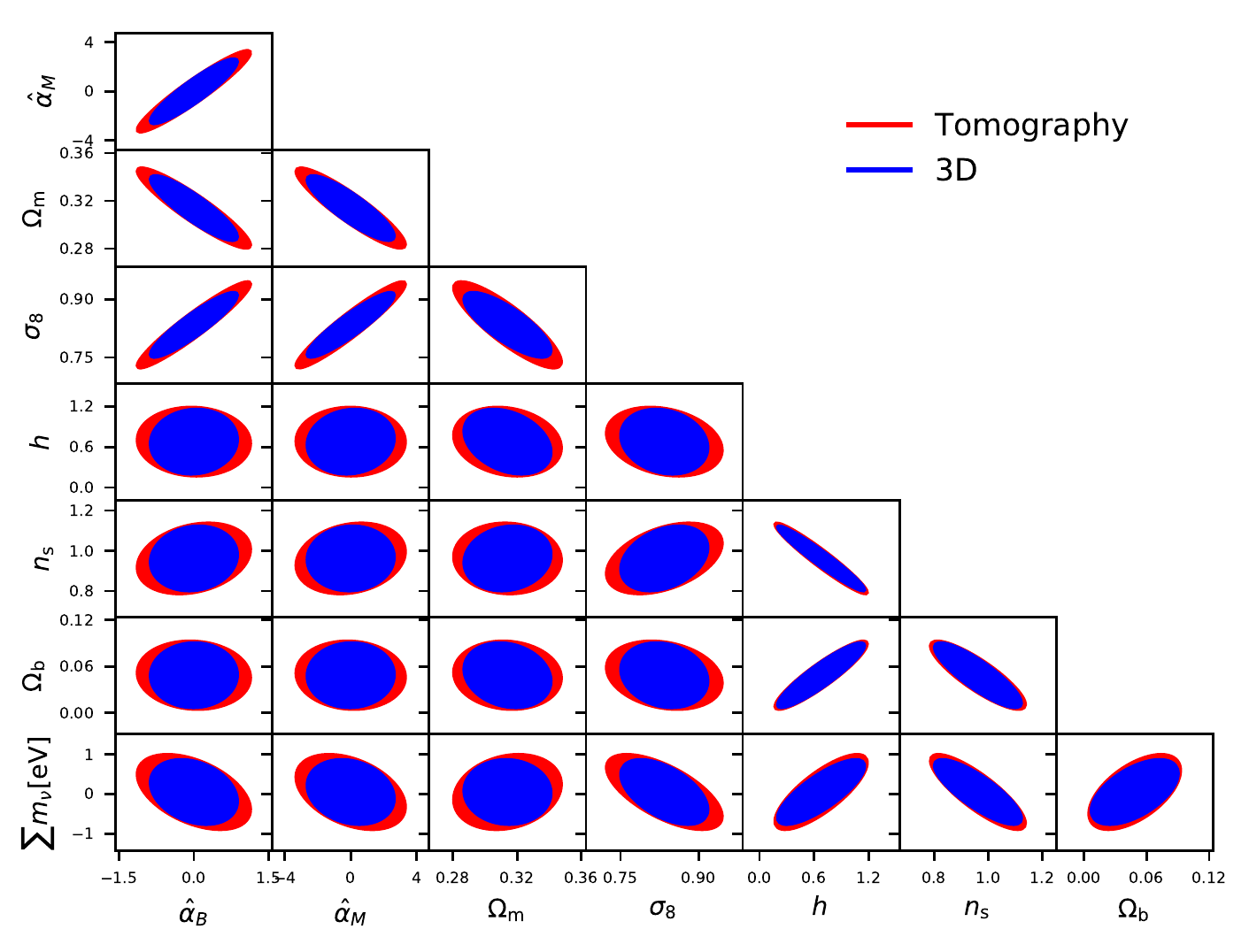}
\includegraphics[scale=0.8]{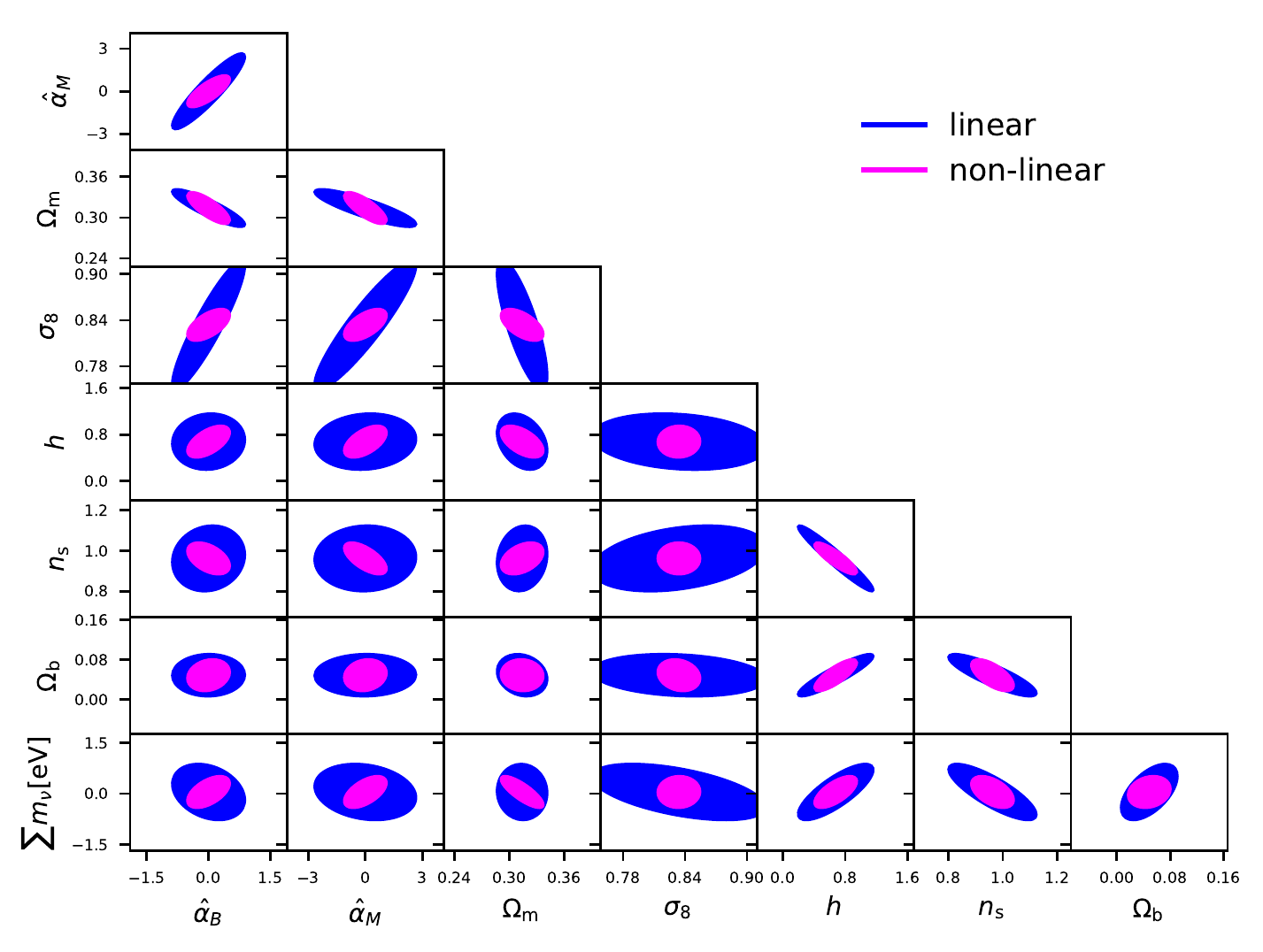}
\caption{Figure 6 and 8 in Ref. \cite{Mancini:2018qtb}. 68\% C.L. of Fisher forecasts from a Euclid-like survey for {\it pure} Horndeski models with $c_t=1$ parametrized with the {\it linear-de} form (eq. \ref{param_de0}) including massive neutrinos. The {\it top} panel compares the constraints obtained from the 3D and tomographic approaches while {\it bottom} compares the use of linear versus non-linear power spectra.}
\label{fig:hornforecast2} 
\end{center}
\end{figure}

The future Stage IV photometric redshift surveys LSST, the radio galaxy survey SKA and CMB-S4 experiments have been used to obtain forecasts for the Horndeski models in the $\alpha$-basis parametrized as follows \cite{Alonso:2016suf}
\be\label{scalealpha}
\alpha_i(t,k)=\alpha_i^0 \,\f{\omde(t)}{\omdeo}\,e^{-\f{1}{2}\l(\f{k}{k_V}\r)^2}\,,
\ee
where $\Omega_{\rm DE}(t)$ is chosen to follow $w$CDM and $\alpha_i^0$ are constant. This modeling includes a phenomenological assumption for screening on small scales. The scale of the Vainshtein screening mechanism $k_V$, above which GR is recovered, has been set to be 0.1 h/Mpc according to numerical simulations \cite{Barreira:2013eea}. Including screening effects helps, for instance, not to overestimate the surveys capacity to test gravity on such scales. The results of the analysis display an improvement in the constraints by a factor of 5 with respect to previous results based on present day surveys \cite{Bellini:2015xja}. In particular, in Figure \ref{fig:hornforecast1} (top panel) one can observe the effects of BAO and RSD data by an independent DESI-like experiment on top of CMB-S4, LSST and SKA measurements on the parameter constraints. Tighter constraints are obtained by setting $c_t=1$ \cite{Reischke:2018ooh} as shown in Figure \ref{fig:hornforecast1} (bottom panel). Such constraints exclude variations of the effective Newtonian constant larger than 10\%   over the age of the Universe \cite{Reischke:2018ooh}. However, the contours obtained from the MCMC analysis are larger than those obtained from a Fisher analysis. This points to the danger of using Fisher forecasts for non-Gaussian likelihoods in forecast investigations. Finally, the model is investigated using two cosmic shear methods \cite{Mancini:2018qtb}: the tomographic method, where the correlations between the lensing signal in different redshifts bins allow to keep track of the redshift information, and the 3D approach, where all the redshift information is carried throughout the analysis. For an Euclide-like experiment, the 3D analysis is shown to constrain the model better than the tomographic approach by about 20\% thanks to the increased redshift information (see Figure \ref{fig:hornforecast2}). The size of the constraints on $\Omega_{\rm m}$, the sum of neutrino mass and EFT functions is further improved by including non-linear corrections in the power spectrum as displayed in Figure \ref{fig:hornforecast2} (bottom panel).

The synergy between future CMB-S4 and DESI has been explored on both the \eftcamb and $\alpha$-basis for the Horndeski class of models \cite{Abazajian:2016yjj}. Two parameterizations of the EFT functions are considered: constant and a time varying behavior with a smooth transition between early and late time values (see eq. \eqref{edform}). When modeling the $\alpha$-functions as constants, CMB-S4 surveys are found to provide the tightest bounds on the effective mass Planck and tensor speed excess respectively by a factor 1.5 with respect to DESI and 2.5 with respect to Planck, while DESI and Planck show the same sensitivity. On the contrary, the Planck measurements are slightly stronger than CMB-S4 on the constant $\alpha_B$. When considering the time varying parameterization, CMB-S4 measurements are sensitive to both early and late time values while the former are, as expected, more efficiently constrained by the CMB surveys than LSS ones. In the \eftcamb basis and for the constant parameterization, Ref. \cite{Abazajian:2016yjj} finds that the sensitivity of CMB probes are unmatched when constraining the parameter $\Omega_\mathrm{0}^\mathrm{EFT}$. Additionally, CMB-S4 measurements better constrain $\gamma_0^{(2)\mathrm{EFT}}$ and $\gamma_0^{(3)\mathrm{EFT}}$. For the time varying parameterization, late time values of the $\gamma_i$ parameters do not change as compared to the constant case, while the forecast bounds on $\Omega_\mathrm{0}^\mathrm{EFT}$ slightly degrades. At early time, the constraints are mostly constrained by physical viability requirements. 

Different time behaviors for the EFT functions $\{\Omega,\gamma_1,\gamma_2\}$ of the \eftcamb basis have been explored further by considering also the constraints on GWs which suggest $\gamma_3=0$ \cite{Frusciante:2018jzw}. The remaining three EFT functions are modeled firstly with the {\it scaling-a} parameterization and then with the {\it de-density} form (respectively eq. \eqref{param_scalinga} and eq. \eqref{dedensity}) both on a CPL background. A constraint analysis is performed using present day observational data (Planck+BOSS DR12+$H_0$+JLA+KiDS) and forecasts from combinations of GC and WL for a prototype of next generation galaxy surveys with specifications like DESI and SKA2. These future surveys will be able to increase the precision on constraints of model parameters by one order of magnitude in both the parameterizations \cite{Frusciante:2018jzw}.

\begin{figure}[!]
\begin{center}
\includegraphics[scale=0.4]{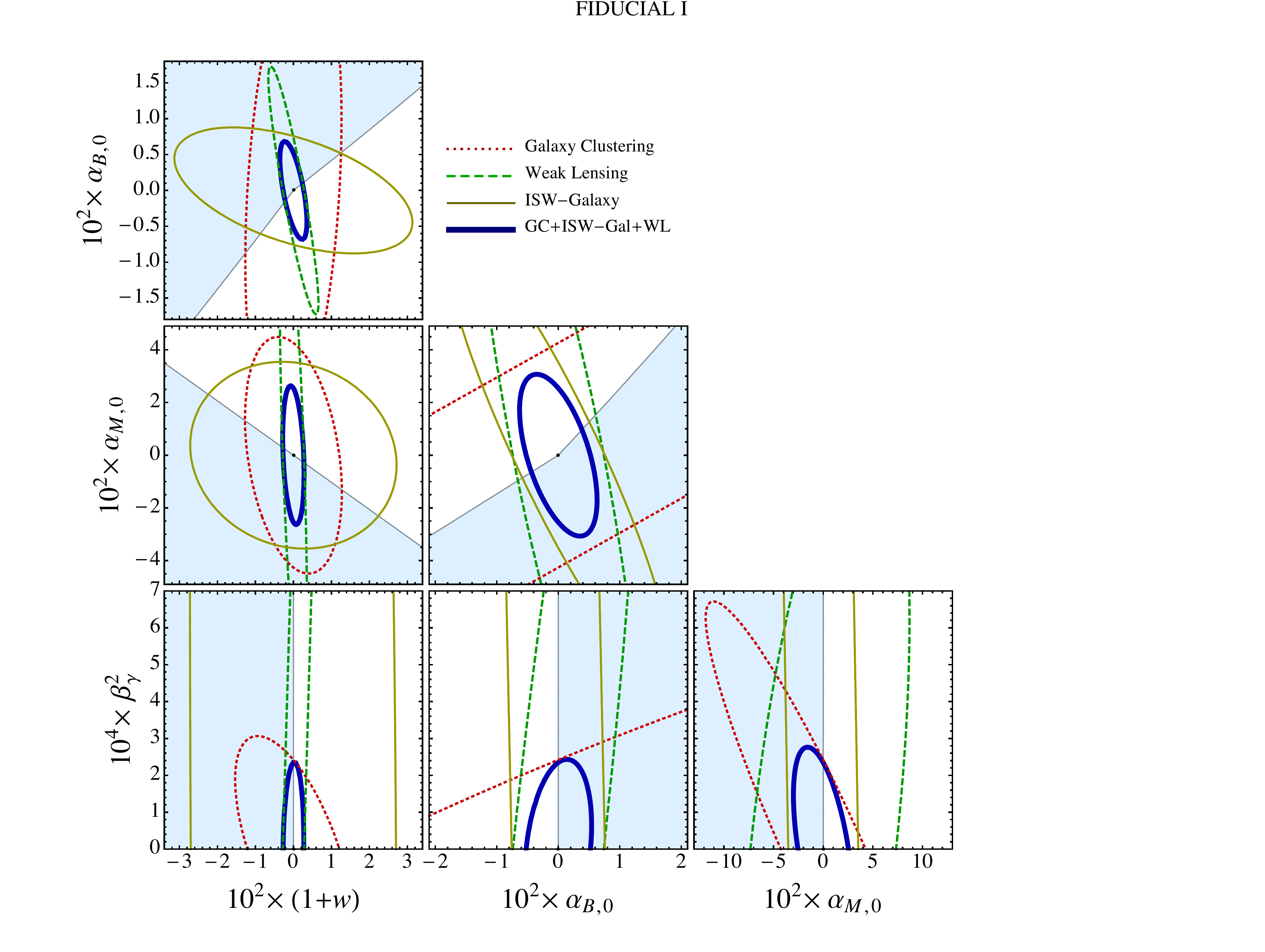}
\caption{Figure 3 in Ref. \cite{Gleyzes:2015rua}. 68\% C.L. contours for the Fisher Euclid-like forecasts on the {\it pure} Horndeski $c_t=1$ model with a direct coupling between DE and DM (see eq. \eqref{eq:dmdecoup}) and the $\Lambda$CDM fiducial background. $\alpha_M$ and $\alpha_B$ are modeled with the {\it linear-de} form (eq. \eqref{param_de0}). The blue region corresponds to the unstable region.}
\label{fig:disformal} 
\end{center}
\end{figure}

We conclude the section discussing the constraints on modifications of gravity when a direct coupling between DE and DM is present. In Ref. \cite{Gleyzes:2015rua}, the {\it linear-de} form (eq. \eqref{param_de0}) of the $\alpha$-functions is considered on top of a $w$CDM expansion history. The dark matter coupling function is parametrized as follows:
\be\label{eq:dmdecoup}
\gamma_c(t)=\f{\beta_\gamma}{2\sqrt{2}}c_s(t)\sqrt{\alpha(t)}\,,
\ee 
where $\beta_\gamma$ is a constant and $c_s(t)$ and $\alpha(t)$ correspond to the DE speed of propagation and kinetic coupling. The forecasts analysis is performed using the specifications of an Euclid-like survey for which three probes are considered: the galaxy power spectrum in redshift space, tomographic weak-lensing shear power spectrum and the correlation spectrum between the ISW effect and the galaxy distribution. This is done selecting three fiducial models: $\Lambda$CDM, a braiding and an interacting model. We display the results for the $\Lambda$CDM fiducial in Figure \ref{fig:disformal} where one can observe how the parameters are degenerate, yet some degeneracies can be effectively broken by combining all three observational probes. $1\sigma$ constraints on the MG parameters are of order $\sim10^{-2}-10^{-3}$ (68\%C.L.) for the first two fiducial models and one order better in the interacting fiducial model. The error on the dark matter coupling parameter is $\sim 10^{-4}$ in all cases. One can also appreciate from Figure \ref{fig:disformal} how an Euclid-like survey gives an order of magnitude tighter constraints than CMB-S4 experiments. Finally, the non-minimal coupling of DE enhances the effects of modification of gravity and reduces the statistical errors accordingly \cite{Gleyzes:2015pma}.

%----------------------------------------------------------
\subsection{Gravitational waves }\label{sec:impactsGW}
%----------------------------------------------------------

Alternatives to GR can lead to two specific modifications in the propagation of GWs: the first is the friction term affected by the running Planck mass, $i.e.$ $\am$, the second is the speed of propagation, $c_t^2=1+\at$. The former alters the amplitude of the GWs and the latter its phase. In detail, the evolution equation of the GWs on cosmological scales reads in Fourier space \footnote{A mass term can be included in the equation of GWs \cite{Ezquiaga:2018btd} but this does not hold for the classes of theories considered in this review.} 
\be\label{eq:gweq}
\ddot{h}^T_{ij}+(3+\am)H\dot{h}^T_{ij}+(1+\at)\f{k^2}{a^2}h^T_{ij}=0\,,
\ee
where $h_{ij}^T$ is the tensor component of the spatial metric. 

$\am$ and $\at$ can lead to observable effects on CMB, on both the temperature and the polarization spectra at any time. In particular, information on primordial GWs is encoded in the B-modes at larger scales \cite{Bennett:2012zja,Liu:2014mpa,Adam:2014bub,Ade:2014zja}. The modification in the speed of propagation at early time can affect the position of the inflationary and of the reionization peaks in the B-modes \cite{Amendola:2014wma,Raveri:2014eea}. Furthermore, along with the expansion rate $H$, the running Planck mass can damp GWs: for the class of Horndeski models, if $\am < 0$ the GW amplitude is smaller than that predicted by the standard scenario, the opposite holds if $\am > 0$ \cite{Nunes:2018zot}. A modified amplitude then introduces a difference in the GWs and electromagnetic luminosity distance, which can be tested by GWs experiments and standard sirens \cite{Abbott:2017xzu,Nissanke:2013fka,Lagos:2019kds}. The next generation Laser Interferometer Space Antenna (LISA) \cite{Audley:2017drz} would constrain a constant $\am$ at redshift 1.5 with a precision varying between 0.03 and 0.13 independently of the underlying cosmological model \cite{Amendola:2017ovw}. A recent analysis combining LISA with CMB+BAO+SNIa data shows that $\am$ can be measured to an accuracy reaching 1.1\% \cite{Belgacem:2019pkk}. Joint measurements with standard sirens will not only directly constrain the running Planck mass but also give constrains on $H_0$ which might also help in resolving the $H_0$ tension~\cite{Riess:2011yx,Riess:2016jrr,Abbott:2017wau,Abbott:2017smn,Delubac:2014aqe}. We refer the reader to \cite{Ezquiaga:2018btd} for a review on GWs astronomy.

The tensor-to-scalar ratio $r$ is degenerate with $\am$ as they both affect the amplitude of the primordial peak \cite{Pettorino:2014bka}. Considering the Background Imaging of Cosmic Extragalactic Polarization 2 (BICEP2) data \cite{Ade:2014xna} in order to have $r$ close to zero, $\am$ must assume negative values, in particular it goes towards $-2$. However, negative values of the running Planck mass enhance the CMB BB spectra at large $\ell$, where the BICEP2 data would favor instead smaller values. This feature needs to be considered in light of other probes on scalar perturbations, indeed as extensively discussed in previous Sections $\am$ enters in the growth of structure, lensing and ISW effect. For example, considering the Brans Dicke theory, the background and scalar perturbations give $0\leq \am\leq0.01$ and the degeneracy with $r$ is removed \cite{Pettorino:2014bka}. General cases do not allow for such tight bounds hence the B-modes can provide useful constraints on early-time MG. Planck and BICEP2 datasets constrain $c_t^2 =1.30\pm0.79$ and $c_t^2 <2.85$ at 95\%C.L. by assuming a power law primordial tensor power spectrum and $c_t^2 < 2.33$ at 95\% C.L. if the running of the spectral index is allowed \cite{Raveri:2014eea}. Forecasts for the next generation CMB satellites Cosmic Origins Explorer mission (COrE) \cite{Bouchet:2011ck} and Polarized Radiation Imaging and Spectroscopy Mission (PRISM) \cite{Andre:2013nfa}, will be able to constrain the speed of GWs at percent level \cite{Raveri:2014eea}. 

As discussed previously, the modification of the speed of propagation of GWs generates a gravitational slip $\eta$ different from unity. The relationship between an anomalous speed of GW and $\eta$ is investigated in three cases \cite{Sawicki:2016klv}: Bi-metric, Einstein-Aether gravity and Horndeski theories. Horndeski theories are studied in their EFT formulation and are shown to a have enough freedom to hide dynamically the modification in the gravitational slip whenever $\at\neq0$, but at the cost of making the perturbations evolve towards a divergent kinetic term. On the contrary, the other theories do not offer this possibility. This result can be used as theoretical argument for the interpretation of future observations if they favor $\eta=1$.

Finally, the anomalous speed of GWs can be used to break the degeneracy between MG and DE behaviors within Horndeski theories, $i.e.$ singling out self-accelerating models \cite{Lombriser:2015sxa}. The linear shielded Horndeski model \cite{Lombriser:2014ira} retains this property but recovers a $\Lambda$CDM background expansion history and exhibits $\mu (t, k) = \eta (t, k) = 1$ at the linear perturbations level in the QS regime. Within this scenario, the region allowing for self-acceleration is very narrow. In particular, for these models to have cosmic acceleration attributed to genuine MG, the present tensor speed has to be $<95\%$ than that of light and $\am$ $ 5\%$ less efficient than in $\Lambda$CDM \cite{Lombriser:2015sxa}. Note that when imposing $c_t^2=1$, these models produce a $3\sigma$ poorer fit to cosmological observables as compared to $\Lambda$CDM \cite{Lombriser:2016yzn}, challenging the concept of self-acceleration within Horndeski theories. This model leads to the conclusion that surveys probing LSS and the background expansion alone are not sufficient to ultimately discriminate between $\Lambda$CDM and such modifications of gravity. Nevertheless it highlights how the addition of current and future GWs detections can break crucial degeneracies induced by MG and offer complementary means to test gravity. 

%-----------------------------------------------------
\subsection{Constraints on $\mu$, $\Sigma$ and $\eta$}\label{sec:musigmaconstraints}
%-----------------------------------------------------

\begin{figure}[!]
\centering
\includegraphics[scale=0.45]{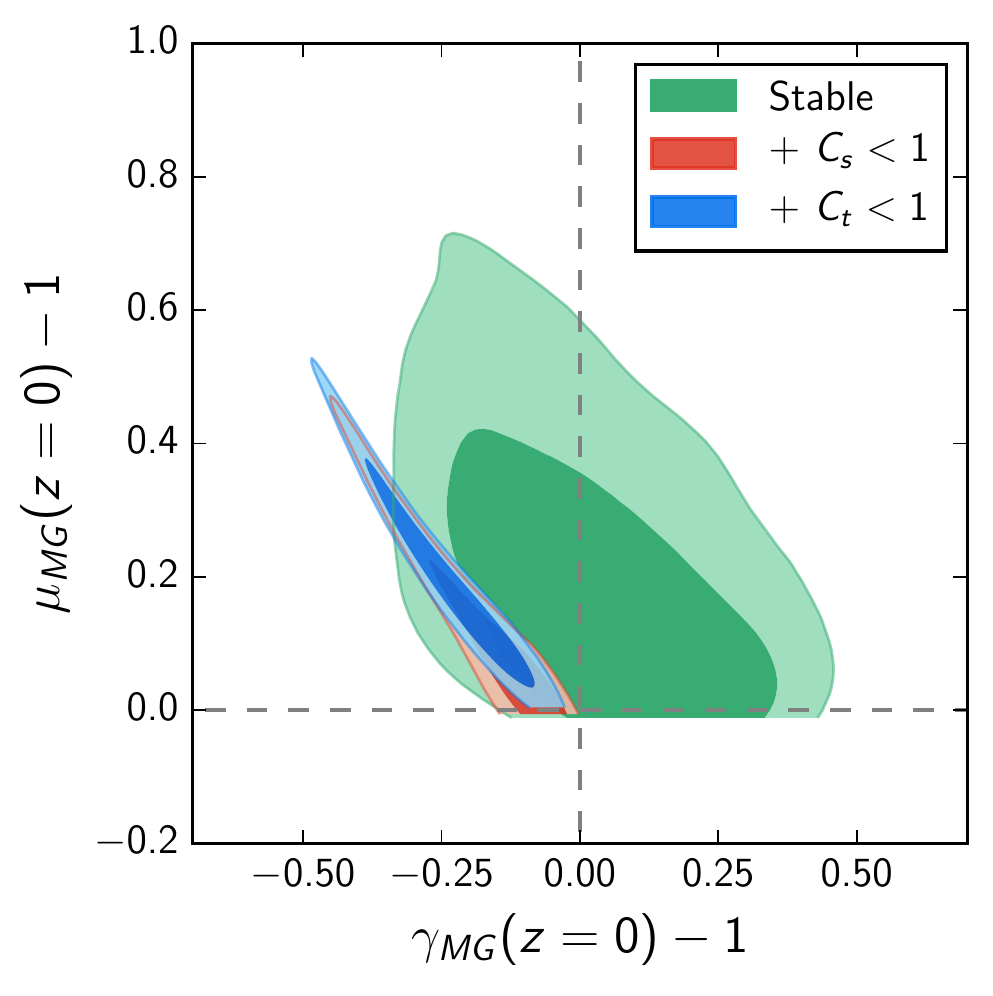}
\hskip6mm
\includegraphics[scale=0.45]{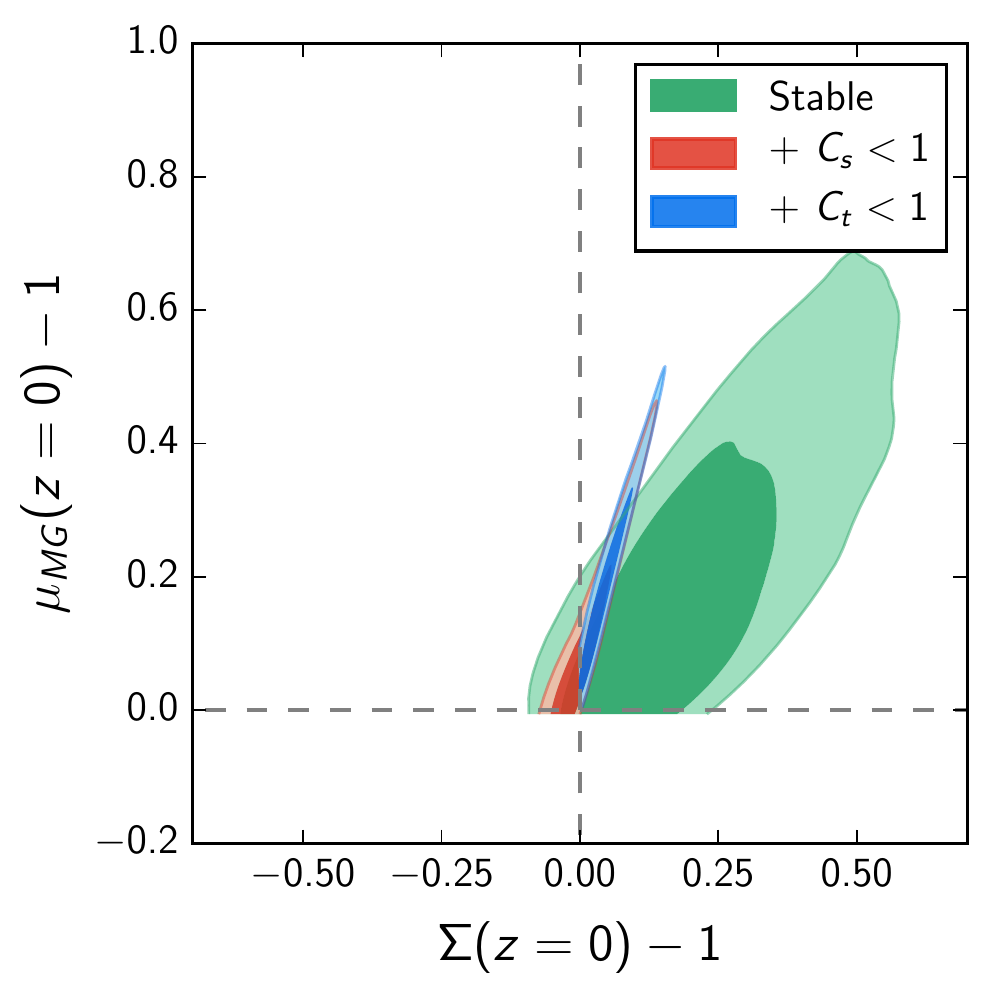}
\vskip4mm
\includegraphics[clip, trim = 0cm 11cm 0cm 0cm,scale=0.34]{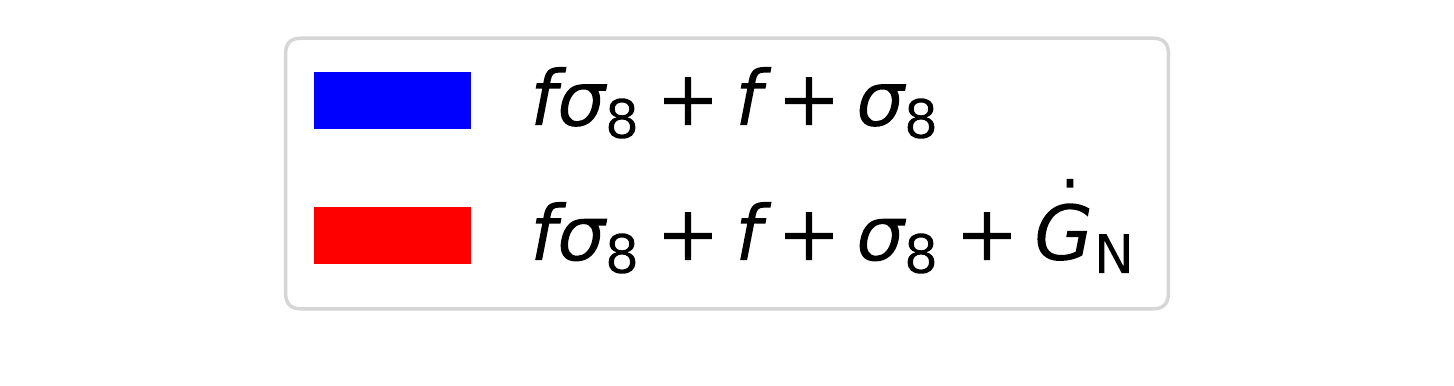}\\
\includegraphics[scale=0.3]{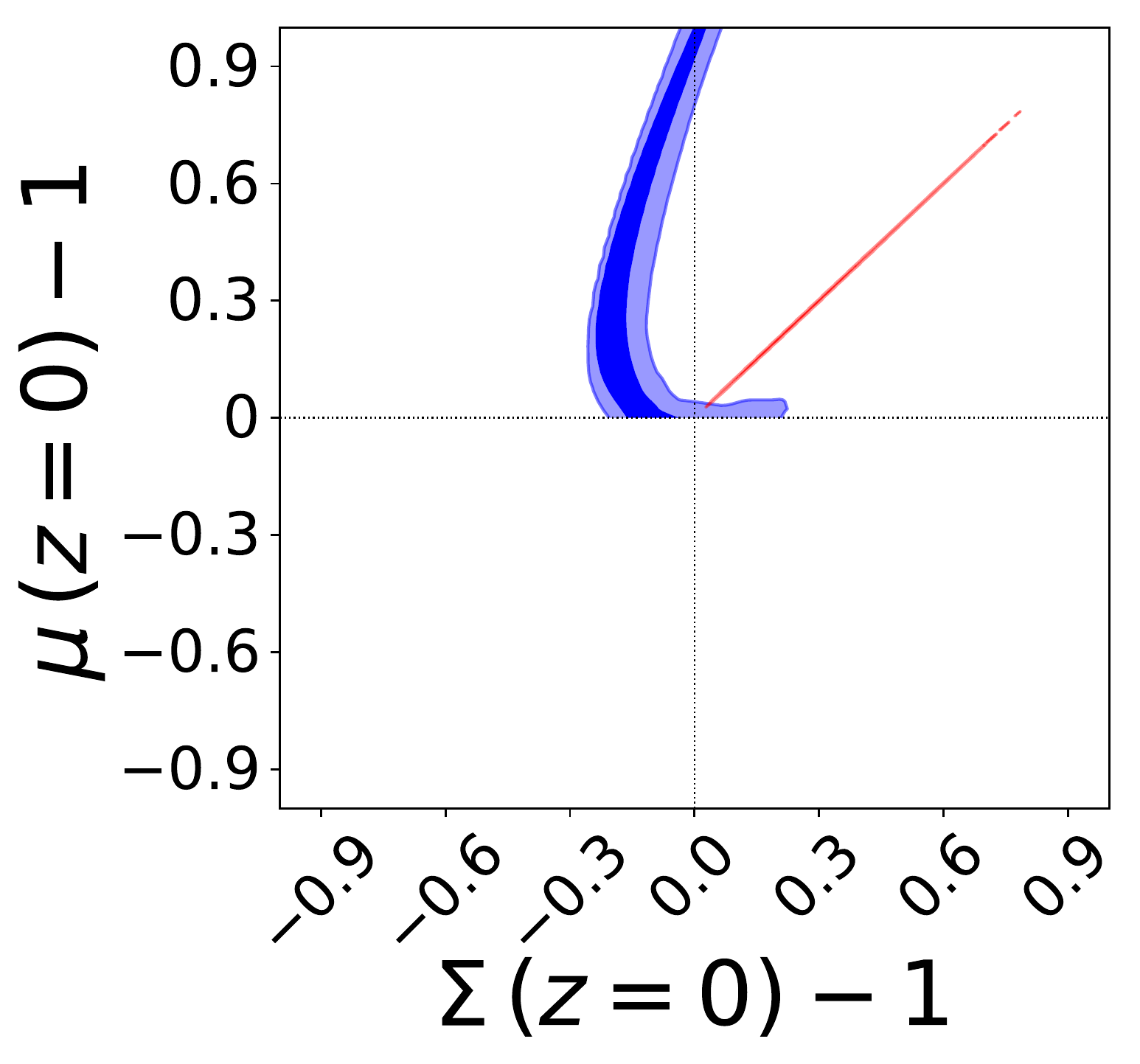}
\includegraphics[scale=0.3]{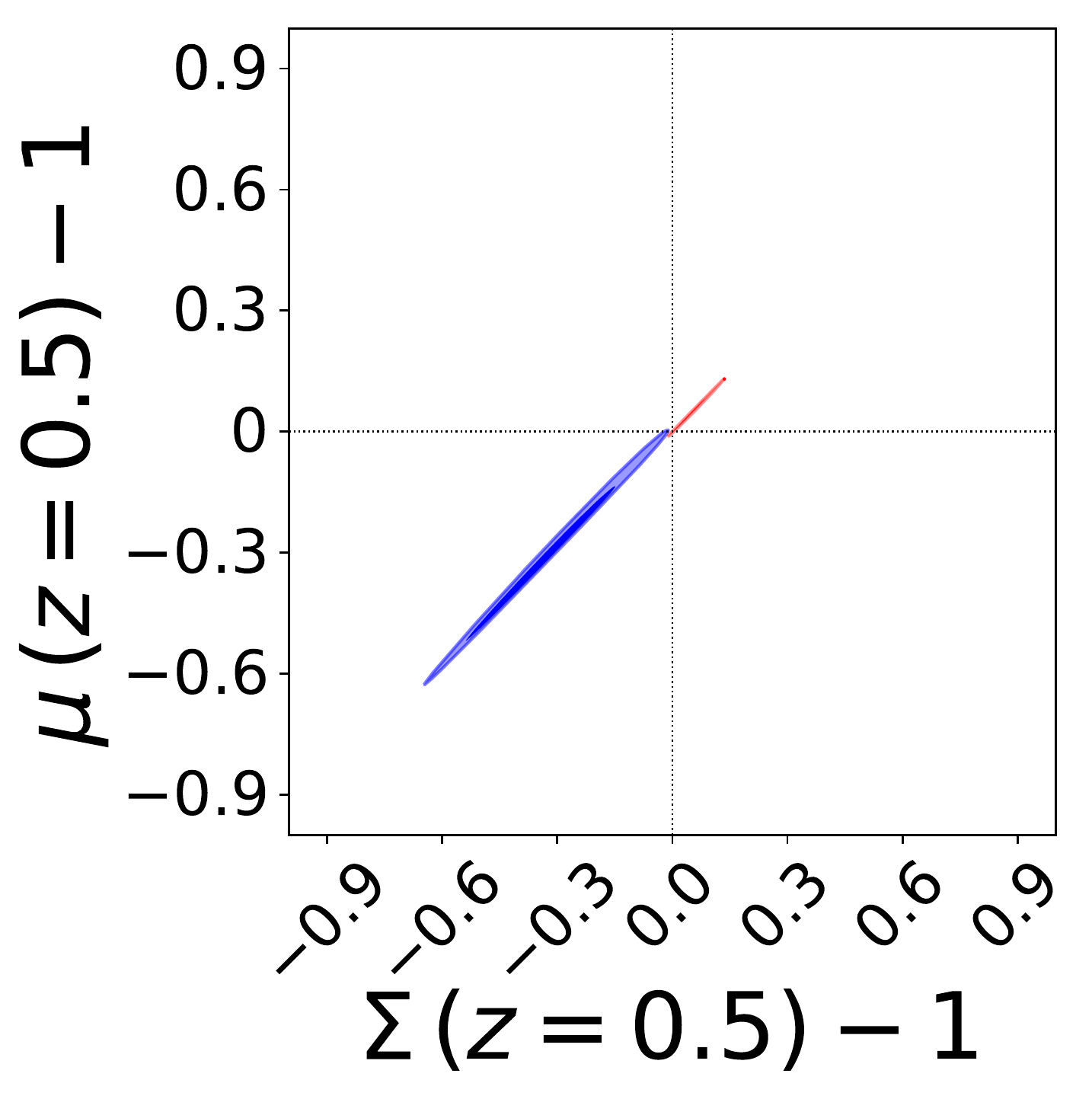}
\includegraphics[scale=0.3]{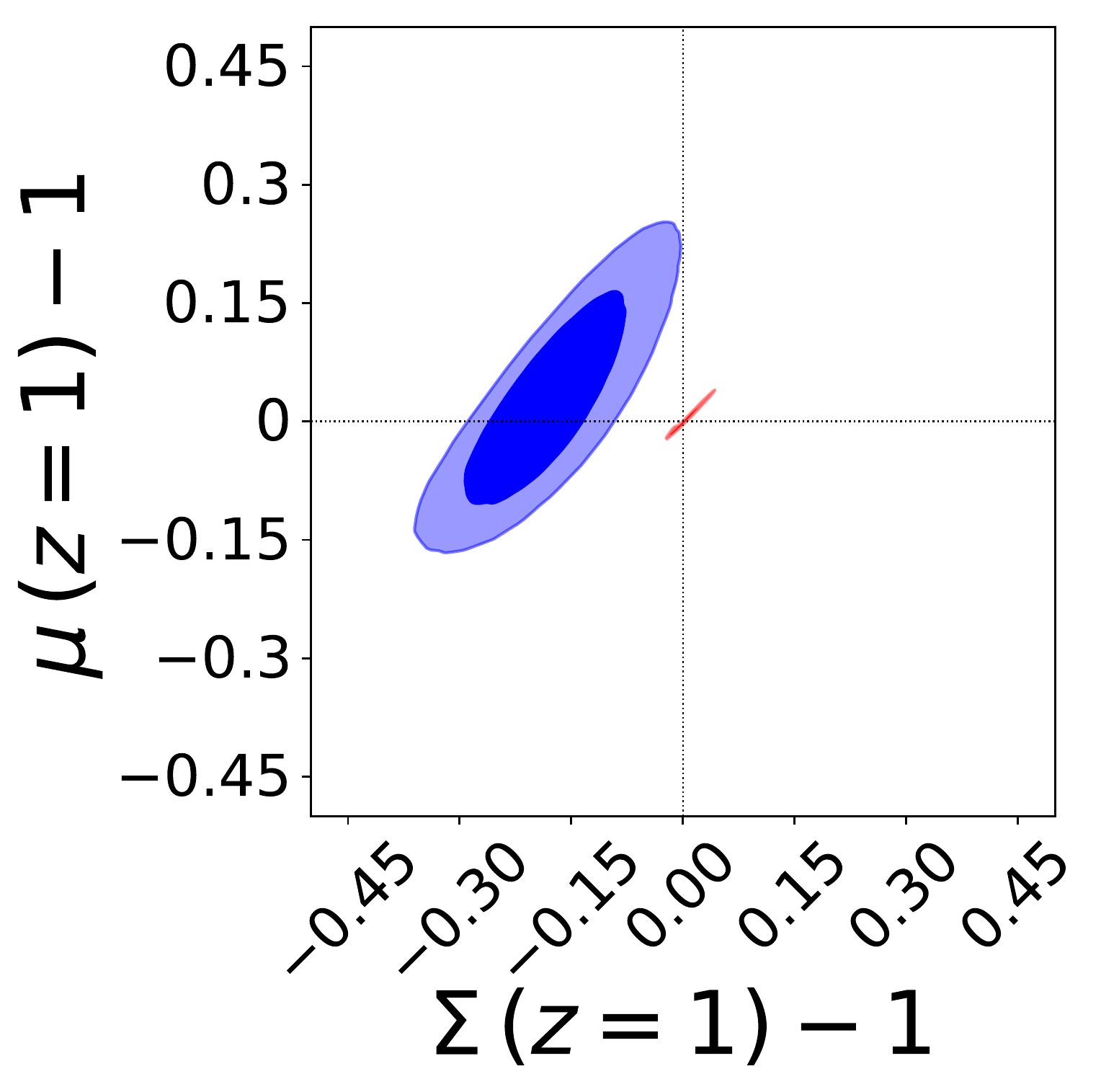}
\includegraphics[scale=0.3]{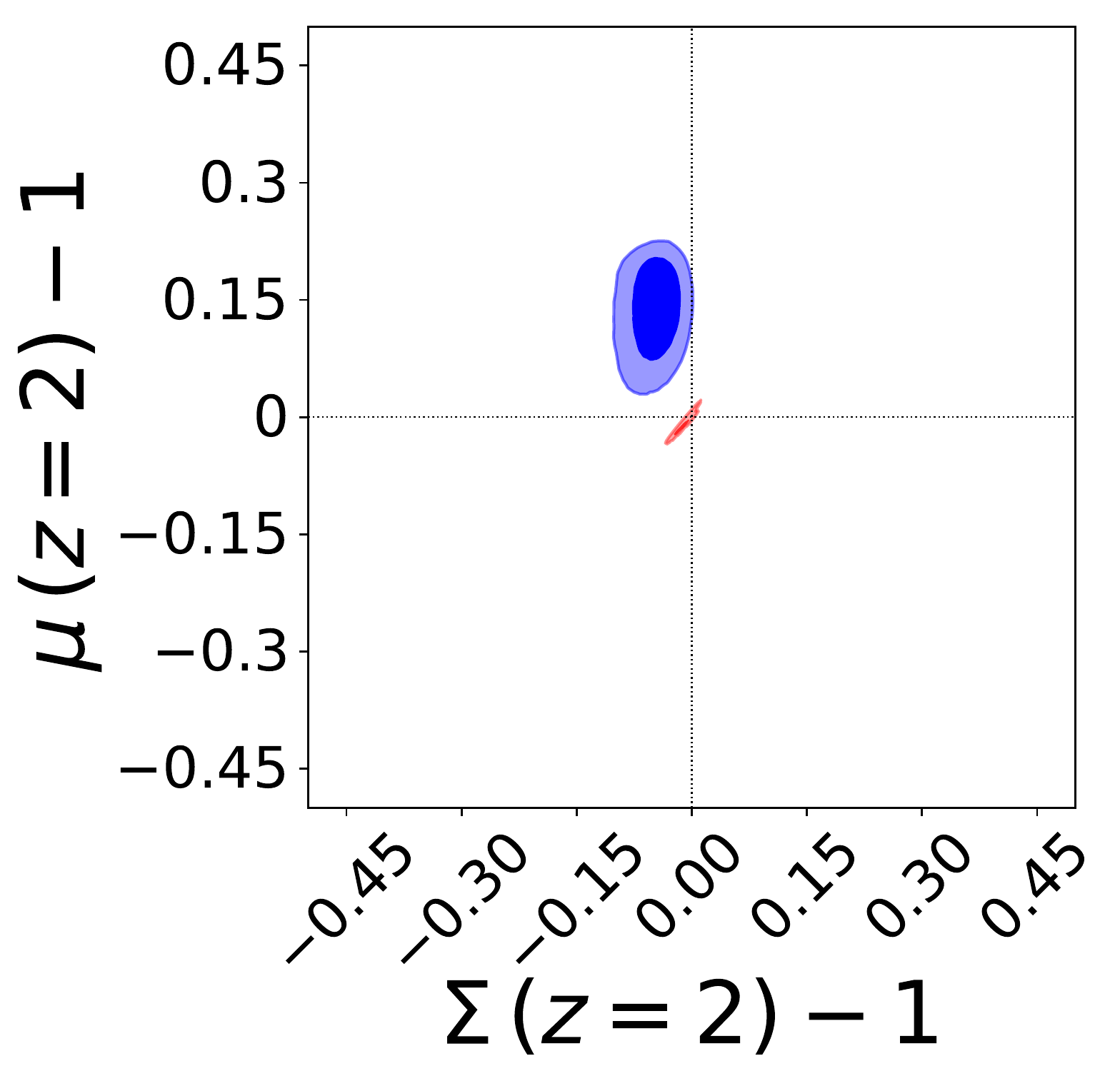}
\caption{\emph{Top panels}: Figure 7 in Ref. \cite{Salvatelli:2016mgy}. 68\% and 95\% C.L CMB constraints on $\mu_{MG}(z=0)$, $\Sigma_{MG}(z=0)$ and $\gamma_{MG}(z=0)$ (in this review $\mu,\Sigma,\eta$ respectively). The EFT functions follow the {\it linear-de} form (eq. \eqref{param_de0}) on $\Lambda$CDM background. Different combinations of stability conditions are imposed: stable (no ghost and no gradient) (green), plus the sub-luminal propagation for scalar mode ($c_s<1$) (red) and plus sub-luminal propagation of tensor mode ($c_t<1$) (blue). \emph{Bottom panels}: Figure 5 in Ref. \cite{Perenon:2019dpc}. RSD constraints are compared with (red) and without (blue) the prior on the variation of the Newton constant for {\it pure} Horndeski models with $c_t=1$ parametrized using the {\it de-1} form (eq. \eqref{param_de}) on a $\Lambda$CDM background.}
\label{fig:hornlsscons}
\end{figure}

A systematic investigation of possible DE/MG extensions to GR can be performed using the $\mu,\Sigma,\eta$ phenomenological approach~\cite{Amendola:2007rr,Bean:2010zq,Silvestri:2013ne,2010PhRvD..81j4023P} reviewed in Section~\ref{sec:gravcouplings}. Signatures of deviations from the cosmological standard model can then be captured by constraining these functions directly.  Despite this approach being model-independent, the way in which constraints on these functions are obtained might instead be model-dependent: one has to choose somehow the underlying gravity model in order to obtain cosmological constrains on either $\mu,\Sigma$ or $\eta$. One can indeed make use of the EFT formulation to compute the constraints. In this case, given a parameterization of the EFT functions one can then translate the constraints on the EFT parameters into the phenomenological functions using the relations~\eqref{muSigma}-\eqref{slip} in Section~\ref{sec:gravcouplings}. Alternatively, one can parameterize directly the phenomenological functions. In the latter case, the connection with a specific model is lost, however depending on the chosen parameterization,  $\mu,\Sigma,\eta$ can mimic the predictions of certain class of theories by appropriately choosing their time and scale dependencies. A public EB code which allows to do so is Modified Growth with \texttt{CAMB} (\texttt{MGCAMB})\footnote{ \texttt{MGCAMB} webpage: \url{www.aliojjati.github.io/MGCAMB/}}~\cite{Zhao:2008bn,Hojjati:2011ix}. Let us note that in both cases the constraints are highly sensitive to the particular parameterizations and priors assumed.   A third approach instead considers  $\mu,\Sigma,\eta$ as free functions (e.g. using a principal component analysis \cite{Zhao:2009fn}) and as such the resulting constraints are completely model-independent. In this Section we focus on the first two approaches.

Firstly let us consider the approach where the constraints are derived within the EFT formulation. In principle, starting from each of the results discussed in Sections \ref{sec:horncons}-\ref{sec:forecasts} one can deduce the corresponding constraints on $\mu,\Sigma$ or $\eta$. Major advantages in using the EFT approach to constrain these phenomenological functions come from the possibility to: impose the appropriate stability conditions to guarantee the viability of the chosen model (see Section~\ref{sec:impactstab}); use observational and experimental priors such as those on the effective Planck mass, speed of GWs and Hubble parameter. Their impact on the $\mu-\Sigma$ or $\mu-\eta$ planes is largely demonstrated in Section \ref{sec:novpred} and when such requirements are used as priors in MCMC analysis they show a strong constraining power \cite{Raveri:2014cka,Frusciante:2015maa,Salvatelli:2016mgy,Perenon:2019dpc}. We show an example in Figure \ref{fig:hornlsscons} (top panel) \cite{Salvatelli:2016mgy} where the $c^2_s\le 1$ and $c^2_t\le 1$ priors push notably the constraints for {\it pure} Horndeski models in the $\eta(z=0)<1$ and $\Sigma (z=0)<1$ quadrant. This is in agreement with the sign conjecture on $\mu$ and $\Sigma$ \cite{Pogosian:2016pwr} discussed in Section \ref{sec:phenomenology}. The conjecture is also confirmed for {\it pure} Horndeski models with $c^2_t=1$ using the Solar-System prior on the variation of the present day value of the effective Planck mass combined with RSD data and a high redshift CMB prior \cite{Perenon:2019dpc}. The Solar-System prior allows notably to obtain a stringent bound on the present value of the gravitational slip parameter $\gsp = 1.0000 \pm{9.3 \times 10^{-4}} $ (95\% C.L.). Having the effective Planck mass stringently bounded at low redshifts by the Solar-System prior and at high redshifts by the CMB one, the redshift evolution of $\gsp$ is constrained by RSD data to be close to unity across matter domination. This implies $\mu$ and $\Sigma$ to be almost equal across redshifts as shown in Figure \ref{fig:hornlsscons} (bottom panels). The fifth force contribution $\muff$ is favored at more than $2\sigma$ at present time from the bound $\mu(z=0) = \muff(z=0) = 1.321^{+0.370}_{-0.284}$ (95\% C.L.). 

\begin{figure}[!]
\centering
\includegraphics[scale=0.5]{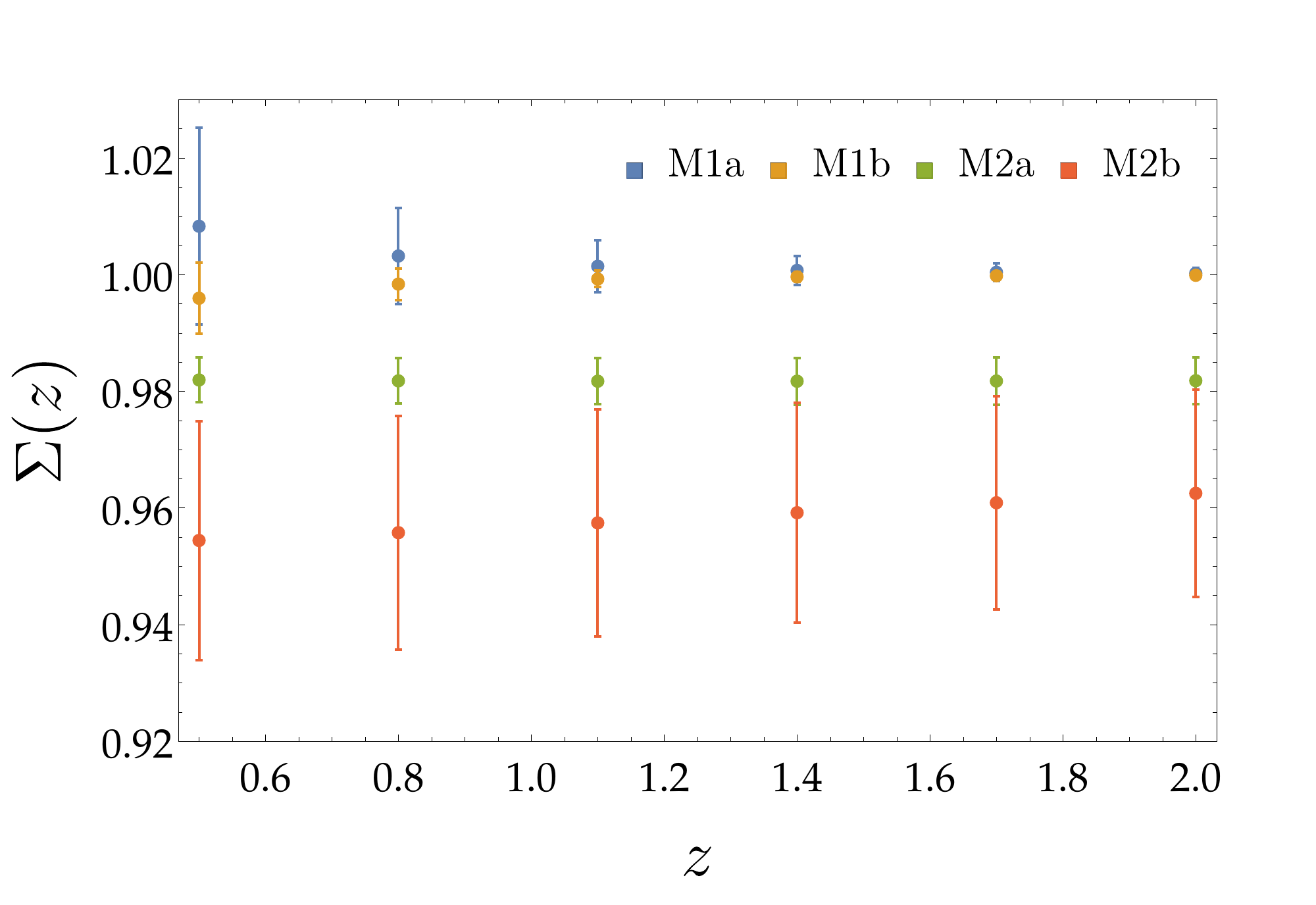}
\caption{Figure 8 in Ref. \cite{Frusciante:2018jzw}. Forecasts 2$\sigma$ errors from DESI and SKA2 like-surveys on the $\Sigma$ function as function of redshift. The models M1 and M2 are {\it pure} Horndeski models with $c_t^2=1$ parametrized respectively as eqs. \eqref{param_scalinga}-\eqref{dedensity} on a CPL background. The labels a and b differentiate for the number of EFT functions active: (a) $\{\Omega,\gamma_i=0\}$, (b) $
\{\Omega,\gamma_i\}$.}
\label{fig:hornlssforecasts}
\end{figure}

In terms of next generation galaxy surveys with specifications like DESI and SKA2, forecasts on $\mu$ and $\Sigma$ obtained from {\it pure} Horndeski models with $c_t^2=1$ parametrized as in eqs. \eqref{param_scalinga}-\eqref{dedensity} on a CPL background are found to reach the 1\% level \cite{Frusciante:2018jzw}. In Figure \ref{fig:hornlssforecasts} \cite{Frusciante:2018jzw}, we show the forecasted 2$\sigma$ errors on $\Sigma(z)$. For the {\it scaling-a} models (M1), the errors are of $\mathcal{O}(10^{-3})$, and decrease towards $\mathcal{O}(10^{-4})$ for $z > 1.5$. This is because $\Sigma(z)$ asymptotically tends to unity, independently of the cosmological parameters, which implies very small errors. For the {\it de-density} form models (M2), the errors are constant in $z$. One can notice that future data will allow to distinguish these EFT models from $\Lambda$CDM at more than 3$\sigma$, assuming the best-fit values obtained with present day data (Planck+BOSS DR12+$H_0$+JLA+KiDS) hold \cite{Frusciante:2018jzw}. 

Let us now concentrate on the second approach, $i.e.$ a direct parameterization of the phenomenological functions. A parameterization commonly used in the literature is \cite{Simpson:2012ra,Ade:2015rim,Ferte:2017bpf,Aghanim:2018eyx} 
\begin{equation}\label{eq:modindeplss}
\Xi(z)=1+\Xi_0 \times \frac{\omde(z)}{\omde(z=0)}\; ,
\end{equation}
where $\Xi$ stands for either $\{\mu,\Sigma,\eta\}$. The latter has been considered by the Planck Collaboration \cite{Ade:2015rim,Aghanim:2018eyx} and the best fit resulting from the combined analysis of several cosmological probes deviates from $\Lambda$CDM by more than $2\sigma$. The constraints obtained suggest that $\mu (z=0)<1$ and $\eta (z=0)>1$. This is explained by the preference of data for lower values of $\sigma_{8,0}$ relative to the standard model (see discussion in Ref. \cite{Salvatelli:2016mgy} and related references therein). Note that using a different data set combination this tension was not recovered \cite{Ferte:2017bpf}. Other examples of direct parameterizations and corresponding constraints can be found in Refs. \cite{Song:2010fg,Casas:2017eob}. An important aspect to note is that  each chosen direct parameterization is differently sensitive to the redshift evolution according to the time scaling of the model thus the tightness of the constraints obtained can indeed vary significantly \cite{Song:2010fg}.

The direct parameterization of the phenomenological functions has some advantages: it provides a null test of the standard cosmological model and enables to clearly identify the tendencies of data with a low amount of free parameters. However, being not based on a gravitational theory, stability and normalization requirements or additional observational priors cannot be considered or are hard to implement. These instead are easily included in the EFT approach. Comparing the results from Planck \cite{Ade:2015rim,Aghanim:2018eyx} and the results obtained from the Monte-Carlo generation of viable {\it pure} Horndeski models discussed in Sections \ref{sec:phenomenology}-\ref{sec:impactstab}, it is clear that the marginalized posterior distributions obtained with the model in eq.~\eqref{eq:modindeplss} dwell in a region where Horndeski theories are mostly absent. Further imposing the normalization of the effective Planck mass to the Planck mass, $i.e.$ $\musc(z=0)=1$ prevents to recover models with  $\mu (z=0)<1$ since $\muff(z)>1$ as shown in Figure \ref{fig:hornlsscons}. Additionally, in a given theory, $\mu,\Sigma,\eta$ are not independent as eq.~\eqref{eq:linkphenofunc} highlights and are hence bound to be correlated. Direct parameterizations of these phenomenological functions do not allow to retain the link they share affecting the constraints. Using instead the EFT approach allows to overcome this issue. However in both approaches the time evolutions of these phenomenological functions might result in simplified behaviors which poorly capture the ones obtained from full covariant theories, $e.g.$ scalar-tensor theories \cite{Linder:2015rcz,Perenon:2015sla,Linder:2016wqw}. The risk here is to miss information in the data. A way to bypass this issue would be to extract model-independent information about the phenomenological functions from current and future data. An example to minimize the model dependence is to consider them as unknown functions of both redshift and scale, and bin them into a large number of narrow bins on a grid in the $(z, k)$ space \cite{Huterer:2002hy,Zhao:2009fn,Hojjati:2011xd,Crittenden:2005wj}. This approach would however miss most of the advantage described above and has its own limitations.

%--------------------------------------------
\subsection{Specific modified gravity models} \label{sec:constraintMG}
%--------------------------------------------

The EFT formalism has a twofold face when implemented in EB codes, indeed the background dynamics can be solved either with a \textit{pure} approach  as discussed in the previous Sections or by implementing a specific background solver for a chosen theory, $i.e.$ the \textit{mapping} approach. In the latter case thanks to the {\it mapping} procedure discussed in Section \ref{sec:mapping}, the EFT functions are fully specified. Let us note that the \textit{mapping} approach can also involve a \textit{designer} procedure to solve the background. 
In the following we review specific DE/MG models implemented in EB codes using the {\it mapping} approach. 

\begin{itemize}
\item \textit{Quintessence} \cite{Tsujikawa:2013fta}: the action for quintessence reads
\begin{align}
S_Q = \int{}d^4x\sqrt{-g}\l[\frac{\mp^2}{2}R-\frac{(\nabla\phi)^2}{2}-V(\phi)\r] \,,
\end{align}
where $V(\phi)$ is the potential of the scalar field $\phi$. It is possible to show that the freedom in choosing the functional form of the potential can be replaced by the choice of the equation of state, indeed in the EFT formalism one gets
 \be
c = \frac{1}{2} \rho_{\rm DE}(1+w_{\rm DE})\, , \qquad \Lambda = V(\phi_0) = \rho_{\rm DE}\, .
\ee
Then, one has only to specify the expansion history and use a \textit{designer} approach to investigate minimally coupled quintessence models. The model is implemented in \texttt{EFTCAMB}. Using the CPL background, the stability conditions alone drastically limit the allowed range of variation of the CPL parameters $\{w_0,w_a\}$ (removing the $\w<-1$ space) and drive the observational constraints \cite{Peirone:2017lgi}. As a result, the region of the CPL plane that would be favored by WL data is eliminated, thus quintessence becomes significantly disfavored with respect to the standard cosmological model. An extension of the viability region of CPL could be performed with multifields quintessences models which might provide a better fit to data. Furthermore, the no-ghost, no gradient and mathematical\footnote{Mathematical conditions are a set of requirements implemented in \eftcamb  worked out at the level of the dynamical equation for the perturbation of the scalar field $\pi$. They guarantee the stability of the theory in absence of the full set of physical conditions. Such condition can be replaced by the no-tachyonic condition which instead is a well physical motivated condition \cite{Frusciante:2018vht}.} help reducing the tension between KiDS and Planck data sets \cite{Peirone:2017lgi}. 

\item \textit{Designer $f(R)$-gravity}~\cite{Song:2006ej,Bean:2006up,Pogosian:2007sw,DeFelice:2010aj}: the action for $f(R)$-gravity is shown in eq. \eqref{fRaction} with the corresponding mapping into the EFT framework. The model is implemented in \eftcamb using the \textit{designer} approach \cite{Raveri:2014cka}. On a $w$CDM background, the free parameters are $B_0,w_0$, $i.e.$ the present day values respectively of the \textit{scalaron} and the equation of state for the DE. The results show that using a combination of data from Planck temperature and lensing potential spectra, WMAP low-$\ell$ polarization spectra (WP), and BAO, the constraints on the DE equation of state are $w_0 \in (-1, -0.9997)$ at 95\% C.L.. In particular, the lower bound is the result of the requirement of stability conditions which induce a strong correlation between $B_0$ and $w_0$ (see Figure \ref{fig:stabregion2} top panel), indeed when $Log_{10} B_0<-4$ follows $w_0\rightarrow -1$. Furthermore, the combination of the viability priors and the Planck lensing data allowed to obtain a stringent constraint also on $B_0$: $Log_{10} B_0 = -3.35^{+1.79}_{-1.77}\; (95\,\%\rm C.L.)$. This result was confirmed later in Ref. \cite{Hu:2015rva}.

\item \textit{$f(R)$ Hu-Sawicki model} \cite{Hu:2007nk}: this model is characterized by a specific form of the $f(R)$- function which reads
\be
f(R)=-m^2\f{c_1(R/m^2)^n}{c_2(R/m^2)^n+1}\,,
\ee 
where $m=\rho/3\mp^2$, $n>0$ and $c_{1,2}$ are constant. The above functional form is introduced to mimic $\Lambda$CDM in the high-energy regime and to give rise to an accelerated expansion which is not driven by a true cosmological constant at low-energy. The model is implemented in \eftcamb using the \textit{mapping} approach \cite{Hu:2016zrh}. In the latest analysis \cite{Hu:2016zrh}, three cases are considered: $n=1$, $n=4$, $n$ free, with only one extra free parameter, $i.e.$ the scalaron Compton wavelength $log(-f_R^0 )$, where $f_R^0 \equiv df/dR (z=0)$. The constraints for the baseline datasets, D1 (Planck15, JLA, BAO) are weak in the case $n=1$ ($log (-f_R^0)<-2.7$ at 95\% C.L. ), in the other cases the constraints on $log (-f_R^0)$ are not statistically significant. The inclusion of the WiggleZ data drives the bound on this parameter away from $\Lambda$CDM, for $e.g.$ in the $n=1$ case $log (-f_R^0) = -3.4^{+1.4}_{-1.2}$, at 95\% C.L., while considering D1+CFHTLenS the value of $\log_{10}(-f_R^0)$ is driven back to its $\Lambda$CDM limit, $log (-f_R^0)<-4.5$ at 95\% C.L. (n=1). This analysis reveals for the first time a degeneracy between $\sigma_8$ and $f_R^0$. This was possible thanks to a full implementation of the background solver which is not forced to be $\Lambda$CDM. Interestingly, when $n = 4$ and $log(-f_R^0 ) > -2$, this degeneracy changes in direction, a feature also found in the posterior distribution of $\sigma_8$ and $H_0$. Furthermore, for $log(-f_R^0 ) > -2$, this parameter shows a degeneracy with $H_0$ also, due to the fact that the effect of the background modification is not negligible within this bound.

\item \textit{Hybrid-metric Palatini $f(\hat{\mathcal{R}})$ gravity}~\cite{Harko:2011nh}: the action is constructed by adding a Palatini correction $f(\hat{\mathcal{R}})$ to the usual Hilbert-Einstein term as 
\be
S_{Hf}=\int{}d^4x\sqrt{-g}\f{\mp^2}{2}\l[R+f(\hat{\mathcal{R}})\r]\,,
\ee
where $\hat{\mathcal{R}}\equiv g^{\mu\nu}\hat{\mathcal{R}}_{\mu\nu}$ is the Palatini curvature and $\hat{\mathcal{R}}_{\mu\nu}$ is defined in terms of an independent connection $\hat{\Gamma}^\alpha_{\mu\nu}$. 
The mapping relations using the $\alpha$-basis read~\cite{Lima:2016npg}
\be
\alpha_M=\f{f^\prime_{\hat{\mathcal{R}}}}{1+f_{\hat{\mathcal{R}}}}\,, \qquad \alpha_K=-\f{3}{2}\f{f^\prime_{\hat{\mathcal{R}}}}{f_{\hat{\mathcal{R}}}}\alpha_M\,, \qquad\alpha_B=-\alpha_M\,,
\ee
where prime is the derivative with respect to $\ln\,a$. The background evolution is solved using the \textit{designer} approach with $\Lambda$CDM background \cite{Lima:2016npg}. In this model, early modifications of gravity become significant after recombination while they decay towards the present. Background probes and Planck measurements do not show evidence for such effects. The constraints for the scalar field value are $\vert f_{\hat{\mathcal{R}}}(z=z_\mathrm{on})\vert \lesssim 10^{-2}$ where $z_{on}$ is the redshift at which the decaying early-time modification is introduced ($z_\mathrm{on}\sim (500-1000)$), and $\vert f_{\hat{\mathcal{R}}}(z=0)\vert \lesssim 10^{-8}$ (95\% C.L.) at present time. 

\item \textit{Jordan Brans-Dicke theory (JBD)} \cite{Brans:1961sx}: the JBD action reads
\be
S_{JBD}=\int{}d^4x\sqrt{-g}\l[\phi R-\f{\omega_{BD}}{\phi}\nabla_\mu\phi\nabla^\mu\phi+V(\phi)\r]\,,
\ee
where $\omega_{BD}$ is the constant Brans-Dicke parameter and $V(\phi)$ is a potential. GR is recovered for $\omega_{BD} \rightarrow \infty$. The theory is very well constrained at all scales: $\omega_{BD}>4 \times 10^4$ from Shapiro time delay \cite{Bertotti:2003rm}, $\omega_{BD}>6 \times 10^2$ from Planck \cite{Avilez:2013dxa}. The model is investigated further in Ref. \cite{Alonso:2016suf} with $V=0$ and a tracker solution for the background 
is found where $\phi=\phi_0a^{1/(\omega_{BD}+1)}$ and $\phi_0=(2\omega_{BD}+4)/(2\omega_{BD}+3)$ \cite{Nariai:1969vh}. In the $\alpha$-basis the JBD model can be written as 
\ba
\alpha_M=\f{d\ln\phi}{d\ln a}\,,\qquad \alpha_B=-\alpha_M\,,\qquad \alpha_K= \omega_{BD}\alpha_M^2\,,\qquad \alpha_T=0\,,
\ea
where the tracker solution implies $\alpha_M=1/(\omega_{BD}+1)$. This background evolution and the mapping relations are implemented in \hiclass and the analysis reveals that the combined future Stage-IV surveys will be able to place the bound $\omega_{BD} > 1.7 \times 10^4$ \cite{Alonso:2016suf}, which is comparable to Solar-System and astrophysical tests.

\item \textit{Covariant Galileon (CG)} \cite{Deffayet:2009wt}: the Galileon field $\phi$ was first introduced in flat space with the Galileon symmetry $\partial_\mu \phi \rightarrow \partial_\mu \phi +b_\mu$ which guarantees that the resulting theory posses second order equations of motion~\cite{Nicolis:2008in}. Its generalization to a dynamical space-time leads to the breakdown of such symmetry when demanding the field preserves second order equation of motion~\cite{Deffayet:2009wt}. The resulting theory is called Covariant Galileon (CG). Even though it does not satisfy the original Galileon symmetry it preserves the shift symmetry: $\phi\rightarrow \phi+c$. The CG action reads 
\ba\label{CGaction}
S_{CG}&=&\int{}d^4x\sqrt{-g}\left\{\f{\mp^2}{2}R-\frac{1}{2}c_2 X+\frac{c_3}{M^3} X \Box\phi 
+ \frac{c_4}{4 M^6} X^2 R \r.\nn\\
&&-\left. \frac{c_4}{ M^6}X\left[ (\Box\phi)^2 - \phi^{;\mu \nu} \phi_{;\mu \nu}\right]+\frac{3 c_5}{4 M^9} X^2 G_{\mu \nu}\phi^{;\mu \nu}\r.\nn\\
&&+\left.\frac{ c_5}{2 M^9}X\left[ (\Box\phi)^3 -3\Box\phi \, \phi^{;\mu \nu} \phi_{;\mu \nu}+2  \phi^{;\mu \nu} \phi_{;\mu \sigma}\phi_{;\sigma}^{;\nu}\right]\r\}\,,
\ea
where $G_{\mu\nu}$ is the Einstein tensor, $X=\phi^{;\mu}\phi_{;\mu}$ and $\{;\}$ stands for the covariant derivative. Moreover, $c_i$ are constant dimensionless parameters and $M^3=\mp H_0^2$.  One can fix a canonical normalization by choosing $c_2=-1$~\cite{Barreira:2014jha}.
Note the original version of the action also includes a $c_1\phi$ term, but if one wants the late time acceleration to be driven by the field kinetic energy, one should set $c_1 = 0$. The above model has been widely studied in literature because of its rich phenomenology which makes it a possible candidate to explain late time acceleration~\cite{Barreira:2013xea,Barreira:2014jha,Renk:2017rzu}. It was recently implemented both in \eftcamb and \hiclass thanks to the \textit{mapping} procedure. Thus the EFT functions are fully determined once the background is solved. Three cases are explored~\cite{Renk:2017rzu,Peirone:2017vcq}: Cubic model ($G_3$), $i.e.$ $c_3\neq 0\,,\{c_4,c_5\}=0$; Quartic model ($G_4$), $\{c_3\,,c_4\}\neq 0\,, c_5=0$; Quintic model ($G_5$), $\{c_3,c_4,c_5\}\neq 0$. Using data from the CMB (including lensing), BAO and ISW, $G_3$ is excluded as a viable candidate because it shows 7.8$\sigma$ tension with the data~\cite{Renk:2017rzu}. The $G_4$ and $G_5$ cases show a reduced viable parameter space due to the ISW data and in this region the goodness-of-fit is comparable to $\Lambda$CDM~\cite{Renk:2017rzu}. When using the combination of CMB (temperature and polarization), BAO, $H_0$ and WL measurements, the $G_3$, $G_4$ and $G_5$ models are also statistically ruled out~\cite{Peirone:2017vcq}. This result is obtained using only cosmological data as such it is independent from any assumption coming from GWs~\cite{Creminelli:2017sry,Baker:2017hug,Ezquiaga:2017ekz}. We further discuss the implication of GW on DE/MG in Section~\ref{Sec:GW}.

\item \textit{Low-energy Ho\v{r}ava gravity \cite{Horava:2008ih,Horava:2009uw}: } Ho\v{r}ava gravity is considered a candidate for the ultra-violet completion of GR \cite{Horava:2008ih,Horava:2009uw}. The idea behind this theory is to add only higher order spatial derivatives in the action in order to modify the graviton propagator. To this purpose the theory is formulated using the 3+1 decomposition of the ADM formalism. Considering the above arguments, the action of Ho\v rava gravity at low-energy can be written as follows~\cite{Blas:2009qj}
\begin{eqnarray} \label{horavaaction}
S_{H}=\f{\mp^2}{(2\xi-\eta)}\int{}d^4x\sqrt{-g}\left(K_{ij}K^{ij}-\lambda K^2 -2 \xi\bar{\Lambda} +\xi \mathcal{R}+\eta a_i a^i \right),
\end{eqnarray}
where $\left\{\lambda,\xi,\eta\right\}$ are dimensionless running coupling constants and $\bar{\Lambda}$ is the ``bare'' cosmological constant. The model parameters are constrained using both cosmological \cite{Carroll:2004ai,Zuntz:2008zz,Audren:2013dwa,Blas:2012vn,Audren:2014hza,Frusciante:2015maa} and astrophysical \cite{Will:2014kxa,Yagi:2013qpa,Yagi:2013ava,Ramos:2018oku} data. Furthermore, the theory is still viable \cite{Gumrukcuoglu:2017ijh} after the detection of GW170817 event~\cite{TheLIGOScientific:2017qsa,Monitor:2017mdv}. The theory is mapped into the EFT formalism and implemented in \eftcamb \cite{Frusciante:2015maa}. Using the combination of CMB (temperature and lensing), galaxy power spectrum, local Hubble measurements, SNIa and BAO data, two cases are analyzed \cite{Frusciante:2015maa}: H3, where the three extra free parameters $\{\xi, \eta, \lambda\}$ are free to vary and H2 where the Parameterize post-Newtonian (PPN) bounds are imposed \cite{Bell:1995jz}, which imply $\xi=\eta/2+1$. Thus only two parameters of the theory are left to vary $\{\eta, \lambda\}$ \footnote{Let us note that after the detection of GW170817~\cite{TheLIGOScientific:2017qsa,Monitor:2017mdv} the viability region identified by the PPN has been revised and the condition $\xi=\eta/2+1$ becomes less relevant \cite{Gumrukcuoglu:2017ijh}.}. In the H3 case the constraints on the parameters are $\xi-1 = -0.01^{+0.01}_ {-0.02}$ and $log_{10}(\lambda - 1) < -4.31$ (at 99.7\%C.L) while $\eta$ is unconstrained. Furthermore, the constraint on derived parameters such as $G_{cosmo}/G_N$, where $G_{cosmo}$ is the cosmological gravitational constant appearing in the Friedmann equation,  improves by one order of magnitude with respect to previous results \cite{Carroll:2004ai}. Indeed, $G_{cosmo}/G_N - 1 < 0.028$ at 99.7\%C.L.. For the H2 case, the results are $log_{10}(\lambda - 1) < -4.39$, $log_{10}(\eta) <-4.51$ and $G_{cosmo}/G_N -1< 6.1\times10^{-5}$ (at 99.7\%C.L.). The latter is several orders of magnitude better than the Big Bang Nucleosynthesis bound \cite{Carroll:2004ai}. 

\item \textit{K-mouflage} \cite{Brax:2014wla,Brax:2015pka}: these theories are characterized by a screening mechanism which acts through the derivative of the scalar field $\phi$. They are constructed by including a universal coupling of the scalar field to matter fields in the K-essence action. The action can be written as follows:
\be
S_{Km}=\int{}d^4x\sqrt{-\tilde{g}}\l[\f{\mp^2}{2}\tilde{R}+\mathcal{M}^4K(\tilde{\chi})\r]+S_M[g_{\mu\nu},\psi_i]\,,
\ee
where $\mathcal{M}^4$ is the energy scale of the scalar field, $g_{\mu\nu}$ obeys to the transformation $g_{\mu\nu}=A^2(\phi)\tilde{g}_{\mu\nu}$ between the Jordan frame and Einstein frame  metric $\tilde{g}_{\mu\nu}$. The kinetic term in the Einstein frame is defined as $\tilde{\chi}=-\tilde{g}^{\mu\nu}\partial_\mu\phi\partial_\nu\phi/2\mathcal{M}^4$, and $K$ is a general function of the scalar field. The quantities with tilde are defined in the Einstein frame. The deviation with respect to $\Lambda$CDM is modeled through two functions:
\be
\epsilon_2=\f{dln\bar{A}}{dlna}\,,\qquad \epsilon_1=\f{2}{\bar{K}_{\tilde{\chi}}}\l(\epsilon_2M_{*}\l(\f{d\bar{\phi}}{dlna}\r)^{-1}\r)^2\,.
\ee

The model is implemented in \eftcamb and two cases are investigated \cite{Benevento:2018xcu}: the first in which the background is solved in the \textit{mapping} approach exploiting the full dynamics of the K-mouflage model and a second one where the scalar field is forced to reproduce a background evolution degenerate with $\Lambda$CDM (K-mimic). Using CMB, CMB lensing, SNIa and different galaxy catalogues  and forecasts for future CMB probes, the $H_0$ bound alleviates the tension between Planck and low-redshift probes. Moreover, because the model predicts a suppression in the growth of matter perturbations relative $\Lambda$CDM, this results in a lower value for $\sigma_8$,  easing the tension between Planck and WL measurements~\cite{Hildebrandt:2016iqg,deJong:2015wca,Kuijken:2015vca,Conti:2016gav,Abbott:2017wau,Abbott:2017smn}. Finally, the constraints on the parameters of the model are at 95\% C.L.: $-0.04 < \epsilon_{2,0} < 0$ for K-mouflage and $0 <\epsilon_{2,0} < 0.002$ for K-mimic. They could improve by approximately one order of magnitude with future surveys such as COrE \cite{Delabrouille:2017rct}.

\item \textit{Galileon ghost condensate (GGC)} \cite{Kase:2018iwp}: the model extends the cubic covariant Galileon \cite{Deffayet:2009wt} by including an additional higher-order field derivative $X^2$ following the Ghost-Condensate model~\cite{ArkaniHamed:2003uy}. The action reads
\be\label{GGCaction}
S_{GGC}=\int {\rm d}^4 x \sqrt{-g} \left[ \frac{M_{\rm pl}^2}{2}R
+a_1 X+a_2 X^2+3a_3X \square \phi \right],
\ee
where $a_{1,2,3}$ are constants. The model is implemented in \eftcamb \cite{Peirone:2019aua} and for convenience the following dimensionless variables are considered on a FLRW background 
\be\label{xiparameters}
x_1=-\f{a_1\dot{\phi}^2}{3M_{\rm pl}^2 H^2}\,,\quad x_2=\f{a_2\dot{\phi}^4}{M_{\rm pl}^2 H^2}\,, \quad x_3=\f{6a_3\dot{\phi}^3}{M_{\rm pl}^2H}\, .
\ee

A phenomenological analysis shows that interestingly the low-$\ell$ ISW tail is lower than in the standard $\Lambda$CDM scenario for $x_3\ll x_2$ \cite{Peirone:2019aua}. Another relevant feature is that the equation of state of DE can be in the region $-2<w_{\rm DE}<-1$ at low redshifts \cite{Kase:2018iwp}. The constraints on the present day values of the above parameters using a combination of CMB, BAO, RSD and SNIa data \cite{Peirone:2019aua} yield: $x_1^{(0)}$ and $x_2^{(0)}$  of order 1, with $x_1^{(0)}$ negative and $x_2^{(0)}$ positive, $x_3^{(0)}$ smaller than $x_2^{(0)}$. Finally, according to the Deviance Information Criterion (DIC) \cite{RSSB:RSSB12062} and the Bayesian evidence factor ($\log_{10}B$)~\cite{Heavens:2017afc,DeBernardis:2009di}, the GGC model is found to be statistically preferred over $\Lambda$CDM even with additional model parameters. In details, the above quantities computed with respect to $\Lambda$CDM give: $\Delta {\rm DIC}= -0.6$ (negative value supports the GGC model) and $\Delta \log_{10}B=5.1$ (value $>2$ favors the GGC model) for the complete dataset \cite{Peirone:2019aua}. 

\item \textit{Beyond Horndeski dark energy model} \cite{Kase:2018iwp}: The model (hereafter BH) belongs to the quartic-order GLPV theories and extends the GGC model in action \eqref{GGCaction} by including a beyond Horndeski term. The action reads
\ba
S_{BH}&=&\int {\rm d}^4 x \sqrt{-g} \left[ 
a_1 X+a_2 X^2+3a_3X \square \phi+ \l(\frac{\mp^2}{2}-a_4X^2\r)R
\r. \nn\\
&+&\l.8a_4\l(\phi^{;\mu}\phi^{;\nu}\phi_{;\mu\nu}\Box\phi-\phi^{;\mu}\phi_{;\mu\nu}\phi_{;\lambda}\phi^{;\lambda\nu}\r) \right],
\label{action}
\ea
where $a_{1,2,3,4}$ are constants. Along with the dimensionless parameters in eq. \eqref{xiparameters} on a FLRW background, one can consider an additional parameter which reads
\be
x_4=\f{10a_4\dot{\phi}^4}{\mp^2}\,.
\ee
The above parameter defines the deviation with respect to Horndeski model, the so-called beyond Horndeski parameter $\alpha_H$, which is given by
\be
\alpha_H=\f{4x_4}{5-x_4}\,.
\ee

The model is implemented in \eftcamb and investigated using CMB, BAO, RSD and SNIa data \cite{Peirone:2019yjs}. A tight upper bound $|\alpha_{H}^{(0)}| \le {\cal O}(10^{-6})$ on the present day value of $\alpha_{H}$ was found \cite{Peirone:2019yjs}. This is mostly due to the shift of CMB high-$\ell$ peaks induced by the early-time modification in the cosmological background and linear perturbations arising from the dominance of $\alpha_{H}$ in the DE density.  Another bound on $\alpha_H$ comes from the GW decay to DE and is of order $10^{-10}$ \cite{Creminelli:2018xsv}. However let us notice that it is still a matter of debate whether GWs bounds from LIGO can be applied to MG at large linear scales \cite{deRham:2018red} (see Section \ref{Sec:GW} for further discussion). The complete analysis of the BH model also reveals it to suppress the low-$\ell$ ISW tail of the CMB TT power spectrum with respect to $\Lambda$CDM due to the existence of the cubic term ($x_3$) and Ghost condensate one ($x_2$). The modified background expansion also affects the high multipoles CMB power spectrum. These features allow for the BH model to fit the data better than $\Lambda$CDM according to the $\chi^2$ statistics, however the DIC criterion slightly favors the latter \cite{Peirone:2019yjs}. From this result follows that since the BH model with $\alpha_{H}=0$ corresponds to the GGC model which is favored over $\Lambda$CDM \cite{Peirone:2019aua} as discussed before, there are no particular signatures for deviations from Horndeski theories in current data.
\end{itemize}

%------------------------------------------
\subsection{Neutrinos and modified gravity}
%------------------------------------------

 Massive neutrinos leave an imprint on cosmological observables: they impact the matter power spectrum and the shape of CMB anisotropies significantly while neutrino masses in the sub-eV to eV range alter the expansion history at the epoch of radiation-matter equality~\cite{Lesgourgues:2006nd,Wong:2011ip}.  Massive neutrinos change the  height of the first acoustic peak of the CMB TT power spectrum due to the  early ISW effect~\cite{Lewis:2002nc}:  during the transition from radiation to matter epoch, the evolution for the metric perturbation changes hence the photon geodesics as well. This happens near the epoch of photon decoupling leaving a signature in the CMB anisotropies. A further contribution of massive neutrinos to the early ISW effect comes from their transition from relativistic to non-relativistic regime if their mass is of order 1 eV. Furthermore, neutrinos have large thermal velocity so they do not fall into the potential wells at $k$ larger than the neutrino free-streaming comoving wavenumber. This weakens the gravitational potential wells and suppresses the growth of structure on small scales. Some of these effects can therefore be degenerate with those of DE and MG and generally the constraints on neutrinos mass depend  on the cosmological model assumed~\cite{Barreira:2014ija,Shim:2014uta,Baldi:2013iza,He:2013qha,Dossett:2014oia,Hojjati:2011ix,Motohashi:2012wc}. The degeneracy between alternatives to GR and massive neutrinos has been revisited in the EFT framework \cite{Hu:2014sea,Bellomo:2016xhl}. 
 
In the context of \textit{designer} $f(R)$-gravity as implemented in \eftcamb  using a combinations of Planck 2013, BAO measurements and LSS data from WiggleZ, a lower degeneracy with respect to previous literature results is found \cite{Hu:2014sea}. This is motivated by the fact that the dynamics of the $f(R)$ model is fully exploited with \eftcamb (the code does not use QS approximation). The bounds on the Compton wavelength parameter and the neutrino mass read: $\log_{10}B_0<-4.1$ for a fixed $\Sigma m_\nu=0.06$ eV, which set a new upper limit on $B_0$; $\log_{10}B_0 < -3.8$ for varying neutrino mass and $\Sigma m_\nu < 0.32$ at 95\% C.L.. The improved bounds on these parameters are driven mostly by the WiggleZ data which are highly sensitive to changes in $B_0$ and thus are able to partially break the degeneracy between these two parameters. The second model considered is a non-minimally gravitational coupling model parametrized through the linear EFT model \eqref{eq:lineft} with a $\Lambda$CDM background. In this model, no sizable degeneracy is found and the bound obtained on the free parameter is $\Omega_0^\mathrm{EFT} < 0.05$ (95\% C.L.). The constraint is slightly improved with respect to what was previously obtained \cite{Raveri:2014cka} where no massive neutrinos were considered. Finally, the constraint on the sum of neutrino mass is $\Sigma m_\nu < 0.26 $ at 95\% C.L.. In this case the CMB lensing drives the constraints on the coupling constant $\Omega_0^\mathrm{EFT}$ and the bound on $\Sigma m_\nu$ is slightly looser. The combined dataset without the CMB lensing results in an slightly improved value for $\Sigma m_\nu (<0.25)$.

The degeneracy has been explored in the $\alpha$-basis with the $\alpha$-functions parametrized as the {\it z-transition} form in eq. \eqref{zform} on a $\Lambda$CDM background \cite{Bellomo:2016xhl}. The neutrino mass is found to be partially degenerate with the free parameters $c_i$. In particular, using forecast CMB and galaxy power spectrum datasets, one observes the parameter $c_B$, characterizing $\ab$, to dominate the correlation with the total neutrino mass. Furthermore, $\ab$ can cancel the power suppression due to the massive neutrinos at a given redshift. The breakdown of such degeneracy depends on the cosmological LSS data used at different redshifts. Next generation surveys such as Euclid would limit but not fully break the degeneracy between these two parameters \cite{Alonso:2016suf}, where no apparent degeneracy between MG and the sum of neutrino masses is obtained given the forecasted precision on the $\alpha_i$ parameters. 

%%%%%%%%%%%%%%%%%%%%%%%%%%%%%%%%%%%%%%%%%%%%%%%%%%%%%
\section{Astrophysical implications}\label{Sec:Astro}
%%%%%%%%%%%%%%%%%%%%%%%%%%%%%%%%%%%%%%%%%%%%%%%%%%%%%

In this Section, we review the implications astrophysical constraints have on the parameter space identified by the EFT functions. The EFT framework discussed in this review holds in the linear regime. In order to connect the EFT functions with physical quantities describing astrophysical processes, one can assume the validity of a linear treatment only if modifications of gravity are not screened  or weakly screened, i.e. in a regime where non-linearities are subdominant. 
In this context, we review the bounds which constrain the strength of gravity inside massive astrophysical bodies such as dwarf, neutron stars, pulsars and galaxy clusters and the constraint derived from the detection of the GWs event GW170817 and its electromagnetic counterpart GRB170817A.
 
%----------------------------------------
\subsection{Massive astrophysical bodies}\label{sec:CObjects}
%----------------------------------------

A common feature of MG models is the property to screen the fifth force on small scales or high density environments, notably where Solar-System and astrophysical tests constrain gravity to be that of GR with astonishing precision \cite{Uzan2011,Will2014}. The GR limit is then recovered in such models thanks to \textit{screening mechanisms} \cite{Joyce:2014kja}. In scalar-tensor theories, the latter can be classified using the type of interaction as a phenomenological criterion. It can depend on the local field value as in the cases of symmetron \cite{Hinterbichler:2010es}, chameleon \cite{Khoury:2003aq,Khoury:2003rn} and dilaton \cite{Brax:2010gi} mechanisms; or on the first derivative $\partial \phi$, $e.g.$ K-mouflage screening mechanism \cite{Babichev:2009ee} and finally if it acts through the second derivative $\partial^2\phi$,  typical of the Vainshtein mechanism \cite{Vainshtein:1972sx,Nicolis:2008in,Koyama:2013paa,Kimura:2011dc}.

The Vainshtein  screening mechanism is characteristic of Horndeski and GLPV theories. A peculiarity of GLPV theories is that the screening of the extra DoF is not complete however. Outside an extended object the Vainshtein mechanism allows to fully reproduce GR equations, yet it exhibits a  ``partial breaking'' inside astrophysical bodies \cite{Kobayashi:2014ida}. The perturbed potentials of the Minkowski metric, $\phi(r)$, $\psi(r)$, inside the objects obey the equations \cite{Kobayashi:2014ida,Sakstein:2016ggl}
\ba\label{upsilon}
&&\f{d\phi}{dr}=\f{G_N\tilde{M}(r)}{r^2}+\f{\Upsilon_1G_N}{4}\f{d^2\tilde{M}(r)}{dr^2}\,,\\
&&\f{d\psi}{dr}=\f{G_N\tilde{M}(r)}{r^2}-\f{5\Upsilon_2G_N}{4r}\f{d\tilde{M}(r)}{dr}\,,
\ea
where $\Upsilon_i$ are dimensionless constants depending on the specific theory and $\tilde{M}(r)$ is the mass inside the object. Inside a matter overdensity the gravitational interaction becomes dependent on local matter density  and the gradient of the gravitational potentials are no longer equal. This is transcribed by the fact that both the $\Upsilon_i$ are different functions of the MG couplings. This ``breaking" of the Vainshtein screening effect opens a new window to constrain bounds on GLPV models through massive objects, such as dwarf, neutron, hyperon and quark stars and galaxy clusters \cite{Sakstein:2015zoa,Saito:2015fza,Koyama:2015oma,Sakstein:2016ggl,Sakstein:2016oel}. In the EFT formulation, the $\Upsilon_i$ are related to the EFT coupling functions by \cite{Koyama:2015oma,Saito:2015fza}
\ba
\Upsilon_1 = \frac{4 \ah^2}{c_t^2(1+\ab)-\ah-1}\,,\qquad \Upsilon_2 = \frac{4\ah(\ah-\ab)}{5(c_t^2(1+\ab)-\ah-1)}\,.
\ea
Note that the breaking occurs if and only if $\ah \neq 0$, hence it does not apply to Horndeski models. Thus, bounds on these two parameters can be seen as  constraints on the EFT functions and eventually they can be used to rule out competitors to $\Lambda$CDM.
\begin{figure}[!]
\begin{center}
\includegraphics[scale=0.212]{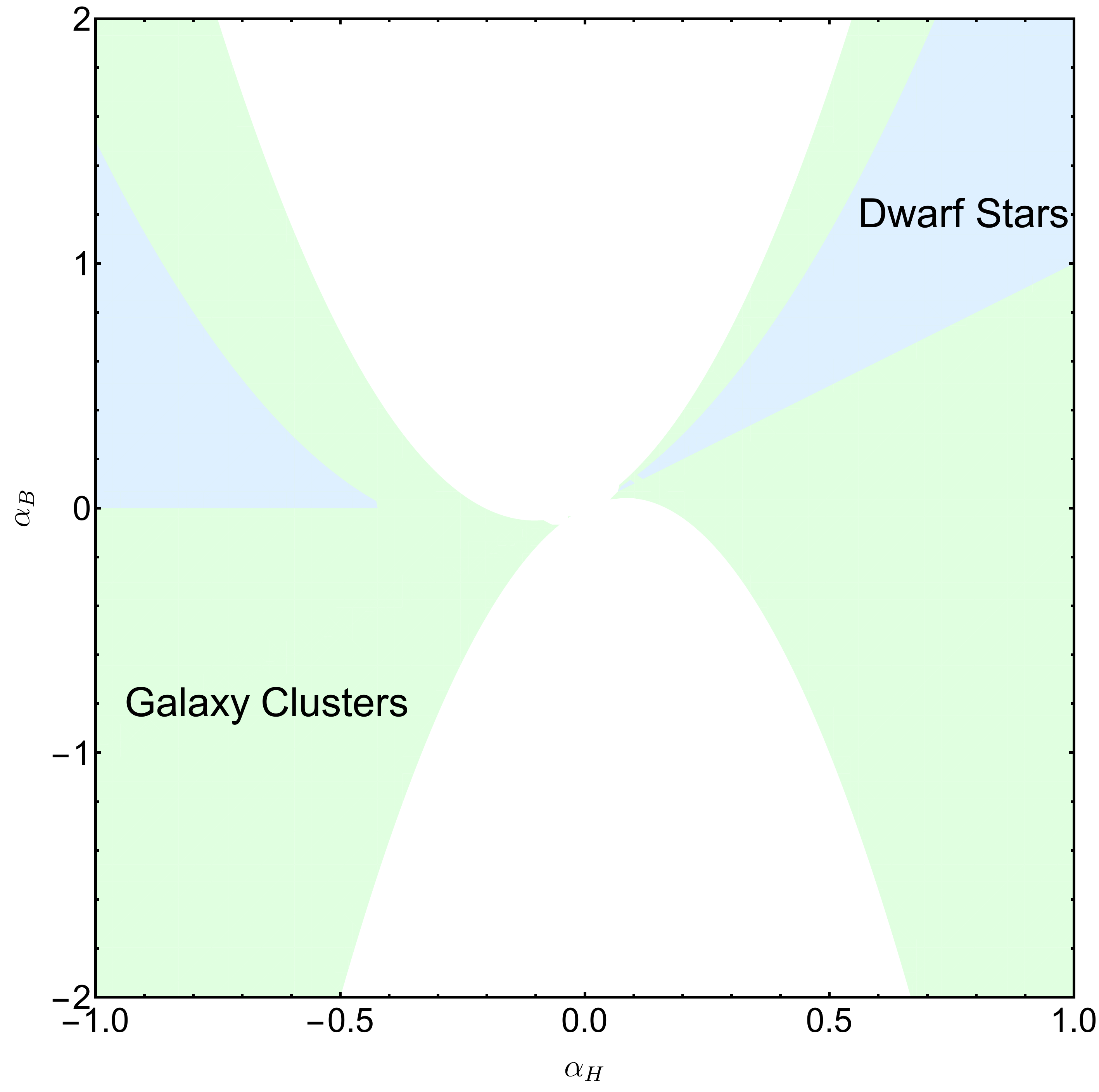}
\hskip2mm
\includegraphics[scale=0.64]{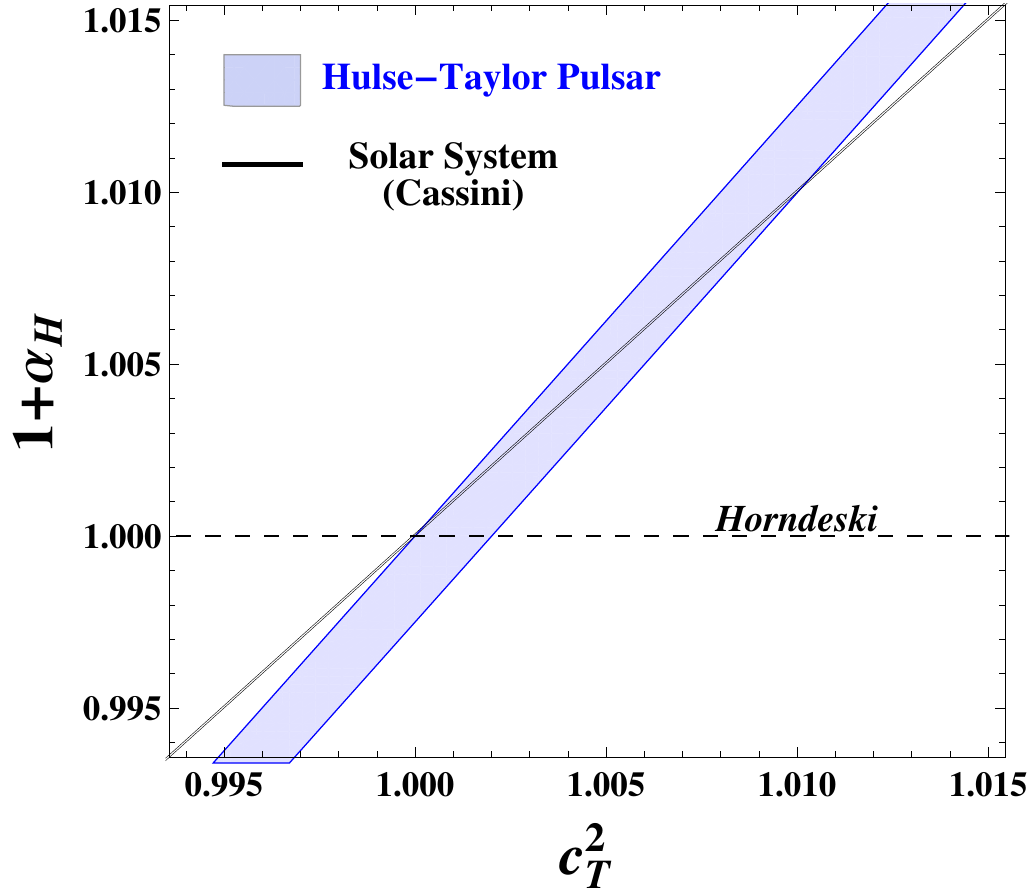}
\caption{\emph{Left panel}: Figure  1 in Ref. \cite{Sakstein:2017xjx}.  Excluded regions on the parameters $\ab$ and $\ah$ from galaxy clusters (green) and dwarf stars (blue) for GLPV models with $c_t^2=1$. \emph{Right panel}: Figure  2 in Ref. \cite{Jimenez:2015bwa}. Hulse-Taylor Pulsar and Cassini constraints on $\ah$ and $c_T^2$ ($=c_t^2$ in this review).}
\label{fig:massiveobjects}
\end{center}
\end{figure}

Observations of several low mass red dwarf stars give a conservative upper bound on the parameter $\Upsilon_1 \lesssim 0.4$  at redshift zero \cite{Sakstein:2015zoa}.  Stable spherically static stellar solutions also require $\Upsilon_1>-2/3$ \cite{Saito:2015fza}, which is satisfied when considering the lower bound $\Upsilon_1>-0.51$ from the consistency of the Chandrasekhar mass with the lowest mass white dwarf \cite{Ohta:2015fpe}. Constraints on $\Upsilon_i$ parameters  can be also obtained from extragalactic measurements of galaxy cluster profiles  \cite{Koyama:2015oma}. Using X-ray and lensing profiles of galaxy clusters from XMM Cluster Survey \cite{Romer:1999qt} and CFHTLenS \cite{Heymans:2012gg}, a stringent constraint on $\Upsilon_1$ and the first estimation of $\Upsilon_2$ are obtained at high redshift ($0.1<z<1.2$), $i.e.$  $\Upsilon_1 =  -0.11^{+0.93}_{-0.67}$ and $\Upsilon_2 = -0.22^{+1.22}_{-1.19}$ at $2\sigma$ \cite{Sakstein:2016ggl}. These constraints directly translate into bounds on the EFT functions. One can further restrict the allowed parameter space for the EFT functions by considering the additional bound on the speed of propagation of GWs  which leads to $c_t^2=1$ ($i.e.$ $\at=0$)~\cite{Monitor:2017mdv}. Then,  only $\ah$ and $\ab$ enter in the definitions of $\Upsilon_i$. The remaining viable regions in the $\ah$-$\ab$ plane are showed in white in Figure \ref{fig:massiveobjects} \cite{Sakstein:2017xjx}. Let us note that the second line in eq. (\ref{upsilon}) is further modified with the inclusion of an additional constant $\Upsilon_3$ if theories such as DHOST are considered \cite{Crisostomi:2017lbg,Dima:2017pwp,Langlois:2017dyl}. 

The Vainshtein mechanism has another peculiar characteristic around massive objects, the so-called \textit{piercing effect} \cite{Jimenez:2015bwa}. Within the screened region, although the background value of the scalar field is suppressed its  gradient is not bound to vanish for theories bearing the shift symmetry, as it is the case of a sub-class of  GLPV theories. Theories with anomalous speed of GWs and coupling to matter for GWs ($i.e.$ an effective Planck mass) can therefore be tested at astrophysical scales. For example, observations of the Hulse-Taylor pulsar led to a model-independent constraint on the local value of $c_t$ to the level of $10^{-2}$ \cite{Jimenez:2015bwa} through the bound
\begin{equation} \label{ppn}
0.995\; \lesssim \;  \frac{G_\mathrm{GW}}{\gn}\, \frac{c}{c_t} \; \lesssim\; 1\;,
\end{equation} 
where $G_\mathrm{GW}$ is the GWs coupling to matter. Combining this bound with the PPN constraint from the Cassini spacecraft experiment on the screened remnant of the gravitational slip parameter $\gsp_\mathrm{sc} -1 = (2.3 \pm 2.1) \times 10^{-5}$ \cite{Bertotti:2003rm} induces the constraints on the EFT functions, $\at=c_t^2-1$ and $\ah$, shown in Figure \ref{fig:massiveobjects} (right panel).

The bounds reviewed in this Section can be used to complement those obtained from cosmological scales in order to further improve our knowledge about the gravitational interaction. 

%-------------------------------------------------
\subsection{GW170817 and GRB170817A}\label{Sec:GW}
%-------------------------------------------------

The first observed merger of a binary neutron star system on August 17, 2017 occurred  through two channels: first the LIGO and Virgo collaborations detected the GWs signal from this event, known as GW170817~\cite{TheLIGOScientific:2017qsa} and after $1.74 \pm 0.05$ s the Fermi and INTEGRAL gamma-ray telescopes observed the gamma-ray burst GRB170817A~\cite{Monitor:2017mdv}. The time delay between the two detections constrained the difference between the speed of GW and the speed of light to be $-3\times 10^{-15}\leq c_t-c\leq 7\times 10^{-16}$~\cite{Monitor:2017mdv}.

This result has a severe impact in selecting viable MG theories compatible with such tiny bound \cite{Creminelli:2017sry,Baker:2017hug,Ezquiaga:2017ekz}. Applying this constraint on the speed of propagation of GWs in the EFT formalism implies $|\alpha_T|<10^{-15}$.   If one imposes exactly such condition at any time, broadly speaking it leads to the conclusion $c_t^2=1$.  Now considering the EFT action (\ref{eftact}), one obtains three separate conditions depending on the chosen sub-class of models:
\ba \label{GWsconstraints}
\bar{M}^2_3&=&0 \; \mbox{(full action)}\,, \nn\\[1mm]
\bar{M}^2_3&=&-\bar{M}^2_2=0\; \mbox{(GLPV)} \,,\nn\\[1mm]
\bar M_3^2&=&-\bar M_2^2=-2\mu_1^2=0\;  \mbox{(Horndeski})\,.
\ea
The cases of GLPV and Horndeski theories received particular attention~\cite{Creminelli:2017sry,Baker:2017hug}. Using the above relations, the GLPV Lagrangian reduces to~\cite{Creminelli:2017sry}
\ba\label{GLPVc1}
L_{c_t=1}=&&G_2(\phi,X)+G_3(\phi,X)+B_4(\phi,X)\nn\\
&&-\f{4}{X}B_{4,X}(\phi,X)(\phi^\mu\phi^\nu\phi_{\mu\nu}\Box\phi-\phi^\mu\phi_{\mu\nu}\phi_\lambda\phi^{\lambda\nu})\,,
\ea
from which one can notice that the quintic GLPV Lagrangian vanishes. Horndeski theories with $c_t^2=1$ can be obtained from the above Lagrangian considering $B_4$ to be solely a function of the scalar field, $i.e.$ $B_4(\phi)$. It is clear that also in this case the quintic Horndeski Lagrangian is completely ruled out.  Note that the  relations leading to action \eqref{GLPVc1} do not represent a fine tuning in the theory because the choice $c_t^2 = 1$ is protected against large quantum corrections \cite{Pirtskhalava:2015nla,Santoni:2018rrx}.
According to action \eqref{GLPVc1} some well known MG models were ruled out \cite{Ezquiaga:2017ekz}, $e.g.$ Quartic and Quintic galileon \cite{Nicolis:2008in,Deffayet:2009wt}, Fab four \cite{Charmousis:2011bf}, de Sitter Horndeski \cite{Martin-Moruno:2015bda}, $G_{\mu\nu}\phi^\mu\phi^\nu$\cite{Gubitosi:2011sg}, $f(\phi)$-Gauss-Bonnet \cite{Nojiri:2005vv}. Considering the beyond Horndeski models, one can exclude quartic and quintic GLPV \cite{Gleyzes:2014dya}, quadratic DHOST (with $A_1\neq 0$) \cite{Langlois:2015cwa} and cubic DHOST \cite{BenAchour:2016fzp}. Additionally, in a recent work the decay of GWs into DE fluctuations in presence of Lorentz breaking was investigated  \cite{Creminelli:2018xsv}. The dominant decay channel is the decay of GWs into two scalar fluctuations $\gamma \rightarrow \pi\pi$ and a second channel leads to $\gamma \rightarrow \gamma \pi$. In both cases the decays are driven by  a  coupling proportional to $\alpha_H$. Values of $\alpha_H\neq 0$ lead to a large  decay rate of the GWs implying no wave would reach the detector. As consequence $\alpha_H$ is forced to be of order $10^{-10}$ and theories such as GLPV and some sub-classes of DHOST models are further ruled out.  Many other models passed the bound on the GWs speed, such as Cubic Horndeski models discussed in Refs. \cite{Nicolis:2008in,Deffayet:2009wt,Albuquerque:2018ymr,Frusciante:2018aew},  the Kinetic Gravity Braiding model \cite{Deffayet:2010qz}, the shift symmetric GLPV model \cite{Kase:2018iwp},  the tracking and scaling DHOST theories discussed in Ref. \cite{Frusciante:2018tvu}, Einstein Aether theory \cite{Oost:2018tcv} and low-energy Ho\v rava gravity \cite{Gumrukcuoglu:2017ijh,Ramos:2018oku}. For a complete review about viable models after GW170817 see  Ref. \cite{Kase:2018aps}.

The constraints in eq. (\ref{GWsconstraints}) assume $\alpha_T=0$ for any background. In other words, they have to be satisfied for any value of $\ddot{\phi}$. However, by requiring the scalar field to satisfy the equation of evolution for the scalar field  ``dynamically'', it is possible to obtain other constraint relations according to which a non trivial quintic Horndeski Lagrangian can be rescued. The latter however is ruled out by the effects of large scale inhomogeneities \cite{Copeland:2018yuh}.  

The range of application of the LIGO bound on MG models is still subject of debate. Let us mention that the applicability of the bound on $\at$ at any time  is questionable since  the source of GWs is at redshift $z \simeq 0.009$. Then  such bound  should  be applied only to constrain the speed of  MG models at recent time, $i.e.$  $z < 10^{-2}$ \cite{Kennedy:2018gtx,Kase:2018aps}. In addition, one of the issues put forward is that LIGO measurement relates to frequencies of $10 - 100$ Hz which correspond to energy scales several orders of magnitude larger than those describing DE. Typically the EFT description of low-energy phenomena breaks down at a cutoff $\sim 100$ Hz. In this regard the measurement of the speed of GWs can be  considered dependent on the  frequency at which it was measured $c_t(k_{LIGO})$ \cite{deRham:2018red}.  Then, the EFT may predict a sub-luminal propagation at low-energy since the speed of GWs is close to unity at LIGO scales, thus, $\bar{M}^2_3(k_{LIGO})=0$ yet  $\bar{M}^2_3(k=0)\neq0$ at cosmological scales  where the transition mechanism would be provided by the partial UV completion of the theory \cite{deRham:2018red}. An experimentum crucis for the GWs in cosmology will be the future  LISA mission \cite{Audley:2017drz} with  sensitivity near $10^{-3}$ Hz. Only at this frequency one can eventually aim at definitively ruling out all the theories with $c_t \neq1$ and restrict the viable parameter space if the LIGO constraint will be confirmed. 

The different scales at which MG phenomena are expected and LIGO measurement happened have been further discussed \cite{Battye:2018ssx}. The former acts on scales $\sim H_0^{-1}$ while one can associate the lookback time of LIGO observation $ 10^{-4}  H_0^{-1}$ to the latter. The corresponding wave number of GWs is $\sim10^{19}H_0$. Defining $K_{grav} = k_{grav}/H_0$, one can use $K_{grav}$ to suppress modifications to gravity on small scales. The phase and group velocity of GWs for a wide range of models can be written as \cite{Battye:2018ssx} 
\be
v_p(K)=\f{1}{a}\sqrt{1+\alpha_T+\f{M^2_{GW}}{K^2}}\,, \qquad   v_g=\f{1+\alpha_T}{a^2v_p}\,,
\ee
where $M^2_{GW}(t)$ is the time dependent, dimensionless graviton mass and $K=k/aH$. In the limit $M_{GW}\ll K_{grav}\sim 10^{19}$ it follows that $v_p=v_g=\sqrt{1+\alpha_T}/a$ which implies $|\alpha_T|<10^{-15}$ as pointed out in previous works. However, the massive graviton could still have significant effects on cosmological scales.

In conclusion, investigation of gravitational models with $c_t^2\neq 1$ would lead to obtain independent constraints from cosmological observations in order to complement the astrophysical observations. Future space-based GW detectors such as LISA will enlarge the reach of multi-messenger GW astronomy. It will then be possible to test gravity to a much higher precision and eventually help in shedding light on the controversy about the applicability of LIGO bounds at cosmological scales.

%%%%%%%%%%%%%%%%%%%%%%%%%%%%%%%%%%%%%%%%%%%%%%%%%%%%%%%
\section{Conclusion and outlook} \label{sec:discussion}
%%%%%%%%%%%%%%%%%%%%%%%%%%%%%%%%%%%%%%%%%%%%%%%%%%%%%%%

The EFT provides a unifying framework for gauging general classes of DE/MG theories at large cosmological  scales  in terms of a variety of free functions of time. One can thereby make predictions and interpret observations directly in the space of class of theories and not within a single paradigm. For example, $\mu$ and $\Sigma$ are known to be a powerful phenomenological parameterization to accurately describe MG effects in the growth of structures and the lensing of light. Observational evidences of any deviations from GR in these functions are hard to connect with  classes of MG models without their interpretation in terms of EFT functions  yet possible for single specific models.  EFT thus provides an appropriate framework to classify different aspects of MG according to their signatures \cite{Pogosian:2016pwr}.  It also helped in obtaining interesting trends in these phenomenological functions which are of large applicability, $e.g.$ the $\mu-\Sigma$ conjecture \cite{Pogosian:2016pwr,Perenon:2016blf,Peirone:2017ywi}. These informations can be translated into specific models since the  EFT framework preserves the  link with covariant theories allowing then  to restrict the forms of the general functions  characteristic of DE/MG Lagrangians.  The use of numerical tools built employing the EFT framework, such as \eftcamb \cite{Hu:2013twa,Raveri:2014cka}, \hiclass \cite{Zumalacarregui:2016pph}, \coop  \cite{Huang:2015srv} and \texttt{EoS\_class} \cite{Pace:2019uow} made  the explorations  of DE/MG effects on observables straightforward. This initiated a systematic analysis of  alternative models against cosmological data. A recollection of these can be found in Appendix \ref{summaryconstraints} summarized in tables. Considerable progress was also made in identifying the stable parameter space of  theories of gravity \cite{Piazza:2013pua,Raveri:2014cka,Perenon:2015sla,Frusciante:2018vht}.  General and theoretically rigorous conditions were derived in the EFT framework encompassing the most significative class of models \cite{DeFelice:2016ucp}. Such conditions are now systematically enforced in numerical codes~\cite{Piazza:2013pua,Salvatelli:2016mgy,Peirone:2017lgi,Raveri:2014cka,Frusciante:2015maa,Peirone:2017lgi}.

Such a unifying description comes at a price: the functional form of the time dependent EFT functions is unknown. Observations generally do not have enough power to fix continuous functions of time but only numbers. One thus resorts to a phenomenological modeling of such functions. In other words, the unknown information contained in the structural functions is compressed into a finite set of parameters, but  one  faces the challenges of general parameterizations against oversimplifications. The risk would be to  miss significant DE/MG signatures or eventually to give a false alert \cite{Linder:2015rcz,Linder:2016wqw}. The chosen parameterization  should thus be universal enough to explore most of the space of stable theories and yet be effectively constrainable by observations. In other words, one must adopt functional behaviors useful for revealing the nature of cosmic acceleration and concurrently avoid including too  many free parameters as they might loosen the constraining power of data \cite{Salvatelli:2016mgy}. Fortunately, cosmological observables seem not to be extremely sensitive to short time-scale variations and therefore smooth parameterizations are in general sufficient to describe the theory space in a satisfactory way \cite{Gleyzes:2017kpi}. A convenient way to fix the EFT functions might be to parameterize directly the stability conditions  and derive, subsequently, the evolution of the EFT functions \cite{Kennedy:2018gtx,Kennedy:2017sof,Lombriser:2018olq,Denissenya:2018mqs}. Alternatively one can use more sophisticated data-driven analysis to reconstruct characteristic functions such as EFT functions, $\mu, \Sigma$ and $\w$ and then derive specific model properties \cite{Espejo:2018hxa,Raveri:2019mxg}.  Beyond this issue, the EFT framework, although not complete  yet, already helped to acquire deeper knowledge about the nature of  gravity force, and derive novel predictions at cosmological scales.  

The attempt for a fundamental understanding of the nature of gravity will require new efforts in both theoretical and observational sides. The next generation of cosmological surveys such as Euclid \cite{Laureijs:2011gra}, DESI \cite{Aghamousa:2016zmz}, SKA \cite{Bacon:2018dui} are specifically tailored to study the impact of DE on the distribution of clustered matter. They will deliver highly accurate data  offering an unprecedented insight into gravity on cosmological scales. The EFT framework is likely to become a sound benchmark to interpret data in the context of MG and extract precious information about fundamental physics. Yet a number of improvements are required to make this approach as complete as possible. For instance, the original EFT description is not properly armed for a correct description of screening effects at intermediate non-linear scales. These are instead very important in the interpretation of data since substantial part of GC, CMB lensing and WL data originates from non-linear scales. A phenomenological way to mimic screening mechanisms has been recently introduced \cite{Alonso:2016suf,SpurioMancini:2019rxy} which highlighted the urgency of the inclusion for non-linearities in the data analysis as these have been proven to increase the constraints on the EFT functions \cite{Mancini:2018qtb}. In this regard, a number of pioneer work contributed to incorporate non-linear effects to the EFT framework \cite{Frusciante:2017nfr,Cusin:2017mzw,Cusin:2017wjg}. Further developments will offer the possibility to include corrections to the power spectrum coming from non-linearities as well as high order correlation functions. In parallel, the interpretation of data in light of stability conditions opened in certain cases  the possibility of multi-fields models to explain the region of the parameter space favored by data  \cite{Peirone:2017lgi}. In this regard, an interesting field of investigation could be  to generalize the EFT construction  to include additional DoFs whose dynamics is relevant at late time following the  example of the EFT of multi-field Inflation \cite{Senatore:2010wk}.

The EFT framework reveals to be an innovating and fascinating research field with very promising prospects opened for improvements and extensions. This not only stems from its flexibility and user friendly approach but also for the era within which it is developed. We believe that in coming years a major breakthrough in the understanding of our Universe will be possible. The window to test gravity is not limited to  cosmological observations but  now extends to multi-messenger probes  which will allow to place constrains on MG/DE models to a much higher accuracy. This becomes even more exciting as our modeling of the Universe perfects, as new ways of making data whisper their secrets are developed from novel machine learning \cite{Peel:2018aei,Ntampaka:2019udw} and gaussian process techniques, without forgetting that evidence for models beyond the standard model is starting to arise at the bayesian level \cite{Peirone:2019aua}.

%%%%%%%%%%%%%%%%%%%%%%%%%%
\section*{Acknowledgments}
%%%%%%%%%%%%%%%%%%%%%%%%%%

We are grateful to E. Bellini, J. Beltr\'an Jim\'enez, C. Marinoni, F. Piazza and A. Silvestri for their detailed comments  and precious feedback on the manuscript. We thank B. Hu, L. Lombriser, R. Maartens, M. Martinelli, F. Pace, G. Papadomanolakis, S. Peirone and D. Vernieri for useful discussions. We acknowledge the authors of \cite{Pogosian:2016pwr,Peirone:2017ywi,Espejo:2018hxa,Lombriser:2015cla,Piazza:2013pua,Raveri:2014cka,Raveri:2017qvt,Perenon:2015sla,Frusciante:2018vht,Perenon:2016blf,Linder:2018jil,Linder:2019bqp,Traykova:2019oyx,Brush:2018dhg,Renk:2016olm,Duniya:2019mpr,Brando:2019xbv,Bellini:2015xja,Noller:2018wyv,Salvatelli:2016mgy,Alonso:2016suf,Reischke:2018ooh,Mancini:2018qtb,Gleyzes:2015rua,Perenon:2019dpc,Frusciante:2018jzw,Sakstein:2017xjx,Jimenez:2015bwa} for the permission to use their figures and we would like to thank also G. Brando and E. Linder for providing us an adapted version of  their figure. The research of NF is supported by Funda\c{c}\~{a}o para a  Ci\^{e}ncia e a Tecnologia (FCT) through national funds  (UID/FIS/04434/2019), by FEDER through COMPETE2020  (POCI-01-0145-FEDER-007672) and by FCT project ``DarkRipple -- Spacetime ripples in the dark gravitational Universe" with ref.~number PTDC/FIS-OUT/29048/2017. LP is supported by the South African Radio Astronomy Observatory (SARAO) and the National Research Foundation (Grant No. 75415).

\appendix
\renewcommand{\thesection}{\Alph{section}}
%%%%%%%%%%%%%%%%%%%%%%%%%%%%%%%%%%%%%%%%%%%%%%%%%
\section{Acronyms and symbols}\label{App:symbols}
%%%%%%%%%%%%%%%%%%%%%%%%%%%%%%%%%%%%%%%%%%%%%%%%%

This Appendix is dedicated to the acronyms and symbols used throughout the review. For the sake of clarity we have collected them respectively in Table \ref{tab:acronym} and Table \ref{tab:symbols}.

\begin{table}[!]
\centering\footnotesize 
\begin{tabular}{|ll|}
\hline\hline
Acronym       & Definition                                     \\ \hline\hline
BAO           & Baryon Acoustic Oscillations                   \\
BICEP         & Background Imaging of Cosmic Extragalactic Polarization\\
BOSS          & Baryon Oscillation Spectroscopic Survey      \\
CDM  (DM)     & Cold Dark Matter (Dark Matter)                              \\
C.L.          & Confidence Level                               \\
CMB           & Cosmic Microwave Background                    \\ 
CMB-S4        & Stage-4 CMB experiment \\
COrE          & Cosmic Origins Explorer mission\\
CPL           & Chevallier-Polarski-Linder                     \\   
CS            & Cosmic shear                      \\ 
DE            & Dark Energy                                    \\
DESI          & Dark Energy Spectroscopic Instrument           \\
DHOST         & Degenerate Higher Order Scalar-Tensor Theories \\ 
DoF           & Degree of Freedom                              \\
EFT           & Effective Field Theory                         \\ 
EB            & Einstein-Boltzmann                             \\ 
FLRW          & Friedmann-Lema\^itre-Robertson-Walker          \\
GBD           & Generalized Brans-Dicke theories               \\
GC            & Galaxy Clustering                              \\ 
GLPV          & Gleyzes-Langlois-Piazza-Vernizzi               \\
GR            & General Relativity                             \\        
GWs           & Gravitational Waves                            \\
ICs           & Initial conditions                             \\
ISW           & Integrated Sachs-Wolfe                         \\
JLA           & Joint Light-curve Analysis                     \\
KiDS          & Kilo-Degree Survey\\
$\Lambda$CDM  & $\Lambda$ Cold Dark Matter                     \\ 
LIGO          & Laser Interferometer Gravitational-Wave Observatory       \\
LISA          & Laser Interferometer Space Antenna    \\
LSS           & Large-scale structure                          \\
LSST          & Large Synoptic Survey Telescope                \\
MCMC & Markov Chain Monte-Carlo\\
MG            & Modified Gravity                               \\   
QS            & Quasi Static                                   \\
RSD           & Redshift-space distortions                     \\
SKA           & Square Kilometer Array                         \\
SNIa          & Supernovae Ia                                  \\    
$w$CDM        & $w$ Cold Dark Matter                           \\  
WEP           & Weak Equivalence Principle                     \\
WL            & Weak Lensing                                   \\
\hline\hline
 \end{tabular}
 \normalsize
  \caption{Table of acronyms and their definitions used throughout the review.}
\label{tab:acronym}
\end{table}

\begin{table}[!]
\centering\footnotesize  
\begin{tabular}{ | ll | ll }
\hline\hline
Symbol                       & Definition                              \\ \hline\hline
$t$, $z$, $a(t)$, $k$        & cosmic time, redshift, scale factor, wavenumber \\
$\partial_\mu,\nabla_\mu (=;)$ & derivative  and covariant derivative\\ 
$\fg(t)$, c(t), $\Lambda(t)$ & (background) EFT functions                \\ 
$M_2^4,\bar{M}_2^2,\bar{M}^2_3,\bar{m}_1^3,m_2^2,\mu_1^2$                        & EFT functions              \\ 
$\mp$                        & Planck mass                             \\ 
$M(t)$                       & effective Planck mass                   \\ 
$\am(t)$                     & running Planck mass                     \\ 
$\ab(t)$                     & braiding function                       \\ 
$\ab^{GLPV}(t)$ &  beyond GLPV function\\
$\at(t)$                     & tensor speed excess                     \\ 
$\ak(t),\alpha_{K_2}(t)$                     & kineticity, extended kineticity                              \\ 
$\alpha_H(t)$ &  beyond Horndeski function\\
$\Omega(t),\gamma_i(t)$ & \eftcamb basis, $i=1,..6$\\
$H(t) \,(H_0)$                        & Hubble function (today)                      \\
$B(t)\, (B_0)$                        & scalaron Compton wavelength (today)                                    \\ 
$\delta_{\rm m}$& linear matter perturbation\\
$\phi(t,x_i)$, $X=\partial_\mu \phi\partial^\mu\phi$                & scalar field, its kinetic term \\
$g_{\mu\nu}(x_\mu)$                 & metric tensor                           \\
$\delta A (t,x_i)$                   & linear perturbation of A                \\
$R$, $R_{\mu\nu}$              & Ricci scalar and tensor                 \\ 
$\Delta_m(t,x_i)$ & Comoving density contrast \\
$\rho_{\rm m}(t),\, \rho_{\rm DE}(t)$ & matter and DE densities\\
$\om(t)\, (\omo)$                & matter density parameter  (today)              \\ 
$\Omega_{\rm DE}(t)\, (\omdeo)$         & DE density parameter  (today)                  \\ 
$\w (t)$                   & DE equation of state                    \\ 
$w_0$, $w_a$ & constant $\w$, first derivative of $\w$ today \\
$c_s(t)^2,\,c_t(t)^2$                   & scalar and tensor speeds of propagation             \\ 
$\alpha(t)$ & kinetic term in Horndeski\\
$\lambda_C\sim M_C^{-1}$     & Compton length scale/Mass               \\ 
$\sigma_8 (\sigma_{8,0})$                   &  matter power spectrum amplitude at 8 h$^{-1}$Mpc  (today)              \\ 
$f(t),\gamma(t)$                       & growth rate and growth index                            \\
$f\sigma_8(t)$               & growth function                         \\
$G_N$                        & Newton constant                         \\ 
$\Phi(t,k),\Psi(t,k)$        & Newtonian and curvature potentials                   \\
$\mu(t,k)$                   & effective gravitational coupling        \\
$\Sigma(t,k)$                & light deflection parameter              \\ 
$\eta(t,k)$                  & gravitational slip parameter            \\ 
$\mu_{\rm sc}(t)/\Sigma_{\rm sc}(t)$                & $\mu/\Sigma$ at super-Compton scale               \\ 
$\mu_\infty(t)/\Sigma_\infty(t)$              & $\mu/\Sigma$ at sub-Compton scale\\
$\mu_{ff}(t)$                & fifth-force contribution to $\mu_\infty$                                 \\
$\pi(t,k)$                   & scalar field perturbation               \\
$K$, $K_{\mu\nu}$               & extrinsic curvature trace and tensor    \\ 
$\mathcal{R},\mathcal{R}_{ij}$& three dimensional Ricci scalar and tensor\\
\hline \hline
\end{tabular}
 \normalsize
  \caption{Table of symbols and their definitions used throughout the review.}
\label{tab:symbols}
\end{table}

\newpage
%%%%%%%%%%%%%%%%%%%%%%%%%%%%%%%%%%%%%%%%%%%%%%%%%%%%%%%%%%%%%%%%%%%%%%%%%%%%%%%%%%
\section{Alternative basis and {\it pure} EFT parameterizations}\label{App:params}
%%%%%%%%%%%%%%%%%%%%%%%%%%%%%%%%%%%%%%%%%%%%%%%%%%%%%%%%%%%%%%%%%%%%%%%%%%%%%%%%%%

In this Appendix, we collect  the different basis used in literature to identify the EFT functions in action \eqref{eftact}.  Table  \ref{tab:basis}  provides the relations among them. The ones that have been largely used in this review are: the EFT basis in action \eqref{eftact}, the \eftcamb basis \cite{Hu:2014oga} and the $\alpha$-basis \cite{Bellini:2014fua}.  The \eftcamb basis  also includes  the $\gamma_6$ function defined as \cite{Hu:2014oga}
\be
\gamma_6=\f{m_2^2}{\mp^2}\,,
\ee  
according to the action \eqref{eftact}. It corresponds to eq. \eqref{alphageneralizeddef} in the $\alpha$-basis. This EFT function is necessary to parameterize  Lorentz violating effects.

\renewcommand{\arraystretch}{1.4}
\begin{sidewaystable}[!]
\centering\footnotesize
\begin{tabular}{c|c|c|c|c|c|c}
\hline 
Ref. & $\boldsymbol{M_{*}^{2}}$ & \textbf{$\boldsymbol{M_{*}^{2}H\alpha_{\textrm{M}}}$} & \textbf{$\boldsymbol{M_{*}^{2}H^{2}\alpha_{\textrm{K}}}$} & \textbf{$\boldsymbol{M_{*}^{2}H\alpha_{\textrm{B}}}$} & \textbf{$\boldsymbol{M_{*}^{2}\alpha_{\textrm{T}}}$}& \textbf{$\boldsymbol{M_{*}^{2}\alpha_{\textrm{H}}}$}  \\ 
\hline 
\cite{Amendola:2012ky} & \textbf{$w_{1}$} & \textbf{$\dot{w}_{1}$} & \textbf{$\frac{2}{3}w_{3}+6Hw_{2}-6H^{2}w_{1}$} & \textbf{$-w_{2}+2Hw_{1}$} & \textbf{$w_{4}-w_{1}$}&-- \\ 
\hline 
\cite{Bloomfield:2012ff,Bloomfield:2013efa} & \textbf{$m_{0}^{2}\Omega+\bar{M}_{2}^{2}$} & \textbf{$m_{0}^{2}\dot{\Omega}+\dot{\bar{M}}_{2}^{2}$} & \textbf{$2c+4M_{2}^{4}$} & \textbf{$-\bar{M}_{1}^{3}-m_{0}^{2}\dot{\Omega}$} & \textbf{$-\bar{M}_{2}^{2}$}& $2\hat{M}-\bar{M}^2_2$\\ 
\hline 
\cite{DeFelice:2011hq} & \textbf{$\mathcal{G}_{T}$} & \textbf{$\dot{\mathcal{G}}_{T}$} & \textbf{$2\Sigma+12H\Theta-6H^{2}\mathcal{G}_{T}$} & \textbf{$-2\Theta+2H\mathcal{G}_{T}$} & \textbf{$\mathcal{F}_{T}-\mathcal{G}_{T}$}&-- \\ 
\hline 
EFT basis  & $\mp^2\fg+\bar{M}^2_2$ & $\mp^2\dot{\fg}+(\bar{M}^2_2)^{\cdot}$ & $2c+4M_2^4$ & $-(\mp^2\dot{\fg}+\bar{m}^3_1)$ & $-\bar{M}^2_2$&$2\mu^2_1-\bar{M}^2_2$ \\ 
\hline 
\cite{Gleyzes:2013ooa,Gleyzes:2014qga} & $M_{*}^{2}f+2m_{4}^{2}$ & $M_{*}^{2}\dot{f}+2\left(m_{4}^{2}\right)^{\cdot}$ & $2c+4M_{2}^{4}$ & $m_{3}^{3}-M_{*}^{2}\dot{f}$ & $-2m_{4}^{2}$ & $2(\tilde{m}_4^2-m_4^2)$  \\ 
\hline 
\eftcamb basis & $m_0^2(1+\Omega+ \gamma_3)$ & $m_0^2(\dot{\Omega}+\dot{\gamma}_3)$& $2c+4H_0^2m_0^2\gamma_1$ & $-m_0^2(H_0\gamma_2+\dot{\Omega})$ & $-m_0^2\gamma_3$ & $m_0^2(2\gamma_5-\gamma_3)$ \\ 
\hline 
{\cite{Piazza:2013pua,Gleyzes:2014qga}} & $M^{2}(1+\epsilon_{4})$ & $\left(M^{2}(1+\epsilon_{4})\right)^{\cdot}$ & $2M^{2}(\mathcal{C}+2\mu_{2}^{2})$ & $-M^{2}(\mu-\mu_3)$ & $-M^{2}\epsilon_{4}$& $M^2(\tilde{\epsilon}_4-\epsilon_4)$ \\ 
\hline 
\end{tabular}
 \caption{Different basis used in literature to identify the  EFT functions. We used  the $\alpha$-basis \cite{Bellini:2014fua} as reference. We note that in this review we have used $M=M_{*}$. -- means the basis has not been extended to the GLPV case yet.}
\label{tab:basis}
\end{sidewaystable}

We summarize the most common {\it pure} EFT parameterizations used to  fix the functional form of the EFT functions. In the following, we denote an EFT function regardless of the basis considered  by $\cf_i$, where $i$ spans on all the EFT functions in a given basis, $e.g.$ in the $\alpha$-basis $\cf_i\equiv \{M,\am,\ab,\ak,\at,\ah\}$. Let us note that in the following we  use $c_i$  to identify general constant parameters. This notation is not a common one, thus the name of the coefficients needs to be adapted to each paper considered in this review.

The {\it pure} EFT parameterizations are:
%---------------------------------------------------------------------------------------
\begin{itemize}
\item {\it linear-de form}:  the EFT functions are chosen to be proportional to  $\omde(z)$ as follows
\begin{equation}\label{param_de0}
\cf_i(z)=c_i\,\frac{\omde(z)}{\omde(z=0)}  \;,
\end{equation}
where  $c_i$ is the constant parameter associated to the coupling $\cf_i$. We note that the normalization with $\omde(z=0)$ is not always used.
%---------------------------------------------------------------------------------------
\item {\it de-N form}: this parameterization  assumes an expansion of the EFT functions in terms of $\omde=1-\om$. In particular it has been built in the formalism of \cite{Piazza:2013pua} for  Horndeski models. The EFT functions in this basis are defined as follows:  
\begin{equation}\label{param_de}
\cf_i(\om)=\frac{1-\om}{1-\omo}H^n\left( c_{i,0}+c_{i,1}\,(\om-\omo) +c_{i,2}\,(\om-\omo)^2+...\right) \;,
\end{equation}
where the  free parameters $c_{i,0}, c_{i,1},c_{i,2}$ are associated to the EFT functions $\cf_i$. The index $n$ is: $n=0$ for $\epsilon_4$, $n=1$ for $\mu$ and $\mu_3$, and $n=2$ for $\mu_2^2$. The correspondence with the basis used in this review is in Table \ref{tab:basis}. Note that the normalization with $1-\omo$ is not always used. In the review we have defined {\it de-1 form} as the expansion up to first order ($c_{i,0}$, $c_{i,1}$ and $c_{2,i}=0$) and {\it de-2 form} the one including the second order ($c_{i,0}$, $c_{i,1}$ and $c_{2,i}$). 
%---------------------------------------------------------------------------------------
\item {\it 6-parameters form}:  the EFT functions resemble  eq. \eqref{param_de} with $\{c_{i,1},c_{i,2}\}=0$. For the running Planck mass function only, the coefficient $c_{M,0}$ is not a constant yet a function of $\om$. Considering  specifically  a $w$CDM background  it assumes the form \cite{Piazza:2013pua}
\ba\label{6form}
c_{M,0}(\om)=(\beta-\alpha)\f{\omo}{\om}+[\alpha-\beta(2+\omo)]\om+2\beta \om^2\,,
\label{6form}
\ea
 where $\alpha,\beta$ are constants. Thus one has two parameters plus three other parameters from the remaining EFT functions and finally the 6th parameter is $\w=w_0$. 
%---------------------------------------------------------------------------------------
\item {\it scaling-a/linear scaling-a form}: this parameterization considers the behavior of each EFT function to scale with the scale factor $a$ as follows
\begin{equation}\label{param_scalinga}
\cf_i(a)=c_{i}\,a^{q_i} \;,
\end{equation}
where there are two free parameters $c_i$ (the amplitude) and $q_i$ (the slope) per coupling $\cf_i$. The {\it linear scaling-a} is obtained when the slope is zero ($q_i=0$).
%---------------------------------------------------------------------------------------
\item {\it de-density form}: the EFT functions are parametrized to be proportional to the DE density ($\rho_{\rm DE}$), {\it i.e.} $\cf_i \propto \rho_{\rm DE}$. In particular using a CPL parameterization for the background it implies:
\ba \label{dedensity}
&&\cf_i(a)=c_i a^{-3(1+w_0+w_a)}e^{-3 w_a (1-a)} \,, 
\ea
where $c_i$ are  constants and $w_0,w_a$ are the CPL parameters. 
%---------------------------------------------------------------------------------------
\item {\it e-fold form or $1+tanh$ form} :
this parameterization allows a transition for the EFT behavior from unity in the past to a constant value in the future as 
\begin{equation}\label{param_efold}
\cf_i(z)=\frac{4 c_i \left(a(z)/a_t\right)^{\tau}}{\left[1+\left(a(z)/a_t\right)^\tau \right]^2}=c_i\l(1-{\rm tanh}^2\l[\f{\tau}{2}{\rm ln}\l(\f{a}{a_t}\r)\r]\r) \;,
\end{equation}
where there are 3 free parameters: $c_i$ giving the amplitude of the transition, $a_t$ the scale factor of the transition and $\tau$ its rapidity. 
%---------------------------------------------------------------------------------------
\item {\it hill form}: this functional form follows the {\it e-fold} form previously presented but  allows for the coupling to have a negative amplitude. It reads
\begin{equation}\label{param_hill}
\cf_i(z)=\frac{4 c_i \left(a(z)/a_t\right)^\tau \left[\left(a(z)/a_t\right)^\tau-1 \right]}{\left[1+\left(a(z)/a_t\right)^\tau \right]^3 } \;, 
\end{equation}
where the definition of the parameters follows the {\it e-fold} form.
%---------------------------------------------------------------------------------------
\item {\it z-transition form}: this  behavior allows to switch on modifications to GR at a given redshift $z_{th}$ (a transition redshift) with a transition given by $\Delta z$.  The expression is given by 
\be\label{zform}
\cf_i(z)=\f{1+{\rm tanh}\l(\f{z_{th}-z}{\Delta z}\r)}{1+{\rm tanh}\l(\f{z_{th}}{\Delta z}\r)}\,.
\ee
For example, the transition redshift $z_{th}$ has been chosen close to the redshift of neutrinos becoming non-relativistic and $\Delta z$ comparable to $z_{th}$ in Ref. \cite{Bellomo:2016xhl}.
%---------------------------------------------------------------------------------------

\item {\it early/late time transition form}: this  behavior allows the EFT functions to shift from  different  early and late time  values with a smooth transition. The expression is given by 
\be\label{edform}
\cf_i(a)=\f{1}{2}(\cf_{i,early}+\cf_{i,late})+(\cf_{i,late}-\cf_{i,early}) \,{\rm arctan} \l[\f{a-a_T}{\Delta a}\r]\f{1}{\pi} \,,
\ee
where $\cf_{i,early}$ and $\cf_{i,late}$ are early and late time values of $\cf_i$, $a_T$ is the scale factor at the time of the transition and $\Delta a$ is the transition sharpness.

%---------------------------------------------------------------------------------------
\item {\it Pad\'e expansion}: the Pad\'e expansion of order [N/M] is the rational function
\be\label{pade}
\cf_i(a)=\f{\sum_{n=0}^{N}c_n(a-a_0)^n}{1+\sum_{m=1}^{M}b_m(a-a_0)^n}\,,
\ee
 where the truncation order is given by $N$ and $M$,  $\{c_n,b_m\}$ are constants coefficients and $a_0$ is the point around which the EFT function is expanded.  Choosing $a_0 = 0$ would select models exhibiting thawing behaviors  and $a_0 = 1$ those having freezing behaviors. This modeling describes well EFT functions  which show a transition from one value at small $a$ to another at large $a$. This parameterization is used in some of  the Monte-Carlo exploration discussed in the review.
%---------------------------------------------------------------------------------------
\item {\it Taylor expansion}: it is a Taylor polynomial expansion defined as follows
\be\label{taylor}
\cf_i(a)=\sum_{n=0}^N\f{c_n}{n!} (a-a_0)^n \,,
\ee
where $N$ is the order at which the expansion is truncated, $a_0$ is the point around which the EFT function is expanded (selecting  $a_0=0$ or $a_0=1$ would give respectively thawing and freezing behaviors)  and $c_n$ is a set of constant coefficients. This parameterization is usually used in the Monte-Carlo exploration and  given $c_n$ coefficients with the same prior distributions would favor the lower order terms.
%---------------------------------------------------------------------------------------
\item {\it Polynomial expansion}: this choice for the parameterization of the EFT functions follows the Taylor expansion with the difference that the term $n!$ in the denominator is not present.  Its form is given as follows
\be\label{poly}
\cf_i(a)= \sum_{n=0}^{N}c_n(a-a_0)^n\,,
\ee
 where $N$ is the order at which the expansion is truncated, $a_0$ is the point around which the EFT function is expanded (selecting  $a_0=0$ or $a_0=1$ would give respectively thawing and freezing behaviors) and $c_n$ are constant parameters. Differently from the Taylor expansion, in this case the high order terms are not suppressed because the $n!$ is absent.  The polynomial expansion is also used in some of the Monte-Carlo exploration.
%---------------------------------------------------------------------------------------
\end{itemize}

%---------------------------------------------------------------------------------------
\section{Constraints summary} \label{summaryconstraints}
%---------------------------------------------------------------------------------------
We summarize the observational constraints and forecasts on the \textit{pure} EFT parameterizations discussed in Sections \ref{sec:running}, \ref{sec:horncons} and \ref{sec:forecasts} in Tables \ref{tab:constr1}, \ref{tab:constr2}, \ref{tab:constr3}. They are organized according to the chosen parameterization making the comparison among different data sets easier. 

\renewcommand{\arraystretch}{1.1}
\begin{table}[!]\footnotesize
\centering
\begin{tabular}{|l|l|l|l|l|}
\hline
\multicolumn{5}{|c|}{Linear-de form eq. \eqref{param_de0}}                                                                                                                                                                                                    \\ \hline \hline
                            Basis    &             EFT functions                                                         &      Constraints                                                            &      Dataset                                                  & Ref. \\ \hline\hline

\multirow{8}{*}{$\alpha$-basis} & \begin{tabular}[c]{@{}l@{}}$\am$\end{tabular}               & \begin{tabular}[c]{@{}l@{}} $ \am>-1.6$ (*)\end{tabular}                & \begin{tabular}[c]{@{}l@{}}CMB + $H_0$ prior \\  95.4\% C.L.\end{tabular}           & \cite{Huang:2015srv} \\ \cline{2-5} 

                                & \multirow{2}{*}{\begin{tabular}[c]{@{}l@{}} $\am$, $\ab$ \\ $\ak$, $c_t^2=1$  \end{tabular} }  & \begin{tabular}[c]{@{}l@{}}$\hat{\alpha}_M=0.25^{+0.19}_{-0.29}$\\ $\hat{\alpha}_B=0.20^{+0.20}_{-0.33}$\\ $\hat{\alpha}_K=0$ (fixed)\end{tabular}     & \begin{tabular}[c]{@{}l@{}}KiDS+GAMA\\ 95\% C.L.\end{tabular} &\cite{SpurioMancini:2019rxy}  \\ \cline{3-5} 
                                                     &       & \begin{tabular}[c]{@{}l@{}}$c_M=0.20^{+1.15}_{-0.82}$\\ $c_B=0.63^{+0.83}_{-0.62}$\\ $c_K= 0.1$ (fixed) \end{tabular}            &\begin{tabular}[c]{@{}l@{}} CMB+BAO\\ +RSD+mPk\\ 95\%C.L.     \end{tabular}     &              \cite{Noller:2018wyv}              \\ \cline{2-5} 
                                
                                & \begin{tabular}[c]{@{}l@{}}$\am$, $\ab$,\\ $\at$, $\ak$\end{tabular}    & \begin{tabular}[c]{@{}l@{}}$-1.36<c_M<-0.06$\\ $0.19<c_B<2.30$\\ $-0.90<c_T<-0.41$\\ $c_K=10$(fixed)\end{tabular} & \begin{tabular}[c]{@{}l@{}}CMB+BAO\\+RSD+PK\\  95\% C.L.\end{tabular}     &  \cite{Bellini:2015xja}\\ \cline{2-5}     
                                                            &  \multirow{3}{*}{\begin{tabular}[c]{@{}l@{}} \begin{tabular}[c]{@{}l@{}}$(\am,\ab)\times S(\f{k}{k_V})$,\\ $\ak\times S(\f{k}{k_V})$,\\$c_t^2=1$,\\ see eq. \eqref{scalealpha}\end{tabular}\end{tabular}} & \begin{tabular}[c]{@{}l@{}}$\hat{a}_K=0.056 $ (fixed) \\$\sigma(\hat{a}_M)=0.065$\\ $\sigma(\hat{a}_B)=0.049$\end{tabular} & \begin{tabular}[c]{@{}l@{}} CMB+GC+CS \\ (forecasts)\end{tabular}& \cite{Reischke:2018ooh} \\ \cline{3-5} 
                                        &  & \begin{tabular}[c]{@{}l@{}}$\hat{a}_K=0.01 $ (fixed) \\$\hat{a}_M=126\%$\\ $\hat{a}_B=41\%$ \end{tabular} & \begin{tabular}[c]{@{}l@{}} CS:\\ 3DWL linear \\ (forecasts)\end{tabular}& \cite{Mancini:2018qtb} \\ \cline{3-5} 
                                        
&  & \begin{tabular}[c]{@{}l@{}}$\hat{a}_K=0.01 $ (fixed) \\$\hat{a}_M=158\%$\\ $\hat{a}_B=54\%$\end{tabular} & \begin{tabular}[c]{@{}l@{}} CS:\\ tomography linear\\ (forecasts)\end{tabular}& \cite{Mancini:2018qtb} \\ \cline{2-5} 
                        
                           &\begin{tabular}[c]{@{}l@{}} \begin{tabular}[c]{@{}l@{}}$(\am,\ab)\times S(\f{k}{k_V})$,\\$(\ak,\at)\times S(\f{k}{k_V})$,\\ see eq. \eqref{scalealpha}\end{tabular}\end{tabular} & \begin{tabular}[c]{@{}l@{}}$\sigma(c_M)=0.056$\\ $\sigma(c_B)=0.123 $\\ $\sigma(c_K)=3.1$ \\ $\sigma(c_T)=0.146$\end{tabular} & \begin{tabular}[c]{@{}l@{}}S4+LSST\\+SKA1-IM+DESI\\(forecasts) \end{tabular}                                                   & \cite{Alonso:2016suf} \\ \cline{2-5} 
                                
                      & \begin{tabular}[c]{@{}l@{}}$\am,\ab,$\\ $\beta^2_\gamma$ (see eq.\eqref{eq:dmdecoup}) \end{tabular}                  & \begin{tabular}[c]{@{}l@{}} $\sigma(\alpha_{M,0})=0.0146 $ \\ $\sigma(\alpha_{B,0})=0.0030 $ \\ $\sigma(\beta^2_\gamma)=0.00135 $\end{tabular}           & \begin{tabular}[c]{@{}l@{}}GC+WL\\ISW-Galaxy\\ (forecasts)\end{tabular}     &     \cite{Gleyzes:2015rua}     \\ \cline{2-5} 
                                & \begin{tabular}[c]{@{}l@{}}$\am$, $\ab$, $\ak$,\\ $\ah$, $c_t^2=1$\end{tabular}  & \begin{tabular}[c]{@{}l@{}}$-0.75<\hat{\alpha}_M<3.75$ (*)\\ $0.2<\hat{\alpha}_B<3$ (*)\\ $0.382<\hat{\alpha}_H<2.457$\\ $\ak$ (fixed)\end{tabular}     & \begin{tabular}[c]{@{}l@{}}CMB+BAO+RSD\\ 95\% C.L.\end{tabular}      & \cite{Traykova:2019oyx} \\ \hline
\end{tabular}
\caption{Summary of the cosmological constraints on the MG parameters discussed in Sections \ref{sec:running}, \ref{sec:horncons} and \ref{sec:forecasts} for the $\alpha$-basis and linear-de form eq. \eqref{param_de0}.	 We use the notation adopted in the original papers. If not specified otherwise the background assumed is $\Lambda$CDM. A (*) denotes ranges we estimated (by eye) from the marginalized contour plots in the respective papers. }\label{tab:constr1}
\end{table}

\renewcommand{\arraystretch}{1}
\begin{table}[!]\footnotesize
\centering
\begin{tabular}{|l|l|l|l|l|}
\hline
Basis                                          & EFT functions                                                                                        & Constraints                                                        & Data sets                                              & Ref.              \\ \hline \hline
\multicolumn{5}{|c|}{Constant form}  \\ \hline\hline
\begin{tabular}[c]{@{}l@{}}$\alpha$-basis\end{tabular}   & \begin{tabular}[c]{@{}l@{}}$M^2$, $\ab$, $\at$\end{tabular}               & \begin{tabular}[c]{@{}l@{}} $ \sigma(\tilde{M}_0)=0.006$\\ $\sigma(\alpha^B_{0})=0.02$\\$\sigma(\alpha^T_0)=0.001$\end{tabular}                & \begin{tabular}[c]{@{}l@{}}CMB-S4+DESI \\ (forecasts)\end{tabular}           & \cite{Abazajian:2016yjj} \\ \hline
\begin{tabular}[c]{@{}l@{}}\eftcamb basis\end{tabular}   & \begin{tabular}[c]{@{}l@{}}$\Omega$, $\gamma_2$, $\gamma_3$\end{tabular}               & \begin{tabular}[c]{@{}l@{}} $ \sigma(\Omega_0)=0.01$\\ $\sigma(\gamma_2^{(0)})=0.05$\\$\sigma(\gamma_3^{(0)})=0.003$\end{tabular}                & \begin{tabular}[c]{@{}l@{}}CMB-S4+DESI \\ (forecasts)\end{tabular}           & \cite{Abazajian:2016yjj} \\ \hline\hline  

\multicolumn{5}{|c|}{early/late transition form eq. \eqref{edform}}  \\ \hline\hline
\begin{tabular}[c]{@{}l@{}}$\alpha$-basis\end{tabular}   & \begin{tabular}[c]{@{}l@{}}$M^2$, $\ab$, $\at$\end{tabular}               & \begin{tabular}[c]{@{}l@{}} $ \sigma(\tilde{M}_{early})=0.05$ \\$\sigma(\tilde{M}_{late})=0.007$ \\ $\sigma(\alpha^B_{early})=0.04$\\$\sigma(\alpha^B_{late})=0.08$\\$\sigma(\alpha^T_{early})=0.02$\\$\sigma(\alpha^T_{late})=0.002$\end{tabular}                & \begin{tabular}[c]{@{}l@{}}CMB-S4+DESI \\ (forecasts)\end{tabular}           & \cite{Abazajian:2016yjj} \\ \hline
\begin{tabular}[c]{@{}l@{}}\eftcamb basis\end{tabular}   & \begin{tabular}[c]{@{}l@{}}$\Omega$ \end{tabular}               & \begin{tabular}[c]{@{}l@{}} $ \sigma(\Omega_{early})=0.03$\\ $\sigma(\Omega_{late})=0.02$\end{tabular}                & \begin{tabular}[c]{@{}l@{}}CMB-S4+DESI \\ (forecasts)\end{tabular}           & \cite{Abazajian:2016yjj} \\ \hline\hline  

\multicolumn{5}{|c|}{Scaling-$a$ form eq. \eqref{param_scalinga}}                                                                                                                                                                                                                                                                         \\ \hline \hline
\multirow{4}{*}{$\alpha$-basis}                & \multirow{2}{*}{\begin{tabular}[c]{@{}l@{}}$\am=-\ab$\end{tabular}}                          & \begin{tabular}[c]{@{}l@{}}$\alpha_{M0}= -0.015^{+0.019}_{-0.017}$\\ $\beta=0.66^{+0.44}_{-0.21}$\end{tabular}  &   \begin{tabular}[c]{@{}l@{}}Planck18+WL\\+BAO/RSD\\ 68\%C.L.\end{tabular}   &        \cite{Aghanim:2018eyx}           \\ \cline{2-5} 
                                               & \begin{tabular}[c]{@{}l@{}}$\am$, $\ab$, \\  $\ak$, $c_t^2=1$\end{tabular}   & \begin{tabular}[c]{@{}l@{}}$c_M=0.27^{+0.54}_{-0.26}$\\ $c_B=0.48^{+0.83}_{-0.46}$\\ $c_K= 0.1$ (fixed) \end{tabular}            &\begin{tabular}[c]{@{}l@{}} CMB+BAO\\ +RSD+mPk\\ 95\%C.L.     \end{tabular}     &              \cite{Noller:2018wyv}     \\ \cline{2-5}  \hline
\multirow{5}{*}{  \begin{tabular}[c]{@{}l@{}}\eftcamb basis\end{tabular} } & $\Omega$     & $\Omega_0^{EFT}<0.061$            &    \begin{tabular}[c]{@{}l@{}}Planck13+WP\\+BAO+Lensing\\ 95\%C.L.\end{tabular}       &        \cite{Raveri:2014cka}           \\ \cline{2-5} 
 & \multirow{2}{*}{\begin{tabular}[c]{@{}l@{}}$\Omega$, CPL \end{tabular}} & \begin{tabular}[c]{@{}l@{}}$\Omega_0=-0.07^{+0.17}_{-0.18} $\\ $s_0> 0.435$\end{tabular}  & \begin{tabular}[c]{@{}l@{}}CMB+BAO\\+SNIa+WL\\ 95\% C.L.\end{tabular}      &        \cite{Frusciante:2018jzw}           \\ \cline{3-5} 
                                               &       & \begin{tabular}[c]{@{}l@{}}$2\sigma(\Omega_0) =110\%$\\ $2\sigma(s_0) = 68\%$ \end{tabular}                                                             &                      \begin{tabular}[c]{@{}l@{}}   GC+WL+CMB \\ (forecasts) \end{tabular}                  &         \cite{Frusciante:2018jzw}             \\ \cline{2-5} 
                                               & \multirow{2}{*}{\begin{tabular}[c]{@{}l@{}}$\Omega$, $\gamma_1$, $\gamma_2$\\  $c_t^2=1$, CPL \end{tabular}} & \begin{tabular}[c]{@{}l@{}}$\Omega_0=0.03^{+0.31}_{-0.25}  $\\ $s_0> 0.215$\\$\gamma_1^0> 0.217 $\\ $\gamma_2^0=-0.9^{+1.3}_{-2.0}  $\\ $s_2>0.330$\end{tabular}  & \begin{tabular}[c]{@{}l@{}}CMB+BAO\\+SNIa+WL\\ 95\% C.L.\end{tabular}      &        \cite{Frusciante:2018jzw}             \\ \cline{3-5} 
                                               &      &  \begin{tabular}[c]{@{}l@{}}$2\sigma(\Omega_0) = 128\%$\\ $2\sigma(s_0) =96\% $    \\$2\sigma(\gamma_2^0) = 240\%$ \\$2\sigma(s_2^0) = 136\%$\\ $\gamma_1^0=5$, $s_1=1.4$ (fixed)    \end{tabular} &                \begin{tabular}[c]{@{}l@{}}   GC+WL+CMB \\ (forecasts) \end{tabular}                                                 &     \cite{Frusciante:2018jzw}                \\ \hline
\end{tabular}
\caption{The same as in Table \ref{tab:constr1} for both the $\alpha$ and \eftcamb basis and constant, early/late transition form eq. \eqref{edform} and scaling-$a$ form eq. \eqref{param_scalinga}.\label{tab:constr2}}
\end{table}

\newpage
 \renewcommand{\arraystretch}{1.1}
\begin{table}[!]\footnotesize
\centering
\begin{tabular}{|l|l|l|l|l|}
\hline
Basis                                          & EFT functions                                                              & Constraints                                                        & Data sets                                         & Ref.              \\ \hline\hline
\multicolumn{5}{|c|}{e-fold form eq. \eqref{param_efold}}                                                                                                                                                                                                                                              \\ \hline \hline
\multirow{2}{*}{$\alpha$-basis}                & \begin{tabular}[c]{@{}l@{}}$\ab=-2\am$\end{tabular}                & \begin{tabular}[c]{@{}l@{}}$0.055<\mu<0.145$ (*) \\ $a_t=0.5$,$\tau=1.5$ (fixed)    \end{tabular} & \begin{tabular}[c]{@{}l@{}} RSD (DESI)\\(forecast)\\ 68\% C.L. \end{tabular} &      \cite{Linder:2018jil}             \\ \cline{2-5} 
                                               & \begin{tabular}[c]{@{}l@{}}$\ab=-2\am$\\CPL  \end{tabular}     &  \begin{tabular}[c]{@{}l@{} }$-0.07995<c_M<0.0$ \\ $0.2615<a_t<1.0$\\$0.8304<\tau<2.19$   \end{tabular}  & \begin{tabular}[c]{@{}l@{}} CMB+BAO\\+RSD+SNIa\\95\% C.L. \end{tabular}  &        \cite{Brando:2019xbv}          \\ \hline\hline
\multicolumn{5}{|c|}{de-1 form eq. \eqref{param_de}}                                                                                                                                                                                                                                                \\ \hline\hline
\multirow{2}{*}{\cite{Piazza:2013pua}}  &\begin{tabular}[c]{@{}l@{}}$\mu$, $\mu_3$, $\epsilon_4$\end{tabular} &\begin{tabular}[c]{@{}l@{}}$p_1=-0.28^{+0.17}_{-0.20}$\\ $p_3=0.04\pm0.17$\\$p_4=-0.030^{+0.068}_{-0.035}$\end{tabular} &   \begin{tabular}[c]{@{}l@{}} CMB\\ (Planck,WMAP)  \\ 68\% C.L.     \end{tabular}                & \cite{Salvatelli:2016mgy}  \\  \cline{3-5} 
                                               &                                                                            &    \begin{tabular}[c]{@{}l@{}}$p_1=0.10^{+0.58}_{-0.37}$\\ $p_3=0.13^{+0.28}_{-0.40}$\\$p_4=-0.18^{+0.28}_{-0.13}$\\$p_3^1=0.41^{+0.39}_{-0.91}$\\ $p_4^1=0.03^{+0.18}_{-0.11}$\end{tabular}                                                                &          \begin{tabular}[c]{@{}l@{}} CMB \\ (Planck,WMAP) \\ 68\%  C.L.    \end{tabular}   &        \cite{Salvatelli:2016mgy}     \\ \cline{2-5}  &\begin{tabular}[c]{@{}l@{}}$\mu_1$, $\mu^2_2$, $\mu_3$, $\epsilon_4$\end{tabular} &   \begin{tabular}[c]{@{}l@{}}$p_{10}=-0.000^{+0.002}_{-0.002}$\\$p_{11}=-0.127^{+0.095}_{-0.096}$\\$p_{20}=1.697^{+2.933}_{-2.157}$\\$p_{21}=-0.926^{+5.852}_{-5.990}$\\$p_{30}=1.022^{+0.930}_{-0.806}$\\$p_{31}=-1.447^{+1.510}_{-1.812 }$ \end{tabular}                                                    &  \begin{tabular}[c]{@{}l@{}} $f\sigma_8+f+\sigma_8+\dot{G}_N$  \\ 95\% C.L.     \end{tabular}     &        \cite{Perenon:2019dpc}           \\ \hline\hline
\multicolumn{5}{|c|}{de-density form eq. \eqref{dedensity}}                                                                                                                                                                                                                                         \\ \hline\hline \multirow{5}{*}{\begin{tabular}[c]{@{}l@{}}\eftcamb basis\end{tabular}}
 & \multirow{2}{*}{\begin{tabular}[c]{@{}l@{}}$\Omega$, CPL \end{tabular}} & \begin{tabular}[c]{@{}l@{}}$\Omega_0=-0.018^{+0.032}_{-0.019} $\end{tabular}  & \begin{tabular}[c]{@{}l@{}}CMB+BAO\\+SNIa+WL\\ 95\% C.L.\end{tabular}      &        \cite{Frusciante:2018jzw}           \\ \cline{3-5} 
                                               &       & \begin{tabular}[c]{@{}l@{}}$2\sigma(\Omega_0) =22\%$ \end{tabular}                                                             &                      \begin{tabular}[c]{@{}l@{}}   GC+WL+CMB \\ (forecasts) \end{tabular}                  &         \cite{Frusciante:2018jzw}             \\ \cline{2-5} 
                                               & \multirow{2}{*}{\begin{tabular}[c]{@{}l@{}}$\Omega$, $\gamma_1$, $\gamma_2$\\  $c_t^2=1$, \\CPL \end{tabular}} & \begin{tabular}[c]{@{}l@{}}$\Omega_0=0.047^{+0.068}_{-0.051}  $\\$\gamma_1^0> 0.295 $\\ $\gamma_2^0=-0.23^{+0.26}_{-0.32}  $\end{tabular}  & \begin{tabular}[c]{@{}l@{}}CMB+BAO\\+SNIa+WL\\ 95\% C.L.\end{tabular}      &        \cite{Frusciante:2018jzw}             \\ \cline{3-5} 
                                               &      &  \begin{tabular}[c]{@{}l@{}}$2\sigma(\Omega_0) = 48\%$   \\$2\sigma(\gamma_2^0) = 40\%$ \\ $\gamma_1^0=4.4$ (fixed)    \end{tabular} &                \begin{tabular}[c]{@{}l@{}}   GC+WL+CMB \\ (forecasts) \end{tabular}                                                 &     \cite{Frusciante:2018jzw}                   \\ \hline
\end{tabular}
\caption{The same as in Table \ref{tab:constr1} for the e-fold form eq. \eqref{param_efold}, de-1 form eq. \eqref{param_de}, de-density form eq. \eqref{dedensity}. \label{tab:constr3} }
\end{table}

\newpage

%%%%%%%%%%%%%%%%%%%%%
\section*{References}
%%%%%%%%%%%%%%%%%%%%%

\bibliographystyle{elsarticle-num} 
\bibliography{References.bib}

\end{document}